%% file: main-arxiv.tex
\documentclass[12pt]{article}

\usepackage{microtype}
\usepackage{graphicx}
\usepackage{amsmath, amsthm, amssymb}
\usepackage{bm}
\usepackage{natbib}
\usepackage[unicode, colorlinks]{hyperref}
\usepackage{enumitem}
\usepackage{booktabs}
\usepackage{multirow}
\usepackage{tabularx}
\usepackage[dvipsnames]{xcolor}
\usepackage[skip=1em]{parskip}

\usepackage{newtxtext}               % requires Times New Roman
\usepackage[varbb]{newtxmath}        % requires Times New Roman
\usepackage{geometry}
\usepackage{tikz}
\usepackage{tikzscale}
\usetikzlibrary{arrows, backgrounds, patterns, matrix, shapes, fit,
  calc, shadows, plotmarks, babel, positioning}

\renewcommand\vec{\bm}
\graphicspath{{./figures/}}

\theoremstyle{remark}
\newtheorem*{remark}{Remark}

\newenvironment{fig}[1]{\par\noindent\begin{minipage}{\linewidth}\begin{center}#1\end{center}}{\end{minipage}\par}

\hypersetup{%
  linkcolor=NavyBlue,%
  citecolor=NavyBlue,%
  filecolor=Red,%
  urlcolor=NavyBlue,%
  menucolor=Red,%
  runcolor=Red,%
  linkbordercolor=NavyBlue,%
  citebordercolor=NavyBlue,%
  filebordercolor=Red,%
  urlbordercolor=NavyBlue,%
  menubordercolor=Red,%
  runbordercolor=Red
}

\hypersetup{%
  pdftitle={Bayesian Graphical Entity Resolution using Exchangeable Random Partition Priors},%
  pdfauthor={Neil G. Marchant, Benjamin I. P. Rubinstein and Rebecca C. Steorts},%
  pdfkeywords={entity resolution, record linkage, Bayesian models, exchangeability, random partitions, Ewens-Pitman random partitions}%
}

\makeatletter
\newcommand\sosname{Statement of Significance}
\if@titlepage
  \newenvironment{sos}{%
      \titlepage
      \null\vfil
      \@beginparpenalty\@lowpenalty
      \begin{center}%
        \bfseries \sosname
        \@endparpenalty\@M
      \end{center}}%
     {\par\vfil\null\endtitlepage}
\else
  \newenvironment{sos}{%
      \if@twocolumn
        \section*{\sosname}%
      \else
        \small
        \begin{center}%
          {\bfseries \sosname\vspace{-.5em}\vspace{\z@}}%
        \end{center}%
        \quotation
      \fi}
      {\if@twocolumn\else\endquotation\fi}
\fi

\newenvironment{keywords}{%
\small
\quotation
\noindent \itshape\textbf{Keywords:}\,\relax%
}
{\endquotation}
\makeatother

\title{Bayesian Graphical Entity Resolution using Exchangeable Random Partition Priors\footnote{
  This is a pre-copyedited, author-produced version of an article accepted for 
  publication in the Journal of Survey Statistics and Methodology following 
  peer review. The version of record 
  \begin{quote}
Marchant, N.~G., Rubinstein, B.~I.~P., and Steorts, R.~C. (2023), ``Bayesian 
  Graphical Entity Resolution using Exchangeable Random Partition Priors,'' 
  \emph{Journal of Survey Statistics and Methodology}
  \end{quote}
  is available online at: \url{https://doi.org/10.1093/jssam/smac030}.
}}
\author{%
  Neil G.~Marchant and Benjamin I.~P.~Rubinstein\\ 
  \normalsize School of Computing and Information Systems\\ 
  \normalsize University of Melbourne
  \and 
  Rebecca C.~Steorts\\
  \normalsize Departments of Statistical Science and Computer Science\\
  \normalsize Duke University
}
\date{}

\begin{document}

\maketitle
\begin{abstract}
  Entity resolution (record linkage or de-duplication) is the process of 
  identifying and linking duplicate records in databases.  
  In this paper, we propose a Bayesian graphical approach for entity resolution 
  that links records to latent entities, where the prior representation on 
  the linkage structure is exchangeable. 
  First, we adopt a flexible and tractable set of priors for the linkage 
  structure, which corresponds to a special class of random partition models. 
  Second, we propose a more realistic distortion model for categorical\slash 
  discrete record attributes, which corrects a logical inconsistency with the 
  standard hit-miss model. 
  Third, we incorporate hyperpriors to improve flexibility. 
  Fourth, we employ a partially collapsed Gibbs sampler for inferential 
  speedups.  
  Using a selection of private and non-private data sets, we investigate the 
  impact of our modeling contributions and compare our model with two 
  alternative Bayesian models. 
  In addition, we conduct a simulation study for household survey data, where 
  we vary distortion, duplication rates and data set size. 
  We find that our model performs more consistently than the alternatives 
  across a variety of scenarios and typically achieves the highest entity 
  resolution accuracy (F1~score). 
  Open source software is available for our proposed methodology, and we 
  provide a discussion regarding our work and future directions.
\end{abstract}

\begin{sos}
  In survey statistics, entity resolution (record linkage) is used to identify 
  responses submitted by the same individual across multiple surveys, even when 
  unique identifiers such as social security numbers are not recorded. 
  This paper advances Bayesian methods for entity resolution by: (i)~thoroughly 
  evaluating a general class of priors on links between responses and 
  individuals, and (ii)~proposing a more realistic model for distortions that 
  appear in individuals' identifying attributes (name, address, etc.). 
  Both of these contributions are evaluated independently and jointly on a 
  variety of data sets, one of which is a longitudinal medical survey.
  The results show that the general class of priors -- the Ewens-Pitman random 
  partitions -- achieve similar accuracy in three previously studied parameter 
  regimes, so long as vague hyperpriors are used. 
  This is an important insight, as it addresses questions in the literature 
  about which parameter regime (if any) is most suitable for entity resolution. 
  In addition, the proposed distortion model is found to significantly improve 
  the accuracy of entity resolution, particularly for string-type attributes. 
  We provide two simulation studies supporting our work and comparisons. 
  The paper is complemented by an R package that implements the proposed 
  model using an optimized partially collapsed Gibbs sampler.
\end{sos}


\section{Introduction}
\label{sec:intro}

As commonly known in the literature, entity resolution (ER; record  linkage or 
de-duplication) is the process of taking large,  noisy (dirty) databases and 
removing duplicate records (often in the absence  of a unique identifier) 
\citep{doan2012principles, elmagarmid_duplicate_2007, 
naumann_introduction_2010, getoor_entity_2012, christen_data_2012, 
Christophides2019,  Ilyas2019, Papadakis2021, binette2022almost}. 
This problem has become increasingly  important in many fields, such as survey 
methodology, official statistics, computer science,  political science, health 
care, human rights, and others. 
In this paper, we are motivated by several applications. For example, we 
consider a longitudinal health care survey, where information  is categorical 
in nature due to privacy restrictions on the data set. This may be of interest 
to those in the survey  methodology community as they may face similar issues. 
In addition, we consider categorical and string or textual based data sets (or 
surveys) such as bibliographic/citation documents, information from 
restaurants, and a traditional benchmark (synthetic) study. 
The goal of analyzing multiple data sets is to make the survey community more 
aware of data sets that are relevant for entity resolution methods. 
Other overarching goals are to extend recent Bayesian graphical ER methodology 
for these data sets, provide comparative analyses, simulation studies, and 
guidance to researchers. 
Moreover, we provide open-source software for the community for our proposed 
method and two recently proposed methods in the literature. 

The idea of entity resolution dates back to \citet{Dunn1946}, who envisioned 
a ``book of life'' that would piece together information about an individual.
\citet{newcombe_automatic_1959} proposed one of the first methods for 
performing ER, based on a heuristic statistical test. 
This method was later formalized by \citet{fellegi_theory_1969}, who 
developed a model based on agreement patterns between pairs of records, 
and a likelihood ratio test for classifying pairs as \emph{linked} 
(referring to the same entity), \emph{possibly linked} or \emph{non-linked}.
Under some strong assumptions, they showed that their method -- now known as 
the Fellegi-Sunter (FS) method -- is statistically optimal.
The FS method has been advanced over many decades in the ER literature, 
especially in survey methodology due to its scalability, ease of use and 
simplicity \citep{winkler_overview_2006, christen_data_2012, 
sadinle_generalized_2013, enamorado_using_2019}. 
However, it has some inherent limitations: it makes inconsistent (intransitive) 
predictions, it does not naturally account for uncertainty, it cannot exploit 
patterns at the entity-level, and it is incompatible with generative modeling 
approaches \citep{liseo_2011}. 

Some of these limitations can be addressed by adapting the FS method to a 
Bayesian setting, while also imposing consistency constraints on the links 
between records \citep{sadinle_detecting_2014, sadinle_bayesian_2017}. 
For example, \citet{sadinle_detecting_2014} proposed a Bayesian extension 
of the FS model for performing ER within a single database. 
It incorporates consistency (transitivity) constraints by requiring that 
records are partitioned into groups that are mutually linked. 
In addition, the model supports multiple levels of agreement, and incorporates 
priors on the links and $m$\slash $u$ probabilities (from the FS model).
In contrast with traditional FS methods, quantification of ER uncertainty is 
possible by computing the posterior distribution on the links. 
However, despite these benefits, the model has not been widely examined in 
the literature, perhaps in part due to the lack of a publicly-available 
implementation. 
One of our goals in this paper is to evaluate Sadinle's model as a 
representative example of a Bayesian FS model, and compare it to another class 
of Bayesian entity resolution models, which we now describe. 

In parallel to developments in Bayesian FS models, others have proposed a new 
class of generative models called Bayesian \emph{graphical entity resolution} 
models \citep{liseo_2011, steorts_entity_2015, steorts_bayesian_2016}. 
In contrast with FS methods, these models do not operate on agreement patterns 
between pairs of records. 
Instead, they model a latent population of entities and the process by which 
records are generated from entities. 
They are known as ``graphical'' models because the fundamental objects in the 
model form a bipartite graph -- the latent entities correspond to one set of 
vertices, the records correspond to another set of vertices, and the links 
are edges that connect the vertices. 
\citet{liseo_2011} proposed one of the first models of this kind for performing 
ER across two databases. 
Subsequently, \citet{steorts_bayesian_2016} proposed an extension to multiple 
databases, while optionally allowing for duplicates within each database. 
However, \citet{steorts_bayesian_2016} discovered a limitation of the their 
model and the model by \citet{liseo_2011} -- the uniform prior on the linkage 
structure is highly informative about the number of entities present in the 
data. 
They noted that future work ought to consider more appropriate priors on the 
linkage structure for ER.

The models by \citet{liseo_2011} and \citet{steorts_bayesian_2016} both assume 
entities are described by a set of latent categorical attributes (e.g., date 
of birth, gender) which are distorted in the records. 
However, their model of the distortion process is simple and is unable to 
capture realistic distortions in string-type attributes, such as names.
\citet{steorts_entity_2015} addressed this problem by proposing a string 
pseudo-likelihood and an empirically-motivated prior in a model known as 
\textsf{blink}. 
The \textsf{blink} model was later used as a foundation by 
\citet{marchant_d-blink_2019} for developing more scalable Bayesian graphical 
ER techniques. 
They proposed an end-to-end method that jointly performs blocking and 
ER, where inference can be distributed or parallelized at the block level. 
Importantly, this enables propagation of blocking uncertainty to the ER 
task. 
They observed a 200$\times$ speed-up, which allowed them to scale 
\textsf{blink} to a data set containing over one million records. 
However, the \textsf{blink} model uses the same uniform prior on the linkage 
structure as \citet{liseo_2011} and \citet{steorts_bayesian_2016}, and suffers 
from the same limitations. 

Motivated by the shortcomings of existing Bayesian graphical ER models, we 
propose and evaluate several modeling refinements in this paper. 
First, we propose a flexible and tractable set of priors for the linkage 
structure that are the Ewens-Pitman (EP) family of random partition models 
\citep{pitman_exchangeable_2006}. 
These are the most general family of priors that satisfy exchangeability (an 
elementary requirement) and they are more flexible than the uniform priors 
used in previous work. 
Second, we incorporate hyperpriors on the EP parameters to further increase 
flexibility and reduce the need for tuning. 
This is motivated by the informativeness of the uniform prior used in previous 
work \citep{steorts_bayesian_2016}.
Third, we propose a more nuanced distortion model for categorical\slash 
discrete attributes, which corrects an inconsistency with the standard 
hit-miss model used by \citet{liseo_2011, steorts_bayesian_2016, 
steorts_entity_2015}. 
Fourth, we design a partially collapsed Gibbs sampler to fit our model which 
incorporates computational optimizations.  

We evaluate our modeling contributions independently and jointly on a 
selection of private and non-private data sets, and compare our model with 
the Bayesian graphical ER model by \citet{steorts_entity_2015} and the 
Bayesian FS model by \citet{sadinle_detecting_2014}. 
We also evaluate our model (and the alternatives) in a controlled simulation 
study, where we generate synthetic household survey data, with varying 
numbers of records, levels of distortion, and rates of duplication. 
Overall, we find our model is more robust across the various scenarios 
tested, and it typically achieves superior ER accuracies. 
We provide open source software for all of the ER methods under evaluation, 
and we provide a discussion of our contributions and directions for future 
work.

The rest of the paper proceeds as follows. 
Section~\ref{sec:prelim} provides background on ER and exchangeable random 
partitions and outlines notation used throughout the paper. 
Section~\ref{sec:model} outlines our proposed Bayesian graphical ER model. 
Section~\ref{sec:inference} presents a partially collapsed Gibbs sampling 
algorithm for approximating the posterior distribution and other computational 
speedups. 
Section~\ref{sec:experiments} presents an empirical study of our proposed 
distortion model and linkage structure priors and includes a comparison to 
two recent Bayesian ER models.
Section \ref{sec:controlled} summarizes a controlled simulation study that is 
in the Appendix. 
Section~\ref{sec:discussion} summarizes our findings.

\section{Background and Notation}
\label{sec:prelim}

In this section, we provide notation, assumptions, and a review of exchangeable 
random partitions which are used as a prior in our model.
Figure~\ref{fig:plate} includes an index of symbols used throughout the paper.

\subsection{Notation and Assumptions}
We review notation and assumptions used throughout the paper. 
We assume the data (from one or more sources) is \emph{structured}, meaning 
that it has been standardized using schema alignment techniques. 
For the purposes of our paper, the data is represented in a tabular format, 
where rows correspond to records and columns correspond to attributes. 
This in contrast to \textit{unstructured} entity resolution, which deals with 
textual descriptions (paragraphs) or images. 
For a full review of these terms, see \citet{Papadakis2021}.

Let $s \in \{1, \ldots, S\}$ be an index over the data sources and 
$i \in \{1, \ldots, N\}$ be an index over the records, which is unique 
across all sources. 
The source of the $i$-th record is denoted by $\varsigma_i \in \{1, \ldots, S\}$ 
and the record's attribute values are represented as a tuple 
$\vec{x}_i = (x_{i1}, \ldots, x_{iA})$ indexed by $a \in \{1, \ldots, A\}$. 
Assume $x_{ia} \in \mathcal{D}_a$ for all $i$ and $a$, where the domain 
of the $a$-th attribute $\mathcal{D}_a$ is a finite set of strings. 
Suppose there exists a (possibly infinite) population of entities indexed by 
$e \in \mathbb{N}$, which is represented in the data. 
Denote the entity referenced in the $i$-th record by $\lambda_{i} \in 
\mathbb{N}$ and define the \emph{linkage structure} as 
$\vec{\Lambda} = (\lambda_1, \ldots, \lambda_N)$.

We consider the most general case where there are no constraints on 
$\vec{\Lambda}$ -- i.e., we permit duplicates within sources and arbitrary 
links across sources.
The linkage structure $\vec{\Lambda}$ induces a partition of the records into 
clusters. 
We label the clusters according to their associated entities, allowing for 
empty clusters.
The size of cluster $e$ is denoted by 
$N_e = \sum_{i} \mathbb{1}[\lambda_{i} = e]$ 
and the number of non-empty clusters is denoted by 
$E = \sum_{e} \mathbb{1}[N_e > 0]$.

We are interested in the \emph{fully unsupervised} setting where no information 
is known about the linkage structure or the 
entities. 
Our goal is to infer the linkage structure based solely on the observed record 
attributes  $\{\vec{x}_1, \ldots, \vec{x}_N\}$ and source identifiers 
$\{\varsigma_1, \ldots \varsigma_N\}$. 
Since we are working in a Bayesian setting, we seek a full posterior (not 
merely a point estimate) over the linkage structure so that uncertainty can be 
propagated to post-ER tasks, which may include regression, multiple systems 
estimation, among other examples \citep{kaplan2022practical, 
tancredi_2015_regression, steorts_generalized_2018, tancredi2020unified, 
sadinle_bayesian_2018}. 
While the post-ER task is not a goal of this paper, the previous references 
propose recent approaches of such tasks.

\subsection{Exchangeable Random Partitions}
\label{sec:exchangeable}
The linkage structure is the primary variable of interest for entity 
resolution, so we pay special attention to it when designing our model. 
Since we are working in a Bayesian setting, we must specify a prior on the 
linkage structure. 
We previously noted that the linkage structure can be interpreted as a 
\emph{partition} of the records into subsets, where the records in each 
subset correspond to the same entity. 
This interpretation is convenient, as we can drawn on related work on random 
partitions when considering potential priors.  
In this section, we review a special class of random partitions called the 
\emph{Ewens-Pitman (EP) family} \citep[p.\ 62]{pitman_exchangeable_2006}, 
which we use as a prior on the linkage structure in our model (see
Section~\ref{sec:model}).

Before defining the EP~family of random partitions, we define key 
concepts and notation. 
Consider a set of $N$ records, where $[N] = \{1, \ldots, N\}$ denotes 
the record identifiers. 
A \emph{partition} of $[N]$ is a collection of disjoint non-empty subsets 
of $[N]$. 
For example, $\{1, 2\}, \{3\}$ and $\{1\}, \{2\}, \{3\}$ are 
partitions of $[N]$ for $N = 3$.
We can equivalently define a partition in terms of the linkage structure 
$\vec{\Lambda} = (\lambda_i)_{i = 1 \ldots N}$ where $\lambda_i$ labels the 
subset (entity) record $i$ is assigned to.\footnote{
  One can use any labels to identify the entities. 
  All that matters is that $\lambda_i = \lambda_j$ if records $i$ and $j$ are 
  assigned to the same entity and $\lambda_i \neq \lambda_j$ otherwise.
}
Using this notation, the above examples for $N = 3$ could be written as 
$\vec{\Lambda} = (1, 1, 2)$ and $\vec{\Lambda} = (1, 2, 3)$.

Let $\mathcal{P}_{[N]}$ denote the set of all partitions of $[N]$. 
A \emph{random partition} of $[N]$ is a random variable whose values 
lie in $\mathcal{P}_{[N]}$. 
The EP family are the most general class of random partitions that satisfy 
the following two desirable properties:
\begin{enumerate}
  \item \emph{Exchangeability.} This means the distribution $P_N$ over the 
  partitions $\mathcal{P}_N$ is invariant under permutations of the record 
  identifiers $[N]$. 
  Or equivalently, the distribution over $\vec{\Lambda}$ is exchangeable. 
  This is a reasonable requirement if the records have no natural 
  ordering -- e.g., it is not known whether one record was generated before or 
  after another.
  \item \emph{Consistency.} This is a property of the distribution as $N$ 
  varies. 
  It ensures the distribution is not altered when more records are observed. 
  This is desirable because the model can be learned sequentially in 
  a consistent manner.
  We say that the sequence of distributions $P_1, P_2, \ldots$ over 
  $\mathcal{P}_{[1]}, \mathcal{P}_{[2]}, \ldots$ is \emph{consistent} if the 
  distribution on $\mathcal{P}_{[N]}$ induced by $P_M$ for $M > N$ is $P_N$. 
  Mathematically, this means
  \begin{equation*}
    P_M(\{\rho: \mathsf{Proj}(\rho, [N]) = \rho'\}) = P_N(\rho'),
  \end{equation*}
  where 
  \begin{equation*}
    \mathsf{Proj}(\rho, [N]) := 
      \{A_e \cap [N] : A_e \cap [N] \neq \emptyset, A_e \in \rho\}
  \end{equation*}
  is the projection of a partition $\rho = \{A_e\}_{e = 1 \ldots E}$ of 
  $[M]$ onto $[N]$ and $\rho'$ is a partition of $[N]$.
\end{enumerate}
In fact, the EP family ensures these properties hold in a limiting sense as 
$N \to \infty$ \citep[p.\ 62]{pitman_exchangeable_2006}.

\begin{figure}
  \def\svgwidth{\linewidth}
  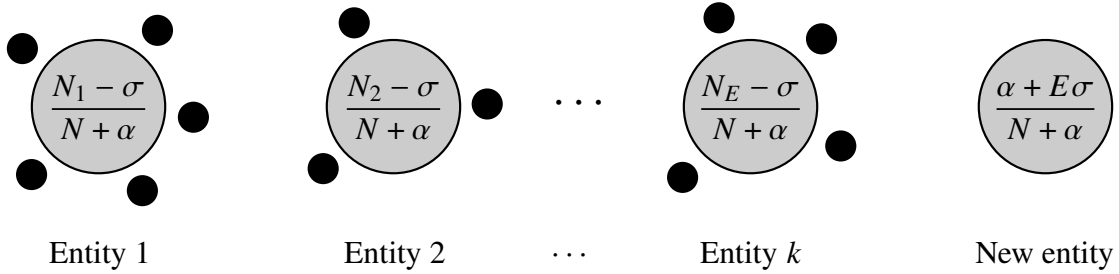
  \caption{%
    Illustration of a sequential construction of a Ewens-Pitman random 
    partition. 
    At step $N$ a record (black circle) is assigned to one of the occupied 
    entities or a new entity (grey circles) conditioned on the assignments of 
    the previous $N - 1$ records. 
    The probabilities assigned to the entities are given inside the grey 
    circles and are dependent on the Ewens-Pitman parameters $\sigma$ and 
    $\alpha$.
  }
  \label{fig:ep-visualize}
\end{figure}

We can develop an intuitive understanding of the EP family by examining 
how a random partition is generated sequentially, one record at a time. 
Let $(\rho_N)_{N = 1, 2, \ldots}$ be a sequence of EP random partitions 
where $\rho_N$ is a random partition of $[N]$. 
We begin at step~1 with $\rho_1 = \{1\}$ -- i.e., a single record assigned 
to a single entity. 
The random partition $\rho_{N}$ at any later step $N > 1$ is generated 
conditional on the random partition $\rho_{N - 1}$ at step $N - 1$. 
Let $E$ be the number of subsets (occupied entities) in $\rho_{N - 1}$ and 
$N_e$ be the size of subset (entity) $e$ in $\rho_{N - 1}$. 
Then $\rho_N$ is generated by assigning record $N$ to:
\begin{itemize}
  \item an existing subset (entity) with probability 
  $\frac{N_e - \sigma}{N + \alpha}$, or
  \item a ``new'' subset (entity) with probability 
  $\frac{\alpha + E \sigma}{N + \alpha}$,
\end{itemize}
where $\sigma$ and $\alpha$ are EP parameters.
This construction is known as a two-parameter Chinese Restaurant Process and 
is visualized in Figure~\ref{fig:ep-visualize}.

The allowable values of the EP parameters fall into two regimes 
depending on the sign of $\sigma$:
\begin{itemize}
  \item $\sigma < 0$ and $\alpha = - m \sigma$ for some $m \in \mathbb{N}$. 
  We refer to this regime as the \emph{generalized coupon partitions}, since 
  they are closely related to the coupon-collector's partition 
  \citep[p.~46]{pitman_exchangeable_2006}. 
  These partitions are generated by sampling with replacement from a finite 
  population of size $m$, where the mixing proportions are drawn from a 
  symmetric Dirichlet distribution with concentration parameter $-\sigma$. 
  The coupon-collector's partition is obtained in the limit 
  $\sigma \to -\infty$.
  \item $0 \leq \sigma \leq 1$ with $\alpha > - \sigma$.
  These are called \emph{Pitman-Yor partitions} after 
  \citet{pitman_two-parameter_1997}, and are generated by sampling with 
  replacement from an infinite population. 
  The resulting partitions demonstrate preferential attachment behavior.  
  The special case $\sigma = 0$ corresponds to the Ewens partition 
  \citep{kingman_random_1978}.
\end{itemize}

To illustrate the varying behavior of the random partitions as a function of 
$\sigma$, we can examine the asymptotic number of subsets in the partition 
(entities) $E_N$ as $N \to \infty$.
\citet[][p.~70]{pitman_exchangeable_2006} shows 
\begin{equation}
  E_N \overset{a.s.}{\asymp} \begin{cases}
    m,                   & \sigma < 0 \text{ and } \alpha = - m \sigma 
                           \text{ for } m \in \mathbb{N}, \\
    \alpha \log N,       & \sigma = 0 \text{ and } \alpha > 0, \\
    S_\sigma N^{\sigma}, & 0 < \sigma < 1 \text{ and } \alpha > - \sigma,
  \end{cases}
  \label{eqn:asymp-K}
\end{equation}
where $S_\sigma$ is a strictly positive random variable. 
Thus, by varying $\sigma$, we can encode a prior belief that the number of 
entities $E_N$ is asymptotically constant, logarithmic, or sub-linear 
in $N$.

\section{Graphical Bayesian ER}
\label{sec:model}
In this section, we propose a generative model for entity resolution that 
incorporates a latent population of entities, each with a set of unknown 
attributes.  
Our model employs the Ewens-Pitman class of priors on the linkage structure 
and a modified record distortion model that deviates from the common 
``hit-miss'' model used by \citet{liseo_2011}, 
\citet{steorts_entity_2015} and \citet{steorts_bayesian_2016}. 
We provide an index of the model's variables and an illustration of their 
dependence structure in Figure~\ref{fig:plate}.
We close the section with a discussion of our model's hyperparameters, 
including recommendations about how to set these values when limited 
prior information is available. 

\begin{figure}
  \centering
  \resizebox{0.8\linewidth}{!}{%
    \input{./figures/plate-diagram.tex}
  }

  \medskip
  {\footnotesize
  \begin{tabularx}{0.48\linewidth}{l X}
    $i$ & index over records \\
    $s$ & index over sources \\
    $a$ & index over attributes \\
    $e$ & index over entities \\
    $x_{ia}$ & attribute $a$ for record $i$ \\
    $z_{ia}$ & distortion indicator for attribute $a$ of record $i$ \\
    $\omega_{ia}$ & distortion propensity for attribute $a$ of record $i$ \\
    $\varsigma_i$ & source of record $i$ \\
    $\lambda_{i}$ & linked entity for record $i$ 
  \end{tabularx} \hfill
  \begin{tabularx}{0.50\linewidth}{l X}    
    $H_{ea}$ & distortion distribution for attribute $a$ of entity $e$ \\
    $\rho_a$ & concentration of $H_{ea}$ \\
    $\vec{\psi}_a$ & base distribution for $H_{ea}$ \\
    $\theta_{s a}$ & distortion probability for attribute $a$ in source $s$ \\
    $y_{ea}$ & attribute $a$ for entity $e$ \\
    $G_a$ & distribution over domain for attribute $a$ \\
    $\vec{\pi}$ & mixing proportions \\
    $\sigma$, $\alpha$ & Ewens-Pitman parameters \\
    $\mathcal{D}_a$ & domain of attribute $a$ \\
    $\operatorname{dist}_{a}$ & distance measure for attribute $a$ 
  \end{tabularx}
  }
  \caption{Plate diagram and index of symbols for our model under a 
  Pitman-Yor prior.}
  \label{fig:plate}
\end{figure}

\subsection{Model Specification}
\label{sec:model-spec}
\paragraph{Entities.}
We assume each entity $e \in \{1, 2, \ldots\}$ is associated with a tuple of 
attribute values $\vec{y}_e = (y_{e1}, \ldots, y_{eA})$, drawn independent and 
identically distributed (i.i.d.) from an unknown distribution $\vec{G}$ with 
support on $\mathcal{D} = \prod_a \mathcal{D}_a$. 
To improve tractability, we assume correlations between attributes are 
negligible, and place independent Dirichlet Process (DP) priors on each 
component of $\vec{G} = (G_1, \ldots, G_A)$:
\begin{align*}
  G_a &\overset{\mathrm{ind.}}{\sim} 
    \operatorname{DP}\!\left(\upsilon_a, \phi_a\right), & \forall a, \\
  y_{ea} \mid G_a &\overset{\mathrm{ind.}}{\sim} 
    \operatorname{Discrete}(G_a), & \forall e, a,
\end{align*}
where $\upsilon_a > 0$ is a concentration parameter and $\phi_a$ is a 
base distribution on domain $\mathcal{D}_a$. 

\paragraph{Links.}
Each record $i \in \{1, \ldots, N\}$ is linked to an entity $\lambda_{i} \in 
\{1, 2, \ldots \}$ which is assumed to be drawn from the population with 
replacement, according to unknown mixing proportions 
$\vec{\pi} = (\pi_1, \pi_2, \ldots)$. 
This process induces a \emph{partition} of the records into clusters according 
to their linked entities.
Following the discussion about exchangeability in 
Section~\ref{sec:exchangeable}, we assume the partition is drawn from the 
Ewens-Pitman (EP) family with parameters $(\sigma, \alpha)$. 
The corresponding distribution on the mixing proportions $\vec{\pi}$ depends 
on the sign of $\sigma$, or equivalently, whether the population of entities 
is finite or infinite.

For the finite regime (generalized coupon partitions) we let 
$\sigma = - \kappa < 0$ and $\alpha = m \kappa$ for some $m \in \mathbb{N}$. 
Our model with hyperpriors on $m$ and $\kappa$ is as follows:
\begin{equation}
  \begin{aligned}
    \kappa 
      &\sim \operatorname{Gamma}(\chi^{(0)}, \chi^{(1)}), 
      & \\
    m 
      &\sim \operatorname{NegativeBinomial}(r, \nu) + 1, 
      & \\
    \vec{\pi} \mid \kappa, m 
      &\sim \operatorname{Dirichlet}(\vec{\kappa}), 
      & \\
    \lambda_i \mid \vec{\pi} 
      &\overset{\mathrm{iid.}}{\sim} \operatorname{Categorical}(\vec{\pi}), 
      & \forall i,
  \end{aligned}
  \label{eqn:finite-regime}
\end{equation}
where $\chi^{(0)}, \chi^{(1)}, r > 0$ and $0 < \nu \leq 1$ are hyperparameters, 
and $\vec{\kappa}$ is a vector of length $m$ with identical entries $\kappa$.
The hyperprior on $m$ is a shifted negative binomial distribution with density 
defined in Appendix~\ref{app-sec:auxiliary-neg-sigma}.

In the infinite regime (Pitman-Yor partitions) the mixing proportions are 
drawn from a two-parameter Poisson-Dirichlet distribution 
\citep{pitman_two-parameter_1997}. 
Our model with hyperpriors on $\sigma$ and $\alpha$ is as follows: 
\begin{equation}
  \begin{aligned}
    \sigma 
      &\sim \operatorname{Beta}(\zeta^{(0)}, \zeta^{(1)}), 
      & \\
    \alpha 
      &\sim \operatorname{Gamma}(\chi^{(0)}, \chi^{(1)}), 
      & \\
    \vec{\pi} \mid \sigma, \alpha 
      &\sim \operatorname{PoissonDirichlet}(\sigma, \alpha), 
      & \\
    \lambda_i \mid \vec{\pi} 
      &\overset{\mathrm{iid.}}{\sim} \operatorname{Categorical}(\vec{\pi}), 
      & \forall i,
  \end{aligned}
  \label{eqn:infinite-regime}
\end{equation}
where $\chi^{(0)}, \chi^{(1)}, \zeta^{(0)}, \zeta^{(1)} > 0$ are 
hyperparameters.
Here we assume $\alpha > 0$ and $0 < \sigma < 1$, which is a subset of the 
admissible parameter space: $0 \leq \sigma \leq 1$ and $\alpha > -\sigma$.
We also consider the case where $\sigma = 0$, which corresponds to the Ewens 
partition.

\begin{remark}
  By placing hyperpriors on the EP parameters, we can improve robustness to 
  misspecified hyperparameters which are difficult to set in a non-informative 
  manner. 
  Special cases of the above priors have been used in other ER models, albeit 
  with fixed hyperparameters.
  \citet{liseo_2011}, \citet{steorts_entity_2015} and 
  \citet{steorts_bayesian_2016} used a coupon-collector's partition with 
  $\kappa \to \infty$ and $m$ fixed, which was shown to be highly informative 
  for the observed population size. 
  \citet{steorts_generalized_2018} used a Pitman-Yor partition with 
  $\sigma$ and $\alpha$ fixed.
\end{remark}

\paragraph{Sources.}
We assume the data source $\varsigma_i \in \{1, \ldots, S\}$ associated with 
record $i$ is drawn i.i.d.\ from a discrete distribution $\vec{\xi}$ over the 
sources $\{1,\ldots, S\}$. 
There is no need to specify $\vec{\xi}$ since it is independent of the other 
model parameters, and the source indicators $\varsigma_i$ are assumed to be 
fully observed.

\paragraph{Distortion.}
We assume the attributes $\vec{x}_i$ for record $i$ are generated by distorting 
the associated entity attributes $\vec{y}_{\lambda_{i}}$. 
For simplicity, we assume the distortion process occurs independently for each 
attribute. 
To decide whether the $a$-th attribute is distorted, a binary indicator 
$z_{ia}$ is drawn which depends on the distortion propensity $\omega_{ia}$ 
scaled by a source\slash attribute-level factor $\theta_{\varsigma_i a}$.
We place a Beta prior on $\theta_{\varsigma_i a}$ and assume the distortion 
propensity $\omega_{ia}$ is deterministic given the true attribute value 
$y_{\lambda_{i} a}$. 
Concretely, we have
\begin{align}
  \theta_{s a} 
    &\overset{\mathrm{ind.}}{\sim} 
      \operatorname{Beta}\!\left(\beta_{s a}^{(0)}, \beta_{s a}^{(1)}\right) 
    & \forall s, a \label{eqn:theta-defn} \\
  \omega_{ia} \mid y_{\lambda_{i} a} 
    &= \mathrm{propensity}\left(
      \min_{x \in \mathcal{D}_a \setminus \{y_{\lambda_{i} a}\}} 
        \operatorname{dist}_{a}(y_{\lambda_{i} a}, x), 
      \max_{x, y \in \mathcal{D}_a} \operatorname{dist}_{a}(y, x)
      \right) 
    & \forall i, a \nonumber \\
  z_{i a} \mid \theta_{\varsigma_i a}, \omega_{ia} 
    &\overset{\mathrm{ind.}}{\sim} 
      \operatorname{Bernoulli}\!\left(\theta_{\varsigma_i a} \omega_{ia}\right) 
    & \forall i, a \label{eqn:z-defn} 
\end{align}
where $\beta_{s a}^{(0)}, \beta_{s a}^{(0)} > 0$ are hyperparameters and 
\begin{equation}
\mathrm{propensity}(d_\mathrm{min}, d_\mathrm{max}) := \begin{cases}
  0, & d_\mathrm{min} = \infty \text{ and } d_\mathrm{max} = \infty, \\
  1, & d_\mathrm{min} = 0 \text{ and } d_\mathrm{max} = 0, \\
  \mathrm{e}^{-\frac{d_\mathrm{min}}{2d_\mathrm{max}}}, & \text{otherwise.}
\end{cases}
\label{eqn:propensity}
\end{equation}

The distortion propensity $\omega_{ia}$ accounts for the fact that some entity 
attribute values $y_{\lambda_{i} a} \in \mathcal{D}_a$ are more likely to be 
distorted than others. 
It makes use of prior information in the attribute distance measure 
$\operatorname{dist}_a(y, x)$ (see Section~\ref{sec:distance-fn}). 
If $y_{\lambda_{i} a}$ is not close to any other values in the domain, it is 
unlikely to be distorted and $\omega_{ia}$ approaches zero.
On the other hand, if $y_{\lambda_{i} a}$ is close to at least one other value 
in the domain, distortion can occur and $\omega_{ia}$ approaches one. 
This logic is not included in a similar model by \citet{steorts_entity_2015}, 
which effectively assumes $\omega_{ia} = 1$.

After drawing the distortion indicator $z_{ia}$, record attribute $x_{ia}$ is 
generated by copying the linked entity attribute $y_{\lambda_{i} a}$ 
directly (if $z_{ia} = 0$) or subject to distortion (if $z_{ia} = 1$). 
If $z_{ia} = 1$, the distorted value is drawn from a distortion distribution 
$H_{\lambda_{i} a}$ associated with the linked entity $\lambda_{i}$. 
We assume $H_{\lambda_{i} a}$ itself is drawn from a Dirichlet Process:
\begin{align}
  \rho_a 
    &\sim \operatorname{Gamma}\! \left(\tau_a^{(0)}, \tau_a^{(1)}\right) 
    & \forall a, \label{eqn:rho-defn} \\
  H_{ea} \mid y_{ea}, \rho_a 
    &\overset{\mathrm{ind.}}{\sim} 
      \operatorname{DP}\! \left(\rho_a; \vec{\psi}_a(y_{ea})\right) 
    & \forall e, a, \label{eqn:distortion-dist}
\end{align}
where $\tau_a^{(0)}, \tau_a^{(1)} > 0$ are hyperparameters, and 
$\vec{\psi}_a(y_{ea})$ is a prior base distribution with support on a subset 
of $\mathcal{D}_a \setminus \{y_{\lambda_{i} a}\}$.
This differs from models by \citet{liseo_2011}, \citet{steorts_entity_2015} and 
\citet{steorts_bayesian_2016,steorts_generalized_2018}, which assume $H_{e a}$ 
is \emph{deterministic} conditional on $y_{ea}$. 

Summarizing this symbolically, we have
\begin{align}
  x_{ia} \mid z_{ia}, y_{\lambda_{i} a}, H_{\lambda_{i} a} 
    &\overset{\mathrm{ind.}}{\sim} \begin{cases}
      \delta(y_{\lambda_{i} a}), & \text{if } z_{i a} = 0, \\
      H_{\lambda_{i} a},         & \text{if } z_{i a} = 1,
    \end{cases} & \forall i, a
  \label{eqn:hit-miss}
\end{align}
where $\delta(y)$ denotes a point mass at $y$.
This is reminiscent of a \emph{hit-miss model} \citep{copas_record_1990}.
However, our construction differs, in that the record and entity attributes 
are forbidden from matching ($x_{i a} \neq y_{\lambda_{i} a}$) if the record 
attribute is distorted ($z_{i a} = 1$).

\begin{remark}
  The hit-miss model of \citet{copas_record_1990} was designed for modeling 
  distortion of continuous attributes. 
  For continuous attributes, the probability of drawing the non-distorted 
  value ($y_{\lambda_i a}$) from the miss component $H_{\lambda_i a}$ is zero, 
  assuming $H_{\lambda_i a}$ is described by a continuous density function. 
  This ensures the record value $x_{ia}$ is always distorted 
  ($x_{ia} \neq y_{\lambda_i a}$) if $z_{ia} = 1$. 
  Our proposal replicates the same behavior for discrete attributes by ensuring 
  $H_{e a}$ has no mass on $y_{e a}$.
  This is especially important if $H_{ea}$ were to place significant mass on 
  $y_{ea}$, as the line between distorted and non-distorted values would become 
  blurred. 
  Apart from the modeling advantages, excluding $y_{ea}$ from the support of 
  $H_{ea}$ also makes inference more tractable as we can collapse $H_{ea}$ 
  (see Appendices~\ref{app-sec:ent-attr-update} and \ref{app-sec:link-update}). 
\end{remark}

\subsection{Choice of Hyperparameters}
\label{sec:hyperparameters}
In this section, we provide recommendations for setting the hyperparameters in 
our model.

\paragraph{Distance Measures.}
\label{sec:distance-fn}
Our proposed distortion model is parameterized by a set of distance measures 
$\{\operatorname{dist}_{a}\}$, one for each attribute $a \in \{1, \ldots, A\}$.
They encode prior knowledge about the likelihood that a record attribute value 
$x$ appears as a distorted alternative to an entity attribute value $y$. 
The larger the distance $\operatorname{dist}_a(y, x)$, the less likely $x$ is 
a distortion of $y$. 
Since the likelihood of distorting $x$ to $y$ may not be the same as the 
likelihood of distorting $y$ to $x$, we do not require that the distance 
measures are symmetric. 
We recommend selecting the distance measures carefully, leveraging prior 
knowledge about the distortion process where possible. 
For instance, one might select edit distance to model typographic distortion 
in a generic string-type attribute. 
For categorical attributes, one could select a constant distance function 
$\operatorname{dist}_a(y, x) \equiv 0$, which encodes the prior belief that all 
values in the domain are equally likely as a distorted alternative to $y$.

\paragraph{Distortion Base Distribution.}
\label{sec:distortion-base-dist}
We recommend using the distance measures to set the base distribution 
$\vec{\psi}_a(y_{e a})$ in Equation~\eqref{eqn:distortion-dist}. 
Specifically, we recommend a softmax distribution 
\begin{equation}
  \psi_a(x \mid y_{ea}) \propto \mathbb{1}[x \neq y_{ea}]
    \exp(-\operatorname{dist}_{a}(y_{ea},x)),
  \label{eqn:distortion-base}
\end{equation}
where the temperature parameter is absorbed in the definition of the distance 
measure, and the indicator function excludes $y_{ea}$ from the support.
This places more weight on values in the domain closer to $y_{ea}$ and less 
weight on values further away. 
Unlike \citet{steorts_entity_2015}, we do not include a factor proportional to 
the empirical frequency of $x$, as distorted values (e.g., typographical 
errors) tend to be infrequent for the applications we consider.
For a categorical attribute with $\operatorname{dist}_a(y, x) \equiv 0$, 
Equation~\eqref{eqn:distortion-base} reduces to the uniform distribution.
In this case, it may be appropriate to incorporate a factor proportional to 
the empirical frequencies by setting $\psi_a(x \mid y_{ea}) \propto 
\mathbb{1}[x \neq y_{ea}] \sum_{i=1}^{N} \mathbb{1}[x_{ia} = x]$.

\paragraph{Other Hyperparameters.}
In the absence of prior knowledge, we recommend setting the remaining 
hyperparameters to yield vague priors -- i.e., priors that provide little 
information relative to the experiment \citep{tiao1973some, 
bernardo2009bayesian}. 
We note that there are different views in the Bayesian community about how 
to specify vague and\slash or uninformative priors. 
For more on this, we refer to \citet{syversveen1998noninformative} and 
\citet{irony1997non}. 
Putting aside such debates, our recommendations are as follows:
\begin{itemize}
  \item For the shifted negative binomial prior on $m$, we set $r$ and $\nu$ 
  so that the prior mean is $N$ and the prior variance is $N^2$.
  \item For the gamma prior on $\alpha$, we set $\chi^{(0)} = 1$ and 
  $\chi^{(1)}$ to be small (e.g., $10^{-2}$).
  \item For the beta prior on $\sigma$, we set $\zeta^{(0)} = \zeta^{(1)} = 1$ 
  to yield a flat prior.
  \item For the Dirichlet prior on the entity attribute distribution, we 
  recommend setting $\upsilon_a = 1$ and using a uniform base distribution 
  $\phi_a$ for all $a$. 
  \item For the gamma prior on the concentration parameter $\rho_a$, we 
  recommend setting $\tau^{(0)} = 2$ and $\tau^{(1)}$ small (e.g., $10^{-4}$) 
  for all $a$. 
  \item For the beta priors on $\theta_{s, a}$, we encode a weak prior belief 
  of low distortion by setting $\beta_{s a}^{(0)} = 1$ and 
  $\beta_{s a}^{(1)} = 4$ for all $s, a$.
\end{itemize}

\begin{remark}
  The hyperparameters can be varied to encourage \emph{over-linkage} (linking 
  records that do not correspond to the same entity) or \emph{under-linkage} 
  (failing to link records that correspond to the same entity). 
  Since perfect linkage is not always possible, practitioners may have to 
  decide whether over-linkage or under-linkage is preferred for a given 
  application. 
  We can encourage over-linkage in our model by setting:
  \begin{itemize}
    \item $\beta_{sa}^{(0)} \gg \beta_{sa}^{(1)}$ (prior belief of high 
    distortion),
    \item $\upsilon_a \ll 1$ (prior belief of low diversity in attribute $a$ 
    among entities),
    \item $\zeta^{(0)} \ll \zeta^{(1)}$ and $\chi^{(0)} \ll \chi^{(1)}$ 
    (prior belief of more links for Pitman-Yor prior), or
    \item $r$ close to 0 and $\nu$ close to 1
    (prior belief of more links for generalized coupon prior).
  \end{itemize}
  Similarly, we can encourage under-linkage by reversing the inequalities 
  above. 
  We measure the extent to which our model over- or under-links in our 
  empirical evaluation (Section~\ref{sec:experiments}) using precision and 
  recall metrics defined in Equations~\eqref{eqn:pair-precision} and 
  \eqref{eqn:pair-recall}.
\end{remark}

\section{Inference}
\label{sec:inference}
To perform entity resolution using our model, we must find the posterior 
distribution over the linkage structure conditional on the observed record 
attributes and their sources. 
Since the posterior is not analytically tractable, we propose an approximate 
inference scheme based on Markov chain Monte Carlo (MCMC). 

MCMC produces approximate samples from the posterior distribution by 
constructing a Markov chain whose equilibrium distribution matches the 
posterior distribution. 
The samples produced by MCMC are approximate in the sense that they are 
autocorrelated, and they may only match the equilibrium (posterior) 
distribution asymptotically. 
Various algorithms exist within the MCMC framework -- we refer the reader 
to \citet{gamerman_markov_2006} or \citet{brooks_handbook_2011} for an 
introduction to the field. 

In this paper, we use an MCMC algorithm called \emph{partially collapsed 
Gibbs (PCG) sampling} \citep{dyk_partially_2008}. 
It is a generalization of Gibbs sampling that reduces the extent of 
conditioning in the variable updates by collapsing (marginalizing out) 
variables and\slash or updating variables in groups. 
This can significantly improve convergence and reduce autocorrelation, so 
long as prescribed rules are followed to ensure the equilibrium distribution 
of the Markov chain is preserved.

Ideally, we would like to reduce the extent of conditioning as much as 
possible, however this must be balanced with computational and mathematical 
constraints.
In our proposed sampling scheme, we \emph{fully collapse} the entity mixing 
proportions $\vec{\pi}$ and the distortion distributions $H_{e a}$. 
We \emph{partially-collapse} the distortion indicators $z_{i a}$ in a joint 
update for the entity attributes $y_{ea}$ and for the distortion distribution 
concentration $\rho_a$.
By collapsing the mixing proportions, we obtain an urn-based scheme for 
updating the linkage structure similar to those used for nonparametric mixture 
models \citep{neal_markov_2000}.
In the remainder of this section, we highlight some less trivial aspects of 
inference -- full details are provided in Appendix~\ref{app-sec:gibbs-updates}. 

\subsection{Nonconjugacy}
\label{sec:nonconjugacy}
While we attempted to maintain conjugacy in our model, we were unable to avoid 
nonconjugate priors in some cases. 
This complicates inference, as the posterior conditional distributions used 
in Gibbs sampling are no longer of a standard form. 
There are several well-established methods for dealing with nonconjugacy, 
including Metropolis-Hastings algorithms \citep{chib_understanding_1995}, 
rejection sampling \citep{gilks_adaptive_1992} and auxiliary variable methods 
\citep{damlen_gibbs_1999}.
We opt to use auxiliary variable methods owing to their simplicity, 
as there is no need to design proposals or monitor acceptance rates. 

There are three sets of parameters in our model for which nonconjugacy is an 
issue:
\begin{enumerate}
  \item The distortion probabilities $\theta_{sa}$ defined in 
  Equation~\eqref{eqn:theta-defn}, where the incorporation of the 
  distortion propensities $\omega_{ia}$ breaks the conjugacy of the 
  beta prior. 
  We propose an auxiliary variable sampling scheme to update $\theta_{sa}$ 
  in Appendix~\ref{app-sec:dist-prob-update}. 
  \item The EP parameters: $\kappa$ and $m$ defined in 
  Equation~\eqref{eqn:finite-regime} or $\alpha$ and $\sigma$ defined in 
  Equation~\eqref{eqn:infinite-regime}, depending on the regime. 
  We use an auxiliary variable scheme proposed by \citet{teh_bayesian_2006}, 
  to update $\alpha$ and $\sigma$ under a gamma and beta prior,
  as summarized in Appendix~\ref{app-sec:auxiliary-pos-sigma}.
  We design an auxiliary variable update for $\kappa$ and $m$ under a 
  gamma and shifted negative binomial prior in 
  Appendix~\ref{app-sec:auxiliary-neg-sigma}.
  \item The distortion distribution concentration $\rho_a$ defined in 
  Equation~\eqref{eqn:rho-defn}. 
  We design an auxiliary variable update for $\rho_a$ in 
  Appendix~\ref{app-sec:distort-dist-conc-update}.
\end{enumerate}

\subsection{Collapsing the Distortion Indicators}
\label{sec:collapse-z}
\citet{marchant_d-blink_2019} demonstrated the importance of collapsing 
the distortion indicators $\{z_{i a}\}$ to improve convergence\slash mixing 
for a hit-miss model similar to Equation~\eqref{eqn:hit-miss}. 
The posterior factors involving $z_{i a}$ factorize over $i$ and $a$, 
so that collapsing $z_{i a}$ yields:
\begin{align}
  P(x_{ia} \mid \theta_{\varsigma_i a}, \omega_{ia}, y_{\lambda_{i} a}, 
    H_{\lambda_{i} a})
    & \propto \sum_{z_{ia} = 0}^{1} P(x_{ia} \mid  z_{ia}, y_{\lambda_{i} a}, 
      H_{\lambda_{i} a}) P(z_{ia} \mid \theta_{\varsigma_i a}, \omega_{ia}) 
      \nonumber \\
    & \propto (1 - \theta_{\varsigma_i a} \omega_{ia}) 
      \mathbb{1}[x_{ia} = y_{\lambda_{i} a}]
        + \theta_{\varsigma_i a} \omega_{ia} H_{\lambda_{i} a}(x_{ia}).
  \label{eqn:collapse-dist-ind}
\end{align}
We use this result to implement a collapsed update for the entity attributes 
$\{y_{e a}\}$.
While it is possible to implement a collapsed update for the linkage structure 
$\{\lambda_{i}\}$, we opt not to do so, since conditioning on the 
distortion indicators allows us to reduce computational complexity 
via indexing (see Section~\ref{sec:computational}). 
This seems to be more efficient empirically~\citep{marchant_d-blink_2019}, 
so long as the level of distortion is not too high.

\subsection{Computational Considerations}
\label{sec:computational}
We now discuss ways of improving the computational complexity.
The main bottleneck is the update for the linkage structure which 
scales na\"{i}vely as $O(N \cdot E)$ where $E$ is the number of 
instantiated entities.\footnote{
  When stating time complexities in this section, we assume a categorical 
  random variate can be drawn in $\Theta(C)$ time where $C$ is the number of 
  categories. 
  The algorithm proposed by \citet{vose_linear_1991} satisfies this 
  constraint.
}
The update for the entity attributes may also be problematic for large 
domains $\mathcal{D}_a$ as it scales as 
$O(E \cdot \lvert \mathcal{D}_a \rvert)$ for the $a$-th attribute.

We are able to reduce the computational complexity of the linkage structure 
update by exploiting constraints imposed by the distortion model.
Close inspection of the update for the entity linked to record $i$ (see 
Appendix~\ref{app-sec:link-update}) reveals that some entities can 
be immediately excluded from consideration. 
Specifically, only those entities whose attributes match the corresponding 
\emph{non-distorted} record attributes ($x_{ia}$ with $z_{ia} = 0$) may be 
linked to record $i$.
In order to efficiently query this set of entities, we maintain inverted 
indices that map an attribute value $x \in \mathcal{D}_a$ to the set of 
entities instantiated with that value $\{e: x = y_{e a}\}$.
This approach is considerably more efficient than iterating over all 
entities sequentially, so long as the level of distortion is relatively low. 
However it is important to note that it relies crucially on \emph{not} 
collapsing the distortion indicators.

To improve the complexity of the entity attribute update, we can impose a 
cut-off on the distance measures. 
Concretely, for attribute $a$ we replace the ``raw'' distance measure 
$\operatorname{dist}_a$ by
\begin{equation*}
  \underline{\operatorname{dist}}_a(y, x) = \begin{cases}
    \operatorname{dist}_a(y, x), 
      & \text{if } \operatorname{dist}_a(y, x) \leq d_a^{(\mathrm{cut})}, \\
    \infty,          
      & \text{otherwise},
  \end{cases}
\end{equation*}
where $d_a^{(\mathrm{cut})} \in (0, \infty)$ is a configurable cut-off.
This approximation eliminates the need to consider unlikely distortions 
from entity attribute $y$ to record attribute $x$, for which 
$\operatorname{dist}(y, x) > d_a^{(\mathrm{cut})}$.
It plays a similar role to blocking in the record linkage literature 
\citep{christen_2012} and resembles an approach proposed by 
\citet{marchant_d-blink_2019}.
To make use of this approximation, we build indices that can efficiently 
answer range queries -- one for each attribute. 
The index for the $a$-th attribute takes a query value $x \in \mathcal{D}_a$ 
and returns the set of entity attribute values that fall below the cut-off: 
$\{y \in \mathcal{D}_a : \operatorname{dist}_a(y, x) \leq 
d_a^{(\mathrm{cut})}\}$.

\section{Model Comparisons}
\label{sec:experiments}

We conduct an empirical study of our ER model using data sets for which the 
true linkage structure is known. 
Section~\ref{sec:datasets} describes the data sets used in the study, which 
are motivated by ER applications in private and non-private settings. 
We explain how our model (and baseline models) are evaluated in 
Section~\ref{sec:model-evaluation}, by computing metrics that assess 
how well the posterior predictions align with the true linkage structure. 
Section~\ref{sec:study} assesses the impact of our modeling contributions 
by varying the distortion model and the prior on the linkage structure. 
Section~\ref{sec:baselines} compares our ER model against  
baselines proposed by \citet{sadinle_detecting_2014} and 
\citet{steorts_entity_2015}. 
Finally, Section~\ref{sec:controlled} summarizes a controlled simulation 
study that can be found in Appendix~\ref{app-sec:sim-study}. 

\subsection{Data Sets}
\label{sec:datasets}
We study entity resolution in private and non-private settings, 
both of which are encountered by practitioners. 
The data sets we use in our study are summarized in Table~\ref{tbl:datasets}.

\paragraph{Private Setting.}
In this setting the practitioner has access to de-identified data, where 
sensitive attributes such as names, addresses, phone numbers, 
etc. are removed. 
This can make ER quite challenging, as the remaining non-sensitive attributes 
may carry limited information about the identity of records.
To study ER in this setting, we use data extracted from the 
National Long Term Care Survey \citep{manton_nltcs_2010}, which we refer to 
as \textsf{nltcs}.

Our extract contains de-identified respondent records from the 1982, 1989 
and 1994 waves of the survey in the U.S.\ state of Alabama. 
We use all of the available attributes for ER, which include date of birth 
(\texttt{DOB\_YEAR}, \texttt{DOB\_MONTH}, \texttt{DOB\_DAY}), registration 
office (\texttt{REGOFF}) and sex (\texttt{SEX}). 
Since the data is well-curated, the only distortion that can occur is when a 
valid attribute value is replaced by another valid attribute value. 
We, therefore, model the attributes as categorical by employing a constant 
distance function.

\paragraph{Non-private Setting.}
In this setting, the practitioner has access to data with sensitive 
attributes, such as names. 
We assume unique identifiers such as social security numbers are not 
available, as ER would otherwise be trivial.
Obtaining real survey data for a non-private setting with ground truth is 
challenging, so we use three publicly-available data sets from the 
ER literature. 
Although these data sets cover other domains, they exhibit characteristics  
one would expect to encounter in real survey data.
Namely, we observe the presence of multiple ``name-like'' attributes, as well 
as different levels of variation and distortion.
Below, we provide a brief description of each data set and the attributes used 
for ER:
\begin{itemize}
  \item \textsf{RLdata} is a synthetic person data set, where 10\% of the 
  records are duplicates with random errors \citep{sariyar_recordlinkage_2010}.
  We model the name attributes (\texttt{fname\_c1} and \texttt{lname\_c1}) 
  using the normalized Levenshtein distance measure. 
  The attributes related to date of birth -- \texttt{bd}, \texttt{bm} and 
  \texttt{by} -- are modeled as categorical attributes with a constant 
  distance measure.\footnote{%
    This is a benchmark data set that is widely used 
    in the literature.
  }
  \item \textsf{cora} is a collection of computer science citation records 
  hosted on the RIDDLE repository \citep{riddle_repository}. 
  It is the ``dirtiest'' of all the data sets we consider, as it was 
  extracted from various online sources with different citation styles.
  As a pre-processing step, we separate hyphenated words and remove 
  punctuation. 
  We also correct several erroneous ground truth labels. 
  The \texttt{title}, \texttt{venue} and \texttt{authors} attributes 
  generally contain multiple words with semantic and character-level 
  variations, and are therefore modeled using a hybrid token\slash edit 
  distance measure described in Appendix~\ref{app-sec:hybrid-dist-fn}.
  The \texttt{year} attribute is modeled using normalized Levenshtein 
  distance.
  \item \textsf{rest} is a collection of restaurant records from the 
  Fodor and Zagat restaurant guides hosted on the RIDDLE repository 
  \citep{riddle_repository}. 
  It is not as ``dirty'' as \textsf{cora} as there are fewer sources 
  and less variation between them.
  We applied the same pre-processing steps as for \textsf{cora}. 
  The \texttt{name} and \texttt{addr} attributes generally contain 
  multiple words and are therefore modeled using the same hybrid 
  distance measure as for \textsf{cora}. 
  The \texttt{city} and \texttt{type} (cuisine) attributes are 
  modeled as categorical with a constant distance measure.
\end{itemize}

\begin{table}
  \small
  \centering
  \begin{tabular}{lllrr}
    \toprule
    Data set        & Setting     & Entity type     
      & \# records ($N$) & \# entities \\
    \midrule
    \textsf{nltcs}  & Private     & People          
      &            5,359 & 3,307 \\
    \textsf{RLdata} & Non-private & People          
      &           10,000 & 9,000 \\
    
    \textsf{cora}   & Non-private & Citations       
      &            1,295 &   125 \\
    \textsf{rest}   & Non-private & Restaurants     
      &              864 &   752 \\
    \bottomrule
  \end{tabular}
  \caption{Summary of data sets.}
  \label{tbl:datasets}
\end{table}

\subsection{Model Evaluation}
\label{sec:model-evaluation}

We evaluate an ER model on a data set by comparing the inferred linkage 
structure $\hat{\Lambda}$ to the true linkage structure $\Lambda$. 
Recall that $\Lambda = (\lambda_1, \ldots, \lambda_N)$ specifies the 
corresponding entity $\lambda_i$ for each record $i$ in the data set.
The agreement between $\hat{\Lambda}$ and $\Lambda$ can be measured using 
pairwise precision and recall. 
The \emph{pairwise precision} is the proportion of record pairs linked in 
$\hat{\Lambda}$ that are also linked in $\Lambda$:
\begin{equation}
  \operatorname{Pr}(\hat{\Lambda}, \Lambda) = 
    \frac{\sum_{i \neq j = 1}^{N} \mathbb{1}[\hat{\lambda}_i = \hat{\lambda}_j] 
      \mathbb{1}[\lambda_i = \lambda_j]}
    {\sum_{i \neq j = 1}^{N} \mathbb{1}[\hat{\lambda}_i = \hat{\lambda}_j]}. 
  \label{eqn:pair-precision}
\end{equation}
It takes on values from 0 to 1, where larger values indicate fewer false 
positive errors.
The \emph{pairwise recall} is the proportion of record pairs linked in 
$\Lambda$ that are also linked in $\hat{\Lambda}$:
\begin{equation}
  \operatorname{Re}(\hat{\Lambda}, \Lambda) = 
    \frac{\sum_{i \neq j = 1}^{N} \mathbb{1}[\hat{\lambda}_i = \hat{\lambda}_j] 
      \mathbb{1}[\lambda_i = \lambda_j]}
    {\sum_{i \neq j = 1}^{N} \mathbb{1}[\lambda_i = \lambda_j]}. 
  \label{eqn:pair-recall}
\end{equation}
It takes on values from 0 to 1, where larger values indicate fewer 
false negative errors.
It is rarely possible to achieve high precision and recall -- one must usually 
make a trade-off depending on which types of errors are more costly in a 
given application. 
If precision and recall are equally important, then one can measure the 
agreement between $\hat{\Lambda}$ and $\Lambda$ using the 
\emph{pairwise F1 score} which is the harmonic mean of precision and 
recall:
\begin{equation}
  \operatorname{F1}(\hat{\Lambda}, \Lambda) = 
    2\frac{\operatorname{Pr}(\hat{\Lambda}, \Lambda) \cdot 
      \operatorname{Re}(\hat{\Lambda}, \Lambda)}
    {\operatorname{Pr}(\hat{\Lambda}, \Lambda) 
      + \operatorname{Re}(\hat{\Lambda}, \Lambda)}. 
  \label{eqn:pair-F1}
\end{equation}

Since the models in our study are Bayesian, the inferred (posterior) linkage 
structure $\hat{\Lambda}$ is a random variable and the metrics in 
Equations~\eqref{eqn:pair-precision}--\eqref{eqn:pair-F1} can be regarded as 
random variables. 
We estimate the distribution of the metrics under the posterior using samples 
generated via MCMC. 
In doing so, we are able to account for posterior uncertainty in our 
evaluation. 
For each metric, we report a point estimate using the median, along with a 
95\% equi-tailed credible interval. 
Further details about the MCMC implementation and configuration for each 
model are provided in Appendix~\ref{app-sec:experimental-setup} and MCMC 
diagnostics are in Appendix~\ref{app-sec:mcmc-diag}.

\subsection{Study of Linkage Structure Priors and Distortion Model}
\label{sec:study}

In this section, we study the effect of two modeling contributions proposed in 
Section~\ref{sec:model-spec} -- the Ewens-Pitman (EP) linkage structure priors 
and the refined distortion model. 
Our objective is to determine the impact of each modeling contribution in 
isolation, using the \textsf{blink} model as a baseline. 
We summarize the results here and refer the reader to 
Appendix~\ref{app-sec:comparison} for comprehensive results covering eight 
combinations of linkage structure priors and distortion models.

\paragraph{Linkage Structure Priors.}
We consider four priors on the linkage structure, which correspond to 
distinct EP parameter regimes (see Section~\ref{sec:exchangeable}):
\begin{enumerate}
  \item \textsf{PY}: Pitman-Yor regime with $\sigma \in (0, 1)$ and hyperpriors 
  on $\sigma, \alpha$ as detailed in Equation~\eqref{eqn:infinite-regime}.
  \item \textsf{Ewens}: Ewens regime with $\sigma = 0$ and a hyperprior on 
  $\alpha$ as detailed in Equation~\eqref{eqn:infinite-regime}.
  \item \textsf{GenCoupon}: generalized coupon regime with 
  $\sigma = - \kappa < 0$ and hyperpriors on $\kappa, m$ as detailed 
  in Equation~\eqref{eqn:finite-regime}.
  \item \textsf{Coupon}: coupon collector's partition used by 
  with $\kappa \to \infty$ and $m = N$.
\end{enumerate}
The first three priors are flexible, in the sense that hyperpriors are placed 
on the EP parameters. 
The last prior is a particular instance of \textsf{GenCoupon} where the 
EP parameters are fixed. 
It is used in models by \citet{liseo_2011}, \citet{steorts_entity_2015} and 
\citet{steorts_bayesian_2016}, and serves as a baseline here.

\begin{figure}
  \centering
  \includegraphics[scale=0.85]{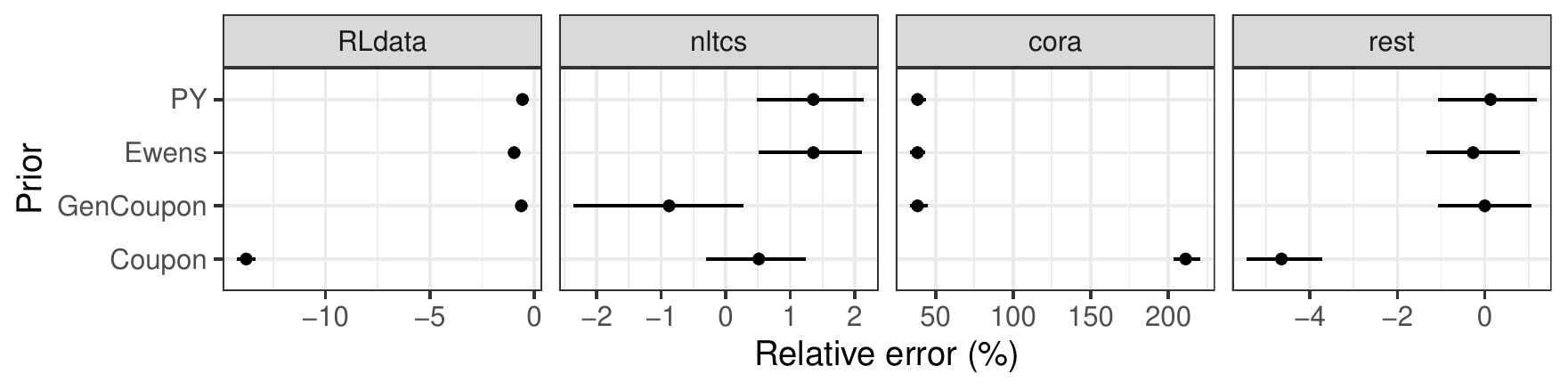}
  \caption{Posterior relative error in the predicted number of entities 
  for all data sets and linkage structure priors. 
  Under-linkage is observed for \textsf{cora}, which is likely due 
  to significant noise that is not well-captured by the distortion model.}
  \label{fig:posterior-num-ents}
\end{figure}

ER evaluation metrics are presented in Table~\ref{tbl:pairwise-measures} for 
the four linkage structure priors, assuming the rest of the model follows the 
specification in Section~\ref{sec:model-spec}.
Another perspective on ER accuracy is provided in 
Figure~\ref{fig:posterior-num-ents}, which plots the relative error in the 
inferred number of entities.
Both results demonstrate the benefit of placing hyperpriors on the 
EP~parameters, as is done for \textsf{PY}, \textsf{Ewens}, and 
\textsf{GenCoupon}. 
These linkage structure priors achieve superior F1~scores compared to 
\textsf{Coupon}, where the EP parameters are fixed.
Figure~\ref{app-fig:post-ep-params} (Appendix~\ref{app-sec:comparison})
is consistent with this finding, demonstrating that vastly different values 
of the EP parameters are inferred for each data set when hyperpriors are 
used. 
Another interesting observation is the fact that the ER accuracy is 
relatively similar among \textsf{PY}, \textsf{Ewens} and \textsf{GenCoupon}.
This was unexpected at first, given the three parameter regimes are known 
to exhibit distinct asymptotic behavior (see equation \ref{eqn:asymp-K}). 
This suggests all three regimes may be flexible enough to model the 
linkage structure of the data sets we consider here.
It would be interesting to see if these observations translate to much larger 
data sets, where the asymptotic behavior of the three regimes would become 
more apparent.

\begin{table}
  \small
  \centering
  \begin{tabular}{llccc}
    \toprule
      &  
        & \multicolumn{3}{c}{Evaluation metric} \\
    \cmidrule{3-5}
    Data set 
      & EP~regime  
        & Precision 
        & Recall 
        & F1 score \\
    \midrule
    \multirow{4}{*}{\textsf{RLdata}}  
      & \textsf{PY} 
        & 0.896 {\scriptsize(0.879,\,0.917)} 
        & 0.961 {\scriptsize(0.952,\,0.972)} 
        & 0.928 {\scriptsize(0.918,\,0.939)} \\
      & \textsf{Ewens} 
        & 0.870 {\scriptsize(0.853,\,0.893)} 
        & 0.970 {\scriptsize(0.961,\,0.978)} 
        & 0.917 {\scriptsize(0.908,\,0.931)} \\
      & \textsf{GenCoupon} 
        & 0.903 {\scriptsize(0.886,\,0.920)} 
        & 0.966 {\scriptsize(0.955,\,0.975)} 
        & 0.933 {\scriptsize(0.923,\,0.941)} \\
      & \textsf{Coupon} 
        & 0.402 {\scriptsize(0.396,\,0.410)} 
        & 0.987 {\scriptsize(0.982,\,0.993)} 
        & 0.572 {\scriptsize(0.565,\,0.580)} \\
    \midrule
    \multirow{4}{*}{\textsf{nltcs}}
      & \textsf{PY}     
        & 0.921 {\scriptsize(0.908,\,0.933)} 
        & 0.924 {\scriptsize(0.915,\,0.934)} 
        & 0.923 {\scriptsize(0.915,\,0.930)} \\
      & \textsf{Ewens}    
        & 0.921 {\scriptsize(0.910,\,0.932)} 
        & 0.925 {\scriptsize(0.915,\,0.934)} 
        & 0.923 {\scriptsize(0.916,\,0.930)} \\
      & \textsf{GenCoupon} 
        & 0.902 {\scriptsize(0.879,\,0.918)} 
        & 0.935 {\scriptsize(0.926,\,0.944)} 
        & 0.918 {\scriptsize(0.906,\,0.927)} \\
      & \textsf{Coupon} 
        & 0.919 {\scriptsize(0.908,\,0.930)} 
        & 0.926 {\scriptsize(0.916,\,0.935)} 
        & 0.923 {\scriptsize(0.915,\,0.930)} \\
    \midrule
    \multirow{4}{*}{\textsf{cora}}
      & \textsf{PY} 
        & 0.971 {\scriptsize(0.963,\,0.979)} 
        & 0.671 {\scriptsize(0.647,\,0.696)} 
        & 0.794 {\scriptsize(0.776,\,0.813)} \\
      & \textsf{Ewens} 
        & 0.974 {\scriptsize(0.965,\,0.981)} 
        & 0.673 {\scriptsize(0.645,\,0.697)} 
        & 0.796 {\scriptsize(0.775,\,0.813)} \\
      & \textsf{GenCoupon} 
        & 0.973 {\scriptsize(0.965,\,0.981)} 
        & 0.657 {\scriptsize(0.632,\,0.683)} 
        & 0.784 {\scriptsize(0.766,\,0.804)} \\
      & \textsf{Coupon} 
        & 0.978 {\scriptsize(0.971,\,0.986)} 
        & 0.173 {\scriptsize(0.164,\,0.181)} 
        & 0.294 {\scriptsize(0.281,\,0.306)} \\
    \midrule
    \multirow{4}{*}{\textsf{rest}}
      & \textsf{PY} 
        & 0.770 {\scriptsize(0.735,\,0.824)} 
        & 0.812 {\scriptsize(0.759,\,0.884)} 
        & 0.795 {\scriptsize(0.755,\,0.828)} \\
      & \textsf{Ewens} 
        & 0.770 {\scriptsize(0.711,\,0.823)} 
        & 0.830 {\scriptsize(0.781,\,0.875)} 
        & 0.798 {\scriptsize(0.760,\,0.838)} \\
      & \textsf{GenCoupon} 
        & 0.794 {\scriptsize(0.742,\,0.850)} 
        & 0.821 {\scriptsize(0.777,\,0.875)} 
        & 0.807 {\scriptsize(0.773,\,0.849)} \\
      & \textsf{Coupon} 
        & 0.637 {\scriptsize(0.602,\,0.674)} 
        & 0.911 {\scriptsize(0.893,\,0.938)} 
        & 0.750 {\scriptsize(0.722,\,0.781)} \\
    \bottomrule
  \end{tabular}
  \caption{Posterior evaluation metrics for our model under four 
  linkage structure priors corresponding to distinct Ewens-Pitman (EP) 
  parameter regimes. 
  A point estimate for each evaluation metric is reported based on the median, 
  along with a 95\% equi-tailed credible interval.
  Similar performance is observed for the three regimes where the EP parameters 
  are permitted to vary (\textsf{PY}, \textsf{Ewens} and \textsf{GenCoupon}). 
  A significant drop in performance is observed for the \textsf{Coupon} regime 
  on \textsf{RLdata} and \textsf{cora}.}
  \label{tbl:pairwise-measures}
\end{table}

\paragraph{Distortion Model.} 
We compare our proposed distortion model (specified in the latter part of 
Section~\ref{sec:model-spec}) to the distortion model proposed by 
\citet{steorts_entity_2015}. 
For brevity, we refer to our distortion model as \textsf{Ours} and that of 
\citet{steorts_entity_2015} as \textsf{blink}.
Here, we report results for the \textsf{GenCoupon} linkage structure 
prior -- the results for the other linkage structure priors are reported 
in Appendix~\ref{app-sec:comparison} and exhibit similar trends.

\begin{figure}
  \centering
  \includegraphics[scale=0.85]{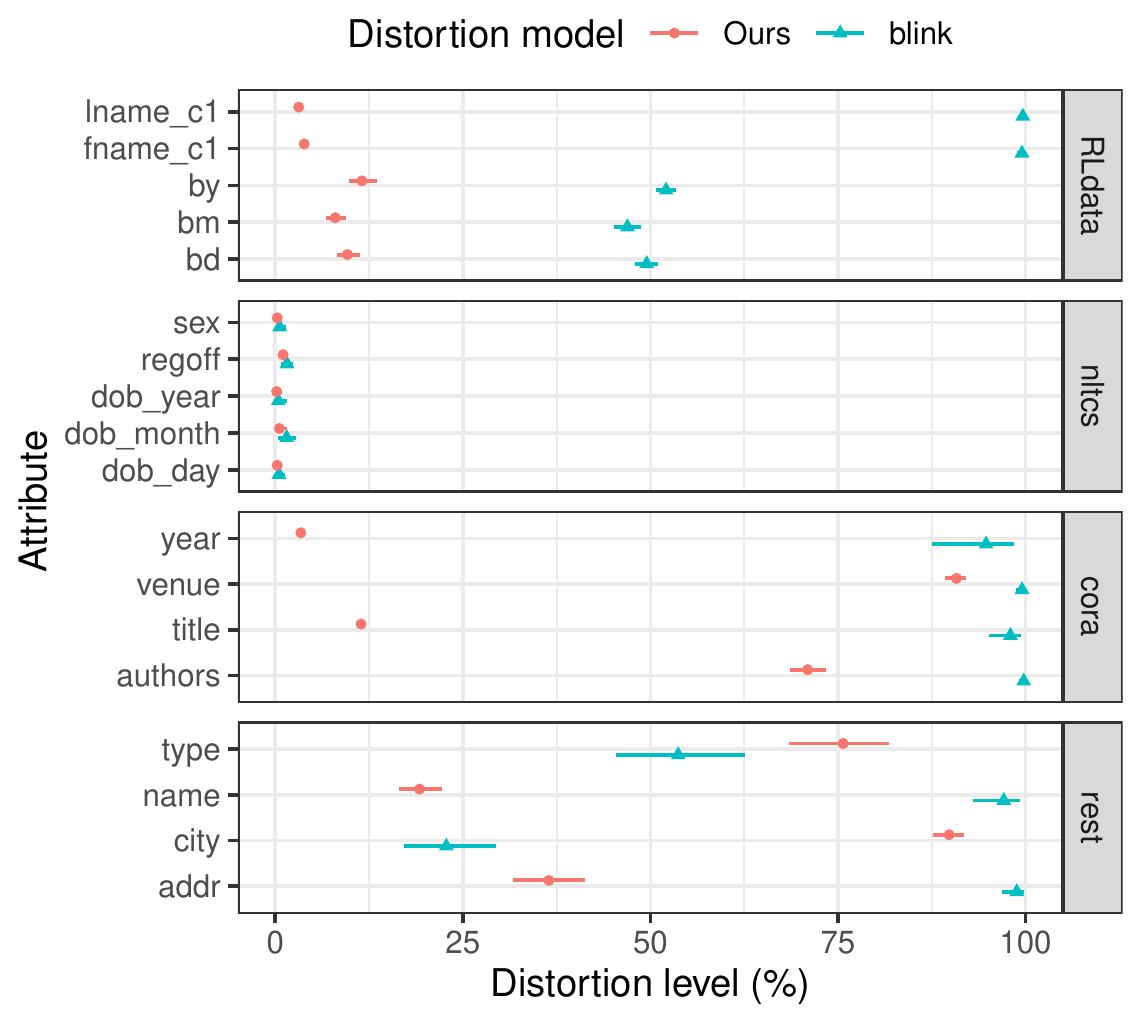}
  \caption{Comparison of the posterior attribute-level distortion under two 
  distortion models: \textsf{Ours} (red top-most intervals) and \textsf{blink} 
  (teal bottom-most intervals). 
  The \textsf{blink} distortion model tends to favor higher levels of 
  distortion -- in some cases approaching 100 percent -- which is not 
  consistent with expectations.}
  \label{fig:posterior-distortion}
\end{figure}

Figure~\ref{fig:posterior-distortion} plots the inferred level of  
distortion for each attribute under both distortion models. 
It shows that \textsf{blink} tends to encourage high distortion, 
particularly for attributes with non-constant distance measures.\footnote{
  The attributes modeled with non-constant distance measures are: all 
  attributes for \textsf{cora}, \texttt{name} and \texttt{addr} 
  for \textsf{rest}, and \texttt{fname\_c1} and \texttt{lname\_c1} 
  for \textsf{RLdata}.
} 
For example, the \texttt{fname\_c1} and \texttt{lname\_c1} attributes 
for \textsf{RLdata} are predicted to be almost 100\% distorted under the 
\textsf{blink} distortion model, which is inconsistent with expectations 
for this data set. 
Our distortion model does not appear to suffer from this problem, as it 
requires disagreement between entity and record attributes in order to 
classify them as ``distorted''. 
Since high distortion makes reliable linkage more challenging, we expect 
that our distortion model is likely to perform better in practice. 
Indeed, it achieves a better balance between precision and recall in 
our full results (see Figure~\ref{app-fig:pairwise-measures-all} in 
Appendix~\ref{app-sec:comparison}).

\paragraph{Summary.}
We return to the original goal of this section and summarize what we have 
learned in this study.
First, we have learned that our proposed linkage structure prior is generally 
more robust due to the use of hyperpriors. 
In addition, our inferences are relatively insensitive to the EP parameter 
regime, which may be due to the fact that the data sets are relatively small 
in size. 
This behavior also holds for the \textsf{blink} distortion model when combined 
with our proposed linkage structure priors. 
Second, when studying the performance of the distortion models (\textsf{blink} 
versus our proposed distortion model), we find that ours predicts more 
reasonable distortion rates and improves the linkage accuracy as measured 
by F1~score. 
Thus, based on this study, we would recommend our distortion model and 
linkage structure priors moving forward for data sets similar to those 
we have considered. 
However, we stress that further exploration is needed to provide more 
general recommendations for other data sets.

\subsection{Comparison with Baseline Models}
\label{sec:baselines}

In this section, we study how our entity resolution model performs in 
comparison with models by \citet{steorts_entity_2015} and 
\citep{sadinle_detecting_2014}. 
The \textsf{blink} model by \citet{steorts_entity_2015} is a natural baseline 
to consider, as it served as inspiration for our model. 
Compared to our model, \textsf{blink} is less Bayesian in its design, as many 
of the parameters are set empirically or arbitrarily. 
Both our model and \textsf{blink}, are examples of direct modeling approaches 
to ER -- i.e., they model how the observed records are generated, incorporating 
the linkage structure as a latent variable. 
In contrast, the model by \citet{sadinle_detecting_2014} (which we refer to 
as \textsf{Sadinle}) adopts a comparison-based approach to ER. 
Instead of modeling the data directly, it models attribute-level comparisons 
between pairs of records, incorporating the presence\slash absence of a link 
between the pair as a latent variable. 
\citet{sadinle_bayesian_2018} compares direct and comparison-based approaches 
from a methodological perspective, however we are not aware of any empirical 
comparisons in the literature. 
Our goal in this section is to provide a comparison for the first time on a 
variety of data sets, where we make no strong claims that our results 
generalize to all applications or data sets.

In order to make the comparison as fair as possible, we use the same distance 
functions to model the distortion in our model, \textsf{blink}, and 
\textsf{Sadinle}.
For instance, if we use edit distance to model distortion for a name attribute 
in our model and \textsf{blink}, then we also use edit distance to compare the 
same name attribute in \textsf{Sadinle}. 
We set the distance cut-offs for our model and \textsf{blink} (see 
Section~\ref{sec:computational}) to align with the blocking design used for 
\textsf{Sadinle}.
Further information about our experimental setup is provided in 
Appendix~\ref{app-sec:experimental-setup}.

ER evaluation metrics are presented in Table~\ref{tbl:baseline-evaluation} for 
all three models. 
For simplicity we only provide results for our model under the 
\textsf{GenCoupon} prior, which is denoted \textsf{Ours} in the table. 
Our model achieves the highest (or equal-highest) F1~score for all four 
data sets. 
We expect the poorer performance of \textsf{blink} is due to its use of 
subjective (inflexible) priors and its distortion model, which tends to favour 
high distortion and over-linkage. 
\textsf{Sadinle} achieves the second highest F1~score in the 
non-private setting (\textsf{RLdata}, \textsf{cora}, \textsf{rest}), 
and the lowest F1~score in the private setting (\textsf{nltcs}). 
The poorer performance for \textsf{nltcs} may be partly related to the 
blocking scheme, which is less aggressive, leaving the model more susceptible 
to over-linkage. 
Another important factor is the sensitivity of \textsf{Sadinle} to the 
truncation points for the priors on the $m$-probabilities. 
We perform coarse-grained tuning of the truncation points in 
Appendix~\ref{app-sec:sadinle-lower-trunc}, however fine-grained tuning could 
result in additional performance gains.

\begin{table}
  \small
  \centering
  \begin{tabular}{llccc}
    \toprule
      &  
        & \multicolumn{3}{c}{Evaluation metric} \\
    \cmidrule{3-5}
    Data set 
      & Model 
        & Precision 
        & Recall 
        & F1 score \\
    \midrule
    \multirow{3}{*}{\textsf{RLdata}}
      & \textsf{Ours} 
        & 0.917 {\scriptsize(0.902,\,0.932)} 
        & 0.966 {\scriptsize(0.953,\,0.973)} 
        & 0.934 {\scriptsize(0.922,\,0.942)} \\
      & \textsf{blink} 
        & 0.336 {\scriptsize(0.327,\,0.344)} 
        & 0.992 {\scriptsize(0.988,\,0.996)} 
        & 0.502 {\scriptsize(0.492,\,0.511)} \\
      & \textsf{Sadinle} 
        & 0.534 {\scriptsize(0.524,\,0.546)} 
        & 0.964 {\scriptsize(0.962,\,0.966)} 
        & 0.687 {\scriptsize(0.679,\,0.697)} \\
    \midrule
    \multirow{3}{*}{\textsf{nltcs}}
      & \textsf{Ours} 
        & 0.901 {\scriptsize(0.878,\,0.917)} 
        & 0.934 {\scriptsize(0.923,\,0.943)} 
        & 0.916 {\scriptsize(0.903,\,0.925)} \\
      & \textsf{blink} 
        & 0.904 {\scriptsize(0.890,\,0.918)} 
        & 0.918 {\scriptsize(0.904,\,0.925)} 
        & 0.910 {\scriptsize(0.903,\,0.918)} \\
      & \textsf{Sadinle} 
        & 0.312 {\scriptsize(0.304,\,0.319)} 
        & 0.975 {\scriptsize(0.969,\,0.979)} 
        & 0.473 {\scriptsize(0.464,\,0.480)} \\
    \midrule
    \multirow{3}{*}{\textsf{cora}}
      & \textsf{Ours} 
        & 0.973 {\scriptsize(0.965,\,0.980)} 
        & 0.657 {\scriptsize(0.627,\,0.683)} 
        & 0.784 {\scriptsize(0.764,\,0.803)}  \\
      & \textsf{blink} 
        & 0.978 {\scriptsize(0.970,\,0.985)} 
        & 0.219 {\scriptsize(0.207,\,0.234)} 
        & 0.358 {\scriptsize(0.341,\,0.378)} \\
      & \textsf{Sadinle} 
        & 0.982 {\scriptsize(0.981,\,0.983)} 
        & 0.359 {\scriptsize(0.357,\,0.362)} 
        & 0.526 {\scriptsize(0.524,\,0.529)} \\
    \midrule
    \multirow{3}{*}{\textsf{rest}}
      & \textsf{Ours} 
        & 0.795 {\scriptsize(0.749,\,0.836)} 
        & 0.830 {\scriptsize(0.776,\,0.871)} 
        & 0.811 {\scriptsize(0.774,\,0.848)} \\
      & \textsf{blink} 
        & 0.635 {\scriptsize(0.586,\,0.671)} 
        & 0.920 {\scriptsize(0.893,\,0.946)} 
        & 0.751 {\scriptsize(0.713,\,0.780)} \\
      & \textsf{Sadinle} 
        & 0.993 {\scriptsize(0.985,\,1.000)} 
        & 0.603 {\scriptsize(0.598,\,0.607)} 
        & 0.750 {\scriptsize(0.744,\,0.756)} \\
    \bottomrule
  \end{tabular}
  \caption{Posterior performance of our model against two baselines: 
    \textsf{blink} \citep{steorts_entity_2015} and \textsf{Sadinle} 
    \citep{sadinle_detecting_2014}. 
    A point estimate for each evaluation metric is reported based on the 
    median, along with a 95\% equi-tailed credible interval.
    Our model achieves the highest (or equal-highest) F1 score within the 
    credible intervals for all data sets.}
  \label{tbl:baseline-evaluation}
\end{table}

\paragraph{Summary.}
This study provides evidence that our model achieves a better balance 
between precision and recall than \textsf{blink} and \textsf{Sadinle}. 
We stress that our results are based on four data sets -- further 
experimentation is required to determine whether our results generalize 
to other data sets and applications. 
We speculate that the better performance of our model is mainly due to 
improved flexibility resulting from the addition of priors and 
hyperpriors, which can be viewed as performing model selection.

\subsection{Controlled Simulation Study} 
\label{sec:controlled}

We conduct a simulation study to evaluate our model under controlled 
conditions, where we vary the size of the data set, the level of distortion, 
and the level of duplication. 
Due to space constraints, we summarize the study here -- full details 
can be found in Appendix~\ref{app-sec:sim-study}.
We design a simulator for household survey data sets, where responses are 
collected for individuals within households. 
Since the attributes of individuals within a household are dependent -- e.g., 
the address is the same, family members may share the same last name, 
the age of individuals may be correlated -- the simulated data follows a more 
complex generative process than our entity resolution model. 
This is intentional, as it allows us to evaluate our model in a more realistic 
setting where it is \emph{misspecified} for the data. 
The dataset simulator also incorporates a record generation process which is 
misspecified for our model. 
Rather than sampling individuals from the population, our data set simulator 
iterates over all individuals, randomly deciding whether to include the 
individual, and if so, how many distorted records to create.

We run entity resolution using our model on 16~simulated data sets, using 
the \textsf{blink} and \textsf{Sadinle} models as baselines. 
We summarize the results across three factors below:
\begin{itemize}
  \item \emph{Duplication level.} 
  The level of duplication has minimal impact on the performance of our 
  model. 
  \textsf{blink} performs well for moderate to high levels of duplication, 
  however it over-links severely when the level of duplication is low. 
  The performance of \textsf{Sadinle} does not seem to follow a consistent 
  trend as the duplication varies -- it achieves a lower F1~score 
  than our model and \textsf{blink} in all cases.
  \item \emph{Distortion level.}
  We find the more distorted data sets are more difficult to link. 
  Specifically, we observe a drop in recall of around 10 percentage 
  points for our model and \textsf{blink} when compared to the data sets with 
  low distortion. 
  Larger drops in recall of 15--20 percentage points are observed for 
  \textsf{Sadinle}.        
  \item \emph{Data set size.}
  We find that our model performs similarly for both data set sizes (1000 
  and 10000 records). 
  \textsf{blink} also performs similarly in most scenarios, however, we observe 
  a drop in precision for the larger data set when the level of duplication 
  is low. 
  \textsf{Sadinle} performs significantly worse for the larger data sets 
  in terms of precision, however, the recall is relatively stable. 
\end{itemize}

In summary, the simulation study shows that our model achieves the most 
consistent performance across all scenarios tested. 
\textsf{blink} is also competitive, but it is has poor performance in the low 
duplication scenario.
\textsf{Sadinle} achieves the lowest F1~score when the level of duplication is 
non-negligible, and is somewhat competitive when the level of distortion 
is low. 

\section{Discussion}
\label{sec:discussion}
In this section, we summarize our contributions and provide a discussion 
regarding future work. 
We have proposed a Bayesian model for entity resolution that addresses 
limitations of previous work \citep{steorts_bayesian_2016, steorts_entity_2015, 
marchant_d-blink_2019}. 
Our model can be viewed as performing graphical entity resolution, where 
observed records are clustered to (unobserved) latent entities. 
To improve upon the scalability of previous work, we designed a partially 
collapsed Gibbs sampler with an optimized implementation that can handle data 
sets of around 10,000 records. 
This allowed us to provide comparisons with models by 
\citet{steorts_entity_2015} and \citet{sadinle_detecting_2014}, which was 
previously only possible for toy-sized data sets. 
We provided comparisons to real and synthetic data sets and a controlled 
simulation study. 
We observed that our model tends to be less sensitive to changes in the 
hyperparameters than competing models for the data sets considered. 
Further analysis is required to make more general conclusions beyond the data 
sets and simulations considered in this paper. 

There are many potential avenues for future work. 
First, it would be of interest to explore scaling for our proposed model and 
the model by \citet{sadinle_detecting_2014}. 
This could be achieved by designing parallel\slash distributed inference 
algorithms, by investigating more efficient MCMC algorithms, or by resorting 
to blocking techniques. 
Another area of interest, is exploring more diverse data sets to understand 
the strengths and weaknesses of each method in practice. 
Although we made recommendations based on four data sets and a simulation 
study, more comparisons would help to alleviate any selection bias regarding 
data sets and provide guidance to users. 
Finally, future work could consider microclustering priors 
\citep{miller_microclustering_2015} to assess their effectiveness compared to 
the infinitely-exchangeable linkage structure priors considered here. 
This would require modifying the sampling scheme and selecting an appropriate 
class of microclustering priors. 

\subsection*{Acknowledgements}
This work was supported by the National Science Foundation [CAREER-1652431], 
the Alfred Sloan Foundation, the Australian Research Council [DP220102269] 
and an Australian Government Research Training Program Scholarship.

\bibliographystyle{asa}
\bibliography{bnp-er}

\appendix
\renewcommand{\theequation}{S\arabic{equation}}
\renewcommand{\thefigure}{S\arabic{figure}}
\renewcommand{\thetable}{S\arabic{table}}
\renewcommand{\bibnumfmt}[1]{[S#1]}
\renewcommand{\citenumfont}[1]{S#1}

\clearpage
\section{Gibbs Updates}
\label{app-sec:gibbs-updates}
In this appendix, we derive updates for the partially collapsed Gibbs 
sampler used to perform approximate inference for the ER model introduced 
in Section~\ref{sec:model}. 
Some of the updates are non-trivial due to non-conjugacy of the proposed model.

\subsection{Update for the Distortion Probabilities}
\label{app-sec:dist-prob-update}
In this section, we provide the update for the distortion probability 
$\theta_{s a}$ (source $s$ and attribute $a$). 
This update is complicated by the presence of the distortion 
propensity variables $\omega_{ia}$, which breaks the conjugacy of the 
beta prior.
To overcome this problem, we introduce the following auxiliary variables:
\begin{align*}
  q_{ia} \mid \omega_{ia} 
    &\sim \operatorname{Bernoulli}(\omega_{ia}) 
    &\forall i, a
\end{align*}
and modify the conditional distribution for the distortion indicators as 
follows:
\begin{align*}
  z_{ia} \mid \theta_{\varsigma_i a}, q_{ia} 
    &\sim \operatorname{Bernoulli}(\theta_{\varsigma_i a} q_{ia}) 
    &\forall i, a.
\end{align*}
It is straightforward to show that one recovers the original model 
in Equation~\eqref{eqn:z-defn} when the auxiliary variables are marginalized out.

Observe that the contribution to the posterior involving $q_{ia}$ is 
\begin{equation*}
  (\theta_{\varsigma_i a} q_{ia})^{z_{ia}} (1 - \theta_{\varsigma_i a} q_{ia})^{1 - z_{ia}}
    \omega_{ia}^{q_{ia}} (1 - \omega_{ia})^{1 - q_{ia}}
  = \left[\theta_{\varsigma_i a}^{z_{ia}} (1 - \theta_{\varsigma_i a})^{1 - z_{ia}} \omega_{ia}\right]^{q_{ia}} 
     \left[1 - \omega_{ia}\right]^{1 - q_{ia}}.
\end{equation*}
Thus, the distribution of $q_{ia}$ conditional on the other variables is:
\begin{align}
  q_{ia} \mid \omega_{ia}, \theta_{\varsigma_i a}, z_{ia} &\sim 
    \operatorname{Bernoulli} \! \left(
      \frac{\omega_{ia} \theta_{\varsigma_i a}^{z_{ia}} (1 - \theta_{\varsigma_i a})^{1 - z_{ia}}}
        {\omega_{ia} \theta_{\varsigma_i a}^{z_{ia}} (1 - \theta_{\varsigma_i a})^{1 - z_{ia}} + 1 - \omega_{ia}} 
    \right) & \forall i, a. 
  \label{app-eqn:distort-prob-aux-update}
\end{align}

Next, observe that the contribution to the posterior involving $\theta_{s a}$ 
is 
\begin{equation*}
  \theta_{s a}^{\beta_{sa}^{(0)} - 1} (1 - \theta_{s a})^{\beta_{sa}^{(1)} - 1} 
    \prod_{i: \varsigma_i = s} (\theta_{s a} q_{ia})^{z_{ia}} 
      (1 - \theta_{s a} q_{ia})^{1 - z_{ia}}.
\end{equation*}
Hence, the distribution of $\theta_{s a}$ conditional on the other variables 
is:
\begin{align}
  \theta_{s a} \mid \vec{Q}, \vec{Z}, \vec{S} &\sim 
  \operatorname{Beta} \! \left(
    \beta_{s a}^{(0)} + \sum_{i: \varsigma_i = s} z_{ia}, \
    \beta_{s a}^{(1)} + \sum_{i: \varsigma_i = s} q_{ia} (1 - z_{ia}) 
  \right) & \forall s, a.
  \label{app-eqn:distort-prob-update}
\end{align}

It is also straightforward to see that the distribution of $z_{ia}$ 
conditional on the other variables is a point mass. 
In particular, we have 
\begin{align}
  z_{i a} \mid x_{ia}, \lambda_i, \vec{Y} = \begin{cases}
    1, & \text{if } x_{i a} \neq y_{\lambda_i a} \\
    0, & \text{otherwise}
  \end{cases}
  \label{app-eqn:distort-ind-update}
\end{align}

In summary, to update the distortion probabilities, one would first compute 
the distortion indicators $\{z_{ia}\}$ using 
Equation~\eqref{app-eqn:distort-ind-update}. 
Then, conditional on the other variables, one would draw auxiliary variables 
$\{q_{ia}\}$ using Equation~\eqref{app-eqn:distort-prob-aux-update}. 
Finally, one can update the distortion probabilities $\{\theta_{sa}\}$ 
using Equation~\eqref{app-eqn:distort-prob-update}.
The updates for the other model parameters are unaffected by the 
introduction of the auxiliary variables $\{q_{ia}\}$.

\subsection{Update for the Entity Attributes}
\label{app-sec:ent-attr-update}
In this section, we provide the update for the entity attributes.
When updating entity attribute $y_{e a}$, we collapse the base 
distribution $H_{e a}$ and distortion indicators $\vec{Z}$. 

The posterior factors involving $y_{ea}$ after collapsing $H_{ea}$ are 
as follows:
\begin{equation*}
  \begin{aligned}
  P(y_{e a} \mid \vec{Z}, \vec{\Omega}, \vec{\Theta}, \vec{S}, G_a, \rho_a) 
  & \propto P(y_{e a} \mid G_a) \\
  & \qquad \times \int \prod_{i: \lambda_{i} = e} 
    P(x_{ia} \mid \theta_{\varsigma_i a}, \omega_{ia}, y_{e a}, H_{e a}) 
      P(H_{e a} \mid y_{e a}, \rho_a) \, \mathrm{d} H_{e a} \\
  & \propto P(y_{e a} \mid G_a) 
    \prod_{\substack{i:\lambda_{i} = e\\x_{ia} = y_{e a}}} 
      (1 - \theta_{\varsigma_i a} \omega_{ia})
        \prod_{\substack{i:\lambda_{i} = e\\x_{ia} \neq y_{e a}}} 
          (\theta_{\varsigma_i a} \omega_{ia}) \\ 
  & \qquad \times \int 
    \prod_{\substack{i:\lambda_{i} = e\\x_{ia} \neq y_{e a}}} H_{ea}(x_{ia}) 
      \frac{ \prod_{v \in \mathcal{D}_{a} \setminus \{y_{ea}\}} 
        H_{ea}(v)^{\rho_a \psi_a(v \mid y_{ea}) - 1} }
      {\mathrm{B}(\rho_a \vec{\psi}_{a}(y_{ea}))} \, \mathrm{d} H_{e a} \\
  & \propto G_a(y_{e a}) \prod_{\substack{i:\lambda_{i} = e\\x_{ia} = y_{e a}}} 
      (1 - \theta_{\varsigma_i a} \omega_{ia})
        \prod_{\substack{i:\lambda_{i} = e\\x_{ia} \neq y_{e a}}} 
          (\theta_{\varsigma_i a} \omega_{ia}) \\
  & \qquad \times \frac{\Gamma(\rho_a)}{\Gamma(\bar{n}_{ea}(y_{ea}) + \rho_a)}
    \prod_{v \in \mathcal{V}_{ea}} 
      \frac{\Gamma(n_{e a}(v) + \rho_a \psi_a(v \mid y_{e a}))}
      {\Gamma(\rho_a \psi_a(v \mid y_{e a}))}, \\
  \end{aligned}
\end{equation*}
where $\mathrm{B}(\cdot)$ is the multivariate beta function,
$\mathcal{V}_{ea} = \left(\bigcup_{i : \lambda_{i} = e} \{x_{ia}\}\right) 
\setminus \{y_{ea}\}$ are the distorted record values for the $a$-th attribute 
associated with entity $e$, 
$n_{ea}(v) = \sum_{i : \lambda_{i} = e} \mathbb{1}[x_{i a} = v]$ is the number 
of records linked to entity $e$ whose $a$-th attribute is equal to $v$ and 
$\bar{n}_{ea}(v) = \sum_{i : \lambda_{i} = e} \mathbb{1}[x_{i a} \neq v]$ 
is the number of records linked to entity $e$ whose $a$-th attribute is 
\emph{not} equal to $v$.

We can rewrite the above expression in a more computationally convenient form 
by repeatedly applying the recurrence relation for the Gamma 
functions\footnote{%
  Repeated application of the recurrence relation for the Gamma function 
  yields 
  \begin{equation*}
    \Gamma(z) = \frac{\Gamma(z + n + 1)}{z (z + 1) \cdots (z + n)}
  \end{equation*}
  for complex $z$ (excluding zero and the negative integers) and 
  non-negative integer $n$.
} 
to yield:
\begin{equation*}
  \begin{aligned}
    P(y_{e a} \mid \vec{Z}, \vec{\Omega}, \vec{\Theta}, \vec{S}, G_a, \rho_a) 
  & \propto G_a(y_{e a}) 
    \frac{ \prod_{v \in \mathcal{V}_{ea}} \prod_{j = 1}^{n_{ea}(v)} 
        \left\{j - 1 + \rho_a \psi_a(v \mid y_{ea})\right\} }
    { \prod_{j = 1}^{\bar{n}_{ea}(y_{ea})} \left\{ j - 1 + \rho_a \right\} } \\ 
  & \qquad \times \prod_{\substack{i:\lambda_{i} = e\\x_{ia} = y_{e a}}} 
      (1 - \theta_{\varsigma_i a} \omega_{ia})
    \prod_{\substack{i:\lambda_{i} = e\\x_{ia} \neq y_{e a}}} 
      (\theta_{\varsigma_i a} \omega_{ia}).
  \end{aligned}
\end{equation*}
Observe that the above distribution may only have support on a subset of the 
full domain $\mathcal{D}_a$ when distance thresholds are applied, as discussed 
in Section~\ref{sec:computational}. 
In particular, one can show that the support is a subset of 
\begin{equation*}
  \bigcap_{i: \lambda_{i} = e} 
    \{y \in \mathcal{D}_a : \operatorname{dist}_a(y, x_{ia}) \leq 
      d_a^{(\mathrm{cut})} \}.
\end{equation*}
This fact can be used to implement the update more efficiently, since 
it is not necessary to construct a pmf over the entire domain $\mathcal{D}_a$.

\subsection{Update for the Linkage Structure} \label{app-sec:link-update}
In this section, we provide the update for the linkage structure. 
When updating the linkage structure, we use an urn-based scheme as described 
by \citet{neal_markov_2000}. 
In doing so, we only need to keep track of entities in the population that are 
linked to records -- any isolated entities not linked to records are ignored. 
This is important, as the population may be infinite in size for some 
Ewens-Pitman parameter regimes (when $\sigma \geq 0$).

To update the linked entity $\lambda_{i}$ for record $i$, we remove the current 
link and allow the record to either join one of the remaining instantiated 
entities (with at least one other record) or instantiate a ``new'' entity.
The conditional distribution has the following form: 
\begin{align}
  P(\lambda_{i} = e \mid \vec{Z}, \vec{X}, \vec{Y}, \vec{\Lambda}_{-i}, 
  \{\rho_a\}) \propto \begin{cases}
    C \frac{\lvert e \rvert - \sigma}{\alpha + N - 1} \prod_a 
      \int P(x_{ia} \mid z_{ia}, y_{e a}, H_{e a}) \mathrm{d} \mathbb{H}_{-i,ea}, \\
        \mspace{100mu} \text{if $e$ is instantiated and $\lvert e \rvert > 0$}, \\
    C \frac{\alpha + \sigma E}{\alpha + N - 1} \prod_a 
      \sum_{y_{ea} \in \mathcal{D}_{a}} P(y_{ea} \mid G_{a}) 
        \int P(x_{ia} \mid z_{ia}, y_{ea}, H_{ea}) \mathrm{d} \mathbb{H}_{0,ea}, \\
        \mspace{100mu} \text{if $e$ is ``new''},
  \end{cases}
  \label{app-eqn:blackwell-macqueen}
\end{align}
where 
\begin{itemize}
  \item $C$ is a normalization constant; 
  \item $\vec{\Lambda}_{-i} = (\lambda_1, \ldots, \lambda_{i - 1}, 
  \lambda_{i + 1}, \ldots, \lambda_N)$ are the linked entities for all records 
  excluding $i$;
  \item $\lvert e \rvert = \sum_{i' \neq i} \mathbb{1}[\lambda_{i'} = e]$ is 
  the number of records (excluding $i$) linked to entity $e$; 
  \item $E = \sum_{e' \neq e} \mathbb{1}[\lvert e \rvert > 0]$ is the number of 
  instantiated entities with at least one linked record; 
  \item $\mathbb{H}_{0,ea}$ is the prior for $H_{ea}$; and
  \item $\mathbb{H}_{-i,ea}$ is the posterior for $H_{ea}$ given the 
  observed distorted record attributes $x_{i' a}$ for which $i' \neq i$ and 
  $\lambda_{i'} = e$ (also conditioned on $y_{ea}$ and $\rho_a$).
\end{itemize}

Recall that the prior $\mathbb{H}_{0,ea}$ for $H_{ea}$ conditioned on $y_{ea}$ 
and $\rho_a$ is $\operatorname{Dirichlet}(\rho_a \vec{\psi}_a(y_{ea}))$.
Since $x_{ia}$ is $\operatorname{Categorical}(H_{ea})$ if $z_{ia} = 1$ (and a 
point mass at $y_{ea}$ if $z_{ia} = 0$), the posterior $\mathbb{H}_{-i,ea}$ is 
also Dirichlet by conjugacy. 
In particular, one can show that $\mathbb{H}_{-i,ea}$ is 
$\operatorname{Dirichlet}(\vec{\alpha}_{-i,ea})$ where 
\begin{equation*}
  \alpha_{-i,ea}(v) = \rho_a \psi_a(v \mid y_{ea}) 
    + \sum_{i' \neq i: \lambda_{i'} = e} z_{i'a} \mathbb{1}[x_{i'a} = v] 
\end{equation*}
for $v \in \mathcal{D}_a \setminus \{y_{ea}\}$.
We can therefore simplify the integral in 
Equation~\eqref{app-eqn:blackwell-macqueen} 
with respect to $\mathbb{H}_{-i,ea}$ as follows:
\begin{align}
  & \int P(x_{ia} \mid z_{ia}, y_{ea}, H_{ea}) \, d \mathbb{H}_{-i,ea} \nonumber \\
  &= \mathbb{1}[x_{ia} = y_{ea}]^{1 - z_{ia}} \int H_{ea}(x_{ia})^{z_{ia}} 
    \Gamma(\rho_a) \prod_{v \in \mathcal{D}_a \setminus \{y_{ea}\}} 
      \frac{H_{ea}(v)^{\rho_a \psi_a(v \mid y_{ea}) - 1}}
      {\Gamma(\rho_a \psi_a(v \mid y_{ea}))} \, \mathrm{d} H_{ea} \nonumber \\ 
  &= \begin{cases}
    \frac{\alpha_{-i,ea}(x_{ia})}
        {\sum_{v \in \mathcal{D}_a \setminus \{y_{ea}\}} \alpha_{-i,ea}(v)}, 
      & z_{ia} = 1 \\
    \mathbb{1}[x_{ia} = y_{ea}], 
      & z_{ia} = 0
  \end{cases} 
  \label{app-eqn:likelihood-int-H}
\end{align}
By a similar argument, the integral in 
Equation~\eqref{app-eqn:blackwell-macqueen} with respect to $\mathbb{H}_{0,ea}$ 
can be simplified to:
\begin{equation}
  \int P(x_{ia} \mid z_{ia}, y_{ea}, H_{ea}) \, d \mathbb{H}_{0,ea} 
  = \begin{cases}
    \psi_a(x_{ia} \mid y_{ea}), 
      & z_{ia} = 1, \\
    \mathbb{1}[x_{ia} = y_{ea}], 
      & z_{ia} = 0.
  \end{cases}
  \label{app-eqn:likelihood-int-H0}
\end{equation}

Putting Equations~\eqref{app-eqn:likelihood-int-H} and 
\eqref{app-eqn:likelihood-int-H0} in \eqref{app-eqn:blackwell-macqueen}, 
then gives:
\begin{align*}
  P(\lambda_{i} = e \mid \vec{Z}, \vec{X}, \vec{Y}, \vec{\Lambda}_{-i}, 
  \{\rho_a\}) \propto \begin{cases}
    C \frac{\lvert e \rvert - \sigma}{\alpha + N - 1} 
      \prod_{a: z_{ia} = 1} \frac{\alpha_{-i,ea}(x_{ia})}
        {\sum_{v \in \mathcal{D}_a \setminus \{y_{ea}\}} \alpha_{-i,ea}(v)}, \\
        \mspace{100mu} \text{if $e$ is instantiated and $\lvert e \rvert > 0$}, \\
    C \frac{\alpha + \sigma E}{\alpha + N - 1} 
     \prod_{a: z_{ia} = 1} \sum_{y \in \mathcal{D}_{a}} 
      G_a(y) \psi_a(x_{ia} \mid y)^{z_{ia}} \mathbb{1}[x_{ia} = y]^{1 - z_{ia}}, \\
        \mspace{100mu} \text{if $e$ is ``new''}.
  \end{cases}
\end{align*}

\subsection{Update for the Ewens-Pitman Parameters}
\label{app-sec:ep-update}
In this section, we describe the update for the Ewens-Pitman parameters.
Since the priors on the Ewens-Pitman parameters $\alpha$ and $\sigma$ are 
non-conjugate, we cannot perform a direct Gibbs update. 
Thus, we describe tractable updates which require the introduction of 
auxiliary variables. 
The updates (and priors) differ depending on the range of $\sigma$.
\citet{teh_bayesian_2006} proposed a scheme for beta\slash gamma priors 
when $0 \leq \sigma < 1$ and $\alpha > 0$, which is summarized in 
Section~\ref{app-sec:auxiliary-pos-sigma}. 
In Section~\ref{app-sec:auxiliary-neg-sigma} we propose a similar scheme 
for gamma\slash shifted negative binomial priors when $\sigma < 0$.

\subsubsection{Case \ensuremath{0 \leq \sigma < 1} and \ensuremath{\alpha > 0}}
\label{app-sec:auxiliary-pos-sigma}
\citet{teh_bayesian_2006} proposed an auxiliary variable scheme for the 
regime $0 \leq \sigma < 1$ and $\alpha >0$ such that the priors
\begin{equation*}
  \sigma \sim \operatorname{Beta}\! \left(\zeta^{(0)}, \zeta^{(1)}\right) 
  \quad \text{and} \quad
  \alpha \sim \operatorname{Gamma}\! \left(\chi^{(0)}, \chi^{(1)}\right)
\end{equation*}
are conjugate.
We provide a summary of the scheme here, but refer the reader to 
\citep{teh_bayesian_2006} for further details.
The scheme introduces the following sets of auxiliary variables conditional 
on the two parameters $\alpha$ and $\sigma$:
\begin{equation}
  \begin{aligned}
    w \mid N, \alpha &\sim 
      \operatorname{Beta} (\alpha + 1, N - 1), \\
    u_k \mid \sigma, \alpha, E &\sim 
      \operatorname{Bernoulli} \!\left(\frac{\alpha}{\alpha + \sigma k}\right), 
        & k \in \{1,\ldots, E - 1\} \\
    v_{ej} \mid \sigma, \vec{\Lambda} &\sim 
      \operatorname{Bernoulli} \!\left(\frac{j - 1}{j - \sigma}\right), 
        & \forall e, j \in \{1, \ldots, N_e - 1\}.
  \end{aligned}
  \label{app-eqn:auxiliary-pos-sigma}
\end{equation}
Here $N_e = \lvert \{i : \lambda_i = e\} \rvert$ denotes the number of records 
linked to entity $e$, and $E = \sum_{e} \mathbb{1}[N_e > 1]$ denotes the number 
of entities linked to at least one record.

It follows that the posterior distributions of $\alpha$ and $\sigma$ 
conditional on the auxiliary variables and other model parameter are given 
by:
\begin{equation}
  \begin{split}
  \sigma \mid \{u_k\}, \{v_{e j}\}, \vec{\Lambda} 
    &\sim \operatorname{Beta}\! \left(
      \zeta^{(0)} + \sum_{k=1}^{E-1} (1 - u_k), 
      \zeta^{(1)} + \sum_{e: N_e > 1} \sum_{j = 1}^{N_e - 1} (1 - v_{e j}) 
    \right), \\
  \alpha \mid \{u_k\}, w, \vec{\Lambda} 
    &\sim \operatorname{Gamma}\! \left(
      \chi^{(0)} + \sum_{k = 1}^{E - 1} u_k, 
      \chi^{(1)} - \log w  
    \right).
  \end{split}
  \label{app-eqn:posterior-alpha-sigma}
\end{equation}

Thus, to update $\alpha$ and $\sigma$, one would first draw auxiliary variables 
$w$, $\{u_k\}$ and $\{v_{ej}\}$ conditional on the linkage structure 
$\vec{\Lambda}$ and the old values of $\alpha$ and $\sigma$ using 
Equation~\eqref{app-eqn:auxiliary-pos-sigma}. 
Then, conditional on the auxiliary variables and the linkage structure, one 
would draw new values for $\alpha$ and $\sigma$ using 
Equation~\eqref{app-eqn:posterior-alpha-sigma}.

\subsubsection{Case \ensuremath{\sigma < 0} and \ensuremath{\alpha = m \kappa} for \ensuremath{m \in \texorpdfstring{\mathbb{N}}{ℕ}}}
\label{app-sec:auxiliary-neg-sigma}
We describe an auxiliary variable scheme for updating the Ewens-Pitman 
parameters in the regime $\sigma < 0$ and $\alpha = m \kappa$, where 
$m \in \mathbb{N}$ and $\kappa > 0$.
This scheme is inspired by \citet{teh_bayesian_2006}. 
The likelihood factor associated with the partition of $N$ records into $E$ 
entities is as follows \citep{pitman_exchangeable_2006}:
\begin{equation}
  P(\text{partition config}) 
    = \frac{(m)_{E \downarrow}}
    {(m \kappa)_{N \uparrow}} \prod_{e = 1}^{E} (\kappa)_{N_e \uparrow} 
    = \frac{\kappa^{E - 1} (m - 1)_{E - 1 \downarrow}}
    {(m \kappa - 1)_{N - 1 \uparrow}} 
      \prod_{e = 1}^{E} (\kappa - 1)_{N_e - 1 \uparrow}, 
  \label{app-eqn:clust-config}
\end{equation}
where ``partition config'' is a representation of the linkage structure 
$\vec{\Lambda}$ as a partition\footnote{%
  Records $i$ and $j$ belong to the same subset of the partition if 
  $\lambda_i = \lambda_j$, and otherwise belong to different subsets.
}, 
$N_e$ is the number of records linked to the $e$-th entity, 
$(x)_{n \uparrow} = \prod_{i = 0}^{n-1} (x + i)$ is the rising 
factorial, and $(x)_{n \downarrow} = \prod_{i = 0}^{n-1} (x - i)$ is 
the falling factorial.
We begin by expressing the denominator in this equation as 
\begin{align*}
  \frac{1}{(m \kappa - 1)_{N - 1 \uparrow}} 
    = \frac{\Gamma(m \kappa + 1)}{\Gamma(m \kappa + N)} 
  & = \frac{\mathrm{B}(m \kappa + 1, N - 1)}{\Gamma(N - 1)} \\
  & = \frac{1}{\Gamma(N - 1)} \int_{0}^{1} w^{m \kappa} (1 - w)^{N - 2} 
    \, \mathrm{d} w,
\end{align*}
which allows us to introduce the following auxiliary variable:
\begin{equation}
  w \mid m, \kappa, N \sim \operatorname{Beta}(m \kappa + 1, N - 1).
  \label{app-eqn:auxiliary-w}
\end{equation}
Expressing each of the latter factors in Equation~\eqref{app-eqn:clust-config} 
as 
\begin{equation*}
  (\kappa - 1)_{N_e - 1 \uparrow} 
  = \prod_{j = 1}^{N_e - 1} (\kappa + j) 
  = \prod_{j = 1}^{N_e - 1} \sum_{v_{ej} \in \{0, 1\}} 
    \kappa^{v_{ej}} j^{1 - v_{ej}}
\end{equation*}
permits us to introduce the following additional auxiliary variables:
\begin{align}
  v_{ej} \mid \kappa 
    & \sim \operatorname{Bernoulli}\! \left(\frac{\kappa}{\kappa + j}\right), 
    & \forall e, j \in \{1, \ldots, N_e - 1\}.
  \label{app-eqn:auxiliary-v}
\end{align}

With this representation, we can place conjugate priors on $\kappa$ and $m$, 
namely:
\begin{equation}
  \kappa \sim \operatorname{Gamma}(\chi^{(0)}, \chi^{(1)}) \ \text{and} \ 
  m \sim \operatorname{NegativeBinomial}(r, \nu) + 1.
  \label{app-eqn:kappa-m-priors}
\end{equation}
The distribution on $m$ is a shifted negative binomial with support on the 
positive integers.
The parameterization we adopt for the negative binomial is in terms of the 
number of failures $x \in \{0, 1, 2, \ldots\}$ in a sequence of trials before 
a given number of successes $r > 0$ occur.
Each trial is an i.i.d.\ draw from a Bernoulli distribution with success 
probability $\nu$.
The density of $x$ is given by
\begin{equation*}
  P(x \mid r, \nu) = \frac{(x + r - 1)!}{(r - 1)! x!} \nu^r (1 - \nu)^{x}.
\end{equation*}

Finally, we combine the priors in Equation~\eqref{app-eqn:kappa-m-priors} 
with the likelihood factors to obtain the following posterior distributions 
for the $m$ and $\kappa$, conditional on the other model parameters:
\begin{equation}
  \begin{aligned}
    m \mid w, \kappa, \vec{\Lambda} 
      & \sim \operatorname{NegBinomial}\! 
        \left(r + E - 1, 1 - (1 - \nu) w^\kappa \right) + E, \\
    \kappa \mid \{v_{ej}\}, w, m, \vec{\Lambda} 
      & \sim \operatorname{Gamma}\! \left(
        \chi^{(0)} + E - 1 + \sum_{e = 1}^{E} \sum_{j = 1}^{N_e - 1} v_{ej}, \ 
        \chi^{(1)} - m \log w 
      \right).
  \end{aligned}
  \label{app-eqn:posterior-m-kappa}
\end{equation}

Thus, to update $\kappa$ and $m$, one would first draw auxiliary variables $w$ 
and $\{v_{ej}\}$ conditional on the linkage structure $\vec{\Lambda}$ and the 
old values of $\alpha$ and $\sigma$ using Equations~\eqref{app-eqn:auxiliary-w} 
and \eqref{app-eqn:auxiliary-v}. 
Then, conditional on the auxiliary variables and the linkage structure, one 
would draw new values for $\kappa$ and $m$ using 
Equation~\eqref{app-eqn:posterior-m-kappa}.

\subsection{Update for the Distortion Distribution Concentration}
\label{app-sec:distort-dist-conc-update}

In this section, we provide the update for the distortion distribution 
concentration $\rho_a$. 
Since we cannot rely on conjugacy for the update, we propose an auxiliary 
variable scheme.
When updating $\rho_a$, we condition on the entity attribute values 
$\{y_{ea}\}_{e=1 \ldots E}$, the record attribute values 
$\{x_{ia}\}_{i=1 \ldots N}$ and the links 
$\vec{\Lambda} = \{\lambda_i\}_{i=1 \ldots N}$. 
We collapse the distortion distributions $\{H_{ea}\}_{e = 1 \ldots E}$.
The contribution to the likelihood involving $\rho_a$ is:
\begin{align}
  & \prod_e \int \prod_{i: \lambda_i = e} 
    P(x_{ia} \mid \theta_{\varsigma_i a}, \omega_{ia}, y_{ea}, H_{ea}) 
      P(H_{ea} \mid \rho_a) \, \mathrm{d} H_{ea} \nonumber \\
  &\propto \prod_e \frac{\Gamma(\rho_a)}{\Gamma(\bar{n}_{ea}(y_{ea}) + \rho_a)} 
    \prod_{v \in \mathcal{V}_{ea}} 
      \frac{\Gamma(n_{ea}(v) + \rho_a \psi_a(v \mid y_{ea}))}
      {\Gamma(\rho_a \psi_a(v \mid y_{ea}))} \nonumber \\
  &= \prod_e \frac{\mathrm{B}(\rho_a, \bar{n}_{ea}(y_{ea}))}
    {\Gamma(\bar{n}_{ea}(y_{ea}))} \prod_{v \in \mathcal{V}_{ea}} 
      \prod_{j = 1}^{n_{ea}(v)} \{j - 1 + \rho_a \psi_a(v \mid y_{ea})\} 
  \label{app-eqn:distort-dist-conc-lklhd}
\end{align}
where $\mathrm{B}(\cdot, \cdot)$ is the beta function, 
$\mathcal{V}_{ea} := \left(\bigcup_{i: \lambda_{i} = e} \{x_{ia}\}\right) 
\setminus \{y_{ea}\}$, 
$n_{ea}(v) = \sum_{i : \lambda_{i} = e} \mathbb{1}[x_{i a} = v]$ and  
$\bar{n}_{ea}(v) = \sum_{i : \lambda_{i} = e} \mathbb{1}[x_{i a} \neq v]$.

Expressing the beta function in Equation~\eqref{app-eqn:distort-dist-conc-lklhd} 
as 
\begin{equation*}
  \mathrm{B}(\rho_a, \bar{n}_{ea}(y_{ea})) = 
    \int_0^1 w_{e}^{\rho_a - 1} (1 - w_{e})^{\bar{n}_{ea}(y_{ea}) - 1} 
      \, \mathrm{d} w_{e}
\end{equation*}
permits us to introduce the following auxiliary variables:
\begin{equation*}
  w_{e} \mid \rho_a, \vec{X}, \vec{Y}, \vec{\Lambda} 
    \sim \operatorname{Beta}\! \left(
        \rho_a, 
        \sum_{e} \sum_{i} \mathbb{1}[x_{ia} \neq y_{\lambda_i a}]
      \right),
  \label{app-eqn:rho-auxiliary-w}
\end{equation*}
for all $e$.
We can also express each of the latter factors in 
Equation~\eqref{app-eqn:distort-dist-conc-lklhd} as 
\begin{equation*}
  j - 1 + \rho_a \psi_a(v \mid y_{ea}) = 
    \sum_{u_{evj} = 0}^{1} \rho_a \psi_a(v \mid y_{ea})^{u_{evj}} 
      (j - 1)^{1 - u_{evj}}
\end{equation*}
which permits us to introduce the following auxiliary variables:
\begin{align}
  u_{evj} \mid \rho_a, \vec{X}, \vec{Y}, \vec{\Lambda} &\sim 
    \operatorname{Bernoulli}\! \left(\frac{\rho_a \psi_{a}(v \mid y_{ea})}
    {j - 1 + \rho_a \psi_{a}(v \mid y_{ea})}\right) 
  \label{app-eqn:rho-auxiliary-u}
\end{align}
for all $e$, $v \in \mathcal{V}_{ea}$ and $j \in \{1, \ldots, n_{ea}(v)\}$.

Now since the prior on $\rho_a$ is 
$\operatorname{Gamma}(\tau_a^{(0)}, \tau_a^{(1)})$, we obtain the following 
posterior distribution for $\rho_a$ conditional on the other parameters:
\begin{align}
  \rho_a \mid \{w_{e}\}, \{u_{evj}\}, \vec{X}, \vec{Y}, \vec{\Lambda} 
  \sim \operatorname{Gamma}\! \left(
      \tau_a^{(0)} + \sum_{e} \sum_{v \in \mathcal{V}_{ea}} 
        \sum_{j = 1}^{n_{ea}(v)} u_{evj}, 
      \tau_a^{(1)} - \sum_{e} \log w_{e} 
  \right). 
  \label{app-eqn:posterior-rho}
\end{align}

Thus to update $\rho_a$, one would first draw auxiliary variables $\{w_{e}\}$ 
and $\{u_{evj}\}$ conditional on the record attributes $\vec{X}$, entity 
attributes $\vec{Y}$, linkage structure $\vec{\Lambda}$, and the previous value 
of $\rho_a$ using Equations~\eqref{app-eqn:rho-auxiliary-w} and 
\eqref{app-eqn:rho-auxiliary-u}. 
Then, conditional on the auxiliary variables and $\vec{X}$, $\vec{Y}$, 
$\vec{\Lambda}$, one would draw a new value for $\rho_a$ using 
Equation~\eqref{app-eqn:posterior-rho}.

\section{Hybrid Distance Measure} \label{app-sec:hybrid-dist-fn}

In this appendix, we describe a hybrid distance measure that is useful for 
comparing text strings containing multiple tokens (words), where individual 
tokens may be subject to distortion. 
We use the measure in this paper for comparing name and address attributes 
in the \textsf{cora} and \textsf{rest} data sets 
(see Section~\ref{sec:datasets}), however it may have wider applications 
beyond this paper. 
Our measure draws inspiration from a hybrid similarity measure proposed by 
\citet{monge_field_1996}.
However, unlike \citeauthor{monge_field_1996}, we attempt to match the 
tokens in each string while incorporating penalties for tokens that are 
``missing'' in one of the strings. 

Suppose we would like to compare a pair of multi-token strings $x$ and $y$. 
As a running example, we consider 
$x = $ ``University of California, San Diego'' and 
$y = $ ``Univ. Calif., San Diego''.
Given a separator character (e.g., a space), we can map each string to 
a set of tokens. 
For example, string $x$ from our running example would be mapped to
\begin{equation*}
  X = \{\text{``California,'', ``Diego'', ``of'', ``San'', ``University''}\}.
\end{equation*}
Note that we have used capital $X$ to denote the token set\footnote{%
  Technically we consider a multi-set, since we allow tokens to appear multiple 
  times.
}
representation of string $x$ -- a convention we adopt throughout this appendix.
Also note that $X$ is a lossy representation of $x$, as it discards 
information about the token order.
This is desirable for our applications to names and addresses\footnote{%
  Specifically, the \texttt{title}, \texttt{venue} and \texttt{authors} 
  attributes in \textsf{cora}, and the \texttt{name} and \texttt{addr} 
  attributes in \textsf{rest}.
}, where permutation of the tokens does not significantly change the meaning 
of the strings.

We propose to measure the distance from $x$ to $y$ via a generalized 
edit distance on the token sets $X$ and $Y$. 
We consider three elementary edit operations:
\begin{itemize}
  \item \emph{token insertions} where a token $b$ is appended to the input set;
  \item \emph{token deletions} where a token $a$ is removed from the input set; 
  and
  \item \emph{token substitutions} where a token $a$ in the input set is 
  replaced by a token $b \neq a$.
\end{itemize}
Each elementary operation takes an input set $Q$ to an output set $Q'$, which 
we write as $Q \to Q'$, and has an associated cost $c(Q \to Q') \geq 0$.
We let
\begin{align*}
  c(Q \to Q') = \begin{cases}
    d_i \, \operatorname{dist}_\mathrm{inner}(\lambda, b), 
      & \text{if $Q = Q' \setminus \{b\}$ (insertion),} \\
    d_d \, \operatorname{dist}_\mathrm{inner}(a, \lambda), 
      & \text{if $Q \setminus \{a\} = Q'$ (deletion),} \\
    d_s \, \operatorname{dist}_\mathrm{inner}(a, b), 
      & \text{if $Q \setminus \{a\} = Q' \setminus \{b\}$ (substitution),}
  \end{cases}
\end{align*}
where $d_i, d_d$ and $d_s$ are non-negative weights; $\lambda$ is the null 
string; and $\operatorname{dist}_{\mathrm{inner}}(\cdot, \cdot)$ 
is an \emph{inner distance measure} on tokens (strings).
We then define the \emph{hybrid distance} between $x$ and $y$ as the minimum 
average cost of transforming $X$ into $Y$ via a sequence of elementary edit 
operations $T_{X,Y} = (X \to Q_1, Q_1 \to Q_2, \ldots, Q_{l - 1} \to Y)$.
Symbolically, we write
\begin{equation*}
  \operatorname{dist}_\mathrm{hybrid}(x, y) = 
    \min_{T_{X, Y}} \frac{1}{\lvert T_{X,Y} \rvert} 
      \sum_{(Q \to Q') \in T_{X, Y}} c(Q \to Q').
\end{equation*}

We can compute the hybrid distance using an off-the-shelf linear sum assignment 
problem (LSAP) solver \citep{crouse_implementing_2016}. 
In order to do so, we need to add null string tokens to $X$ and $Y$ to account 
for all possible insertion and deletion operations.
Concretely, we add $\lvert Y \rvert$ null tokens to $X$ to allow for insertions 
and $\lvert X \rvert$ null tokens to $Y$ to allow deletions.
We then construct a pairwise cost matrix by applying 
$\operatorname{dist}_\mathrm{inner}$ to all pairs of tokens in (the amended) 
$X$ and $Y$. 
The resulting matrix is then passed to the LSAP solver, which returns the 
optimal set of edit operations and their cost.

Returning to our running example, if we set 
$\operatorname{dist}_\mathrm{inner}$ to the Levenshtein distance, the solution 
to the LSAP is 
\begin{gather*}
  \{(\text{``University''} \leftrightarrow \text{``Univ.''}, 5), 
    (\text{``of''} \leftrightarrow \lambda, 2), 
    (\text{``California,''} \leftrightarrow \text{``Calif.,''}, 6), \\
    (\text{``San''} \leftrightarrow \text{``San''}, 0), 
    (\text{``Diego''} \leftrightarrow \text{``Diego''}, 0), 
    (\lambda \leftrightarrow \lambda, 0), 
    (\lambda \leftrightarrow \lambda, 0), \\
    (\lambda \leftrightarrow \lambda, 0), 
    (\lambda \leftrightarrow \lambda, 0)\}.
\end{gather*}
Hence we conclude that $\operatorname{dist}_\mathrm{hybrid}(x, y) 
= \frac{5+2+6+0+0}{5} = 2.6$. 
This distance reflects the semantic closeness between $x$ and $y$ better than 
the Levenshtein distance, which gives a larger value of 14 when evaluated 
directly on $x$ and $y$.

\section{Simulation Study} \label{app-sec:sim-study}

In this appendix, we conduct a simulation study to understand how our model 
performs in controlled scenarios. 
Specifically, we simulate entity resolution data sets where we vary the number 
of records, the level of distortion and the level of duplication. 
Since our model is generative, we could use it to simulate data, however the 
resulting data would have negligible specification error for our model, which 
is not realistic. 
We therefore simulate data that is purposefully misspecified for our model by 
adding additional dependencies between the entities and entity attributes, 
and by using a different process to generate records from entities. 
We were unable to find an existing data set simulator that generated such 
data, so we implemented our own. 

\subsection{Data Set Simulator}
\label{app-sec:sim-study-data}

We provide an overview of our simulator, which generates personal records 
describing a population of households. 
For brevity, we omit low-level details here and refer the reader to the 
included Python script. 
Our simulator operates in two stages: in the first stage it generates a 
population of households, then in the second stage it iterates over 
individuals in all households, generating a random number of distorted records 
for each individual.
By generating households rather than individuals in the first stage, we 
are able to incorporate additional dependencies between individuals (entities) 
that are not present in our ER model (see Section~\ref{sec:model-spec}). 

\paragraph{Generating Households.}
We now describe how households are generated in the first stage. 
In our simplified model, a household may be a couple, a single, a couple 
or single with children, or a group of unrelated adults. 
Individuals within a household are described by the following attributes: 
first and last name (first\_name and last\_name), date of birth (birth\_year, 
birth\_month, birth\_day), gender and zipcode. 
The zipcode is constrained to be the same for all individuals within a 
household, and the first name is conditioned on the gender, however 
the other attributes may vary as described below.
Random values for attributes are generated using the Faker Python 
library\footnote{\url{https://github.com/joke2k/faker}}, which attempts 
to mimic real-world frequency distributions.
We make the distributions more concentrated for the name and zipcode 
attributes to ensure the entities are not too unique (otherwise entity 
resolution would be too easy).

We begin by generating the head(s) of the household, which are a male and 
female couple (for simplicity) or a single male or female. 
If a couple is generated, they have a high chance of sharing the same 
last name and their birth years are likely not too far apart. 
Next we randomly decide whether to generate children. 
If children are generated, they share the same last name as the head(s) of 
the household (the parents) and there is an appropriate gap between their birth 
year and their parents' birth year.
If no children are generated, then we randomly decide whether to generate 
unrelated adults who live with the head(s) of household. 
The unrelated adults are constrained to be of a similar age as the head(s) 
of the household. 
When simulating the household composition, we attempt to follow aggregate 
statistics from the Current Population Survey \citep{uscb_cps_2016}.

\paragraph{Generating Records.} 
In the second stage, records are generated for individuals across all 
households. 
We simulate a single database\slash file with duplicate records by including 
an individual with probability $p_\mathrm{inc} = 0.9$ and sampling the 
number of records according to a Poisson distribution with rate parameter 
$\mu$, truncated to the interval $[1, 4]$.
Each record is obtained by copying the entity attributes subject to distortion. 
This is done by iterating over the attributes in a random order, and deciding 
whether to activate the distortion process with a probability that varies 
for each attribute. 
The distortion process for birth day, birth month, gender and zipcode 
involves drawing a replacement value according to the distribution used 
in the first stage.
The distortion process for birth year involves adding discrete Gaussian 
noise to the true birth year. 
The distortion process for first and last name may proceed in one of three 
ways: (1)~by making a random typographical error (character insertion, 
deletion, substitution or transposition); (2)~by replacing the name with a 
variant drawn uniformly at random; or (3)~by generating a replacement 
according to the distribution used in the first stage. 
Variant names for (2) are sourced from the WeRelate.org Variant Names 
Project\footnote{
  \url{https://www.werelate.org/wiki/WeRelate:Variant_names_project}
}.

\subsection{Results}

\begin{figure}
  \centering
  \includegraphics[scale=0.95]{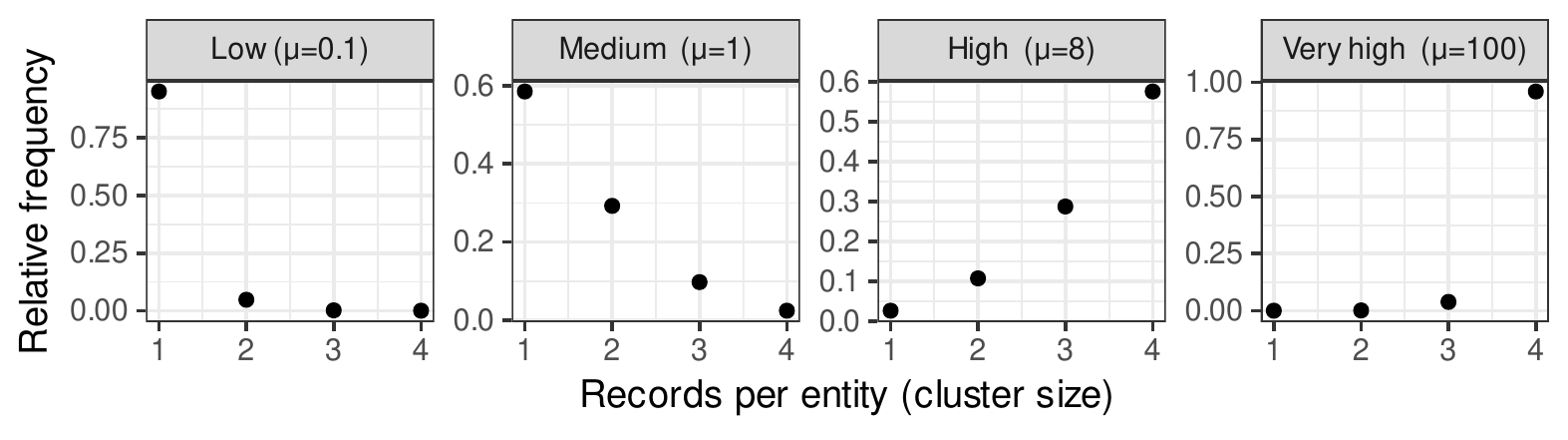}
  \caption{Distribution of records per entity for each level of duplication: 
    low, medium, high and very high. 
    The Poisson rate parameter $\mu$ for each level is given in parentheses.}
  \label{app-fig:sim-link-conf}
\end{figure}

We generate 16 data sets using our simulator for each 
combination of the following variables:
\begin{itemize}
  \item \emph{Number of records.} 
  We consider data sets with 1000 and 10000 records in expectation. 
  The number of records is random and depends on the number of individuals 
  and the Poisson rate parameter. 
  Since the Poisson rate parameter is fixed (see below), we control the 
  number of records by varying the number of individuals generated in the 
  first stage. 
  \item \emph{Level of distortion.} We consider two levels of record 
  distortion which we refer to as ``low'' and ``high''. 
  These correspond to different choices for the probabilities of activating 
  the distortion process as detailed in Table~\ref{app-tbl:sim-dist-conf}.
  \item \emph{Level of duplication.} We consider four levels of duplication 
  which we refer to as ``low'', ``medium'', ``high'' and ``very high''.
  These levels correspond to Poisson rate parameters of 0.1, 1, 8 and 100, 
  respectively. 
  When the duplication is ``low'' ($\mu = 0.1$) over 95\% of the entities 
  represented in the data only appear once. 
  Whereas when the duplication is ``very high'' ($\mu = 100$) over 95\% of 
  the entities represented in the data appear four times.
  The distribution of records per entity is plotted for each level in 
  Figure~\ref{app-fig:sim-link-conf}.
\end{itemize}

\begin{table}
  \centering
  \small
  \begin{tabular}{lrr}
    \toprule
        & \multicolumn{2}{c}{Distortion probability} \\
    \cmidrule{2-3}
    Attribute     & Low distortion & High distortion \\
    \midrule
    first\_name   &           10\% &            40\% \\
    last\_name    &           10\% &            40\% \\
    gender        &            1\% &             1\% \\
    zipcode       &            5\% &            10\% \\
    birth\_year   &            1\% &            10\% \\
    birth\_month  &            1\% &            10\% \\
    birth\_day    &            1\% &            10\% \\
    \bottomrule
  \end{tabular}
  \caption{Attribute-level distortion probabilities for two levels of 
  distortion: low and high.}
  \label{app-tbl:sim-dist-conf}
\end{table}

\begin{figure}
  \centering
  \includegraphics[scale=0.95]{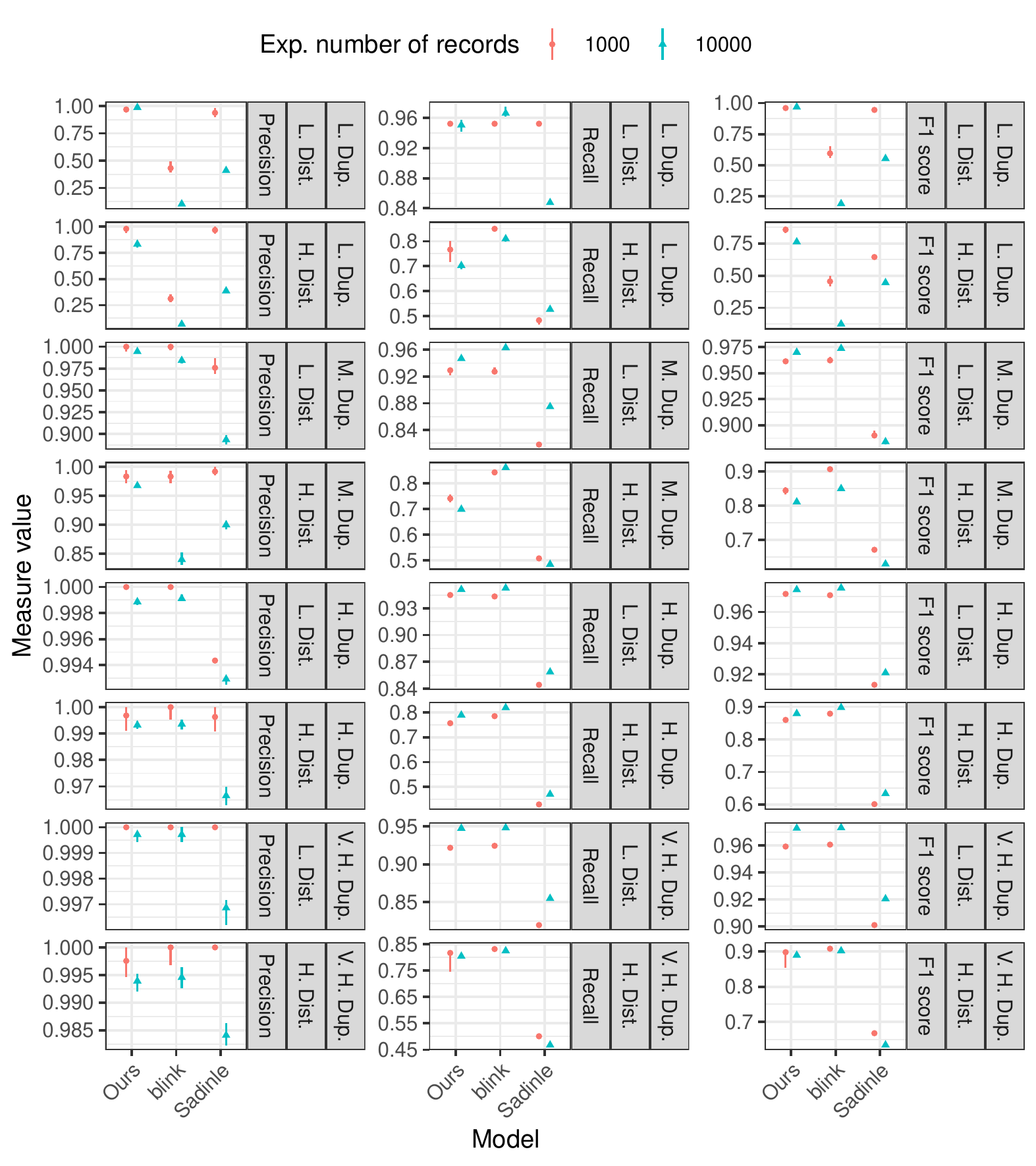}
  \caption{Posterior evaluation metrics for our model, \textsf{blink}, and 
  \textsf{Sadinle} when fitted on simulated datasets with varying levels of 
  distortion and duplication. 
  Low and high distortion levels are abbreviated as ``L. Dist.'' and 
  ``H. Dist.'' respectively. 
  Low, medium, high and very high duplication levels are abbreviated as 
  ``L. Dup.'', ``M. Dup.'', ``H. Dup.'' and ``V. H. Dup.'' respectively. 
  The expected number of records (1000 or 10000) is denoted by the color and 
  shape of the markers.
  A point estimate for each evaluation metric is reported based on the 
  median and 95\% equi-tailed credible interval are represented by intervals 
  around the point.
  }
  \label{app-fig:sim-results}
\end{figure}

We perform a comparative evaluation of our model, \textsf{blink}, and 
\textsf{Sadinle} on the 16 simulated data sets. 
The model evaluation procedure is described in 
Section~\ref{sec:model-evaluation} and the \textsf{blink} and \textsf{Sadinle} 
models are introduced in Section~\ref{sec:baselines}. 
ER evaluation metrics are plotted for each data set in 
Figure~\ref{app-fig:sim-results}. 
We now make several observations about the results. 

First, we observe that our model and \textsf{blink} perform similarly when the 
duplication level is medium, high or very high.  
For these duplication levels, \textsf{blink} has a slight advantage in terms of 
recall when the distortion level is high.
The largest difference is observed for the medium duplication\slash high 
distortion scenario, where the recall for \textsf{blink} is roughly 10 
percentage points higher than for our model.
This difference is due to the priors placed on the concentration parameters 
$\rho_a$ in our model, which favour high concentrations. 
This corresponds to a prior belief that distortions occur in the same way, 
rather than in multiple different ways. 
However, this is not true for distortions in the simulated data -- e.g., an 
individual whose first name is ``JONATHON'' may appear in the data with four 
distinct first names: ``JOHN'', ``JOJN'', ``JONATHON'' and ``ALEX''. 
If we wanted to exploit this knowledge, we could increase $\tau_a^{(0)}$ 
for our model to favour lower concentrations. 

Second, we observe that our model significantly outperforms \textsf{blink} in 
terms of precision when the duplication level is low. 
We believe this is due to the highly informative prior on the linkage 
structure used in \textsf{blink} -- it uses a coupon prior with $m$ fixed to 
the number of records $N$ and $\kappa \to \infty$. 
However our model under the generalized coupon prior selects a value for 
$m$ of approximately $8 \times N$ and $\kappa$ of approximately 100, which 
allows it to more accurately model a low duplication scenario.

Thirdly, we observe that \textsf{Sadinle} achieves the lowest F1~score 
when the duplication level is medium, high or very high. 
For these duplication levels, the performance gap in F1~score is largest when 
the distortion is high -- approximately 20 percentage points.  
The gap is less significant when the distortion is low -- around 5 
percentage points.
These differences in F1~score are mainly due to lower recall in most 
cases -- the precision is generally competitive with the other models.

Fourthly, we comment on the effect of the dataset size, measured in terms 
of the expected number of records (1000 or 10000). 
We observe similar trends for all three models: the precision tends to 
be larger for the smaller dataset, while the recall tends to be larger for 
the larger dataset (however there are exceptions). 
The difference in performance is most pronounced for the low duplication 
setting, where the precision of \textsf{blink} and \textsf{Sadinle} drops 
considerably for the larger dataset, and the recall of \textsf{Sadinle} 
drops also drops considerably for the larger dataset.

\section{Study of Linkage Structure Priors and the Distortion Model}
\label{app-sec:comparison}

In this appendix, we provide additional results for the study of linkage 
structure priors and distortion models presented in Section~\ref{sec:study}. 
Our goal is to study the impact of the modeling contributions 
independently, to determine whether each contribution is beneficial in its 
own right, and\slash or whether one contribution is more beneficial 
than the other.

\begin{figure}
\center
  \includegraphics[scale=0.95]{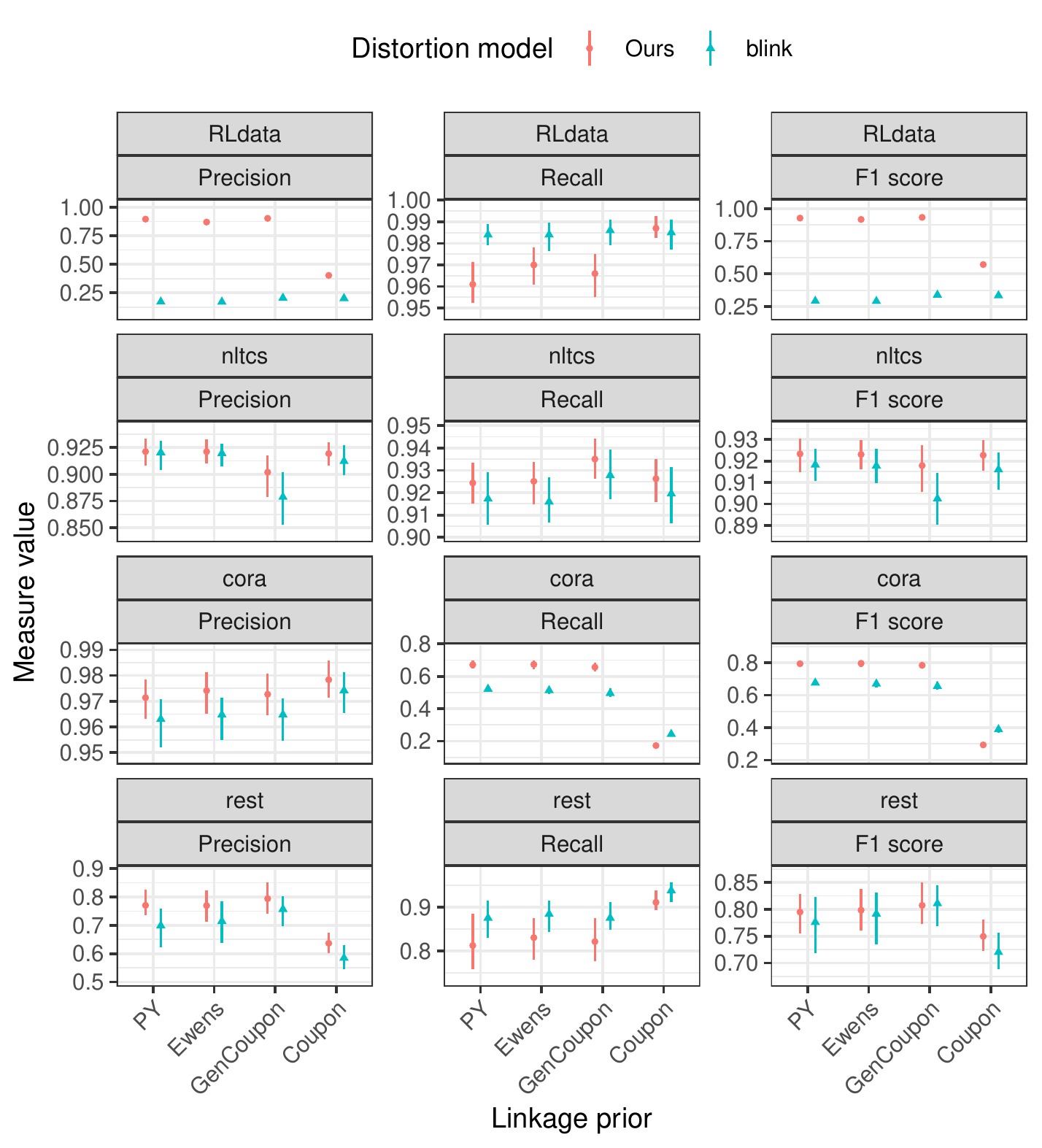}
  \caption{%
    Evaluation of ER quality as a function of the linkage structure prior 
    (plotted on the $x$-axis) and distortion model (indicated by the line 
    color).
    Three pairwise evaluation measures are shown (grouped by column) 
    for four data sets (grouped by row).}
  \label{app-fig:pairwise-measures-all}
\end{figure}

Recall from Section~\ref{sec:study} that we considered four parameter 
regimes for the linkage structure priors -- \textsf{PY}, \textsf{Ewens}, 
\textsf{GenCoupon} and \textsf{Coupon} -- and two distortion 
models -- \textsf{Ours} as proposed in Section~\ref{sec:model-spec} and 
\textsf{blink} as proposed by \citet{steorts_entity_2015}.
Thus, there are eight model variants to test -- one for each linkage 
structure prior and distortion model.
Figure~\ref{app-fig:pairwise-measures-all} presents pairwise evaluation 
metrics (F1~score, precision and recall) for the eight model variants and 
four data sets in a single plot.
We interpret the results for each modeling contribution below.

\begin{figure}
  \centering
  \includegraphics[scale=0.85]{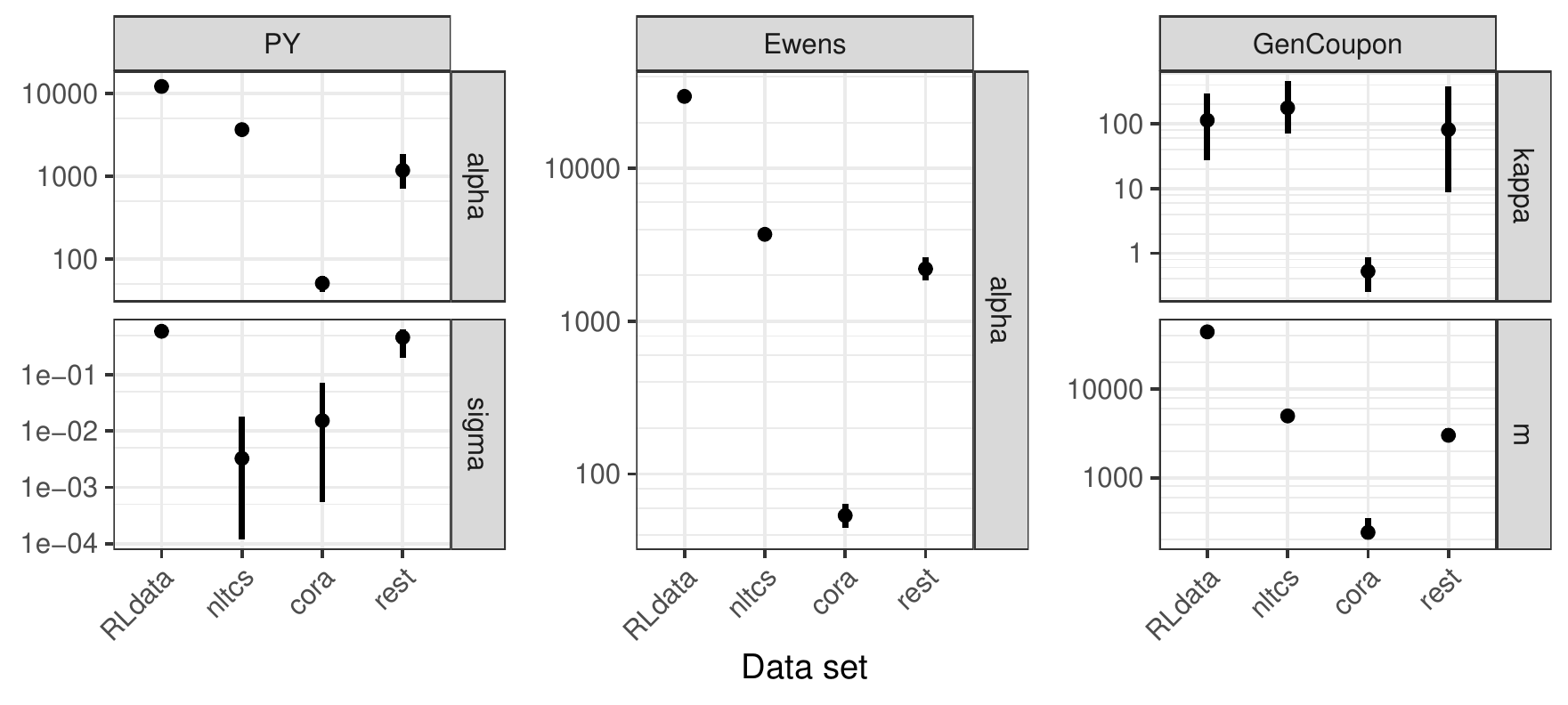}
  \caption{Posterior Ewens-Pitman parameters for three regimes: \textsf{PY}, 
  \textsf{Ewens} and \textsf{GenCoupon} under our distortion model 
  (\textsf{Ours}). Note that the values of the parameters are presented on 
  log-scales.}
  \label{app-fig:post-ep-params}
\end{figure}

\paragraph{Linkage Structure Prior.}
While we discuss the effect of the linkage structure prior in 
Section~\ref{sec:study}, we only present results for our distortion model 
(\textsf{Ours}) due to space constraints. 
In Section~\ref{sec:study}, we draw two main conclusions from the results in 
Table~\ref{tbl:pairwise-measures} which are replicated in 
Figure~\ref{app-fig:pairwise-measures-all} (represented by circular 
vermilion markers):
\begin{enumerate}
  \item Our proposal to place vague hyperpriors on the EP parameters (for 
  \textsf{PY}, \textsf{Ewens} and \textsf{GenCoupon}) improves robustness 
  and yields the highest ER accuracy for three of data sets, as measured by 
  pairwise F1~score (\textsf{nltcs} is the exception). 
  We observe significantly lower F1~scores when hyperpriors are not used 
  (see \textsf{Coupon} in Figure~\ref{app-fig:pairwise-measures-all}), 
  particularly for \textsf{cora} and \textsf{RLdata}. 
  Figure~\ref{app-fig:post-ep-params} provides further justification for 
  this argument, as it shows vastly different values of the EP parameters 
  are selected for each data set, facilitated by the vague hyperpriors.
  \item Our inferences are relatively insensitive to the EP parameter 
  regime (\textsf{PY}, \textsf{Ewens} or \textsf{GenCoupon}) despite 
  the fact that each regime is known to exhibit distinct asymptotic 
  behavior (see Section~\ref{sec:exchangeable}). 
\end{enumerate}
Figure~\ref{app-fig:pairwise-measures-all} shows that these conclusions 
\emph{also hold} for the \textsf{blink} distortion model (represented by 
triangular teal markers). 
We expect the competitive performance for \textsf{nltcs} under the 
\textsf{Coupon} linkage prior may be a coincidence, as the population size 
under the prior is 3,387, which happens to be very close to the true value 
of 3,307 (see Table~\ref{tbl:datasets}).

\paragraph{Distortion Model.}
In Section~\ref{sec:study}, we discuss the effect of the distortion prior 
under the \textsf{GenCoupon} linkage structure prior. 
We now extend the discussion to include results for the three other linkage 
structure priors (\textsf{PY}, \textsf{Ewens} and \textsf{Coupon}), as 
presented in Figure~\ref{app-fig:pairwise-measures-all}. 

We find that our distortion model achieves the highest F1~score for all but 
one of the data sets and linkage structure priors.
The exception is for \textsf{cora} under the \textsf{Coupon} linkage structure 
prior, where the \textsf{blink} distortion model has a slight edge.
An explanation for the improved performance of our model is given in 
Section~\ref{sec:study}, which we summarize here. 
The \textsf{blink} distortion model is susceptible to 
entering a high distortion mode, particularly for attributes with 
non-constant distance measures.
This is because it allows a record attribute value to be marked as 
``distorted'' even if it is not actually distorted.
Our model corrects this inconsistency, and in doing so appears to 
be more robust. 
In general, we expect the \textsf{blink} distortion model to result in 
\emph{over-linkage} (high recall, low precision), while our model is 
expected to be more balanced. Figure~\ref{app-fig:pairwise-measures-all} 
supports this argument, with the difference being most apparent for 
\textsf{RLdata}, where we see a difference of $\sim$0.7 in the F1~score.

\section{Further Details of Experimental Setup}
\label{app-sec:experimental-setup}
In this appendix, we provide further details about the experiments presented 
in Section~\ref{sec:experiments}.

\paragraph{Implementation and Hardware.}
All experiments were conducted in R version 3.4.4, running on a local server
fitted with two 28-core Intel Xeon Platinum 8180M CPUs and 12~TB of 
RAM.\footnote{%
  R scripts are published at 
  \href{https://github.com/cleanzr/exchanger-experiments}
  {github.com/cleanzr/exchanger-experiments}.
}
We developed an open-source R package called \texttt{exchanger}\footnote{
  Package source code published at 
  \href{https://github.com/cleanzr/exchanger}{github.com/cleanzr/exchanger}.
} which implements variants of our model (under different linkage structure 
priors and distortion models) in addition to the \textsf{blink} model 
\citep{steorts_entity_2015}. 
Since an implementation of the model proposed by \citet{sadinle_detecting_2014} 
was not publicly available, we developed our own which we released 
as an open-source R package called \texttt{BDD}\footnote{
  Package source code published at 
  \href{https://github.com/cleanzr/BDD}{github.com/cleanzr/BDD}.
}.
For efficiency reasons, we implemented inference for all models in C++ using 
the Rcpp interface \citep{eddelbuettel_rcpp_2011}. 
The data set simulator described in Appendix~\ref{app-sec:sim-study-data} 
was implemented as a Python script.
A Pipfile is provided to specify the dependencies used when running the script.

\paragraph{Hyperparameter Settings.}
We followed the recommendations in Section~\ref{sec:hyperparameters} 
when setting hyperparameters for our model.
When setting hyperparameters for the two baseline models, we attempted 
to follow the recommendations of the authors. 
For \textsf{blink}, we set $m = N$ for the coupon-collector's prior and 
$\beta_{sa}^{(0)} = N/1000$ and $\beta_{sa}^{(1)} = N/10$ for the 
Beta prior on the distortion probabilities (here $N$ is the total number of 
records). 
For \textsf{Sadinle}, we set the agreement levels by inspecting the 
distribution of distances for each attribute.
We used truncated uniform priors on the $m$-probabilities and a uniform 
prior on the $u$-probabilities, as recommended by the author.
We set the lower truncation points for the $m$-probabilities to 0.95, based 
on tuning experiments presented in Appendix~\ref{app-sec:sadinle-lower-trunc}.

\paragraph{Initialization and MCMC.}
For our model and \textsf{blink}, we initialized the linkage structure 
$\vec{\Lambda}$, entity attributes $\vec{Y}$ and distortion indicators 
$\vec{Z}$ by linking each record to a unique entity and copying the record 
attributes into the entity attributes, assuming no distortion. 
The distortion probabilities $\vec{\Theta}$ and entity attributes distributions 
$\vec{G}$ were initialized by drawing from their conditional distributions.
The Ewens-Pitman parameters and distortion distribution concentration 
parameters were initialized using their prior means.

A similar initialization was used for the \textsf{Sadinle} model. 
We assigned each record to a unique entity (cluster). 
The $m$- and $u$-probabilities were initialized by drawing from their 
conditional distributions.

When fitting each model, we ran Markov chain Monte Carlo (MCMC) for 
$2 \times 10^5$ iterations, discarding the first $10^5$ iterations as 
burn-in, and applying thinning with an interval of 10.\footnote{
  The chain was slower to converge for the \textsf{cora} data set, so we 
  increased the number of iterations to $2.5 \times 10^5$ and the burn-in 
  interval to $1.5 \times 10^5$.
}
This produced $10^4$ approximate posterior samples.

\section{Tuning Hyperparameters for Sadinle~(2014)} 
\label{app-sec:sadinle-lower-trunc}

Our aim in this appendix is to determine reasonable values for the 
hyperparameters in the entity resolution model by 
\citet{sadinle_detecting_2014}, which we refer to as \textsf{Sadinle}.
We assume the distance functions (used to compare attributes) and agreement 
levels (mappings from real-valued distances to discrete levels) are fixed, 
and that flat priors are used, as recommended by Sadinle. 
Given these assumptions, the only hyperparameters that remain unspecified 
are the lower truncation points for the $m$-probabilities. 

\begin{figure}
  \centering
  \includegraphics[scale=0.95]{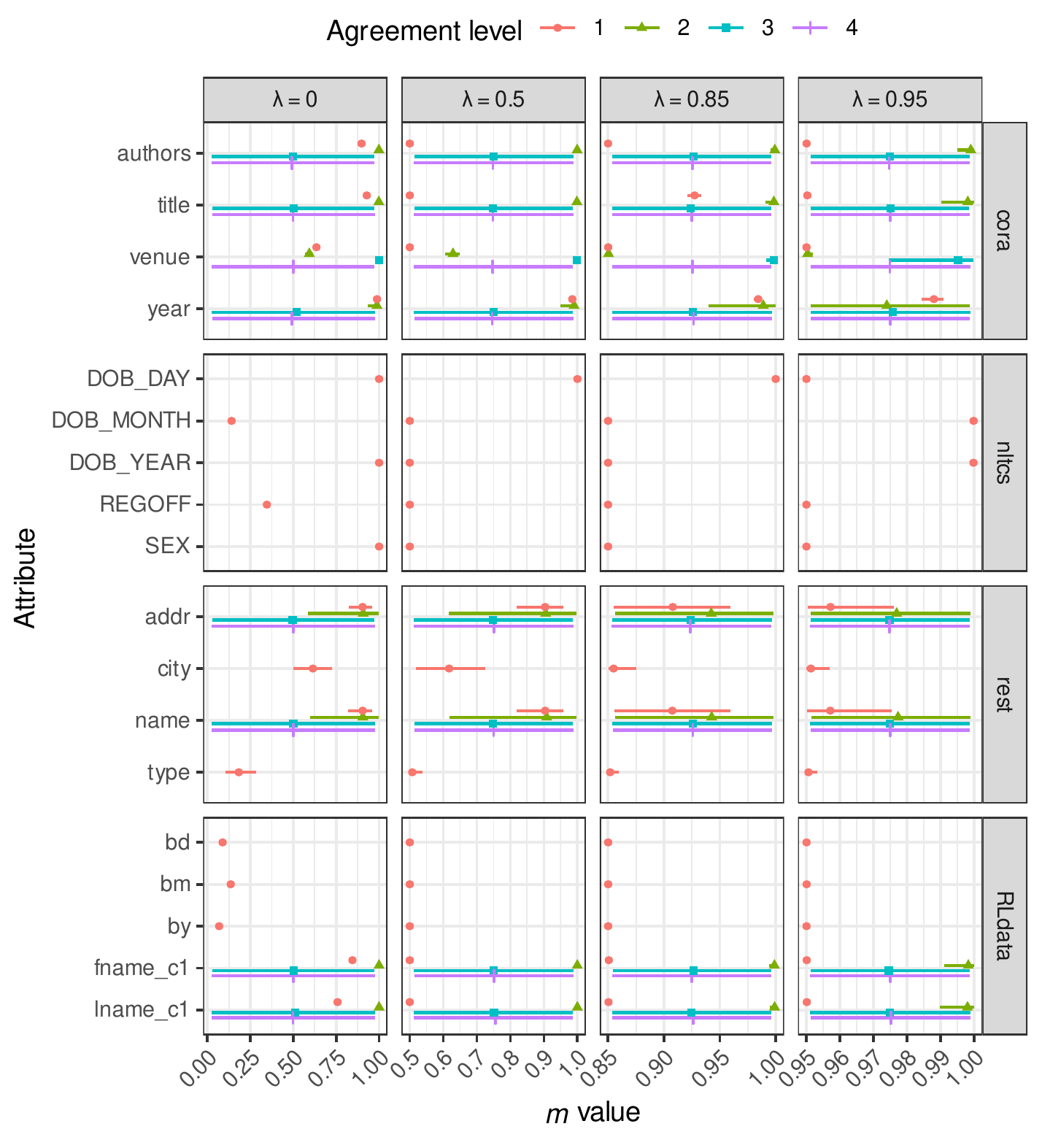}
  \caption{Posterior estimates of the $m$-probabilities (plotted on the 
    $x$-axis) in the \textsf{Sadinle} model as a function of the data set 
    (vertical panels), attribute ($y$-axis), agreement level (color\slash 
    marker) and lower truncation point (horizontal panels). 
    Point estimates are shown based on the median, along with 95\% equi-tailed 
    credible intervals. 
    The model has a tendency to select small values for the $m$-probabilities, 
    close to the lower truncation point, especially for agreement level~1.}
  \label{app-fig:m-sadinle-lower-trunc}
\end{figure}

The $m$-probabilities are a set of parameters $\{m_{al}\}$ where $m_{al}$ is 
the probability that a pair of records referring to the same entity agree at 
level $l$ on attribute $a$, given they do not agree at levels 
$0, 1, \ldots, l - 1$. 
Sadinle recommends using truncated flat priors on $m_{al}$, so that the 
allowed values lie in the interval $[\lambda_{al}, 1]$, where 
$\lambda_{al} \in [0, 1]$ is a hyperparameter typically close to 1. 
More specifically, he recommends setting $\lambda_{al} = 0.95$ if attribute $a$ 
is a ``nearly-accurate'' quasi-identifier and $\lambda_{al} = 0.85$ if 
attribute $a$ is an ``inaccurate'' quasi-identifier. 
Since it is not clear which setting for $\lambda_{al}$ is best for our data 
sets, we run the experiments described in Section~\ref{sec:baselines} 
for four different values: $\lambda_{al} = 0, 0.5, 0.85, 0.95$. 
When setting $\lambda_{al}$, we use the same value for all attributes $a$ and 
agreement levels $l$ for simplicity. 
The results are reported in Figure~\ref{app-fig:m-sadinle-lower-trunc} and 
Table~\ref{app-tbl:sadinle-lower-trunc}. 

Figure~\ref{app-fig:m-sadinle-lower-trunc} plots the posterior values 
of the $m$-probabilities (on the $x$-axis) for each truncation point 
$\lambda_{al}$ (corresponding to the horizontal panels).
We observe that the posterior values of $m_{al}$ tend to be close to 
$\lambda_{al}$, especially for agreement level $l = 0$ and 
$\lambda_{al} \geq 0.5$. 
This suggests that the model favors small values of $m_{al}$, despite 
the fact that $m_{al}$ is expected to be close to 1. 
Consequently, the model has a tendency to ``over-link''---linking records 
that do not refer to the same entity. 
Understanding why the model exhibits this behavior would require further 
exploration and is beyond the scope of this paper. 
However, we speculate that it may be related to the use of flat priors, or 
known stability issues with Fellegi-Sunter-type models 
\citep{goldstein_scaling_2017}. 
As a result, we conclude that the posterior value of $m_{al}$ is highly 
sensitive to the choice of $\lambda_{al}$. 

Table~\ref{app-tbl:sadinle-lower-trunc} shows the impact of the truncation 
point $\lambda_{al}$ on entity resolution performance. 
It shows that the performance is relatively stable for \textsf{cora} and 
\textsf{rest} as a function of $\lambda_{as}$, despite the fact that there is 
some variation in the $m$-probabilities (as shown in 
Figure~\ref{app-fig:m-sadinle-lower-trunc}). 
The reason for the stability may be related to the blocking scheme used for 
these data sets, which rules out a relatively large number of potential links, 
thereby guarding against over-linkage. 
On the other hand, the performance for \textsf{RLdata} and \textsf{nltcs} is 
far less stable. 
We find that the precision drops considerably as $\lambda_{al}$ is reduced, 
while the recall remains relatively stable. 
This is a sign of over-linkage, which is expected since the posterior 
$m$-probabilities are significantly smaller.  
Based on these results, we set $\lambda_{al} = 0.95$ as the default value 
for our other experiments since it seems to achieve balanced performance.

\begin{table}
  \small
  \centering
  \begin{tabular}{lcccc}
    \toprule
      &  
        & \multicolumn{3}{c}{Evaluation metric} \\
    \cmidrule{3-5}
    Data set 
      & Truncation point 
        & Precision 
        & Recall 
        & F1 score \\
    \midrule
    \multirow{4}{*}{\textsf{RLdata}}
      & 0.00 
        & 0.068 {\scriptsize(0.067,\,0.068)} 
        & 0.909 {\scriptsize(0.906,\,0.912)} 
        & 0.126 {\scriptsize(0.125,\,0.126)} \\
      & 0.50 
        & 0.069 {\scriptsize(0.069,\,0.069)} 
        & 0.921 {\scriptsize(0.919,\,0.924)} 
        & 0.128 {\scriptsize(0.128,\,0.129)} \\
      & 0.85 
        & 0.315 {\scriptsize(0.308,\,0.321)} 
        & 0.965 {\scriptsize(0.962,\,0.967)} 
        & 0.475 {\scriptsize(0.467,\,0.481)} \\
      & 0.95 
        & 0.534 {\scriptsize(0.524,\,0.546)} 
        & 0.964 {\scriptsize(0.962,\,0.966)} 
        & 0.687 {\scriptsize(0.679,\,0.697)} \\
    \midrule
    \multirow{4}{*}{\textsf{nltcs}}
      & 0.00 
        & 0.114 {\scriptsize(0.113,\,0.114)} 
        & 0.983 {\scriptsize(0.978,\,0.987)} 
        & 0.204 {\scriptsize(0.203,\,0.205)} \\
      & 0.50 
        & 0.111 {\scriptsize(0.110,\,0.111)} 
        & 0.964 {\scriptsize(0.958,\,0.970)} 
        & 0.199 {\scriptsize(0.197,\,0.200)} \\
      & 0.85 
        & 0.162 {\scriptsize(0.160,\,0.164)} 
        & 0.972 {\scriptsize(0.968,\,0.976)} 
        & 0.278 {\scriptsize(0.275,\,0.281)} \\
      & 0.95 
        & 0.312 {\scriptsize(0.304,\,0.319)} 
        & 0.975 {\scriptsize(0.969,\,0.979)} 
        & 0.473 {\scriptsize(0.464,\,0.480)} \\
    \midrule
    \multirow{4}{*}{\textsf{cora}}
      & 0.00 
        & 0.980 {\scriptsize(0.979,\,0.981)} 
        & 0.378 {\scriptsize(0.377,\,0.378)} 
        & 0.545 {\scriptsize(0.544,\,0.546)} \\
      & 0.50 
        & 0.981 {\scriptsize(0.979,\,0.984)} 
        & 0.390 {\scriptsize(0.389,\,0.391)} 
        & 0.558 {\scriptsize(0.557,\,0.560)} \\
      & 0.85 
        & 0.984 {\scriptsize(0.983,\,0.984)} 
        & 0.383 {\scriptsize(0.382,\,0.384)} 
        & 0.552 {\scriptsize(0.551,\,0.553)} \\
      & 0.95 
        & 0.982 {\scriptsize(0.981,\,0.983)} 
        & 0.359 {\scriptsize(0.357,\,0.362)} 
        & 0.526 {\scriptsize(0.524,\,0.529)} \\
    \midrule
    \multirow{4}{*}{\textsf{rest}}
      & 0.00 
        & 1.000 {\scriptsize(0.985,\,1.000)} 
        & 0.607 {\scriptsize(0.598,\,0.607)} 
        & 0.756 {\scriptsize(0.744,\,0.756)} \\
      & 0.50 
        & 0.985 {\scriptsize(0.985,\,1.000)} 
        & 0.598 {\scriptsize(0.598,\,0.607)} 
        & 0.744 {\scriptsize(0.744,\,0.756)} \\
      & 0.85 
        & 0.985 {\scriptsize(0.985,\,1.000)} 
        & 0.598 {\scriptsize(0.598,\,0.607)} 
        & 0.744 {\scriptsize(0.744,\,0.756)} \\
      & 0.95 
        & 0.993 {\scriptsize(0.985,\,1.000)} 
        & 0.603 {\scriptsize(0.598,\,0.607)} 
        & 0.750 {\scriptsize(0.744,\,0.756)} \\
    \bottomrule
  \end{tabular}
  \caption{Posterior performance of the \textsf{Sadinle} model as a function 
    of the lower truncation point $\lambda_{al}$ on the $m$-probabilities. 
    A point estimate for each evaluation metric is reported based on the 
    median, along with a 95\% equi-tailed credible interval.}
  \label{app-tbl:sadinle-lower-trunc}
\end{table}

\newpage
\section{MCMC Diagnostics}
\label{app-sec:mcmc-diag}

\subsection{Study of Linkage Structure Priors}
Here we present convergence diagnostics for the models fitted in 
Section~\ref{sec:study}. 
We present Geweke diagnostic plots and trace plots for a selection of model 
variables for each data set, linkage structure prior and distortion model. 
Each pair of plots is preceded by a title of the form ``Data set | Linkage 
structure prior | Distortion model''.
The Geweke diagnostic plot (on the left) depicts a Z-score on the x-axis for 
each variable on the y-axis. 
The Z-score tests for equality of the means of the first 10\% and final 50\% of 
the Markov chain, and is typically expected to be in the range $[-2,2]$ 
\citep{geweke_evaluating_1992}.
The trace plot (on the right) depicts the value of variables (labeled in the 
right panel) for each step in the chain (on the x-axis). 
Note that variable $E$ denotes the number of instantiated entities. 
We replace integer indices for the attributes by named indices. 
For instance, $\theta_{0, \mathrm{city}}$ refers to the distortion probability 
in source $0$ of the attribute called ``city''.

\begin{fig}{\textsf{nltcs} | \textsf{PY} | \textsf{Ours} }
\includegraphics[width=0.48\linewidth]{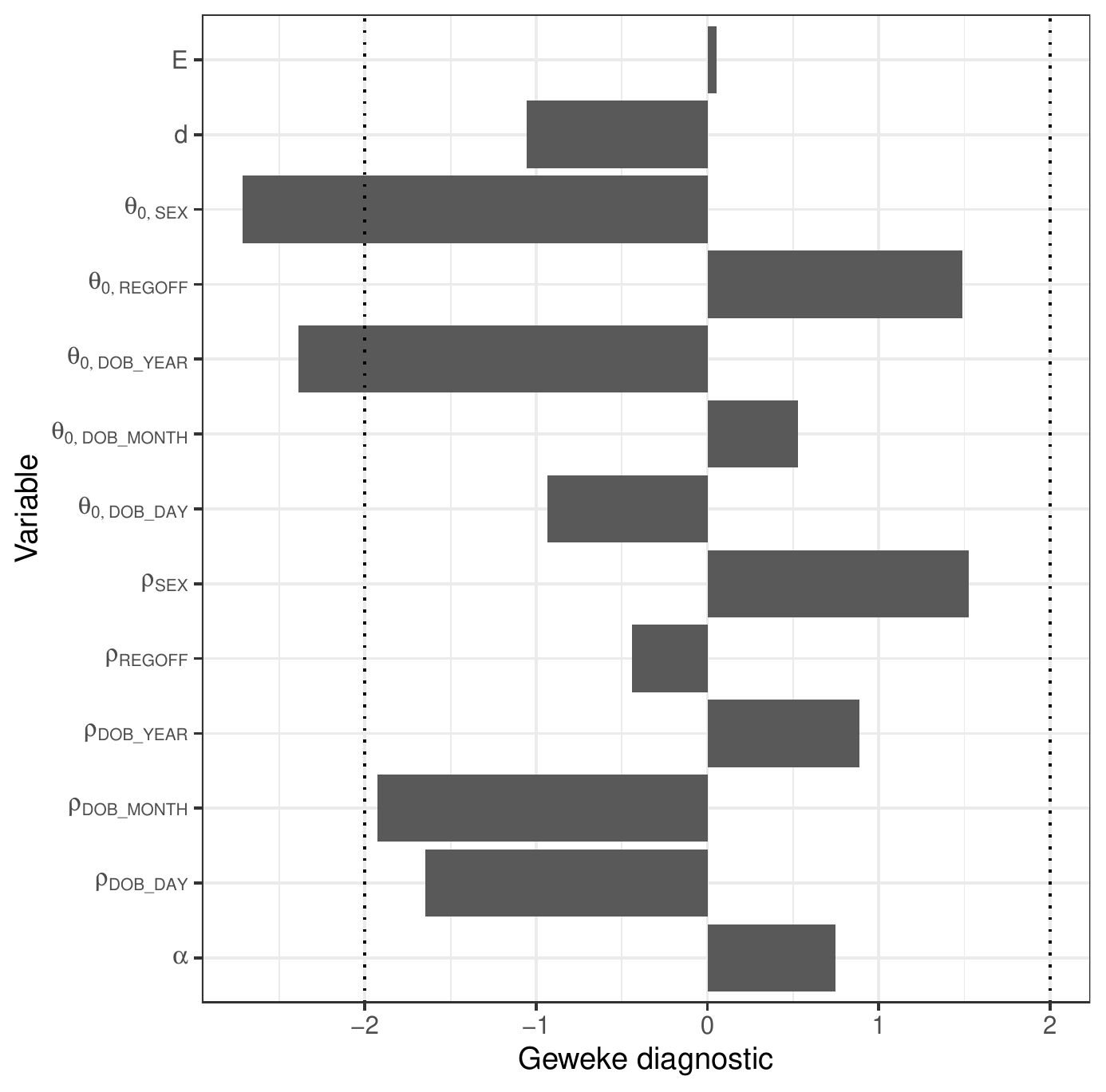} \hfill
\includegraphics[width=0.48\linewidth]{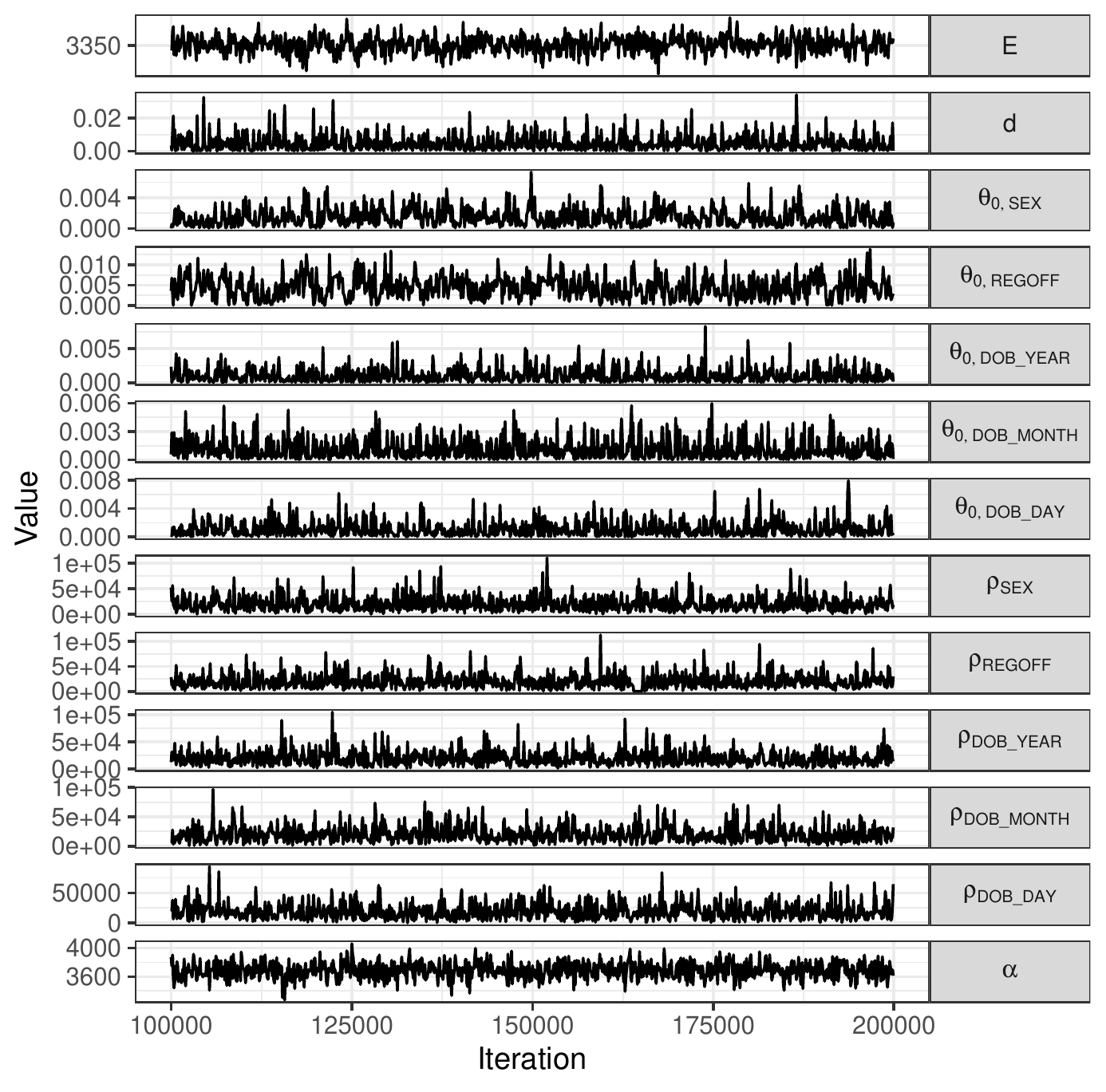} 
\end{fig}

\begin{fig}{\textsf{nltcs} | \textsf{Ewens} | \textsf{Ours}}
\includegraphics[width=0.48\linewidth]{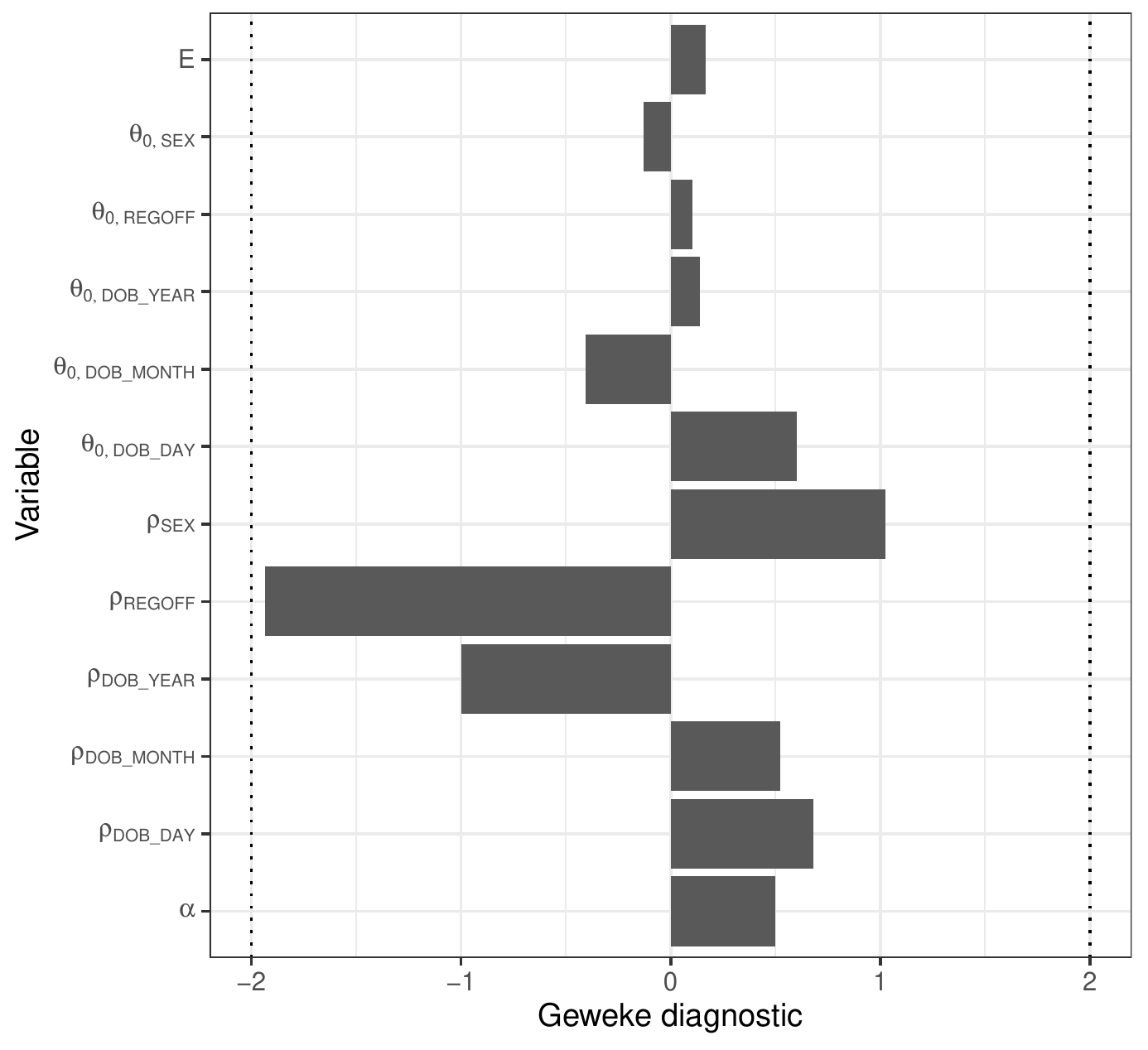} \hfill
\includegraphics[width=0.48\linewidth]{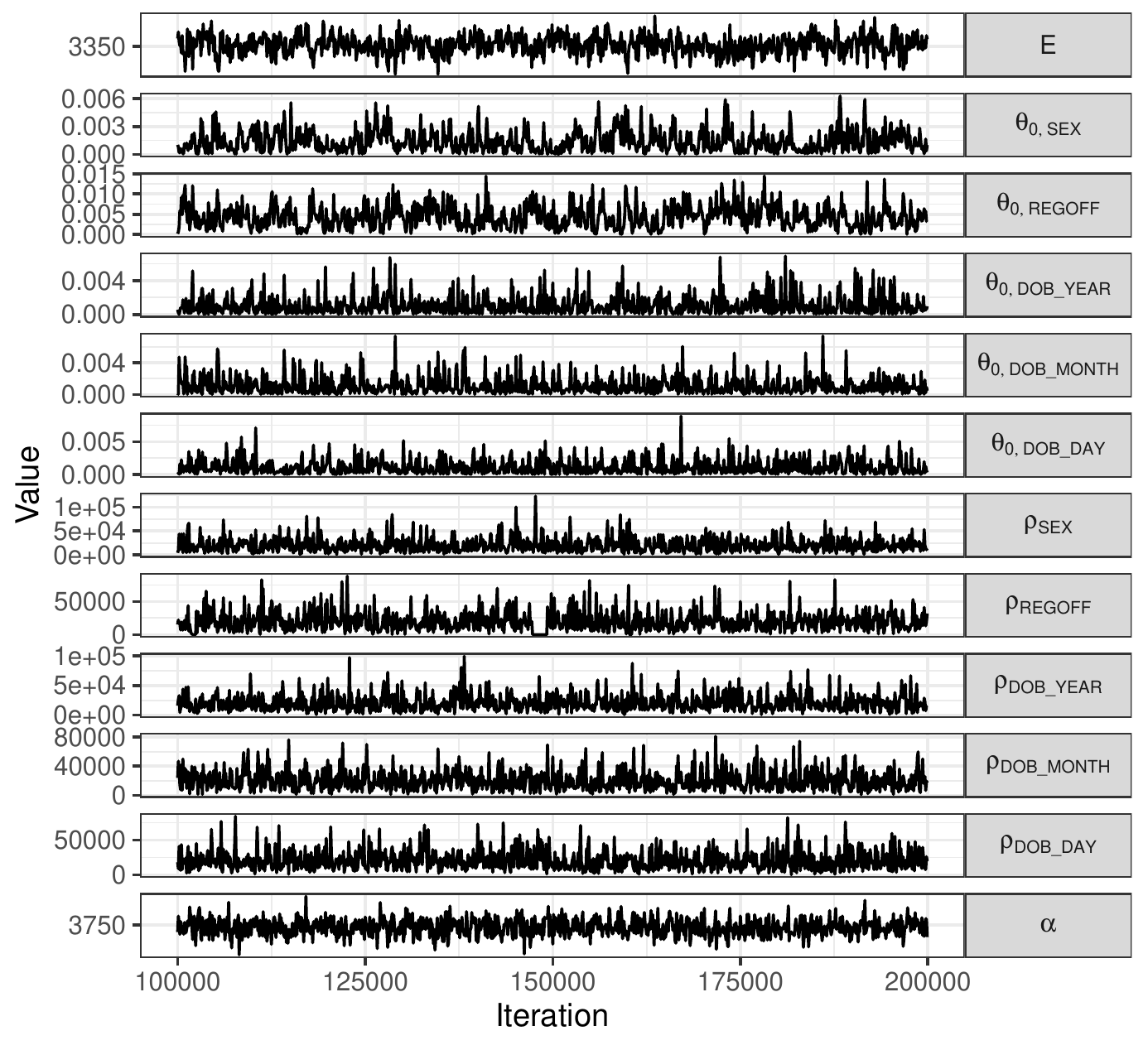} 
\end{fig}

\begin{fig}{\textsf{nltcs} | \textsf{GenCoupon} | \textsf{Ours}}
\includegraphics[width=0.48\linewidth]{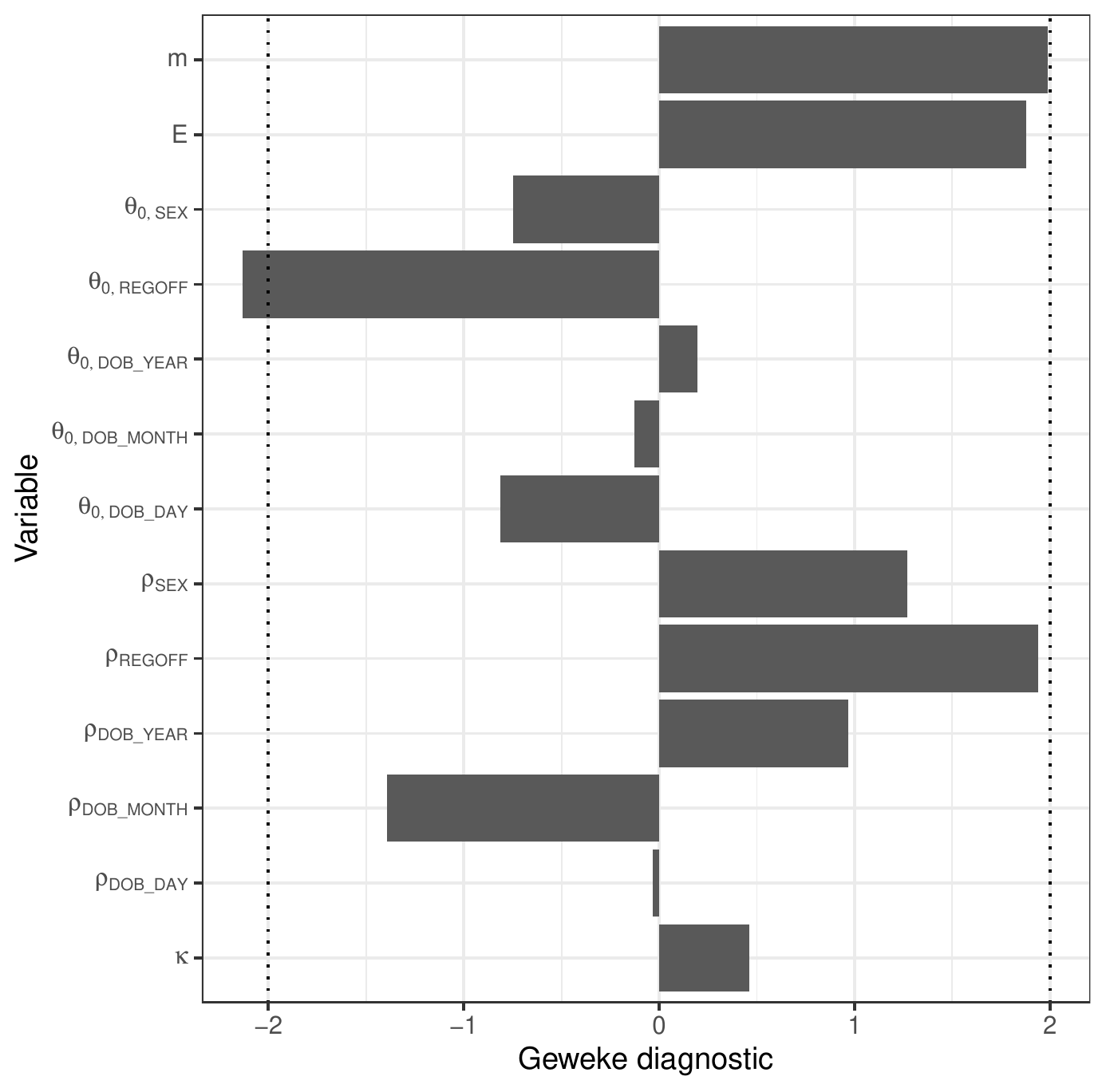} \hfill
\includegraphics[width=0.48\linewidth]{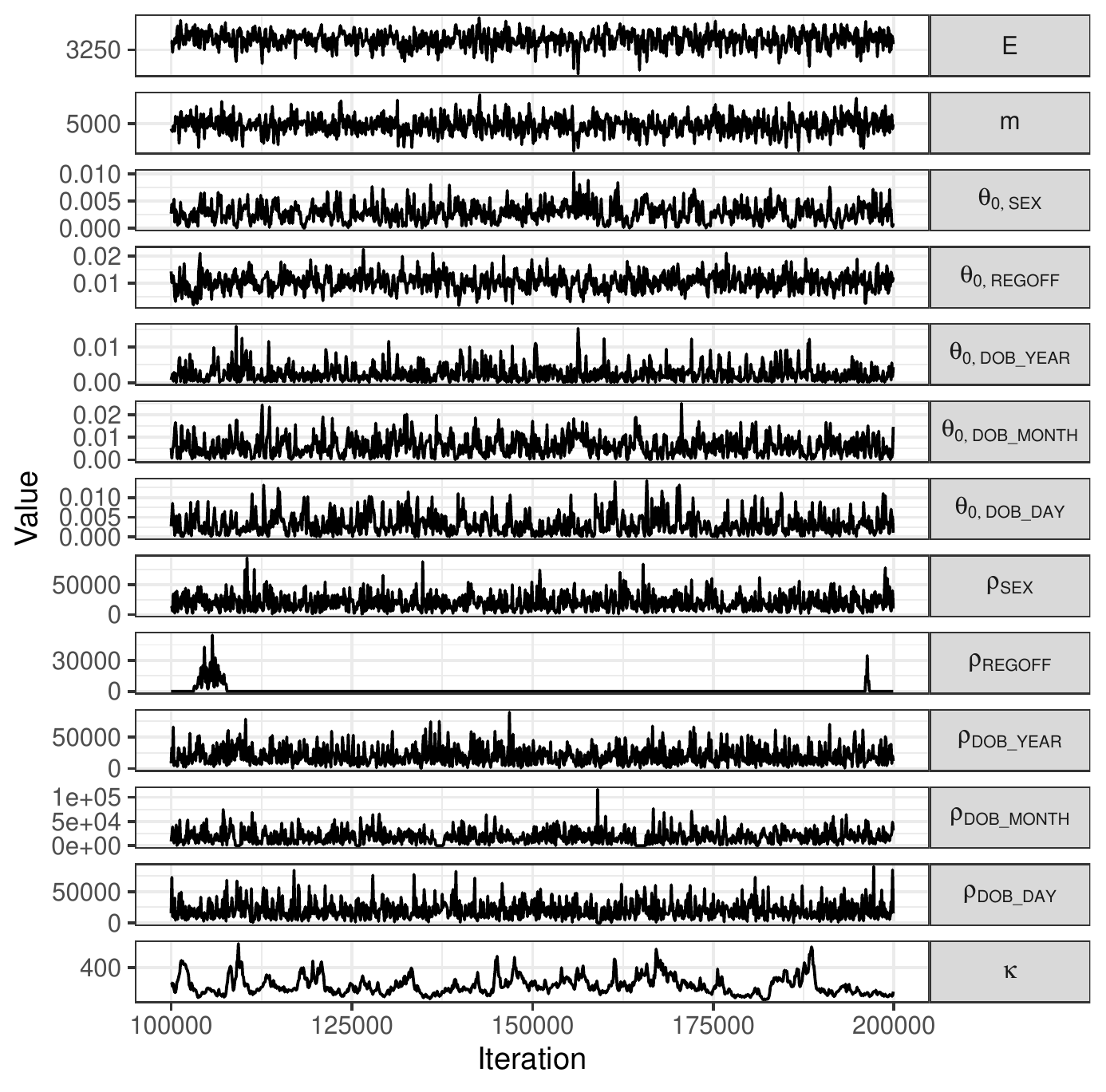} 
\end{fig}

\begin{fig}{\textsf{nltcs} | \textsf{Coupon} | \textsf{Ours}}
\includegraphics[width=0.48\linewidth]{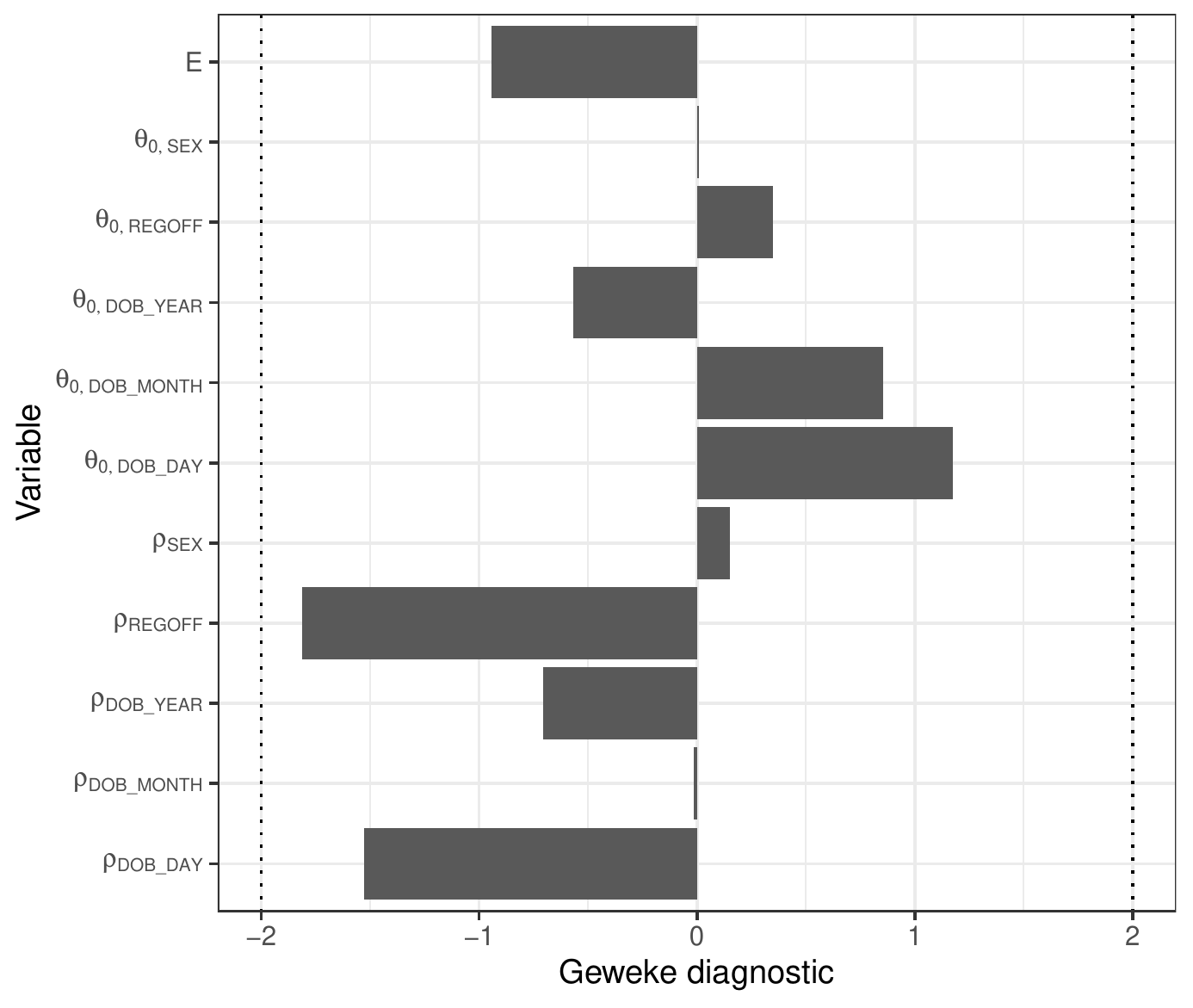} \hfill
\includegraphics[width=0.48\linewidth]{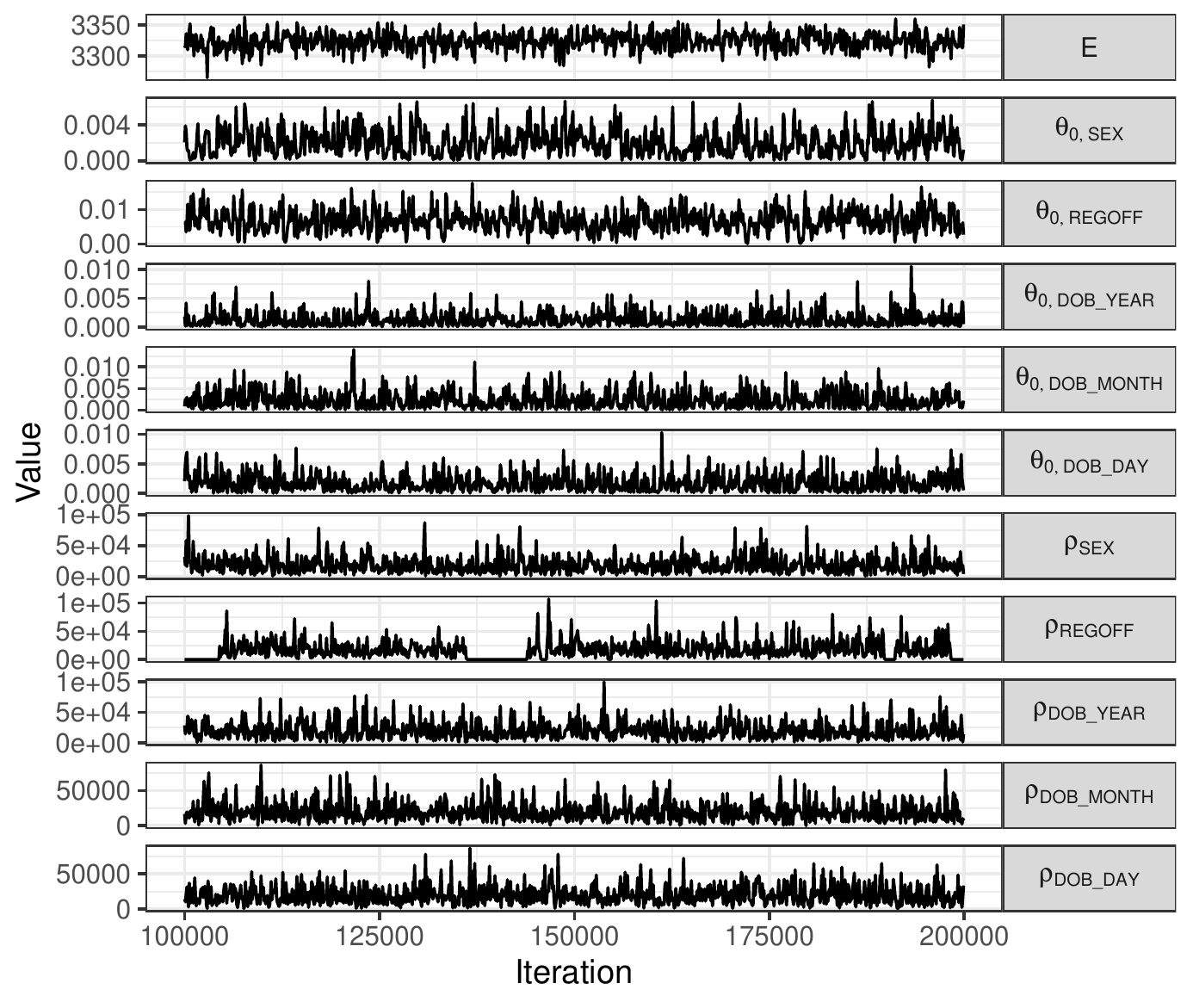} 
\end{fig}

\begin{fig}{\textsf{nltcs} | \textsf{PY} | \textsf{blink}}
\includegraphics[width=0.48\linewidth]{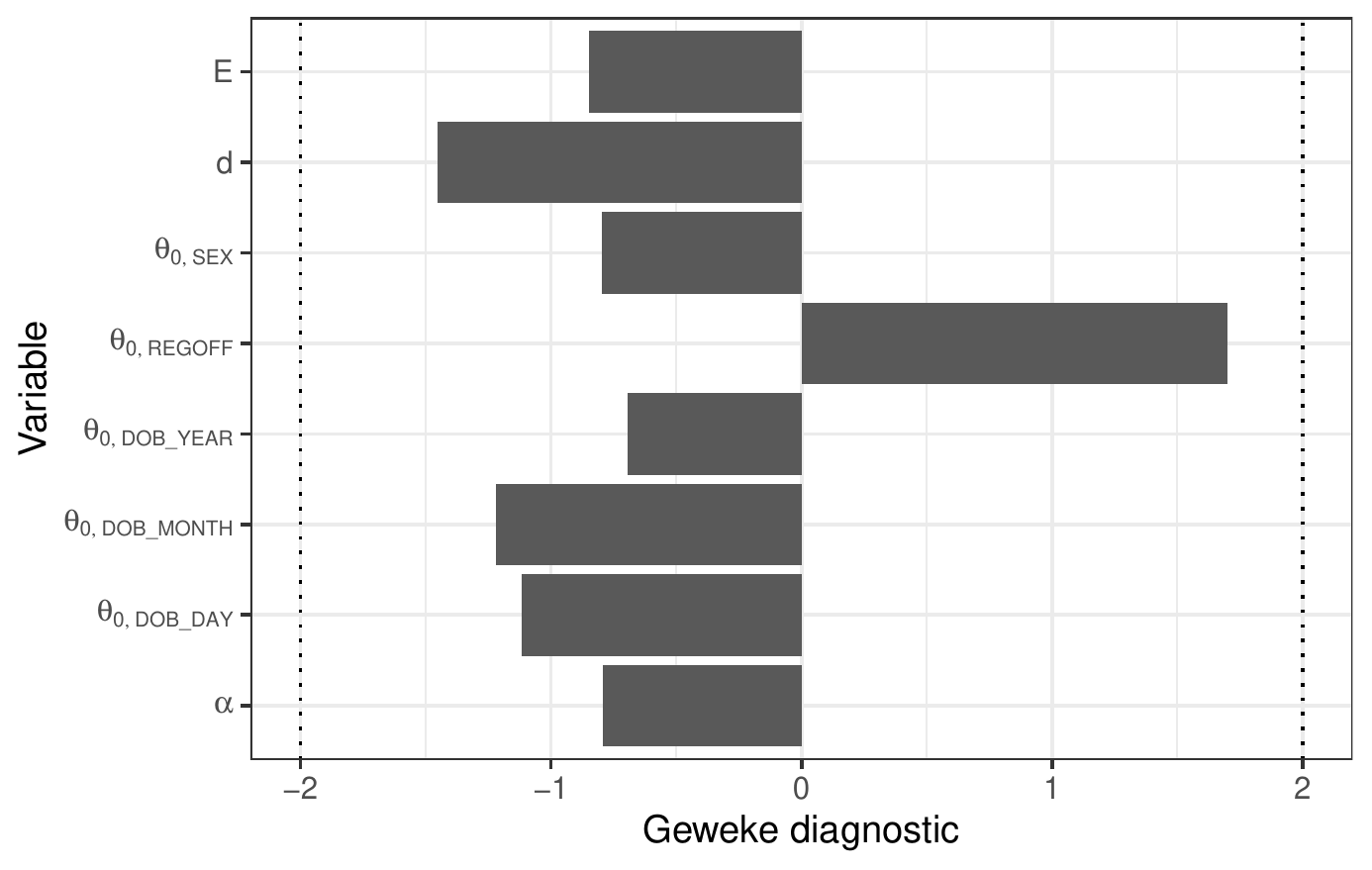} \hfill
\includegraphics[width=0.48\linewidth]{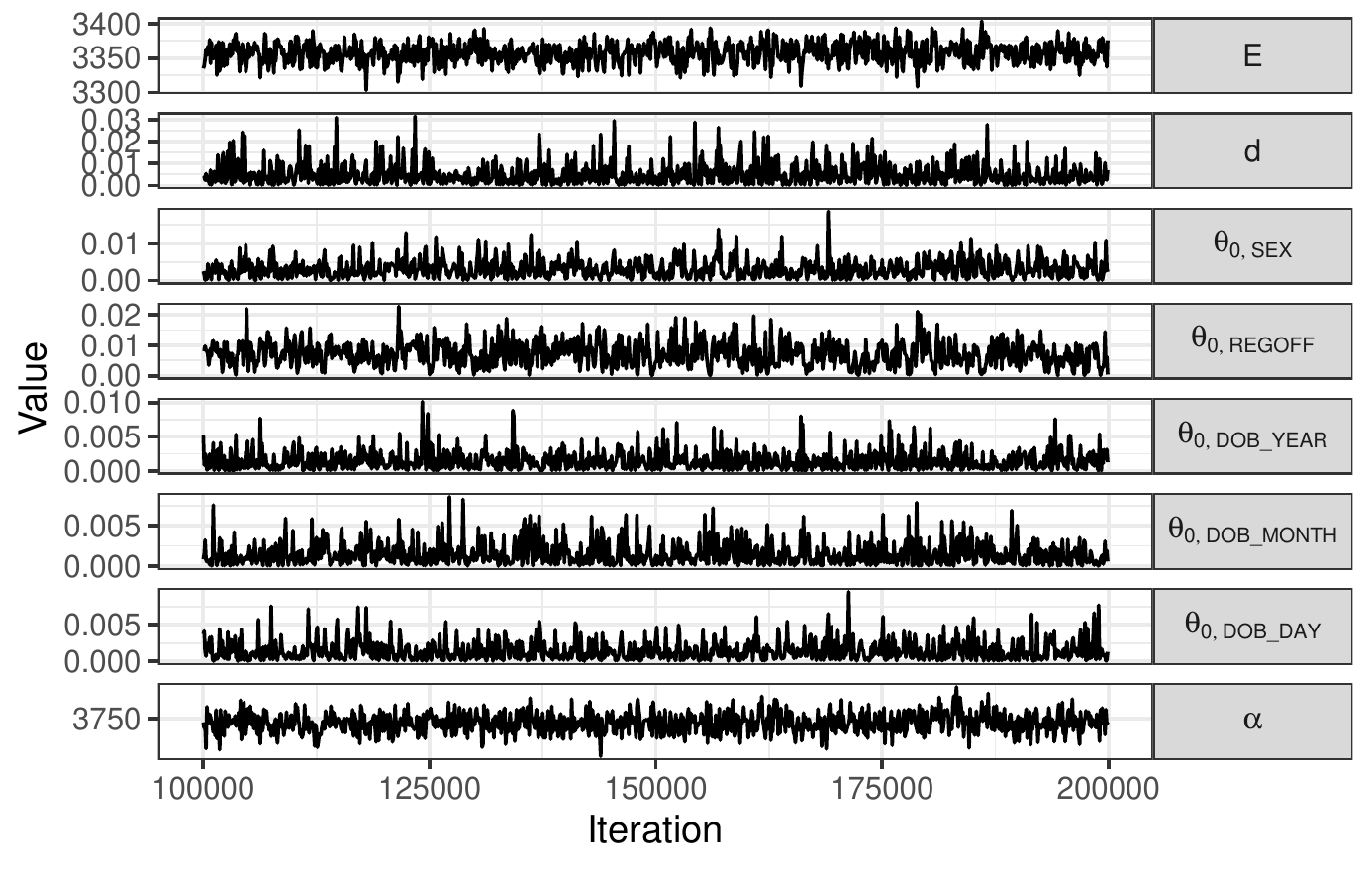} 
\end{fig}

\begin{fig}{\textsf{nltcs} | \textsf{Ewens} | \textsf{blink}}
\includegraphics[width=0.48\linewidth]{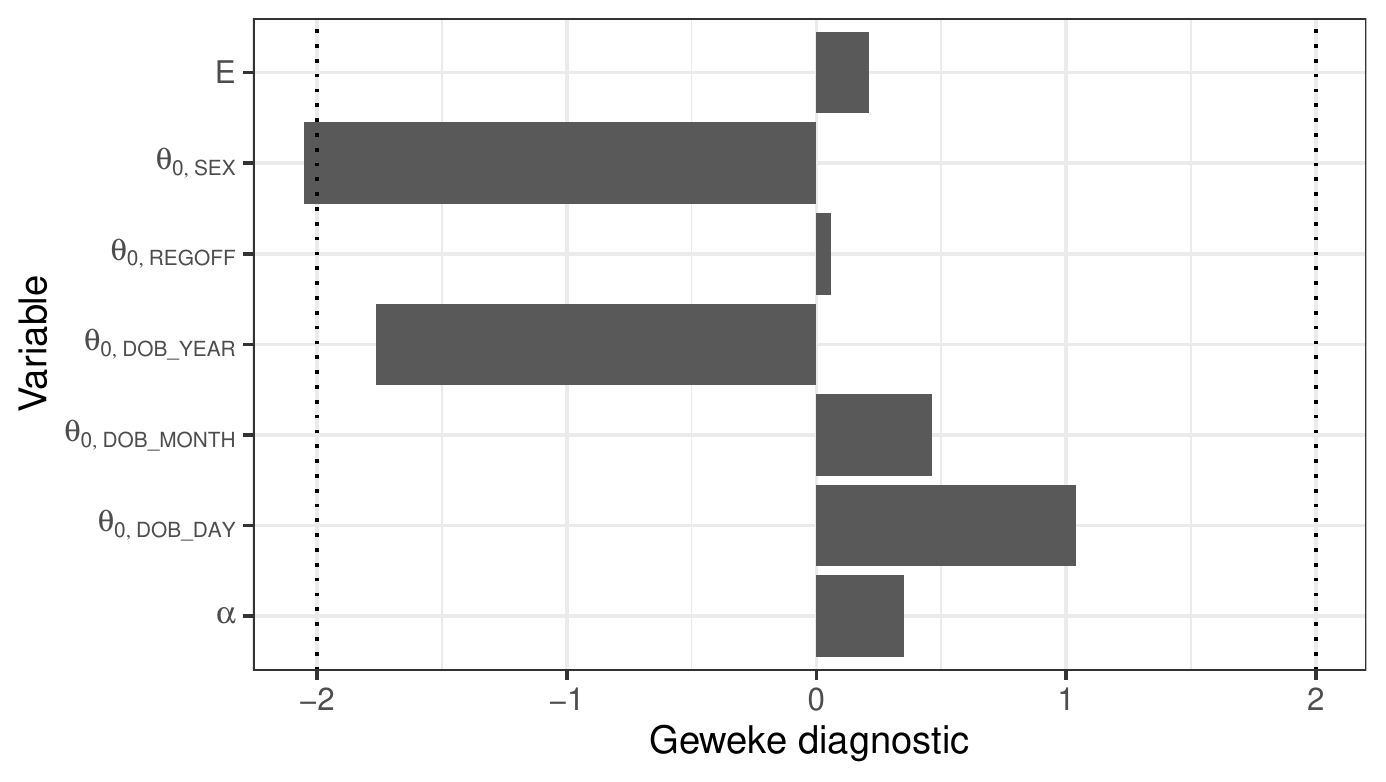} \hfill
\includegraphics[width=0.48\linewidth]{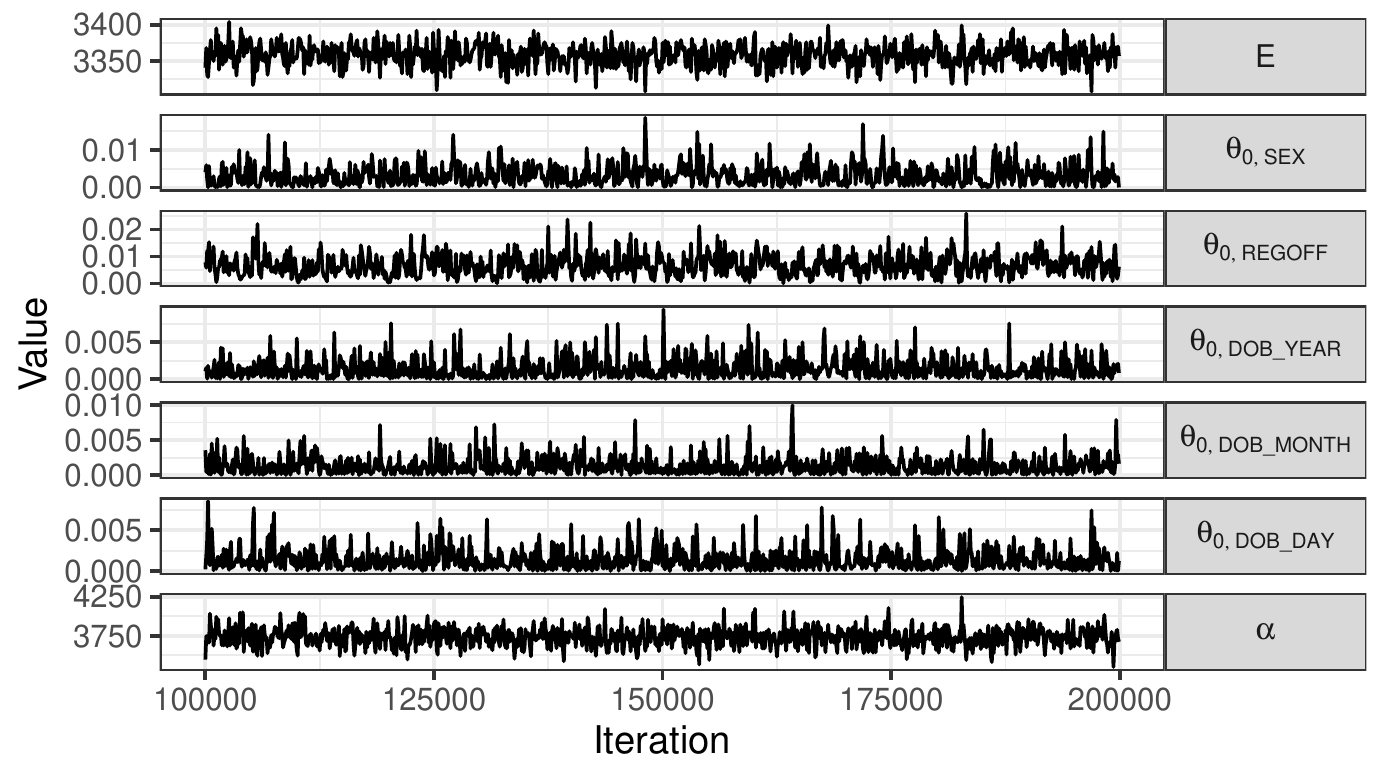} 
\end{fig}

\begin{fig}{\textsf{nltcs} | \textsf{GenCoupon} | \textsf{blink}}
\includegraphics[width=0.48\linewidth]{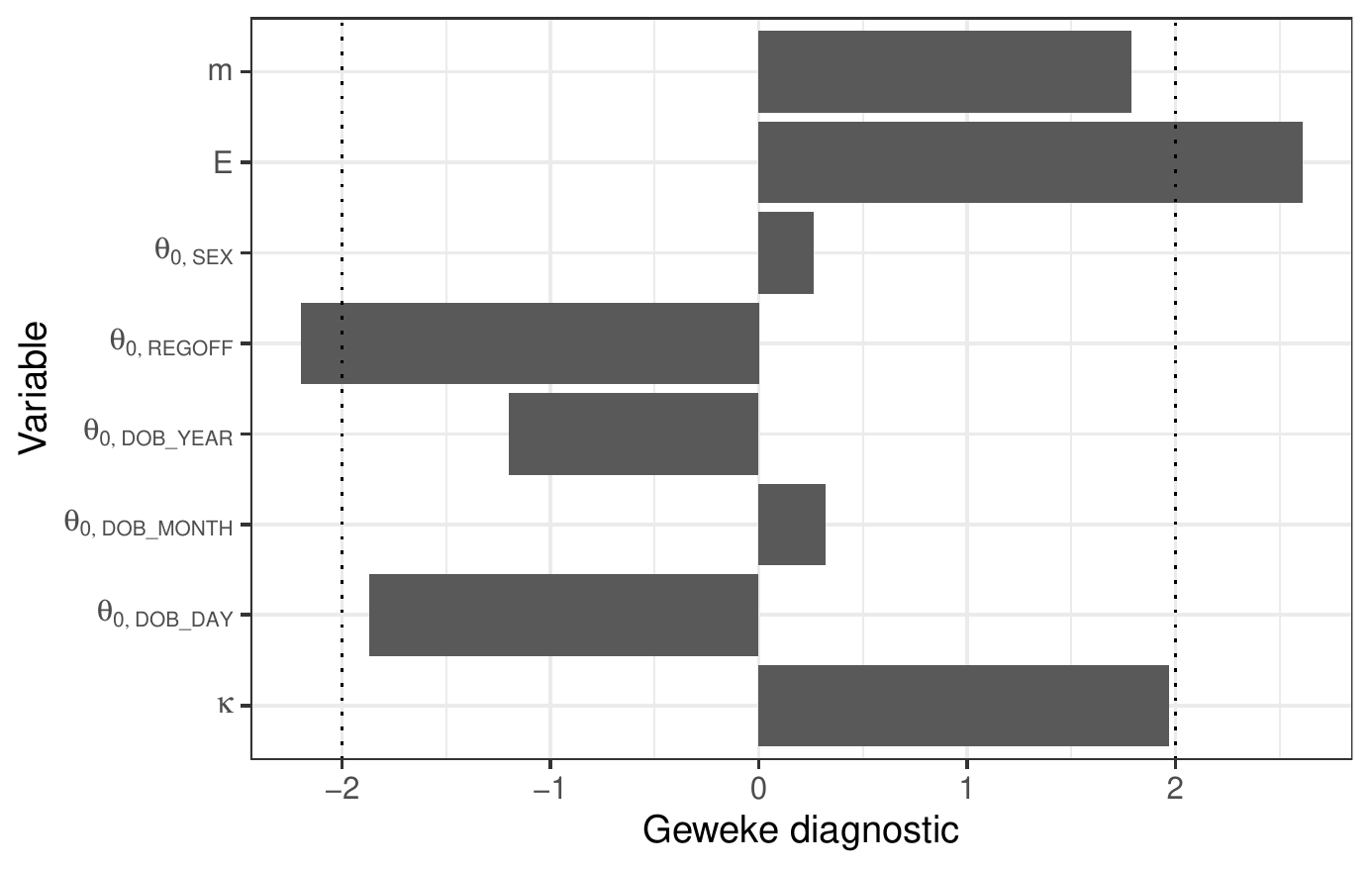} \hfill
\includegraphics[width=0.48\linewidth]{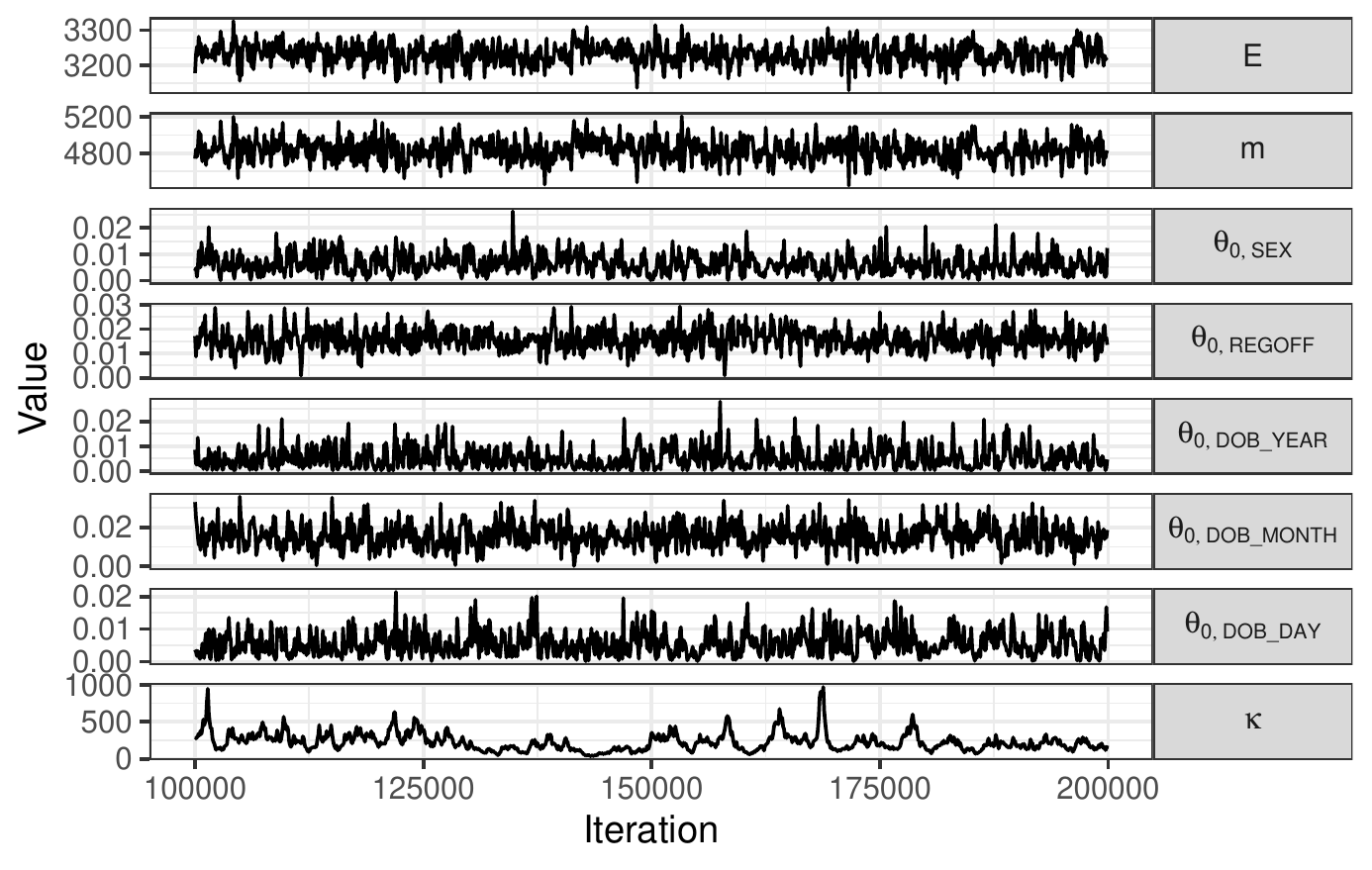} 
\end{fig}

\begin{fig}{\textsf{nltcs} | \textsf{Coupon} | \textsf{blink}}
\includegraphics[width=0.48\linewidth]{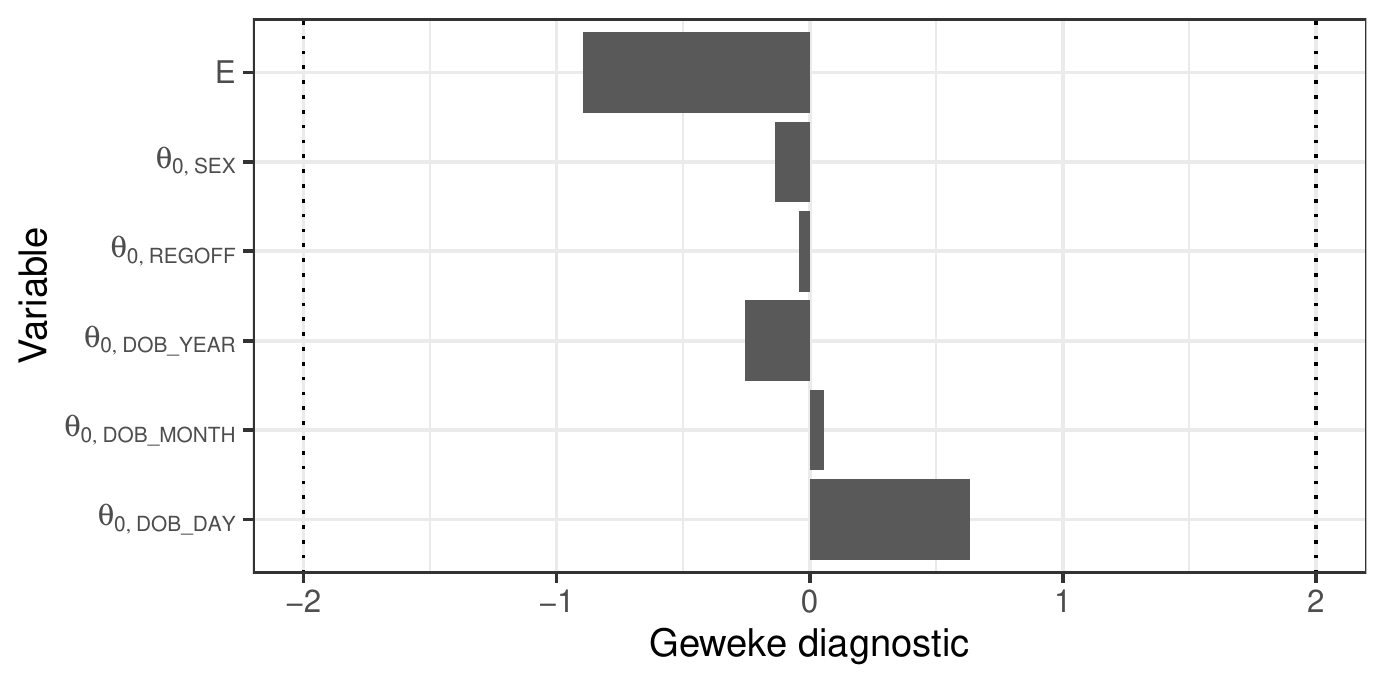} \hfill
\includegraphics[width=0.48\linewidth]{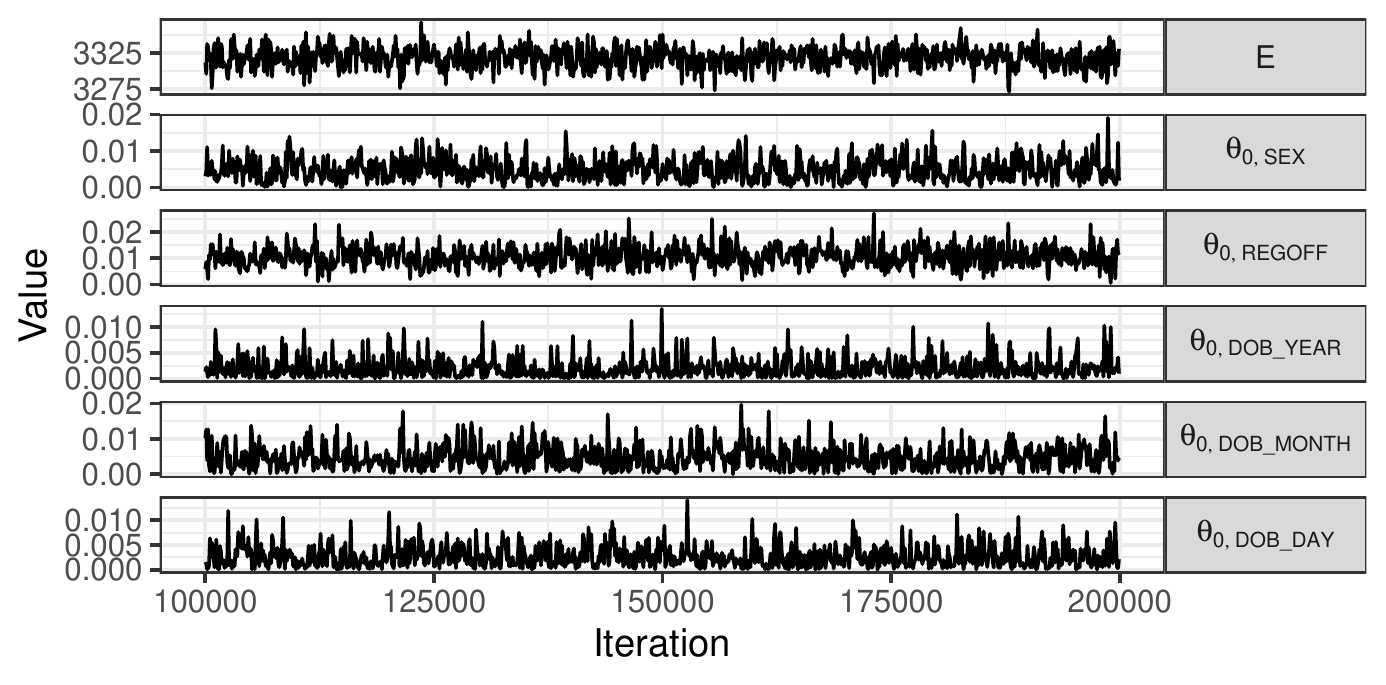} 
\end{fig}

\begin{fig}{\textsf{RLdata} | \textsf{PY} | \textsf{Ours}}
\includegraphics[width=0.48\linewidth]{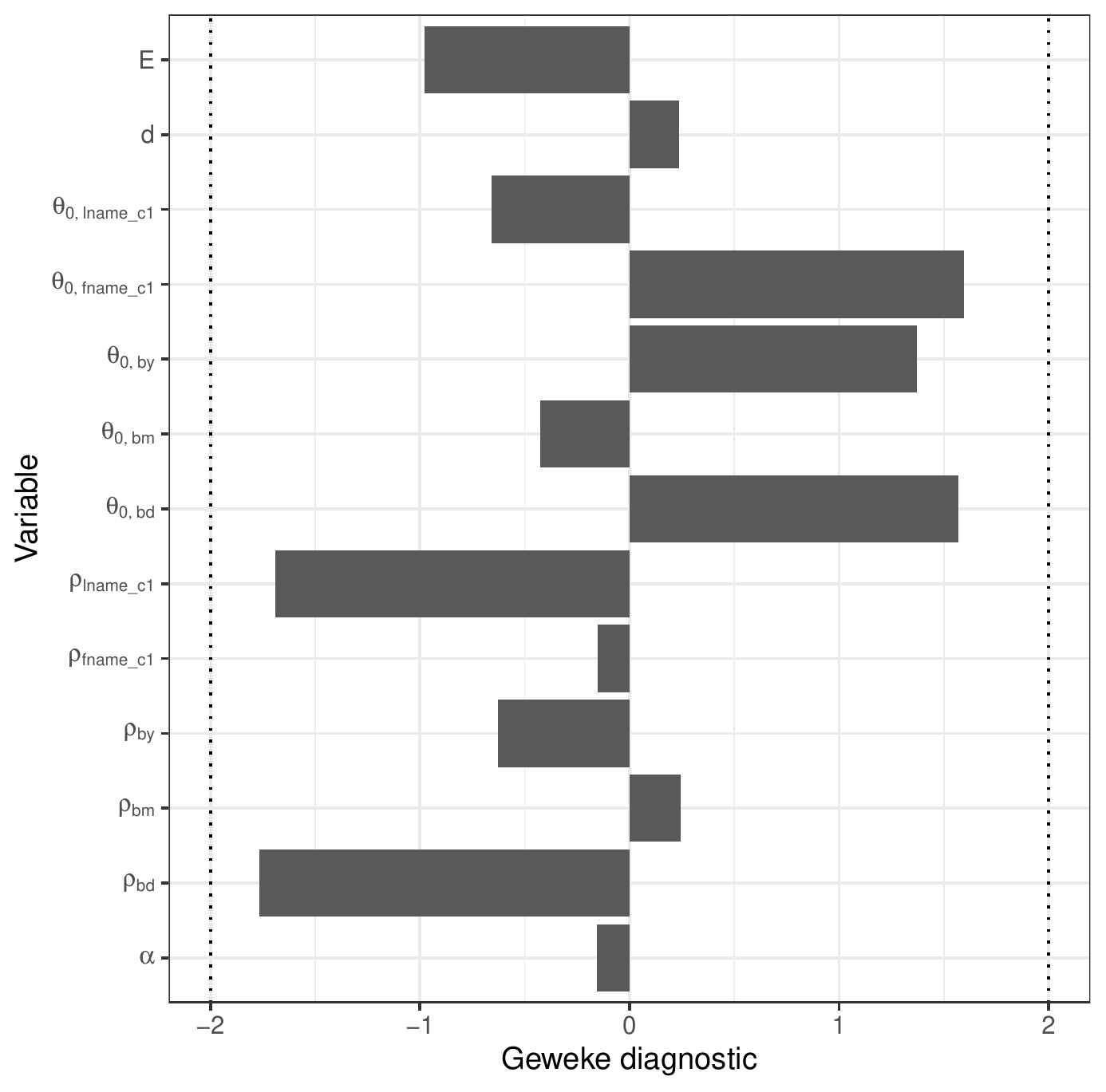} \hfill
\includegraphics[width=0.48\linewidth]{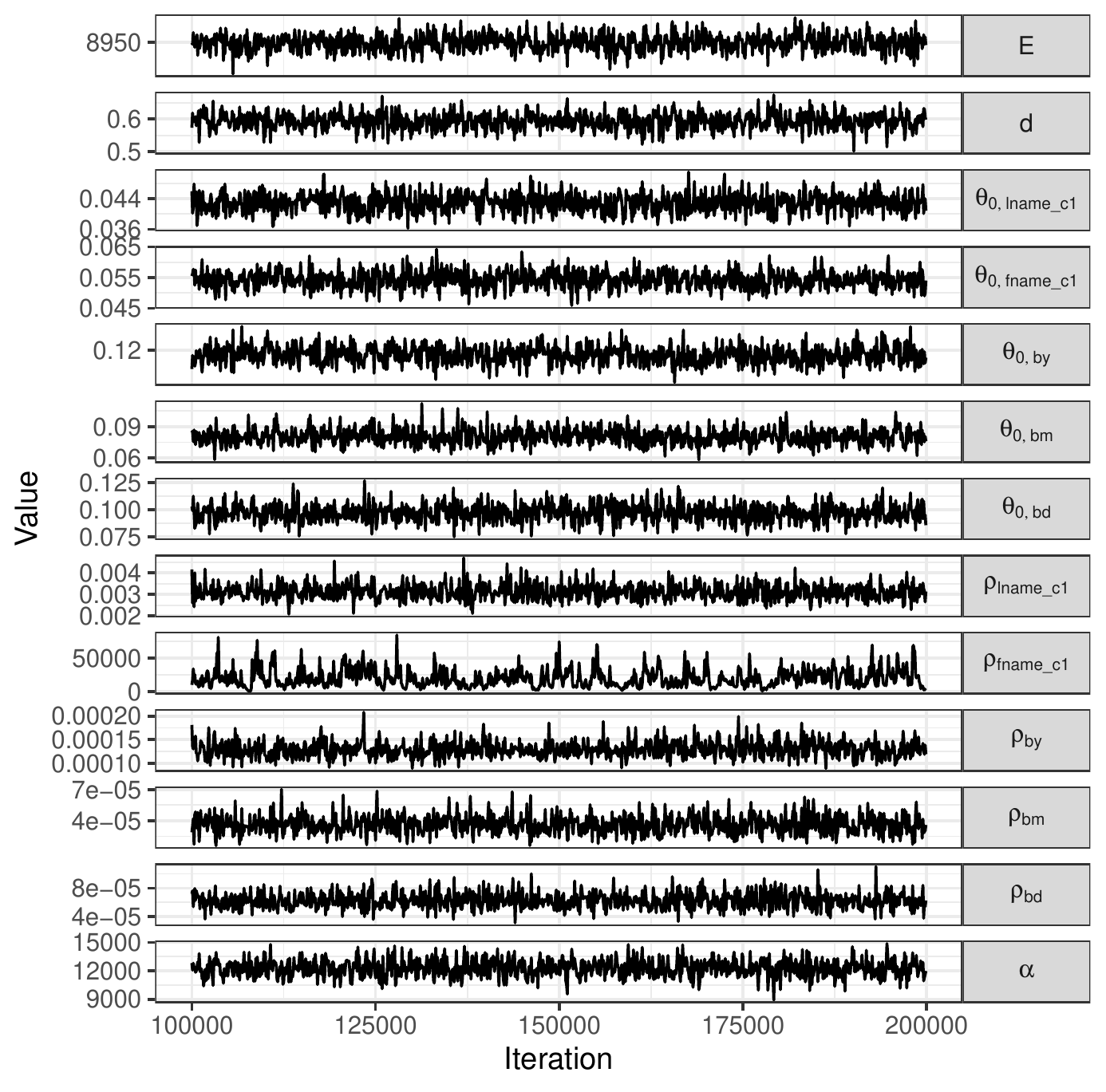} 
\end{fig}

\begin{fig}{\textsf{RLdata} | \textsf{Ewens} | \textsf{Ours}}
\includegraphics[width=0.48\linewidth]{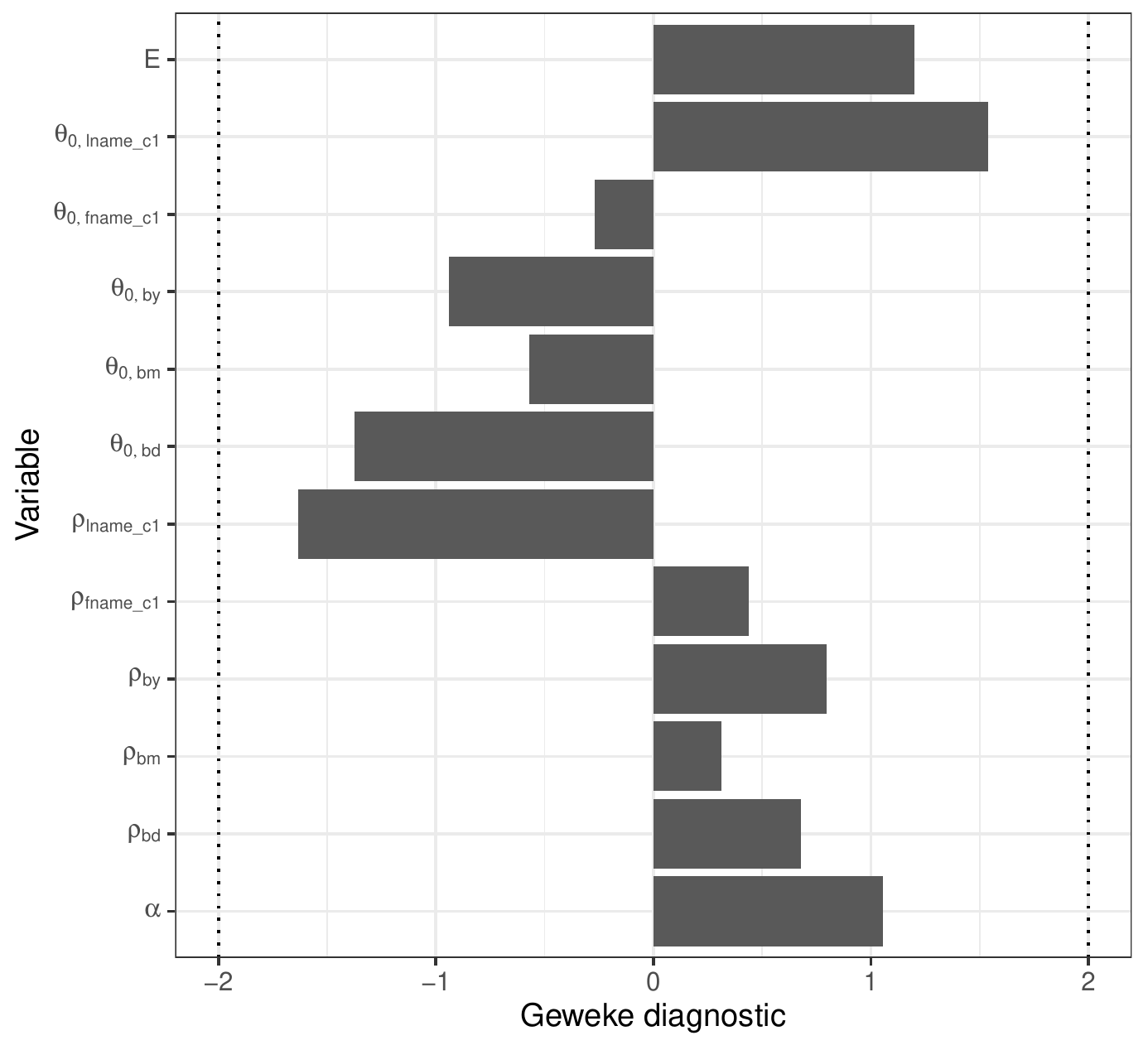} \hfill
\includegraphics[width=0.48\linewidth]{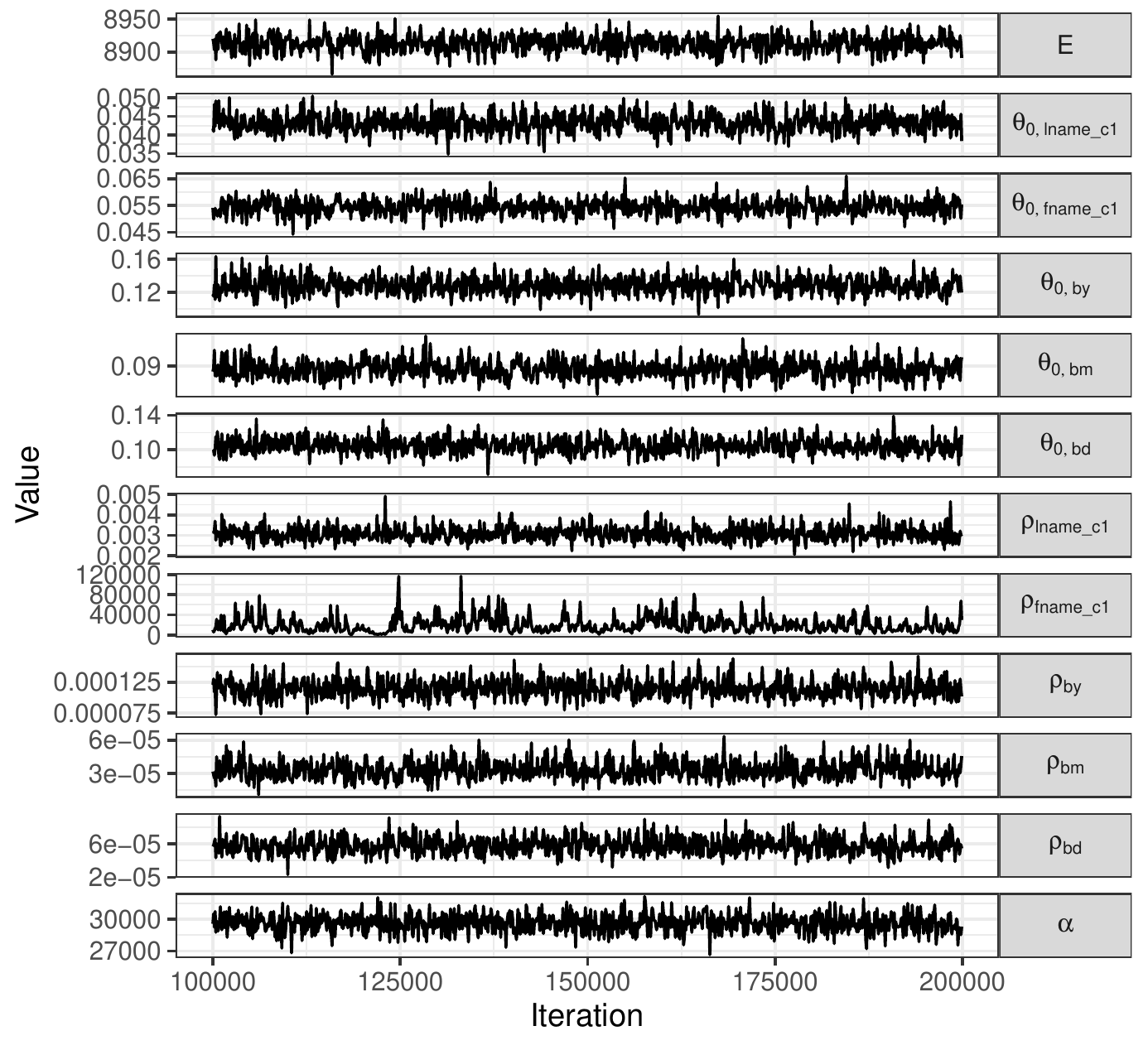} 
\end{fig}

\begin{fig}{\textsf{RLdata} | \textsf{GenCoupon} | \textsf{Ours}}
\includegraphics[width=0.48\linewidth]{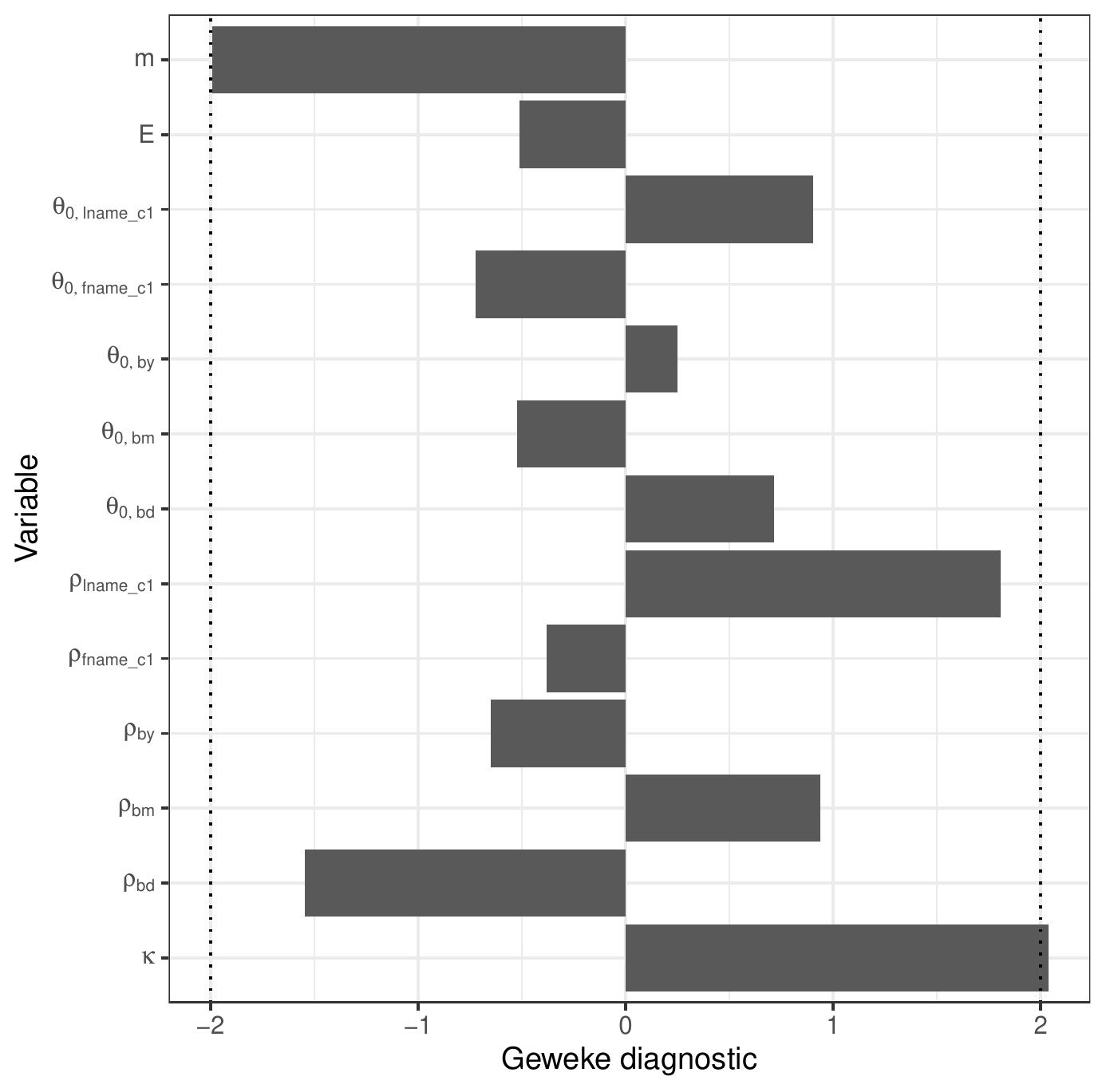} \hfill
\includegraphics[width=0.48\linewidth]{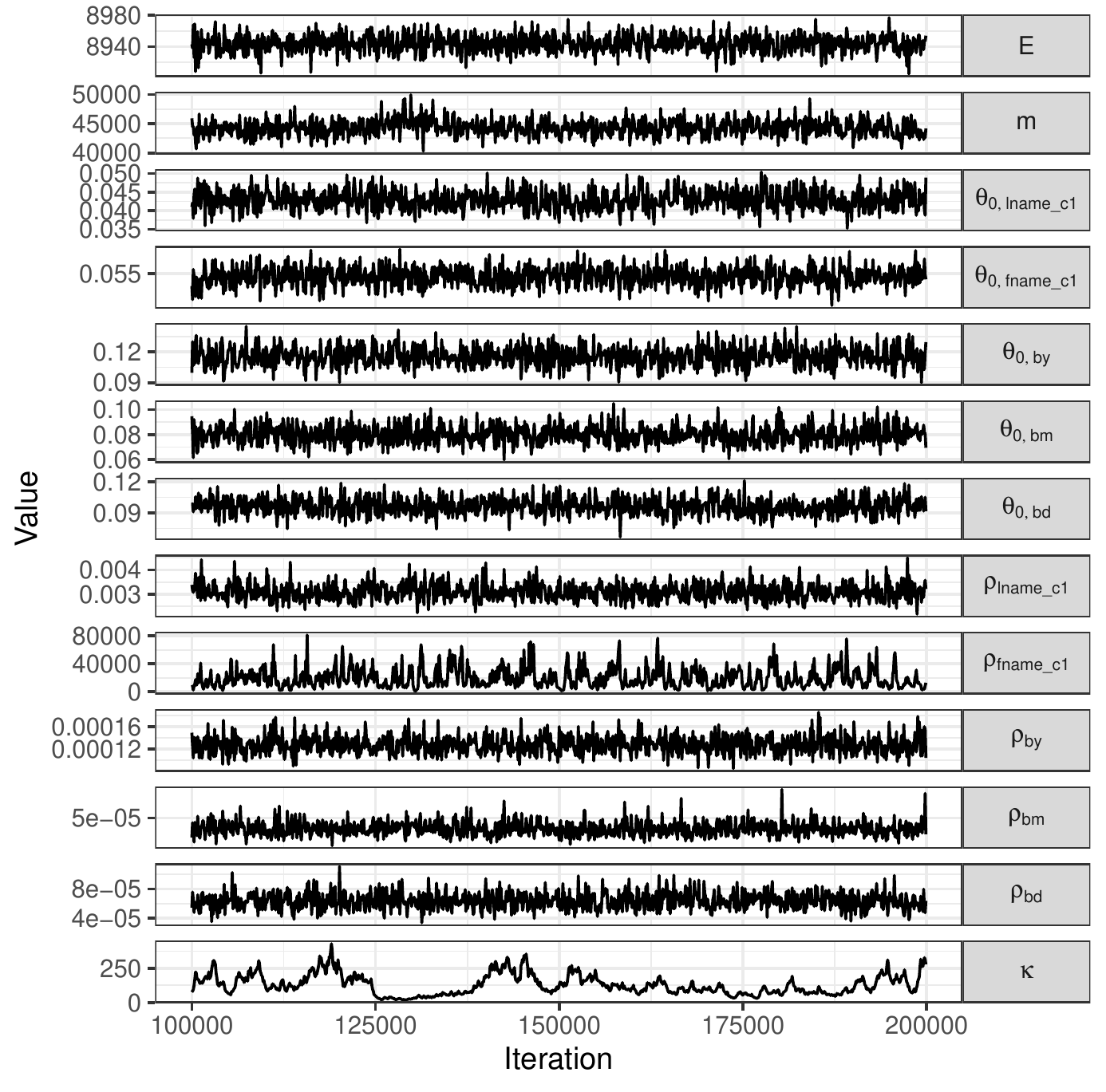} 
\end{fig}

\begin{fig}{\textsf{RLdata} | \textsf{Coupon} | \textsf{Ours}}
\includegraphics[width=0.48\linewidth]{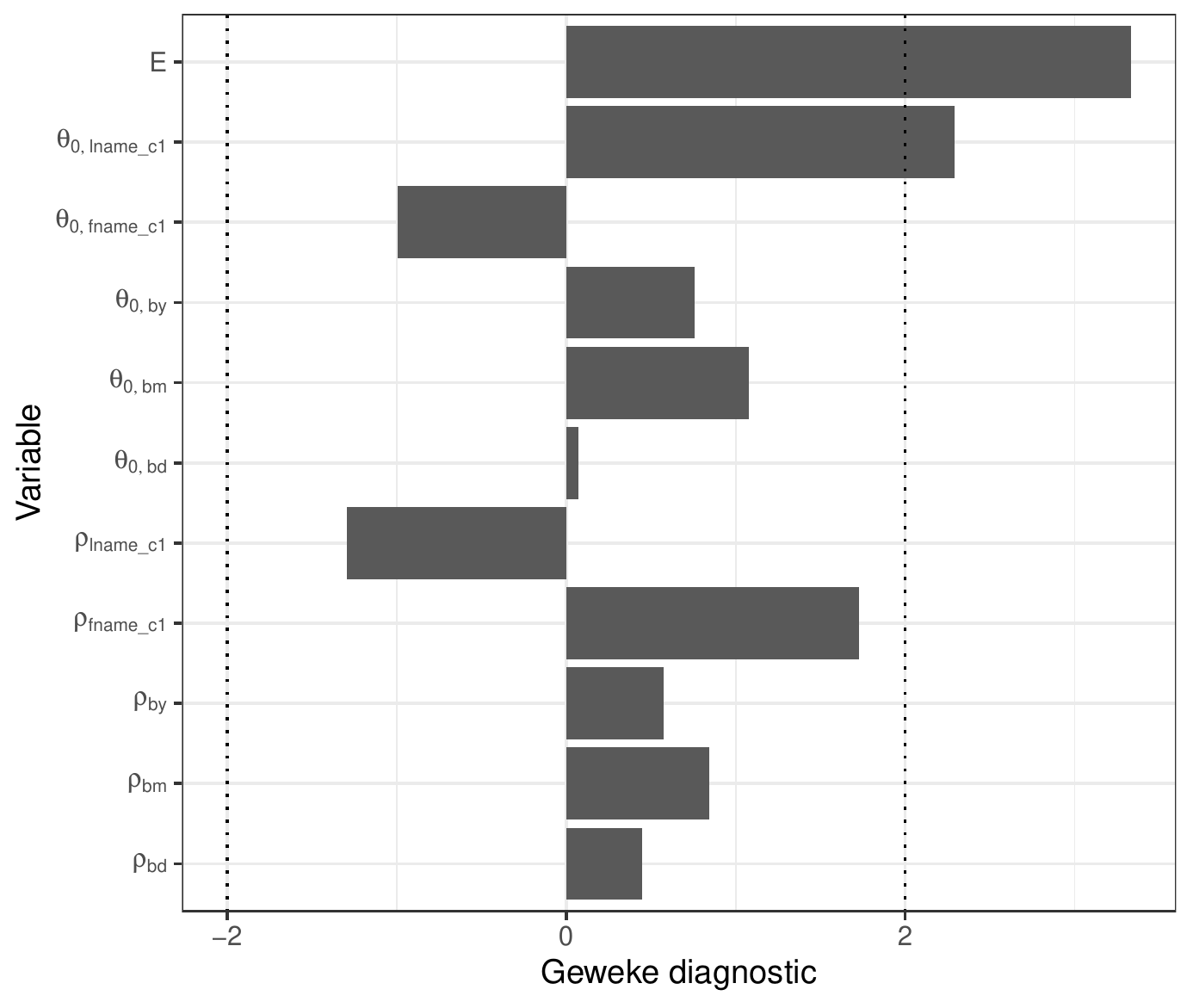} \hfill
\includegraphics[width=0.48\linewidth]{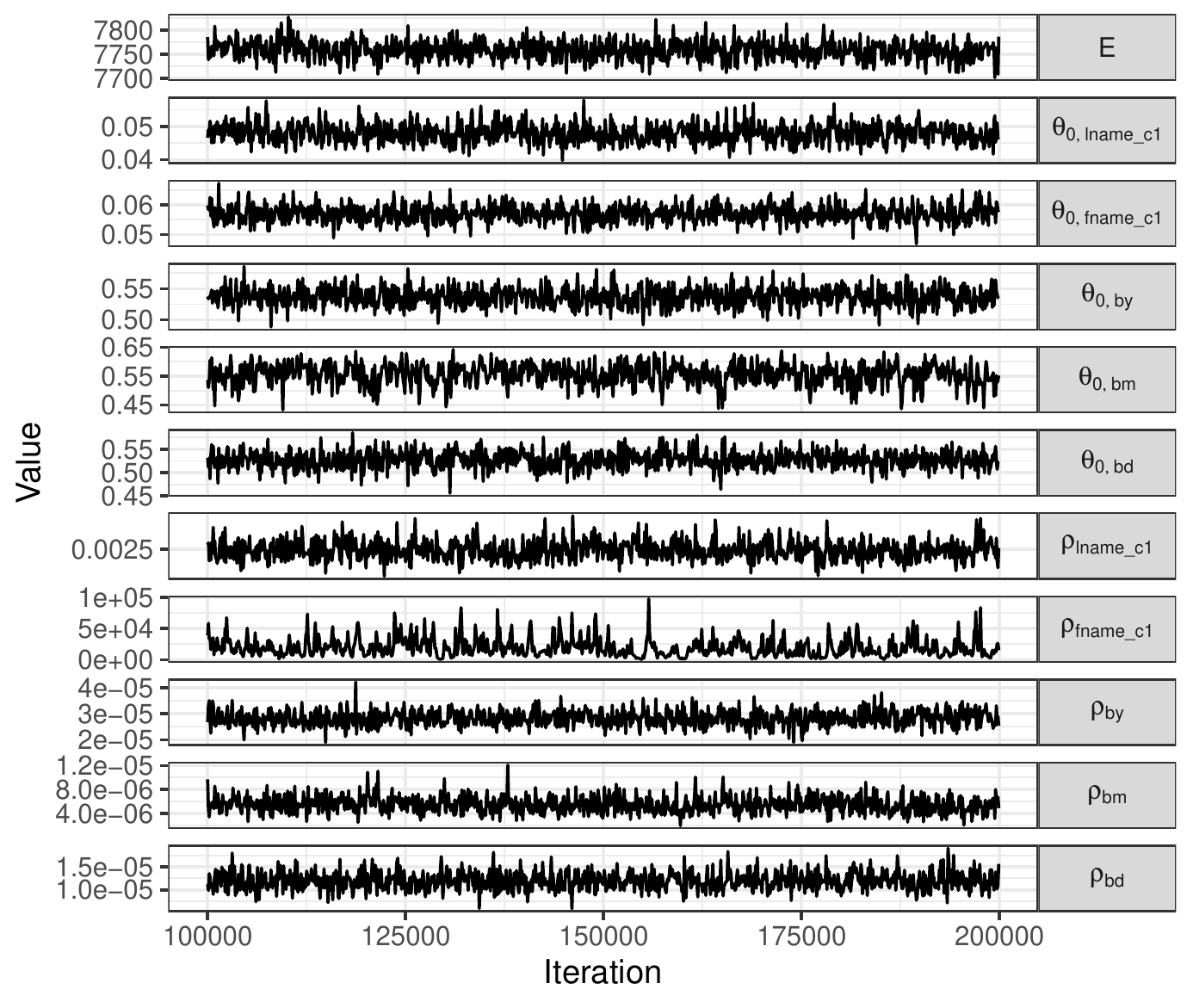} 
\end{fig}

\begin{fig}{\textsf{RLdata} | \textsf{PY} | \textsf{blink}}
\includegraphics[width=0.48\linewidth]{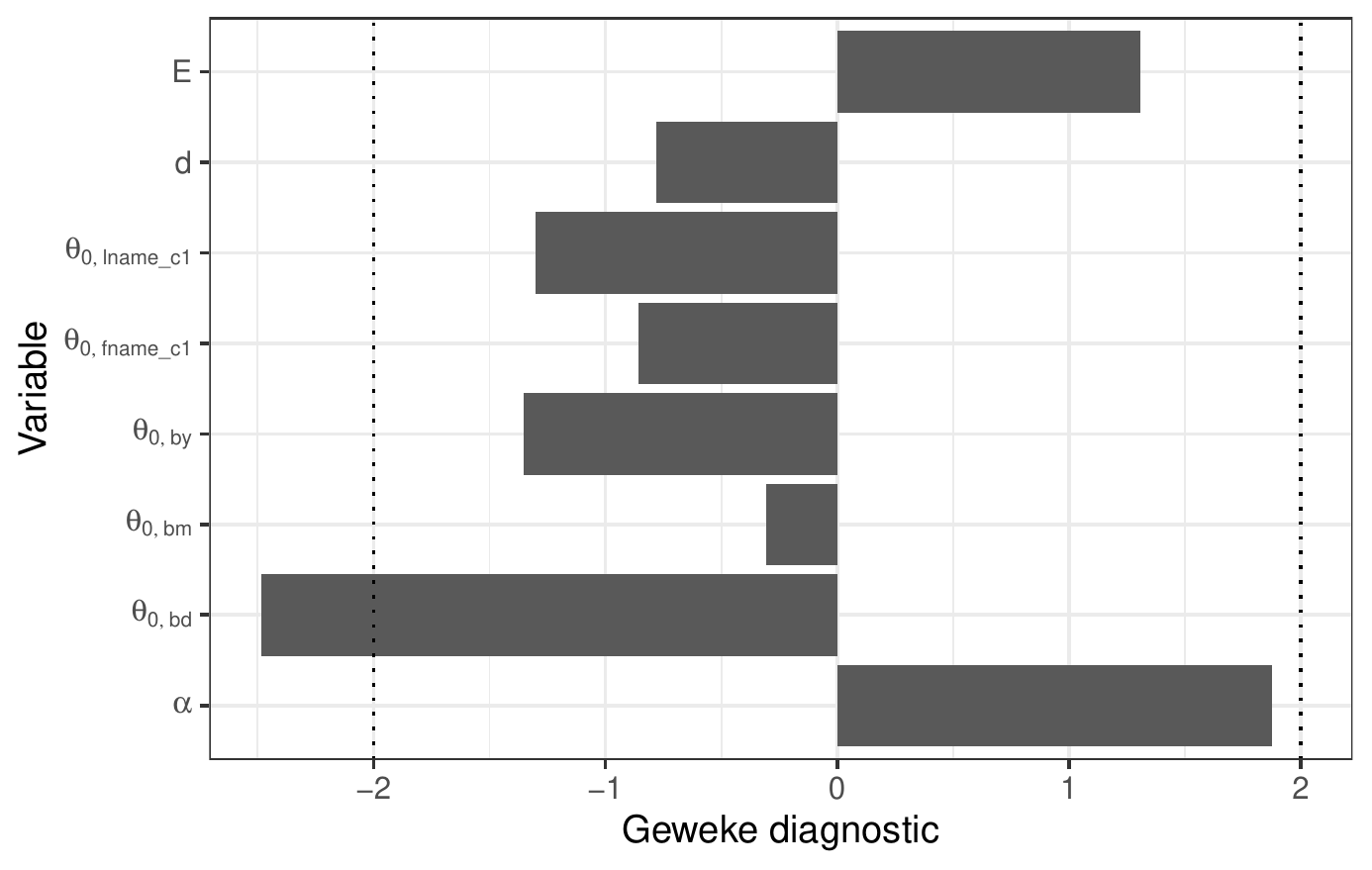} \hfill
\includegraphics[width=0.48\linewidth]{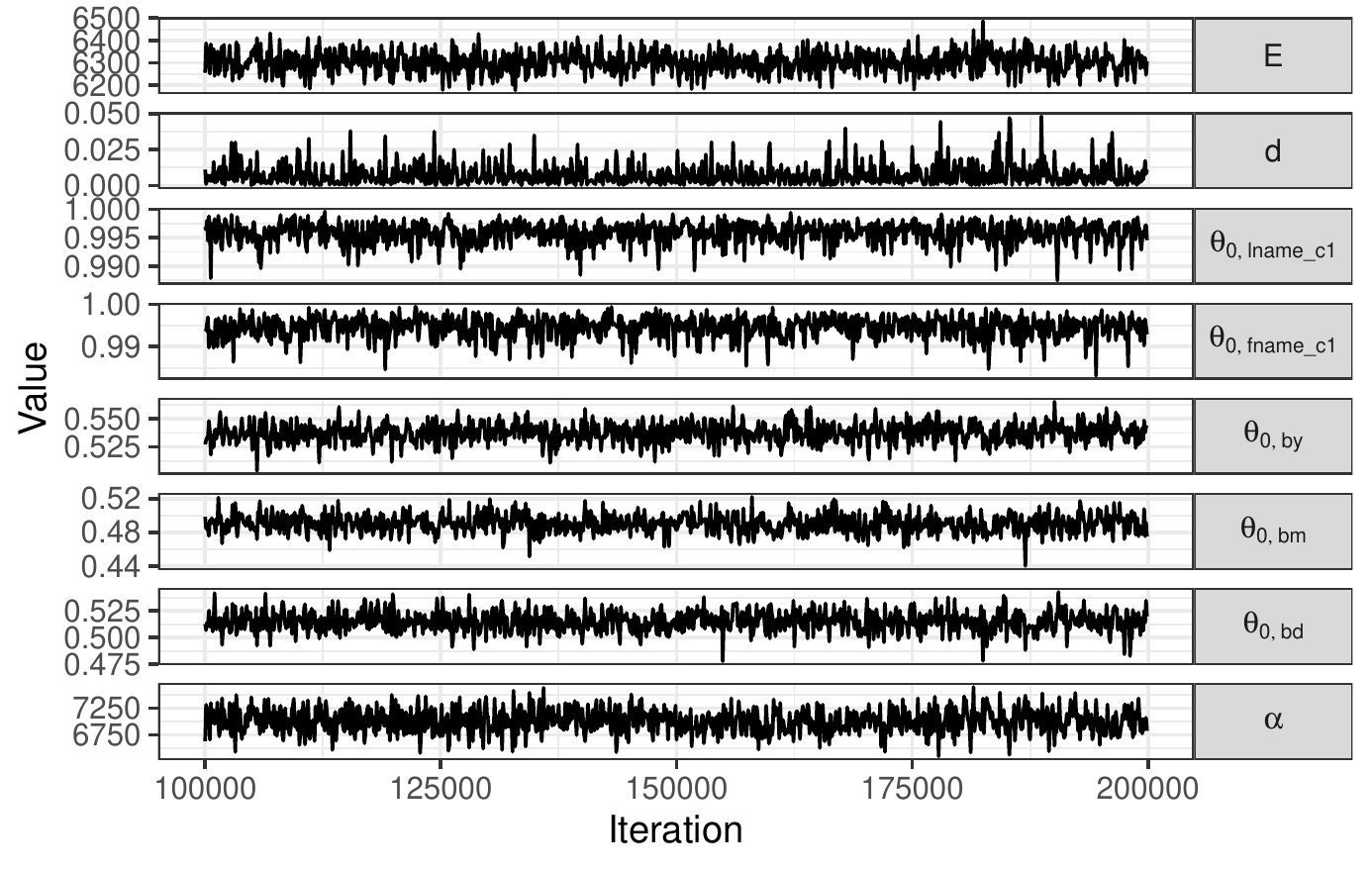} 
\end{fig}

\begin{fig}{\textsf{RLdata} | \textsf{Ewens} | \textsf{blink}}
\includegraphics[width=0.48\linewidth]{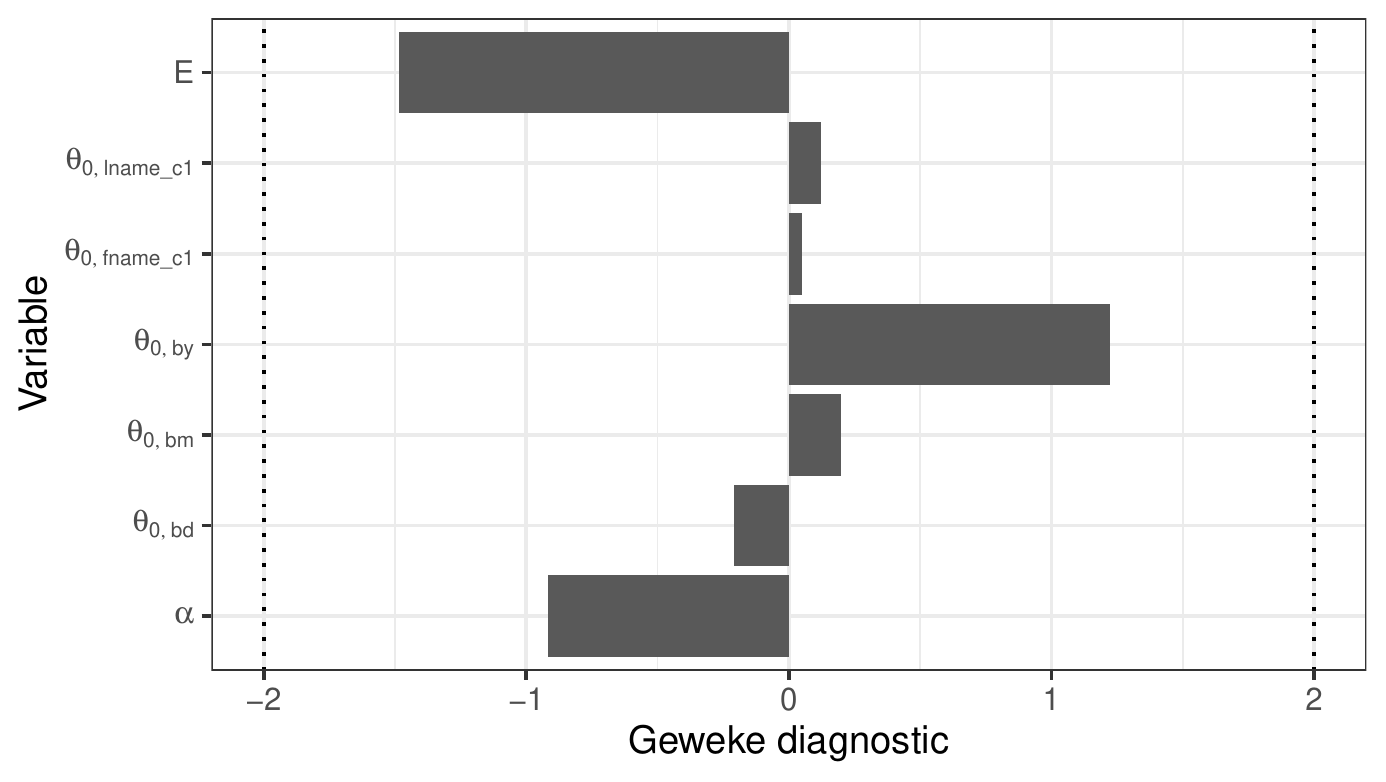} \hfill
\includegraphics[width=0.48\linewidth]{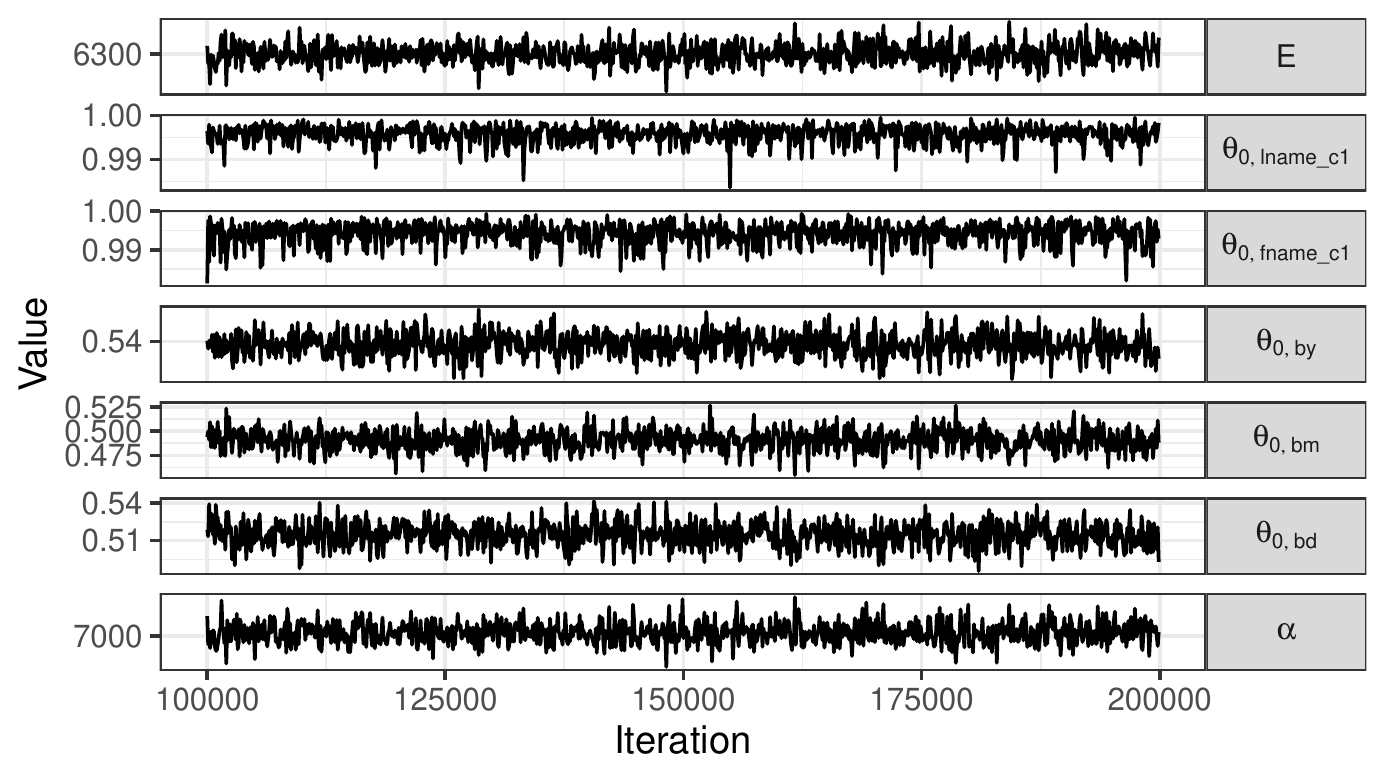} 
\end{fig}

\begin{fig}{\textsf{RLdata} | \textsf{GenCoupon} | \textsf{blink}}
\includegraphics[width=0.48\linewidth]{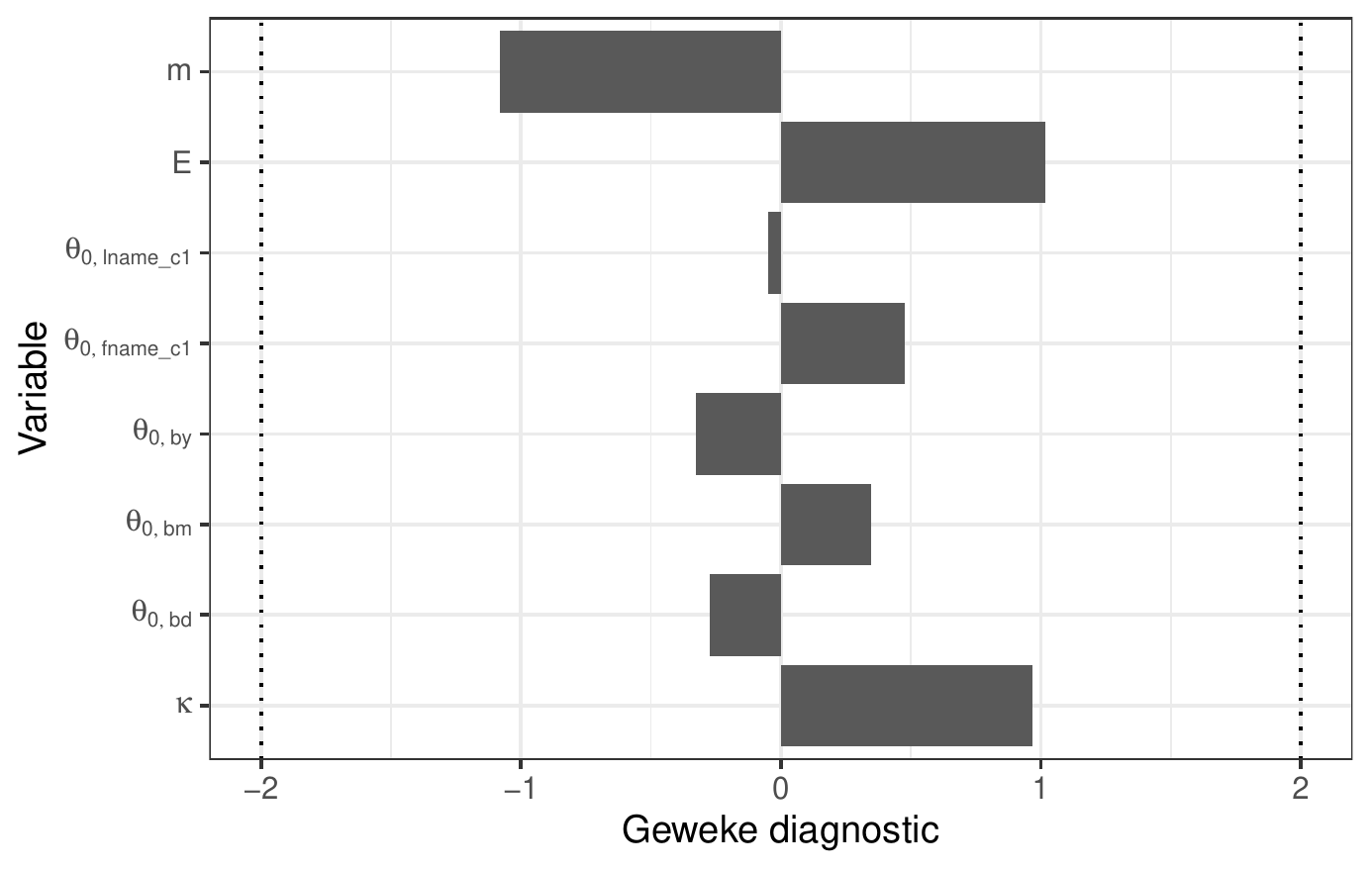} \hfill
\includegraphics[width=0.48\linewidth]{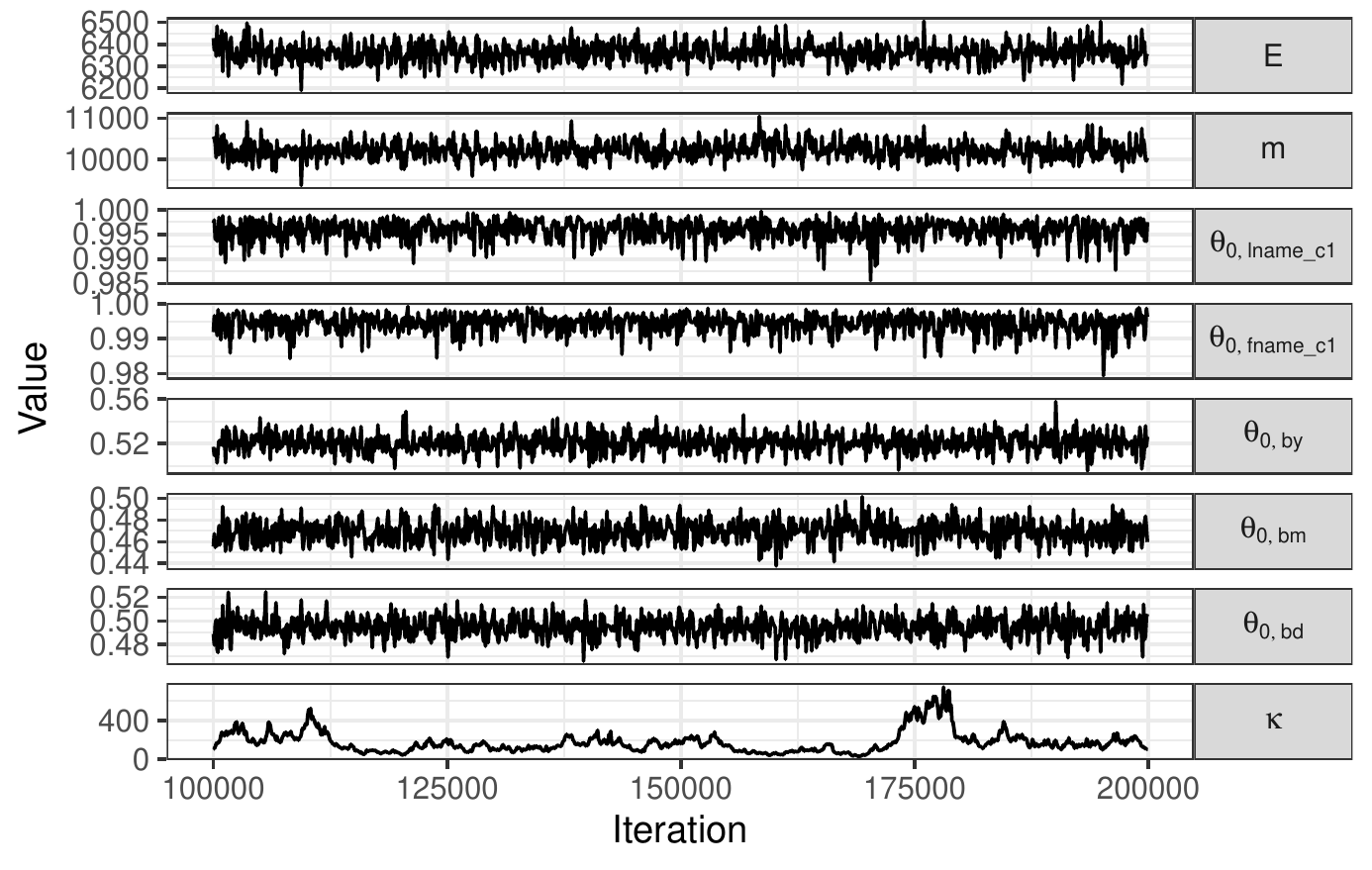} 
\end{fig}

\begin{fig}{\textsf{RLdata} | \textsf{Coupon} | \textsf{blink}}
\includegraphics[width=0.48\linewidth]{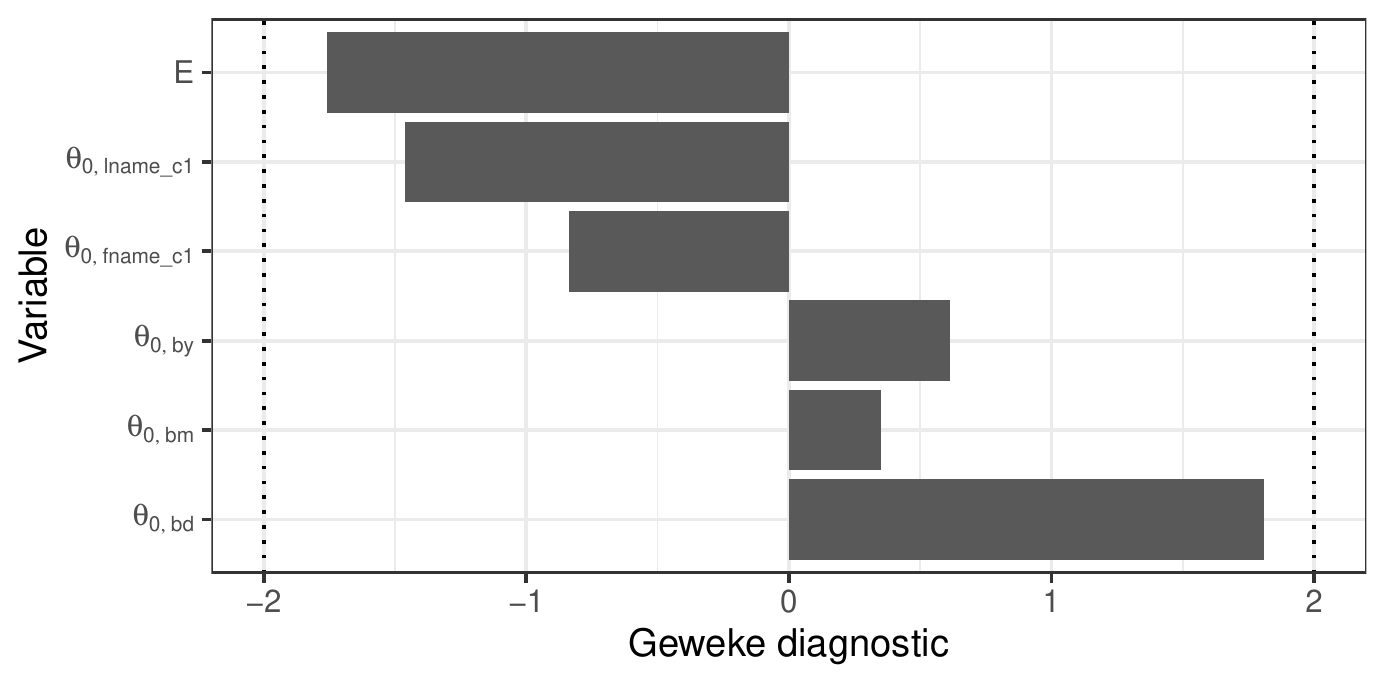} \hfill
\includegraphics[width=0.48\linewidth]{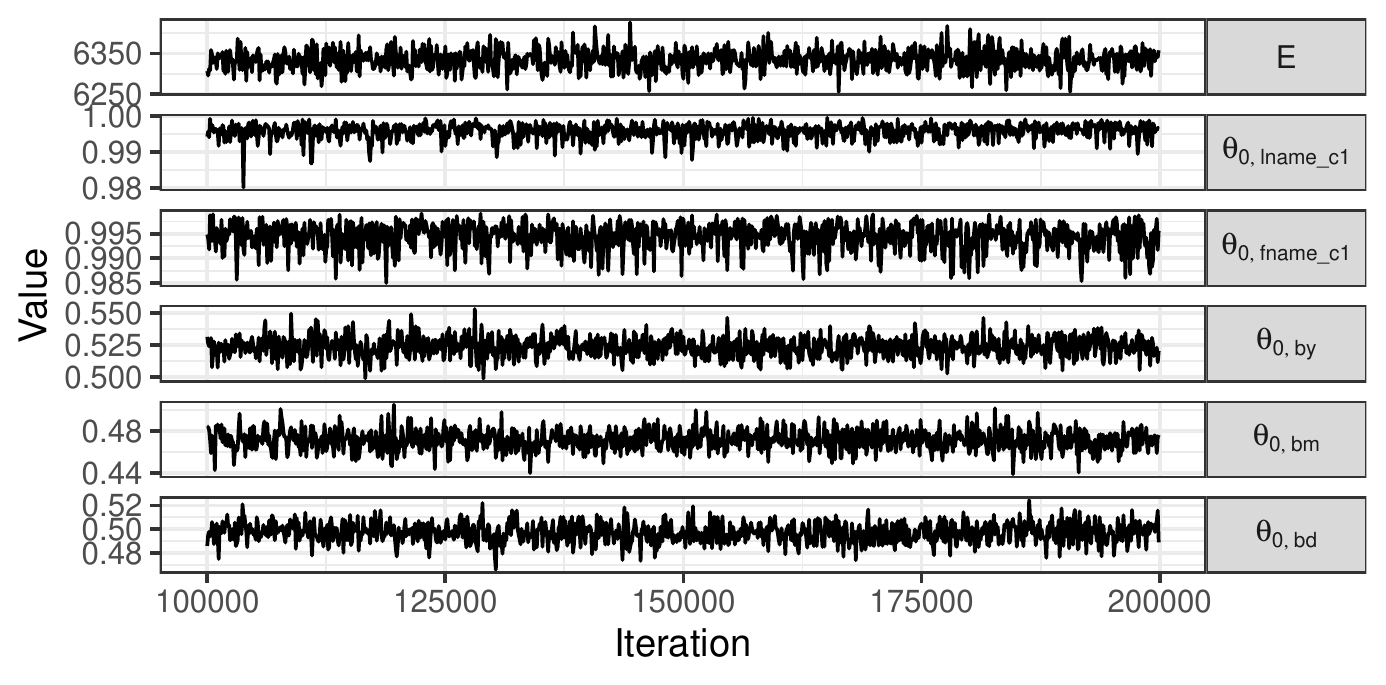}
\end{fig}

\begin{fig}{\textsf{cora} | \textsf{PY} | \textsf{Ours}}
\includegraphics[width=0.48\linewidth]{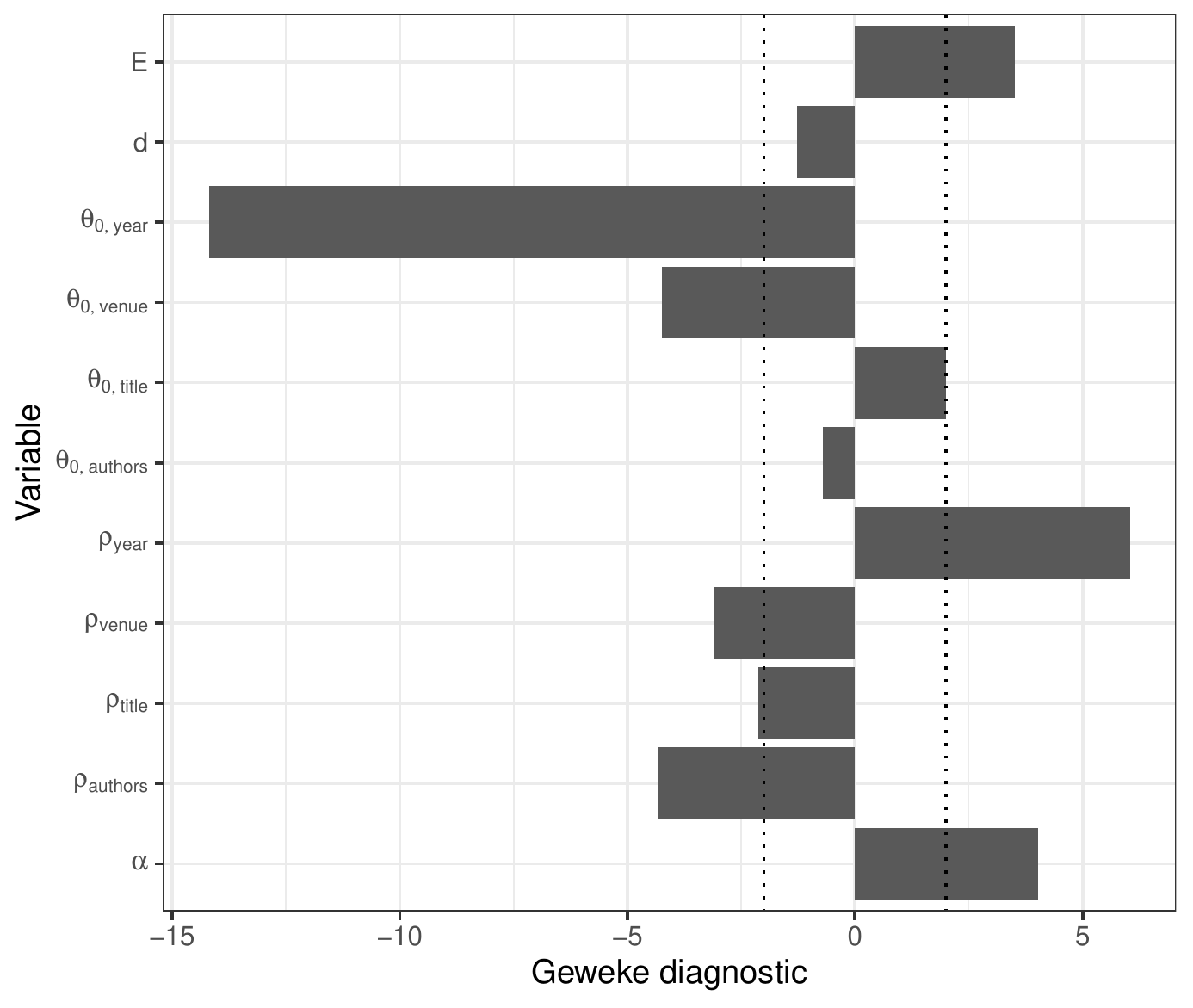} \hfill
\includegraphics[width=0.48\linewidth]{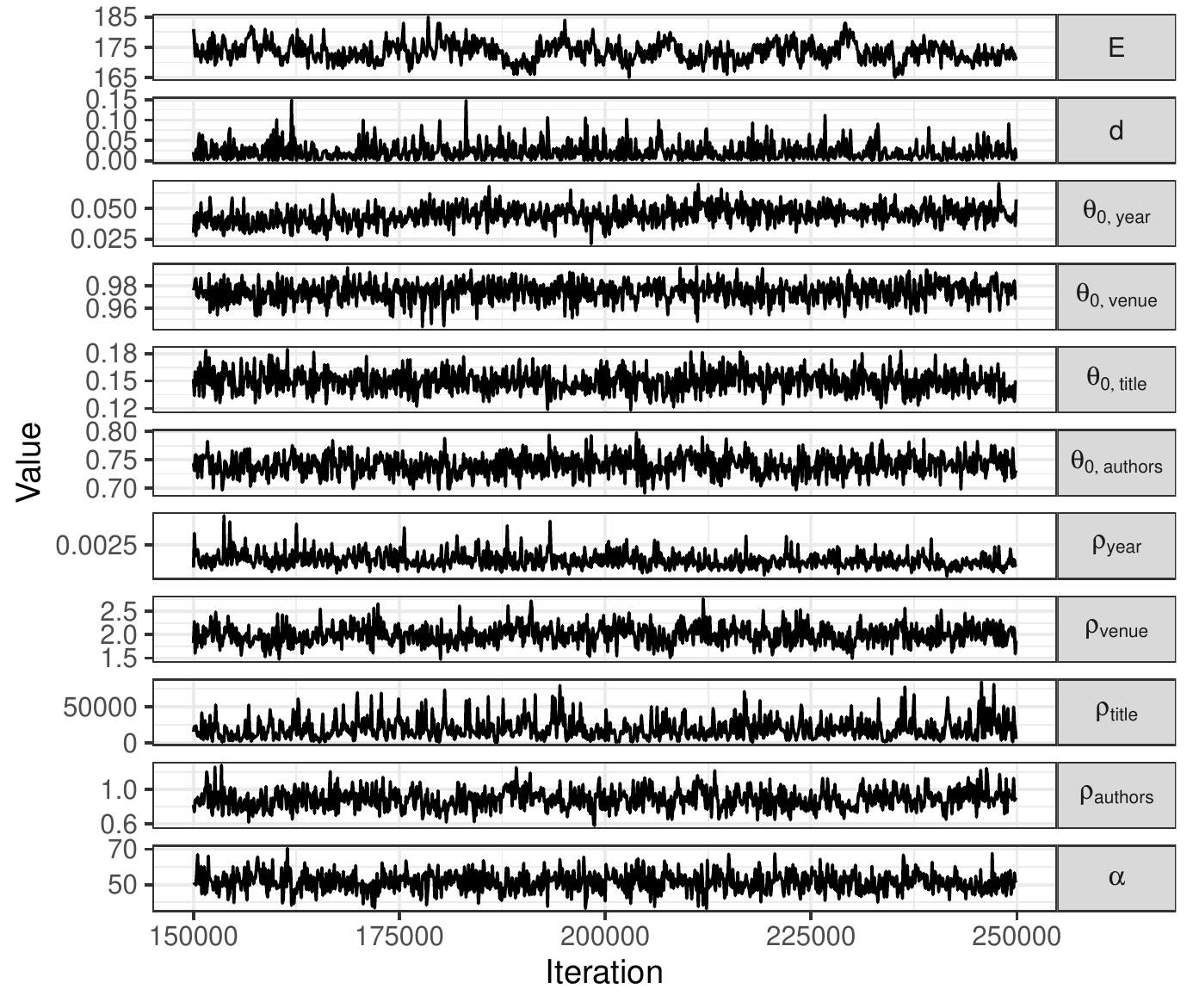} 
\end{fig}

\begin{fig}{\textsf{cora} | \textsf{Ewens} | \textsf{Ours}}
\includegraphics[width=0.48\linewidth]{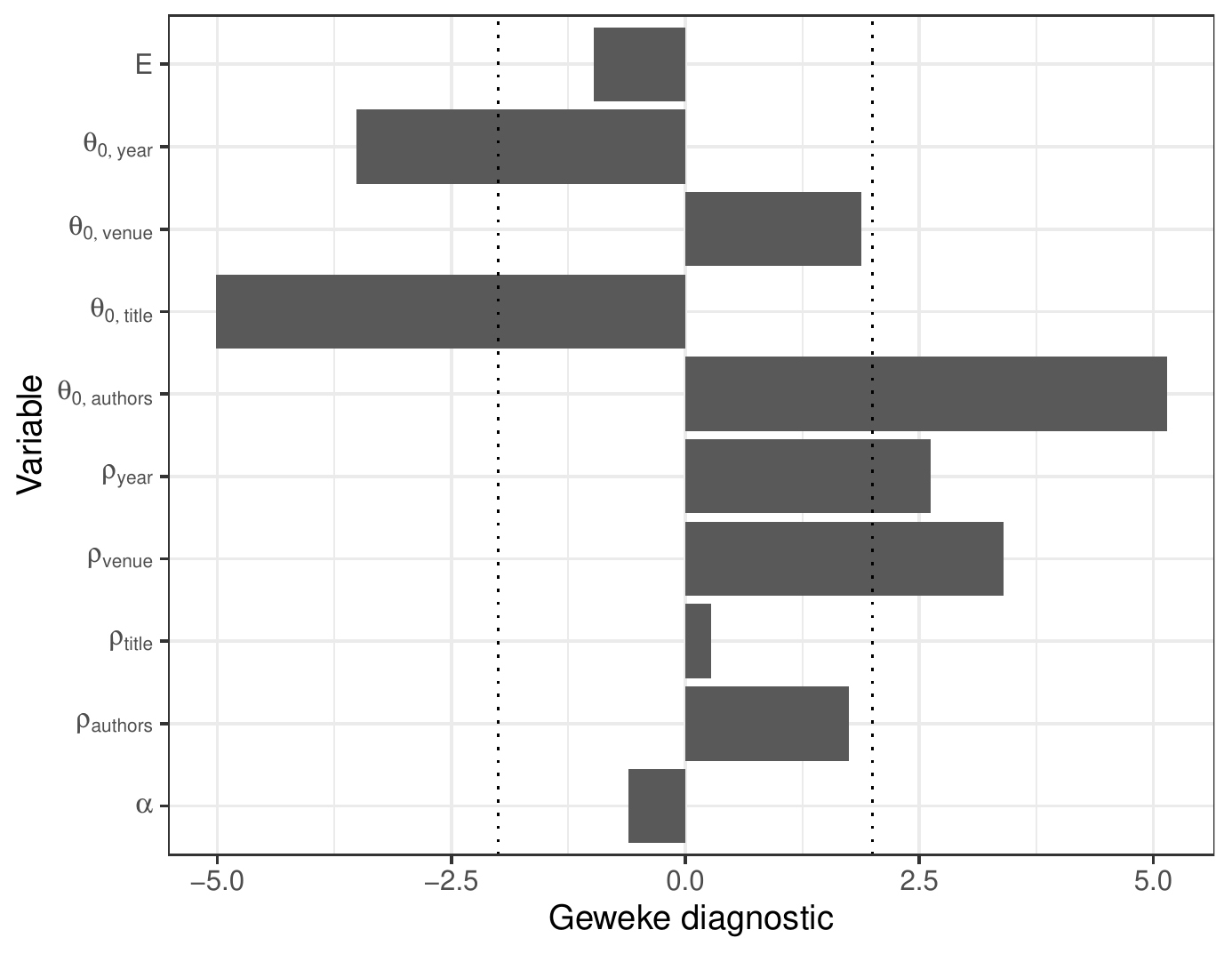} \hfill
\includegraphics[width=0.48\linewidth]{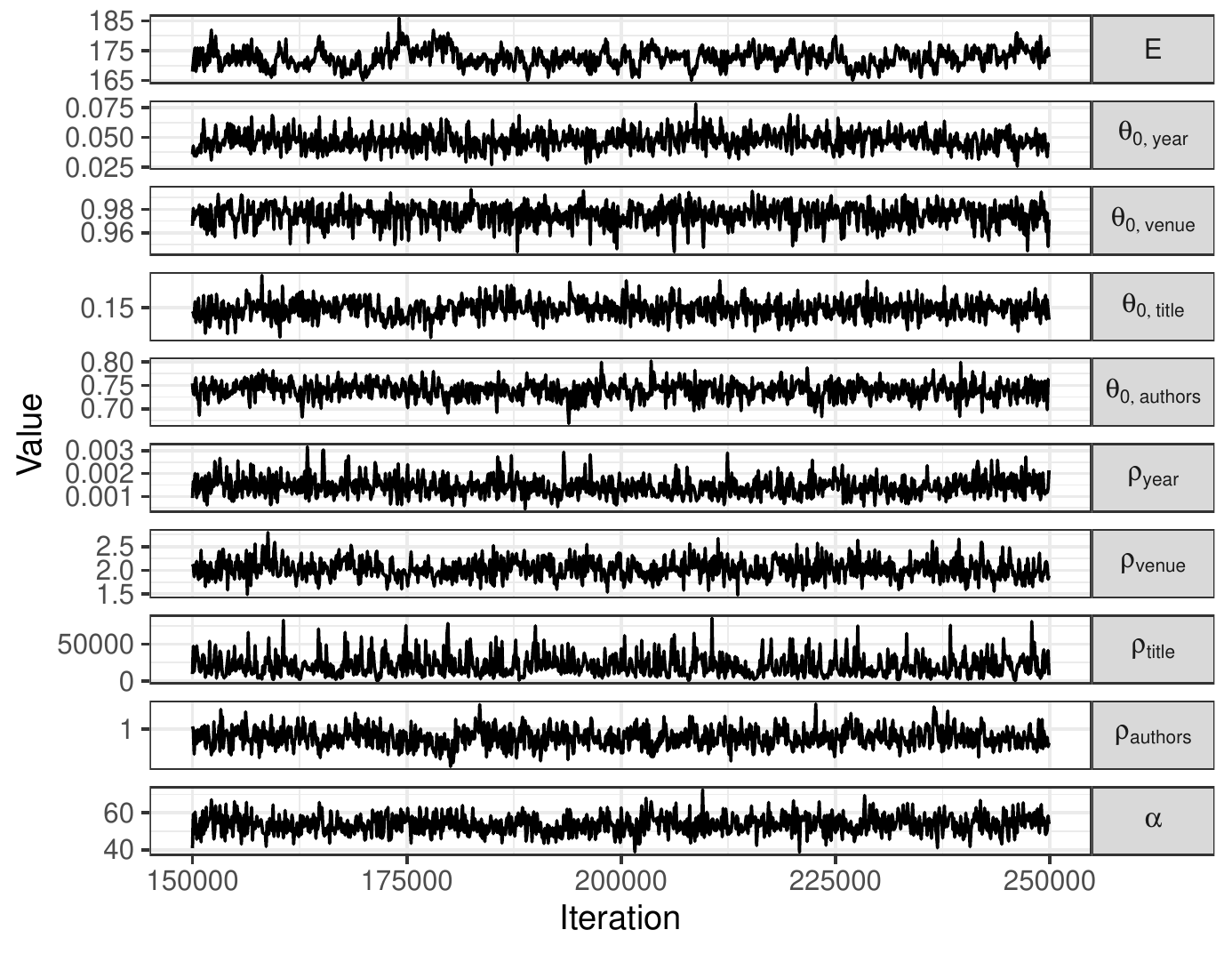} 
\end{fig}

\begin{fig}{\textsf{cora} | \textsf{GenCoupon} | \textsf{Ours}}
\includegraphics[width=0.48\linewidth]{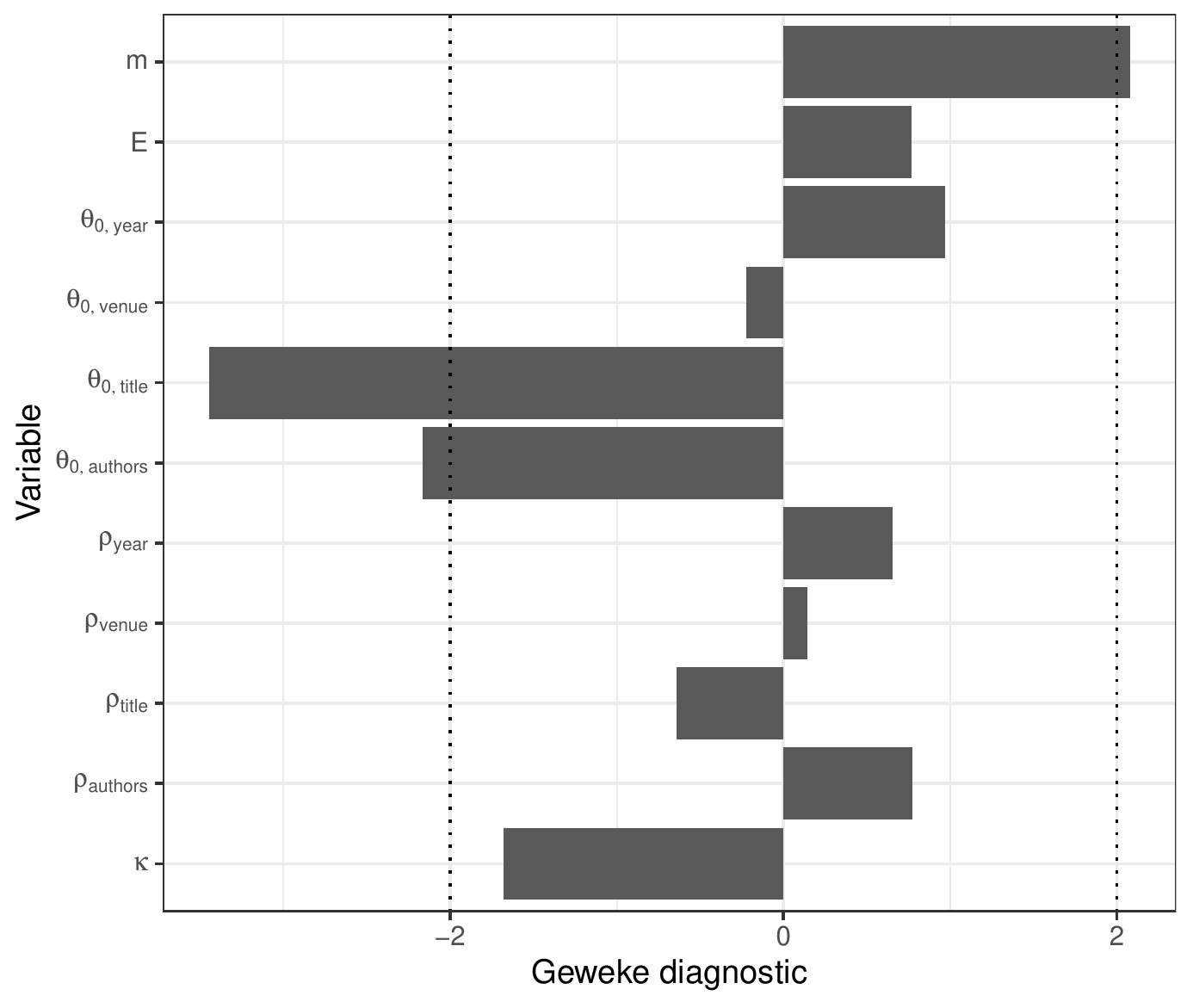} \hfill
\includegraphics[width=0.48\linewidth]{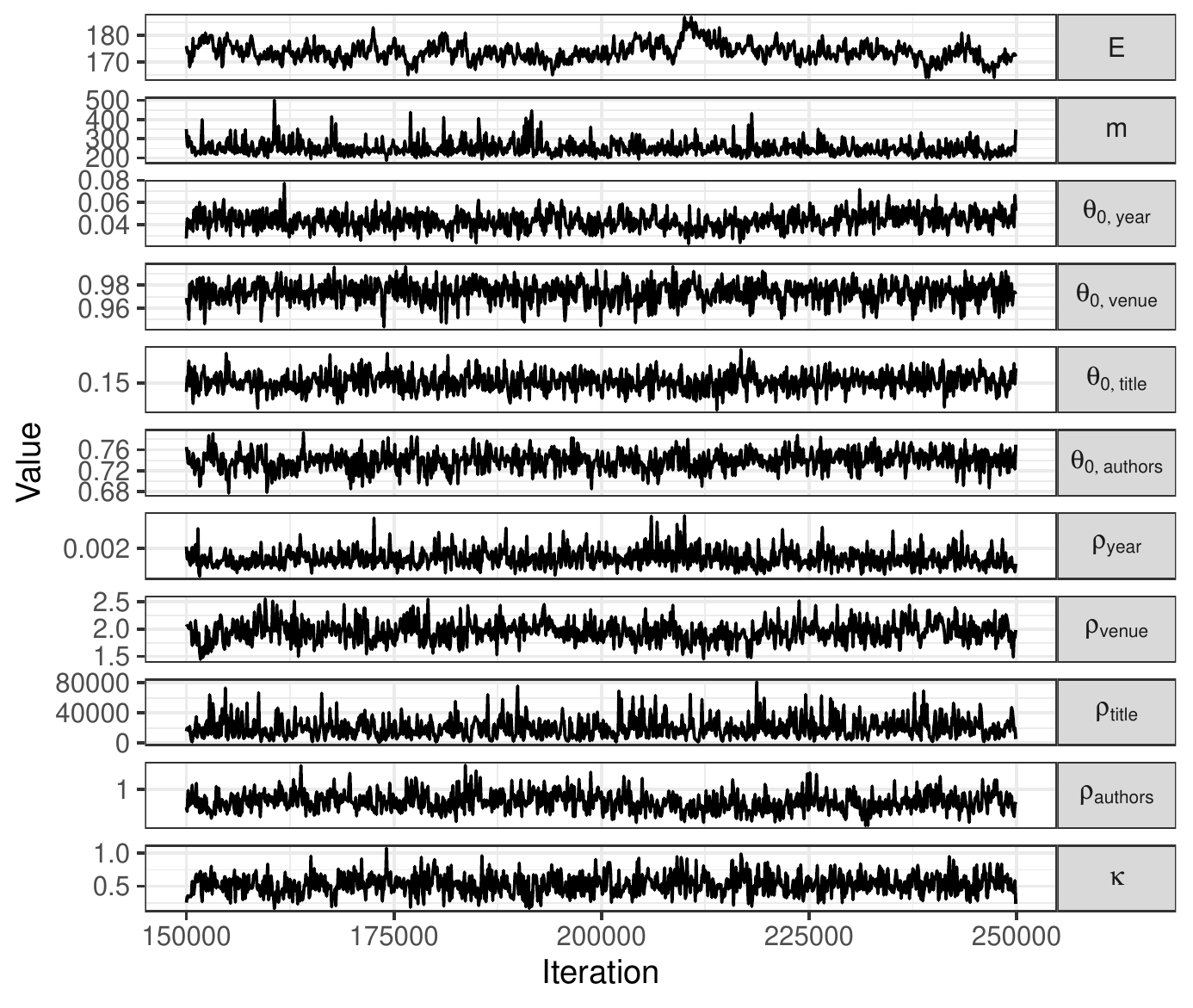} 
\end{fig}

\begin{fig}{\textsf{cora} | \textsf{Coupon} | \textsf{Ours}}
\includegraphics[width=0.48\linewidth]{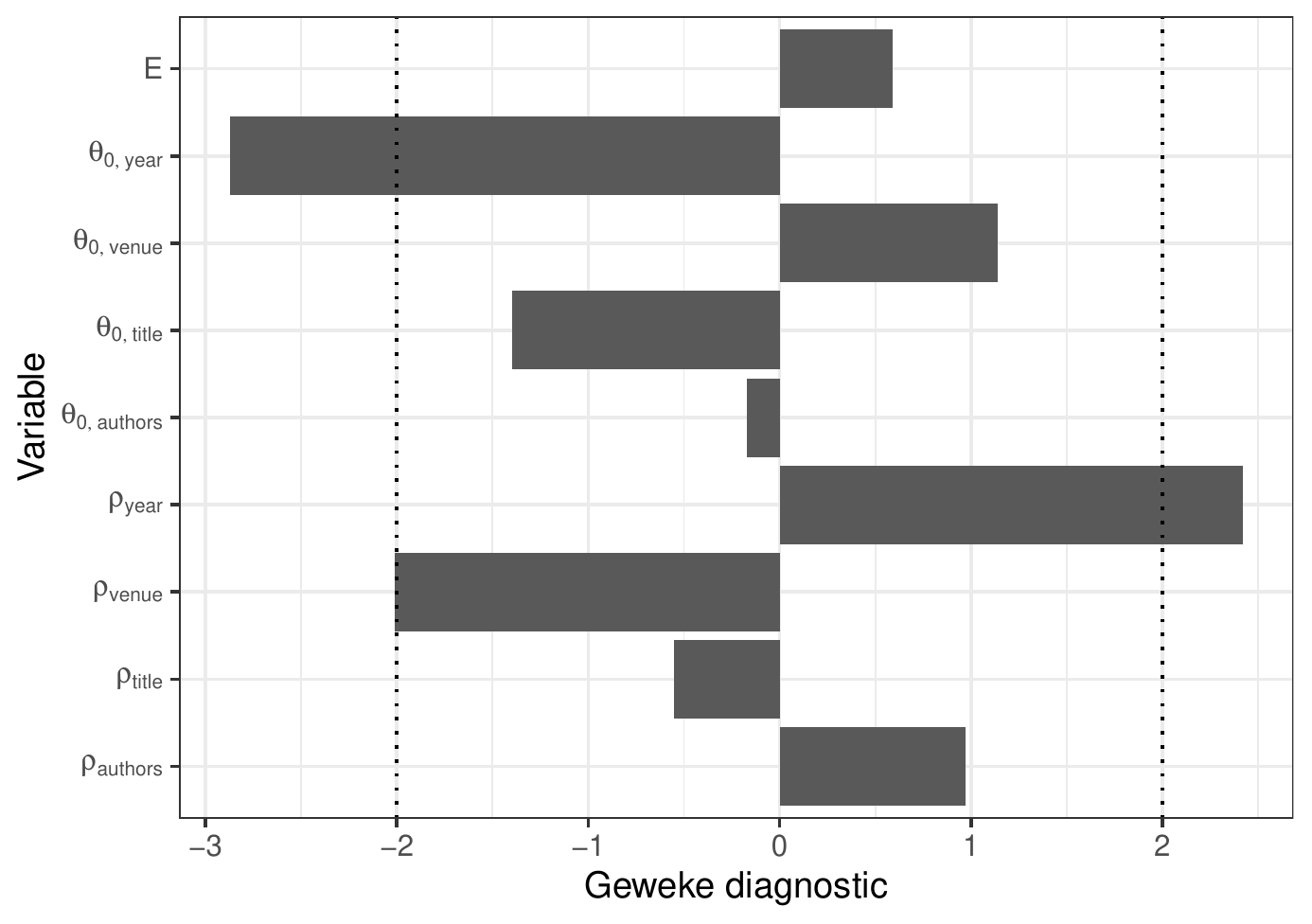} \hfill
\includegraphics[width=0.48\linewidth]{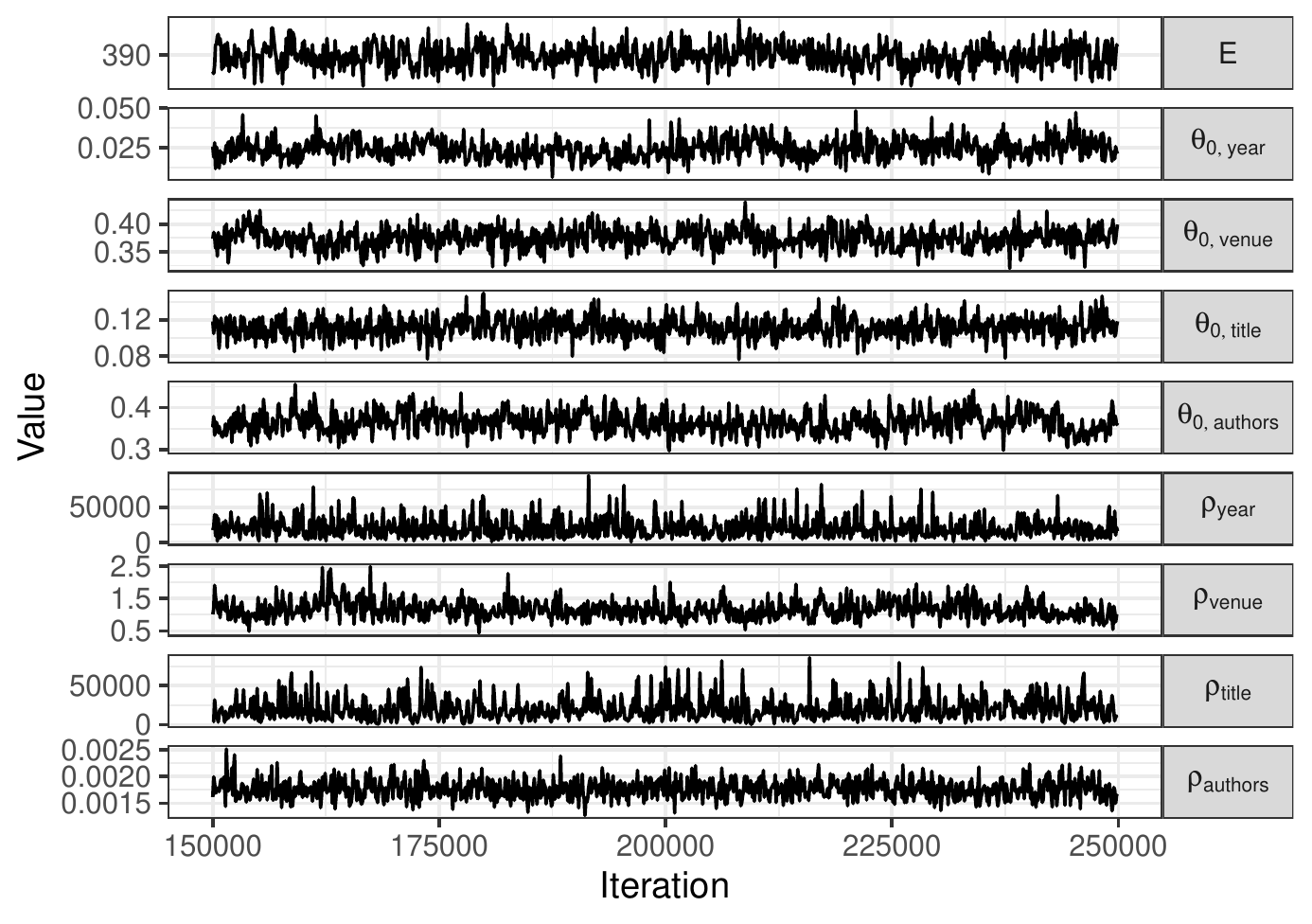} 
\end{fig}

\begin{fig}{\textsf{cora} | \textsf{PY} | \textsf{blink}}
\includegraphics[width=0.48\linewidth]{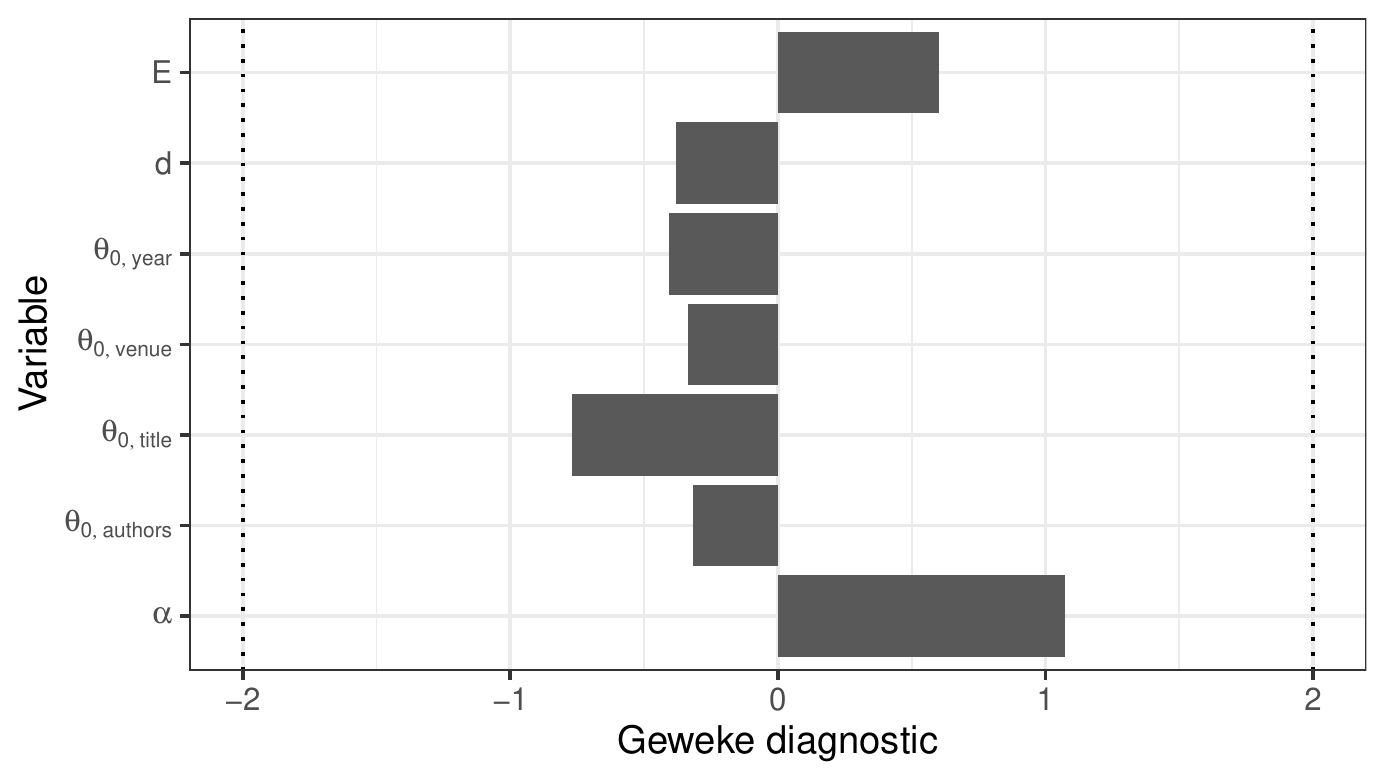} \hfill
\includegraphics[width=0.48\linewidth]{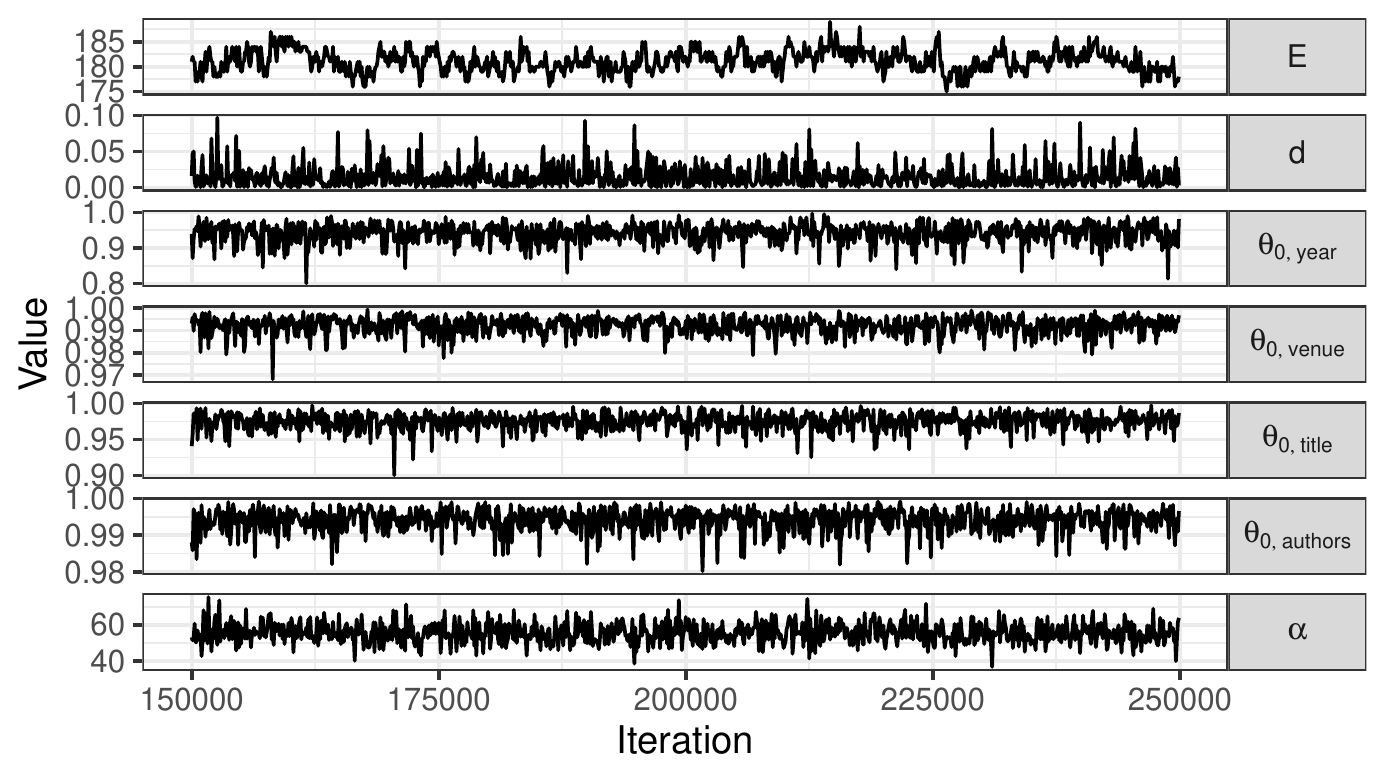} 
\end{fig}

\begin{fig}{\textsf{cora} | \textsf{Ewens} | \textsf{blink}}
\includegraphics[width=0.48\linewidth]{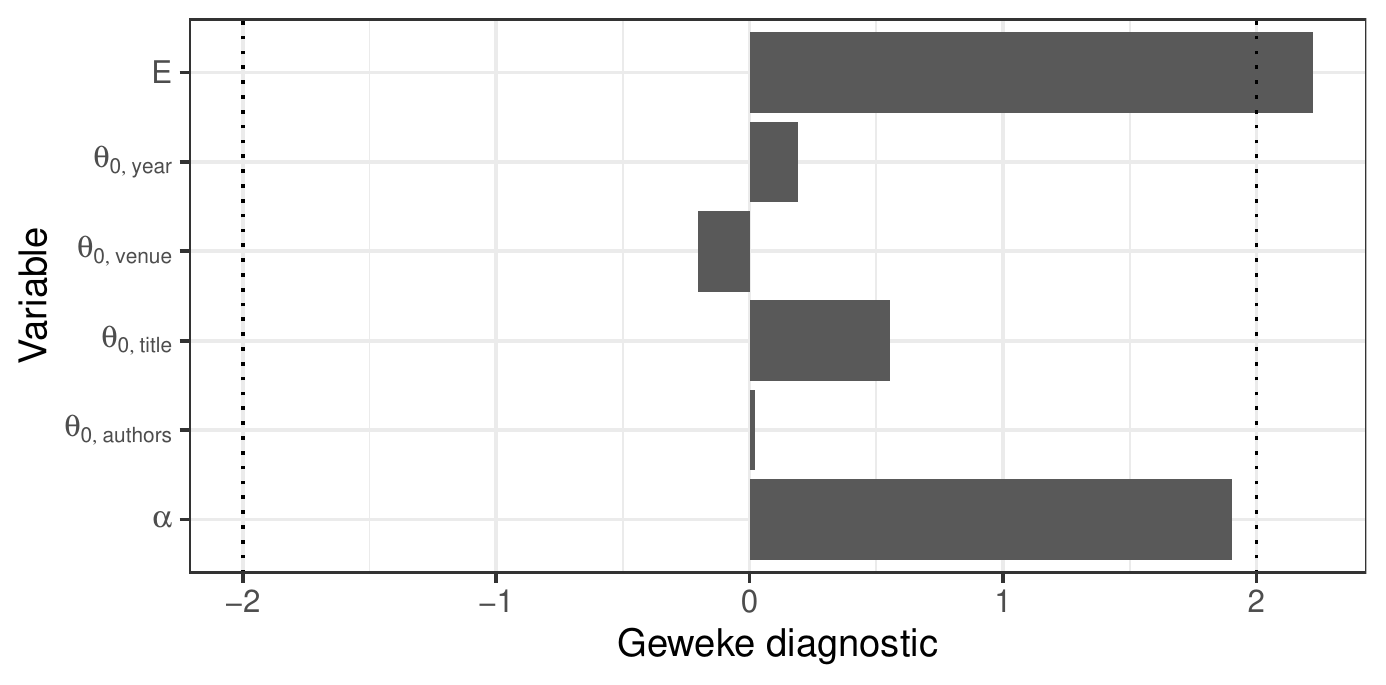} \hfill
\includegraphics[width=0.48\linewidth]{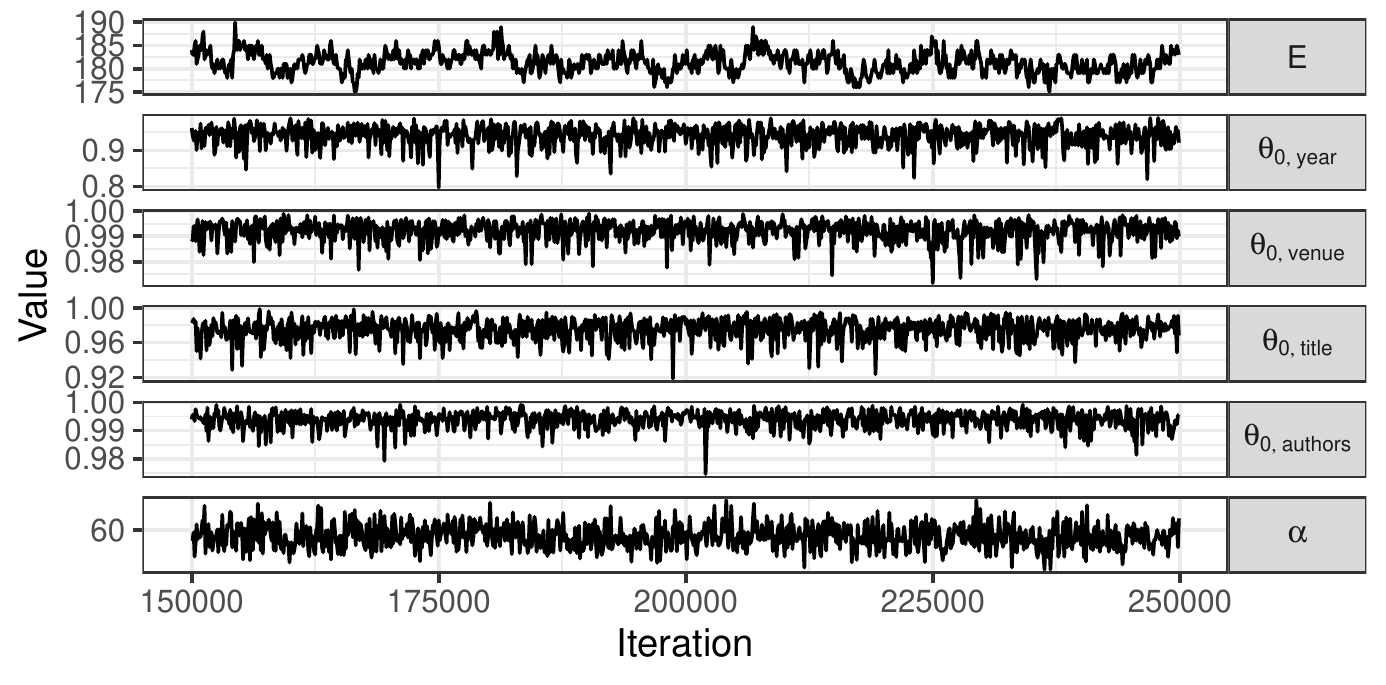} 
\end{fig}

\begin{fig}{\textsf{cora} | \textsf{GenCoupon} | \textsf{blink}}
\includegraphics[width=0.48\linewidth]{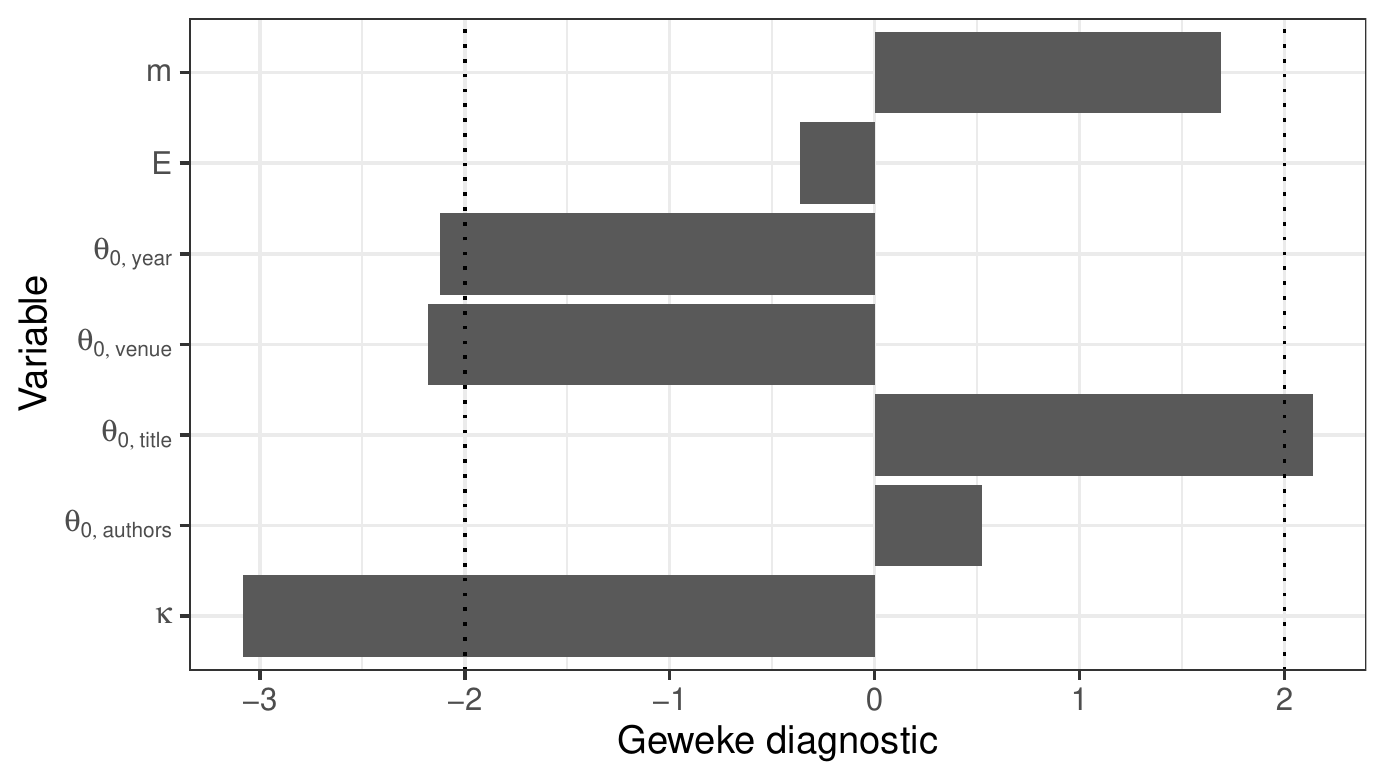} \hfill
\includegraphics[width=0.48\linewidth]{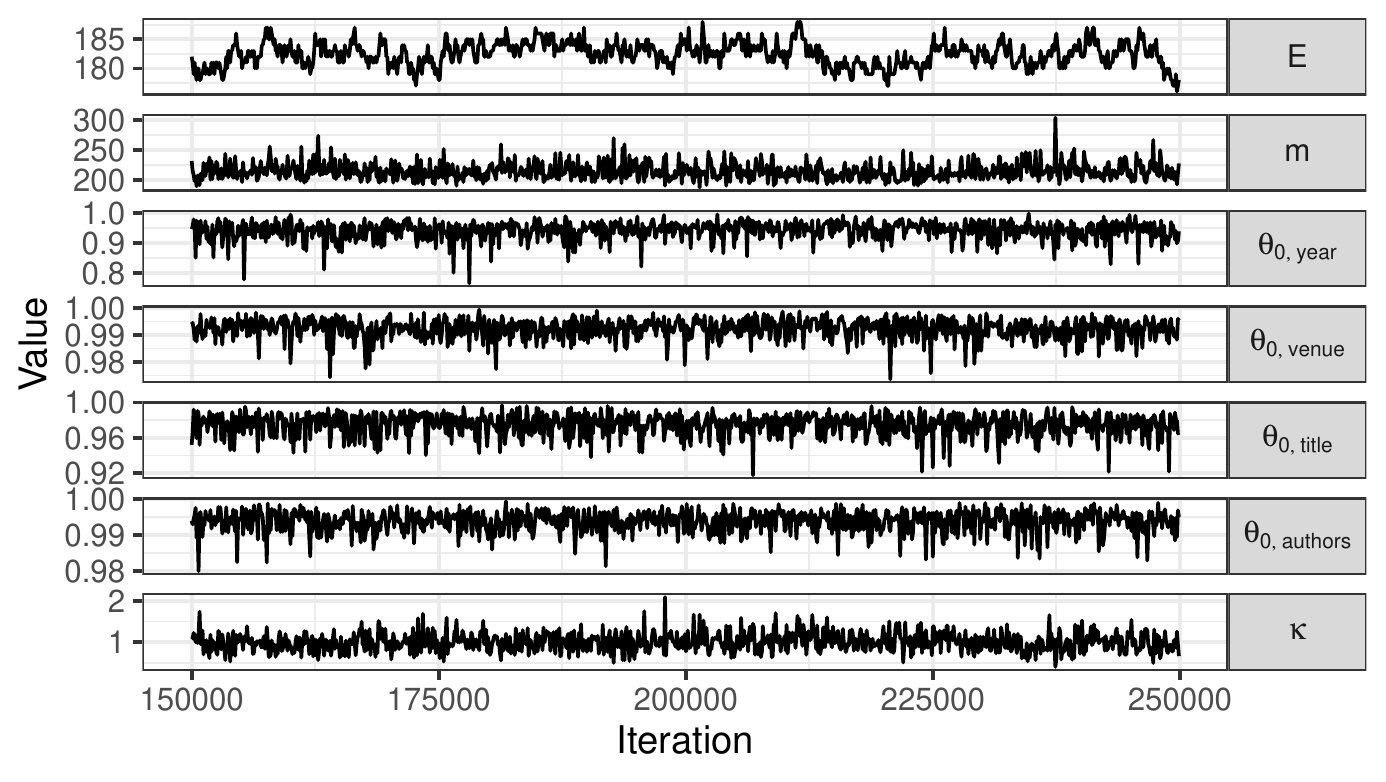} 
\end{fig}

\begin{fig}{\textsf{cora} | \textsf{Coupon} | \textsf{blink}}
\includegraphics[width=0.48\linewidth]{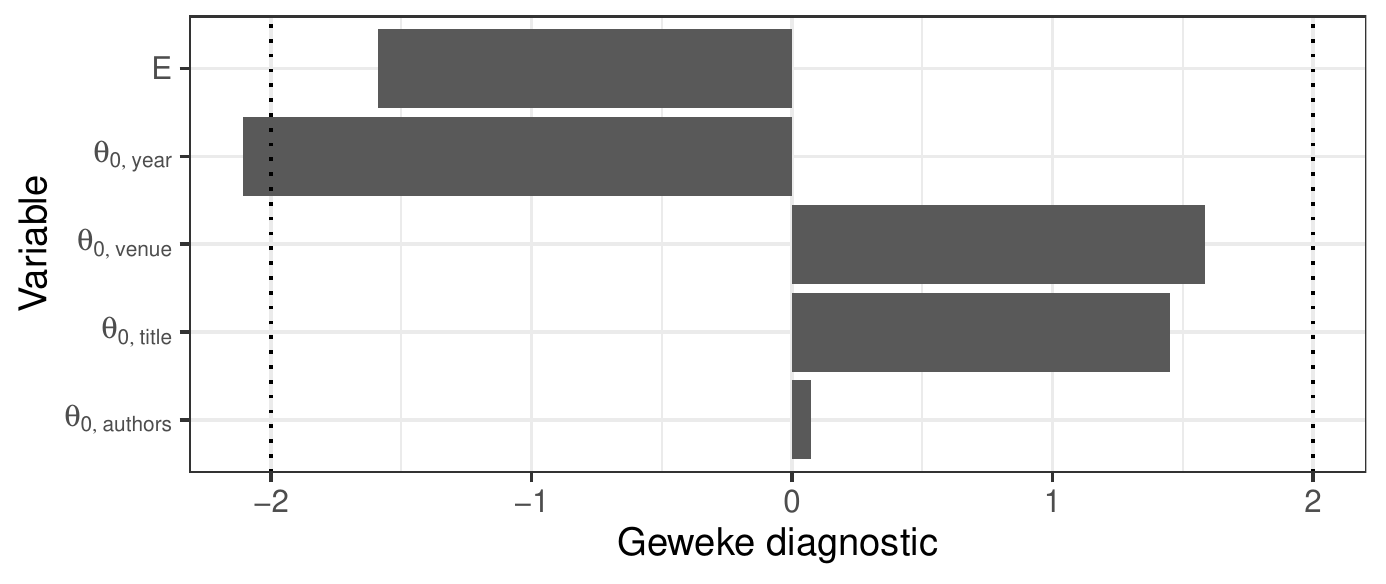} \hfill
\includegraphics[width=0.48\linewidth]{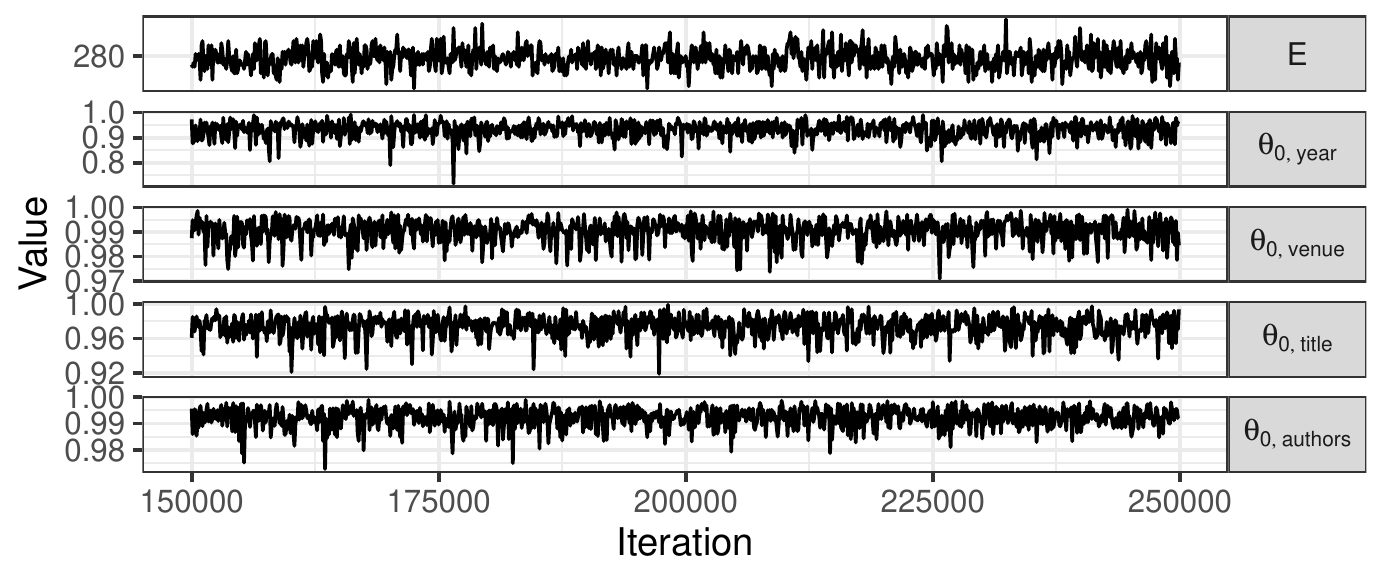}
\end{fig}

\begin{fig}{\textsf{rest} | \textsf{PY} | \textsf{Ours}}
\includegraphics[width=0.48\linewidth]{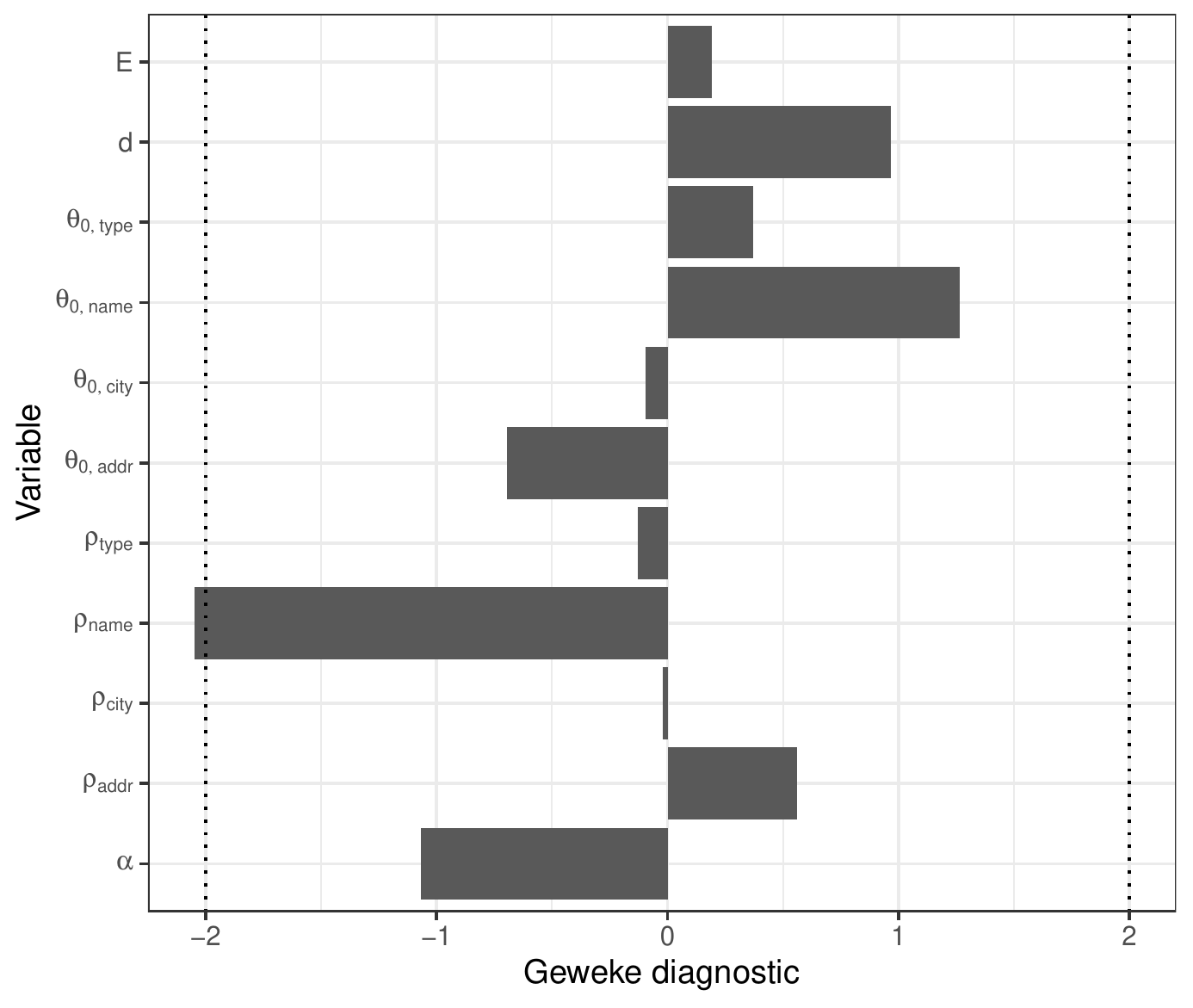} \hfill
\includegraphics[width=0.48\linewidth]{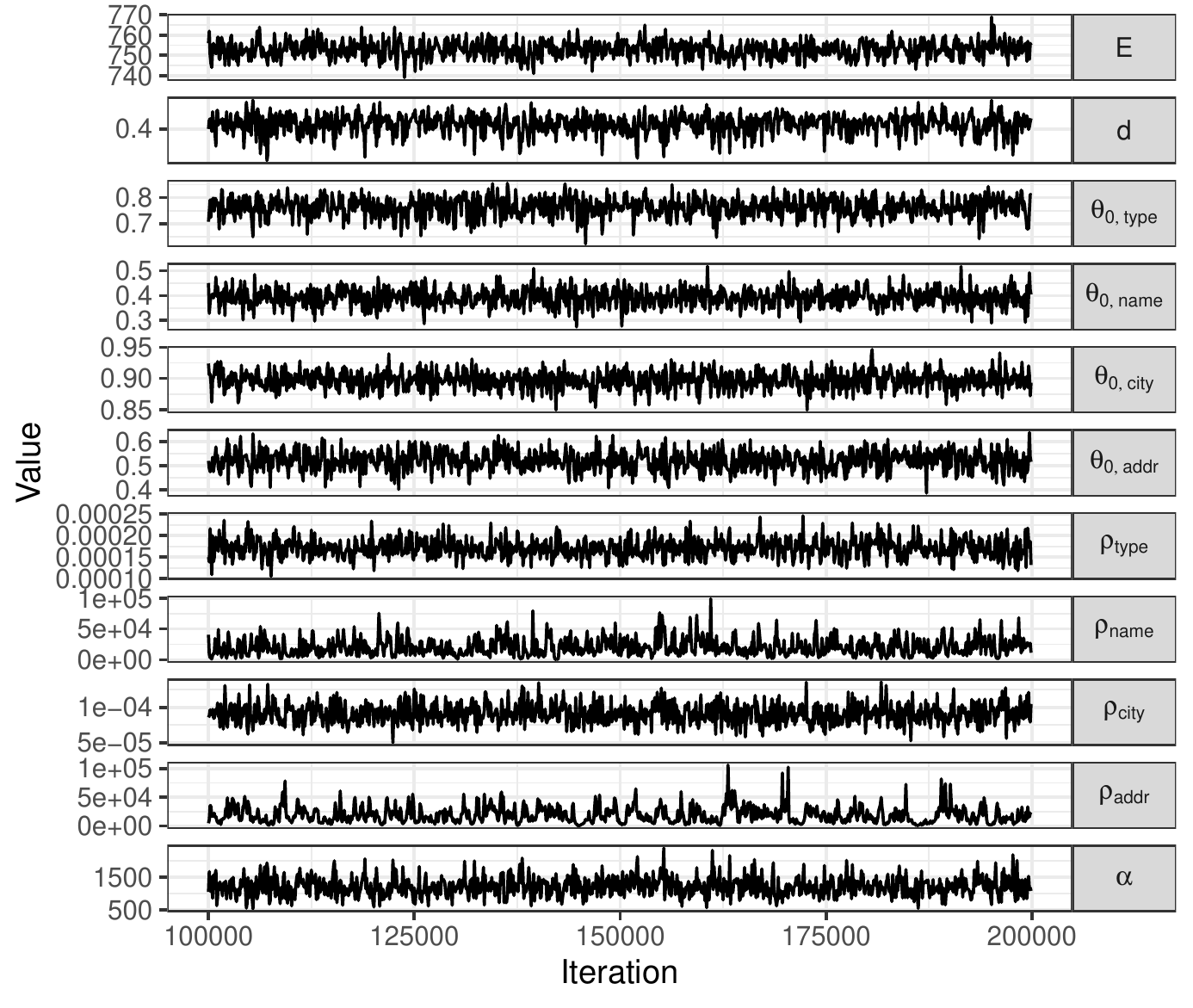} 
\end{fig}

\begin{fig}{\textsf{rest} | \textsf{Ewens} | \textsf{Ours}}
\includegraphics[width=0.48\linewidth]{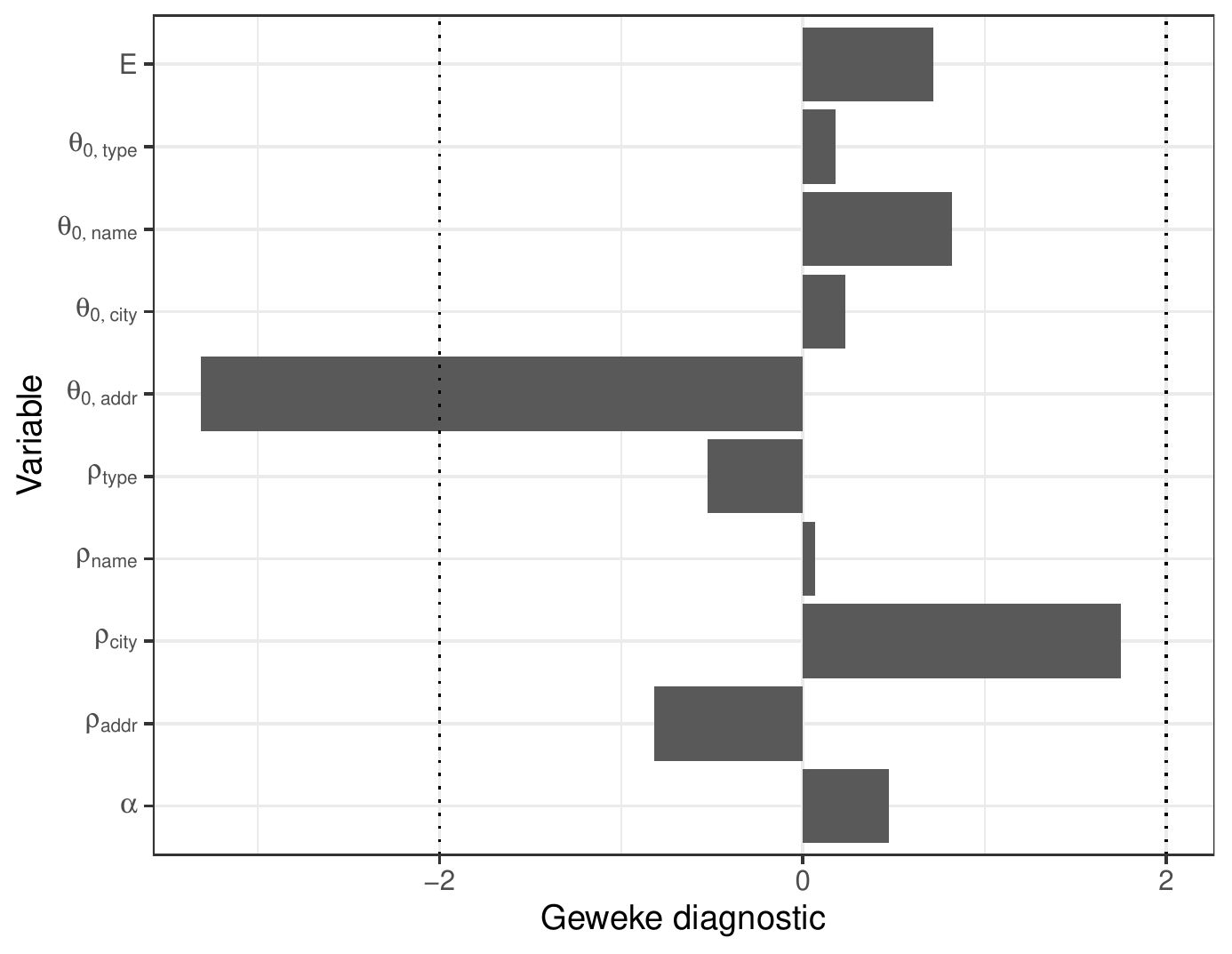} \hfill
\includegraphics[width=0.48\linewidth]{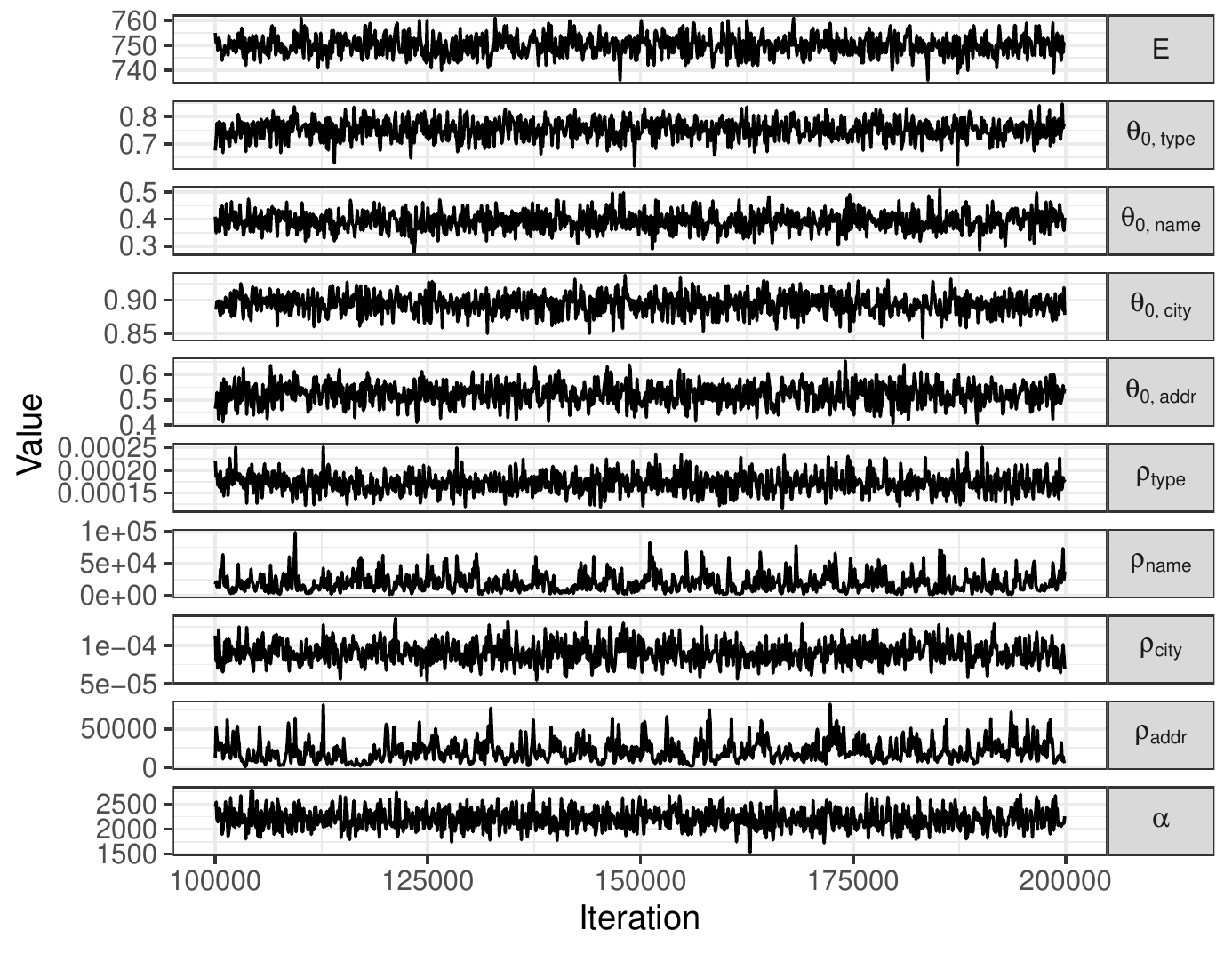} 
\end{fig}

\begin{fig}{\textsf{rest} | \textsf{GenCoupon} | \textsf{Ours}}
\includegraphics[width=0.48\linewidth]{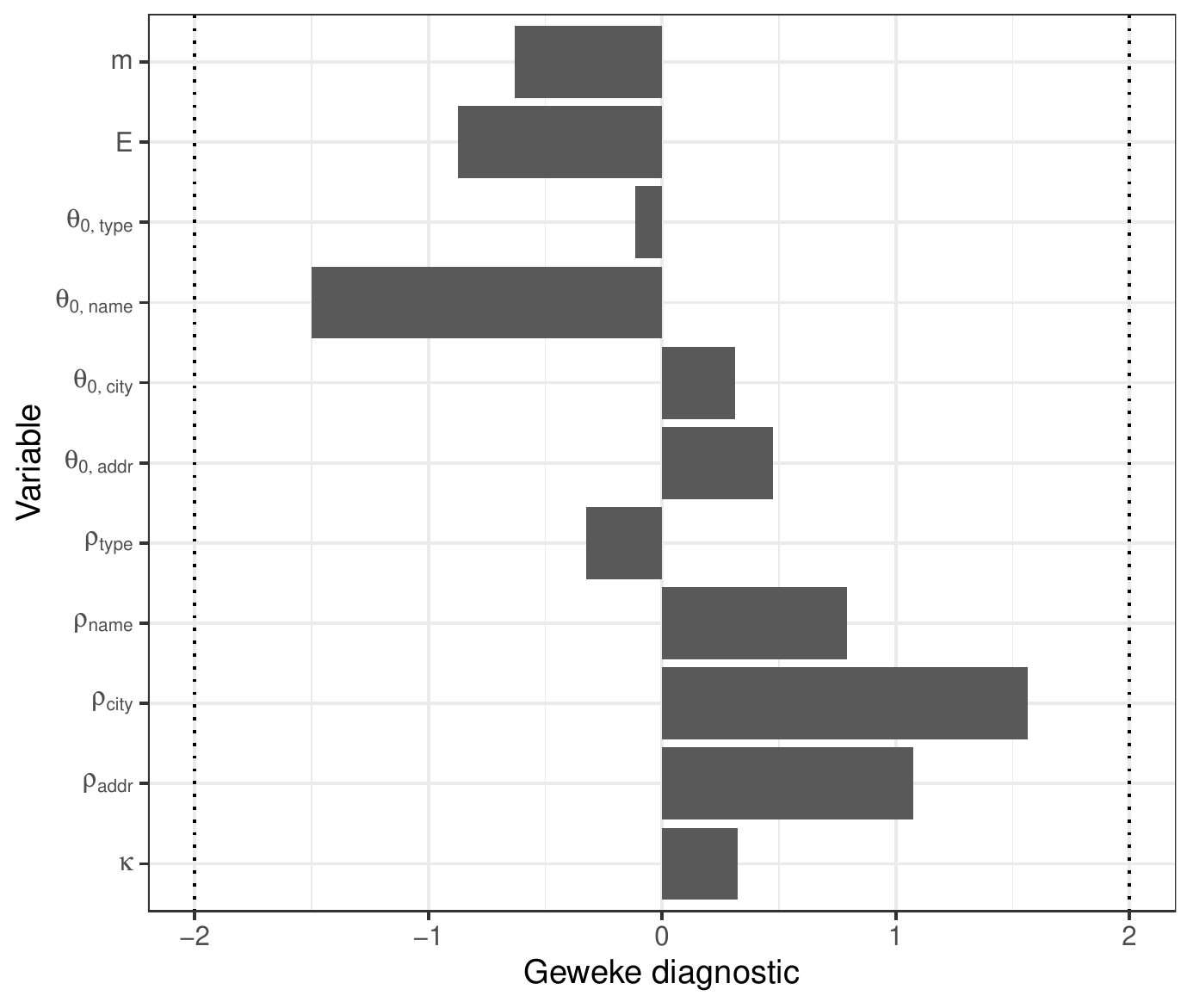} \hfill
\includegraphics[width=0.48\linewidth]{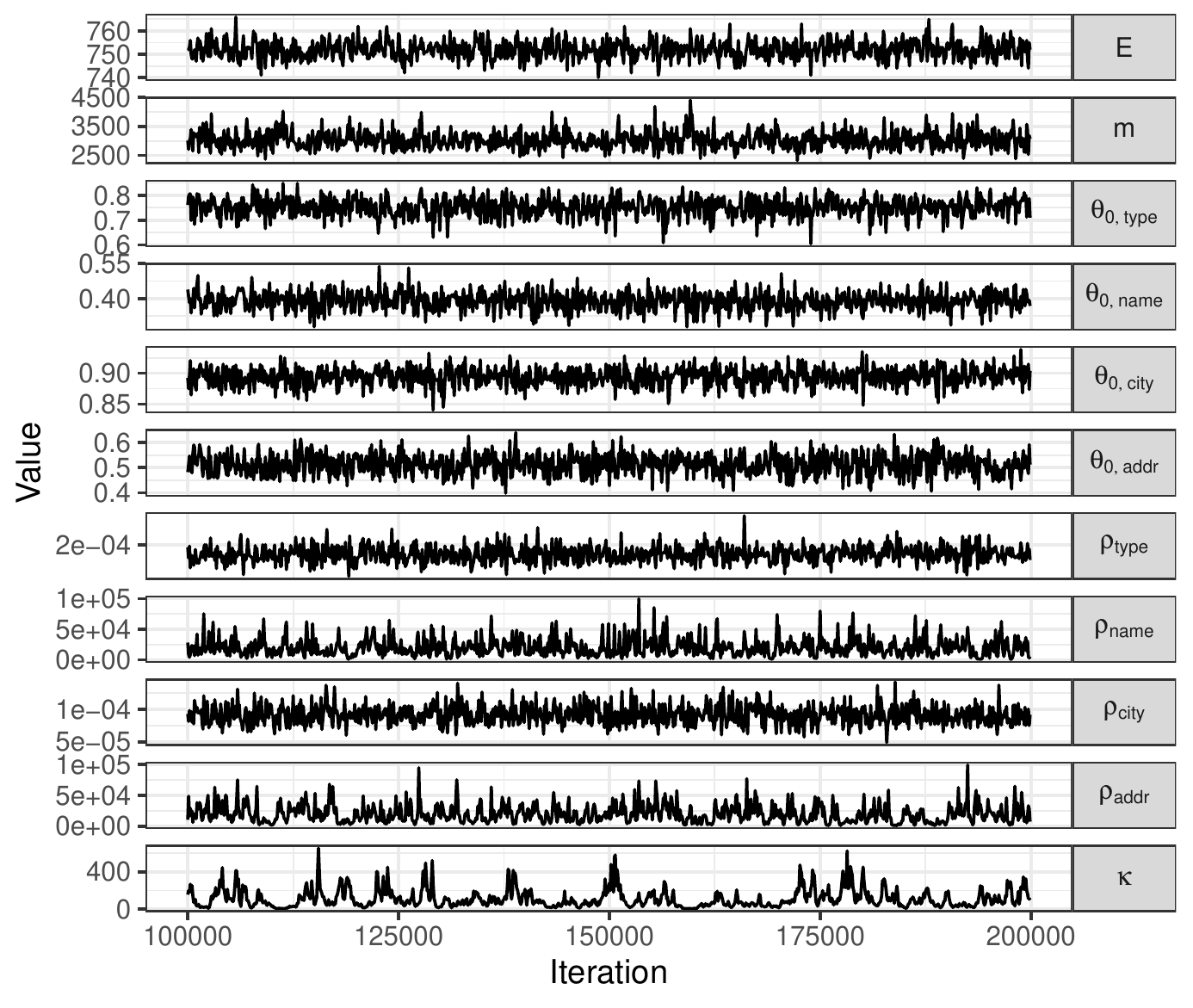} 
\end{fig}

\begin{fig}{\textsf{rest} | \textsf{Coupon} | \textsf{Ours}}
\includegraphics[width=0.48\linewidth]{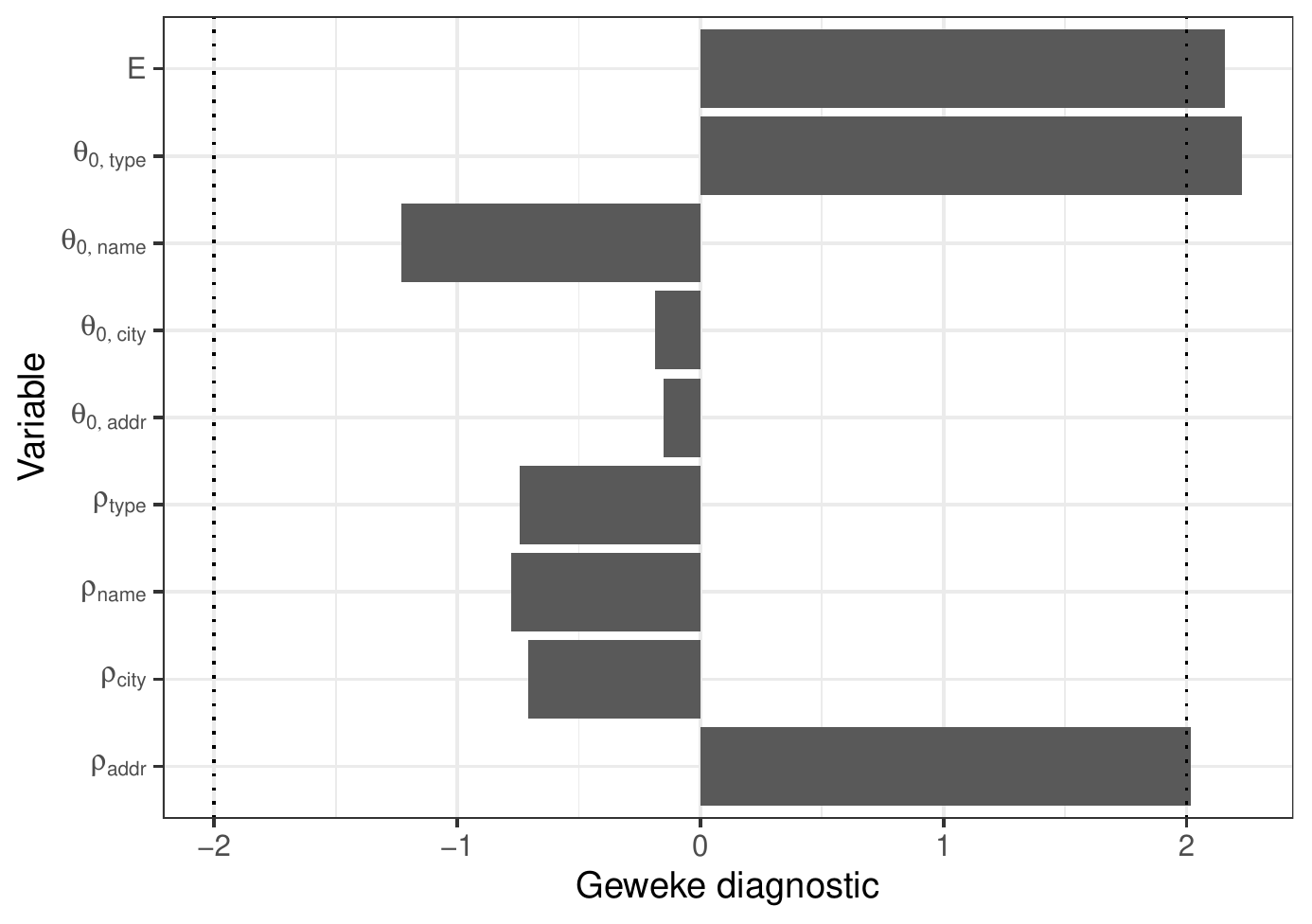} \hfill
\includegraphics[width=0.48\linewidth]{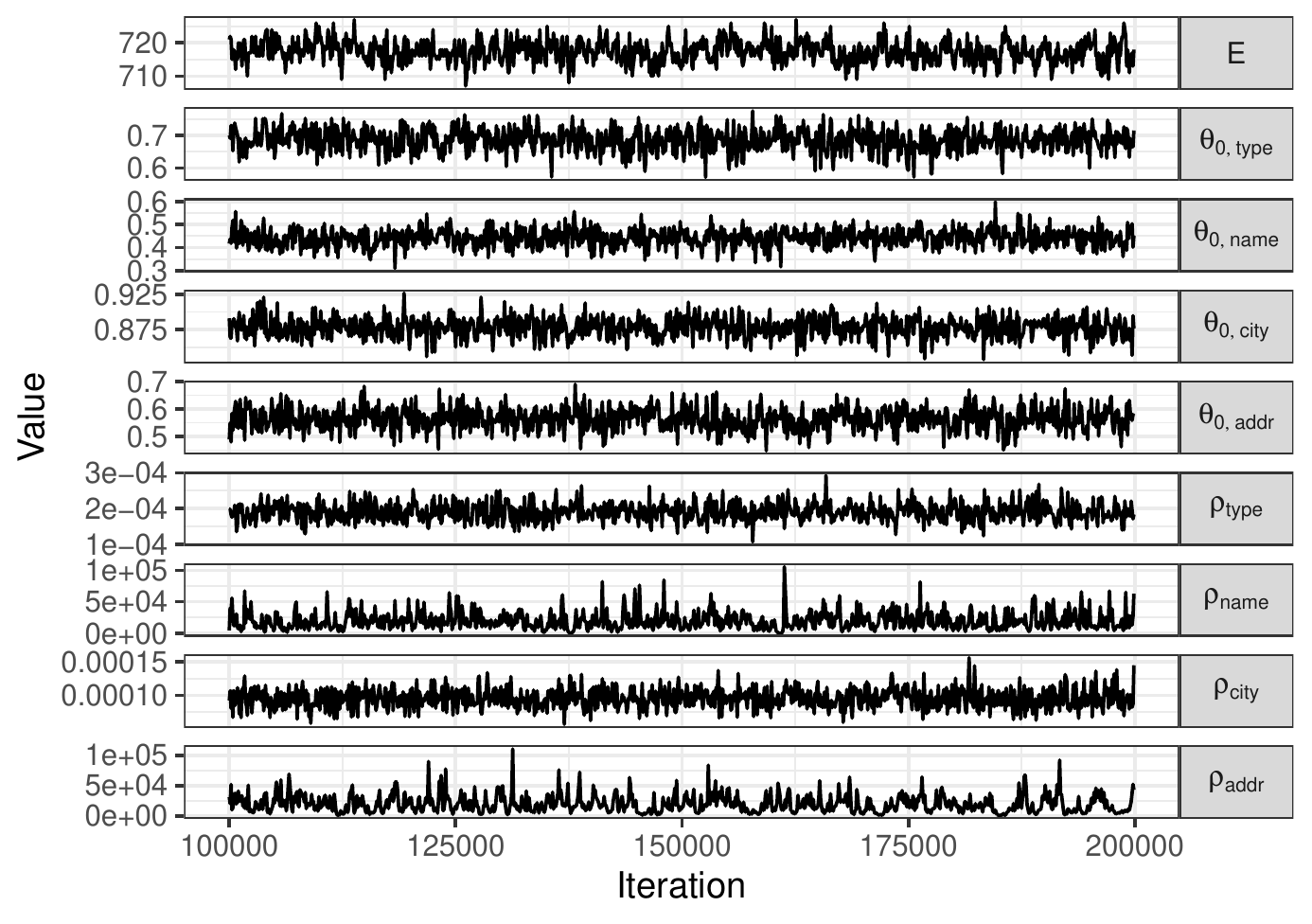} 
\end{fig}

\begin{fig}{\textsf{rest} | \textsf{PY} | \textsf{blink}}
\includegraphics[width=0.48\linewidth]{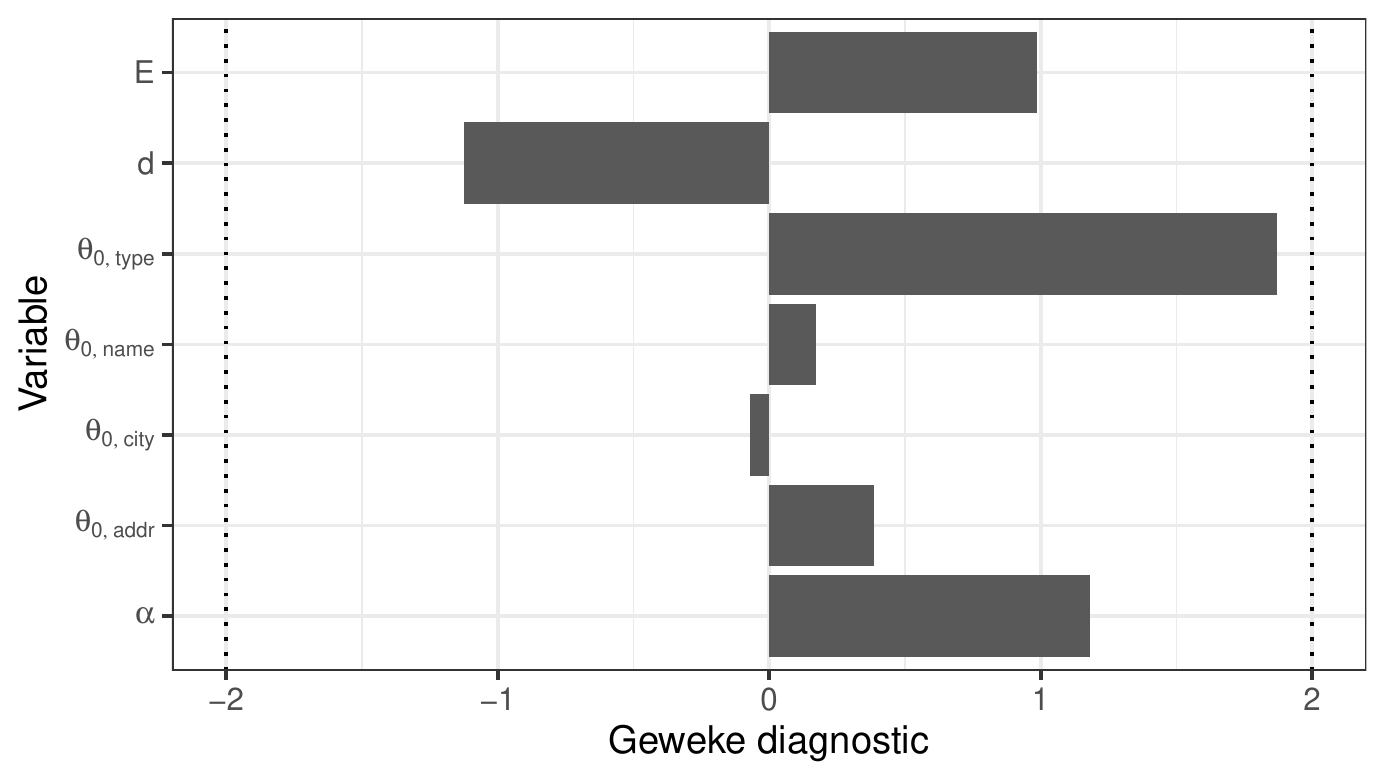} \hfill
\includegraphics[width=0.48\linewidth]{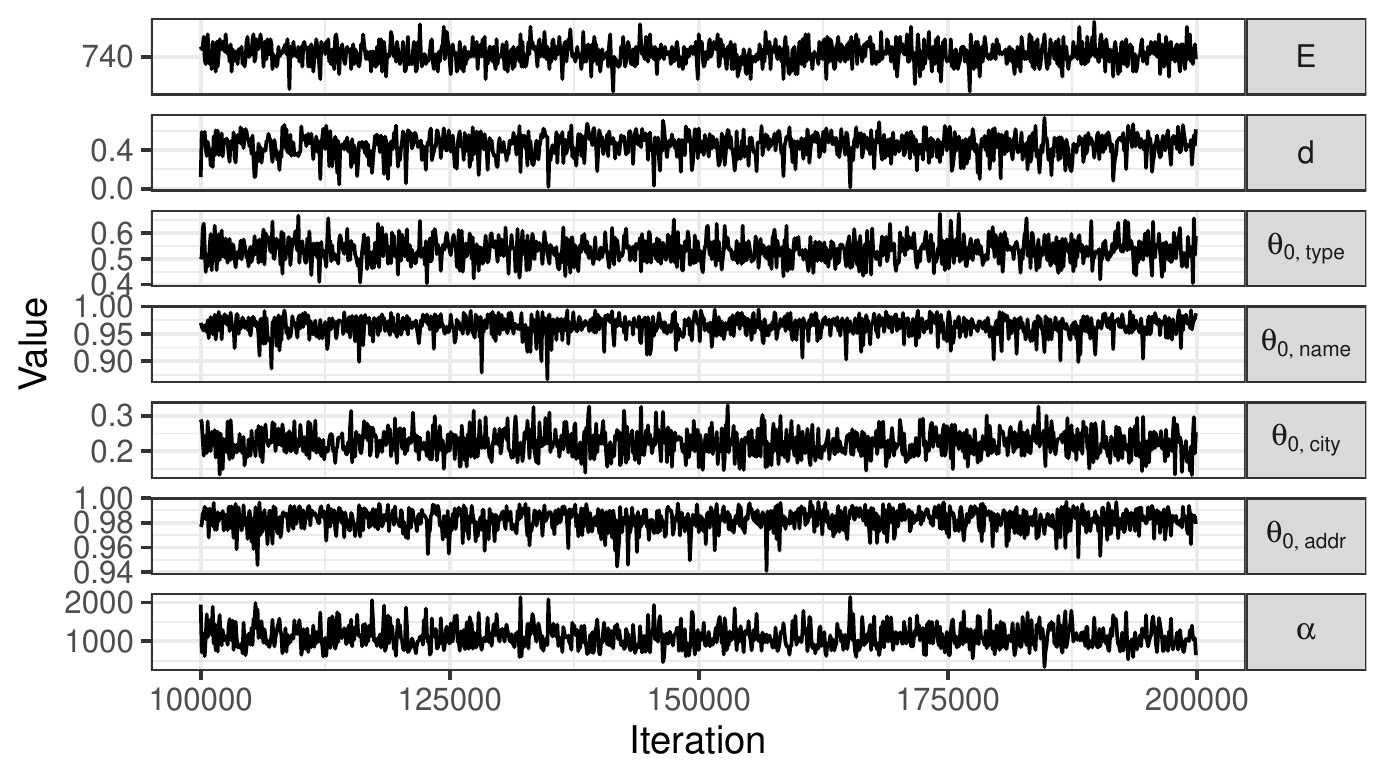} 
\end{fig}

\begin{fig}{\textsf{rest} | \textsf{Ewens} | \textsf{blink}}
\includegraphics[width=0.48\linewidth]{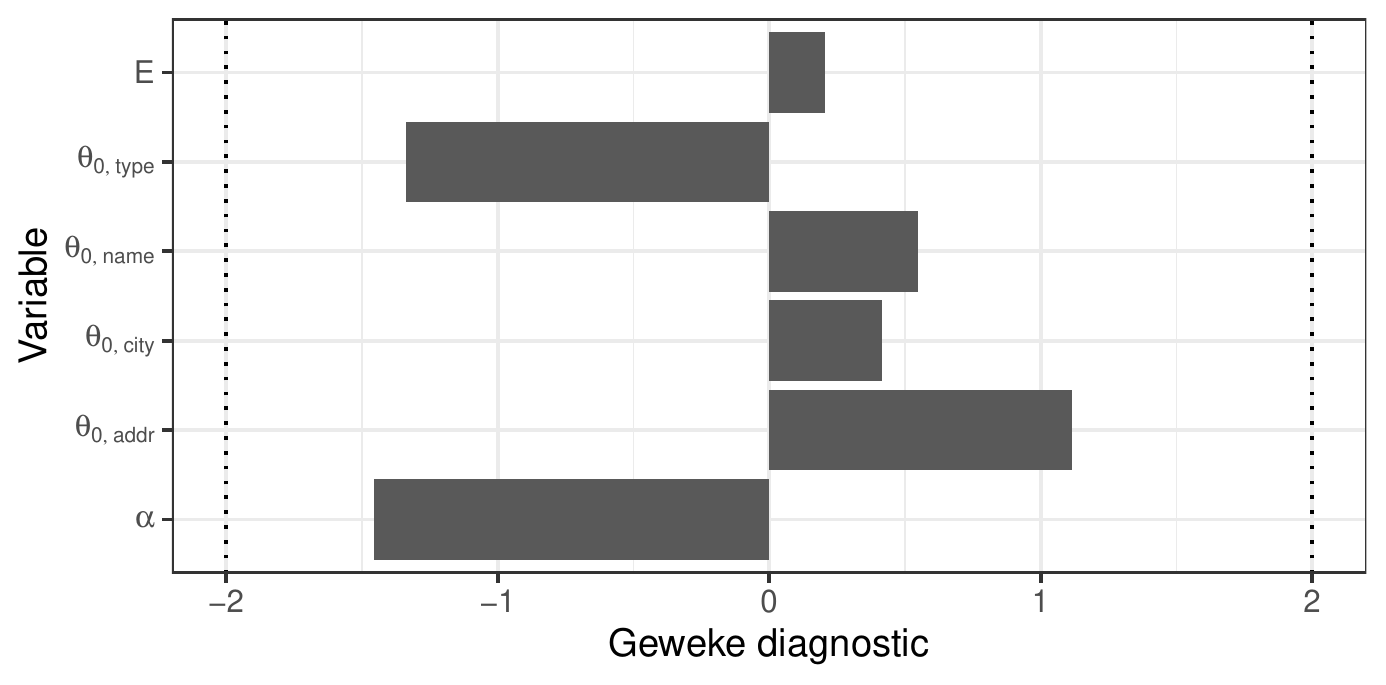} \hfill
\includegraphics[width=0.48\linewidth]{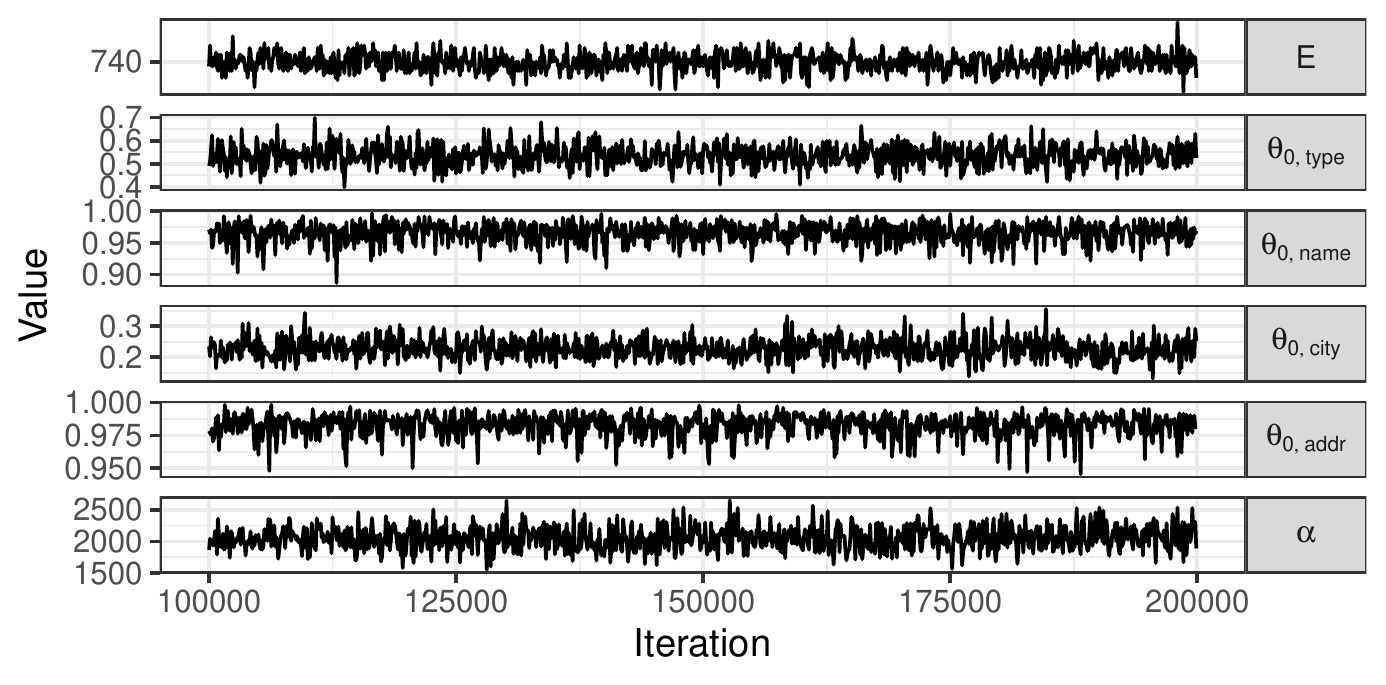} 
\end{fig}

\begin{fig}{\textsf{rest} | \textsf{GenCoupon} | \textsf{blink}}
\includegraphics[width=0.48\linewidth]{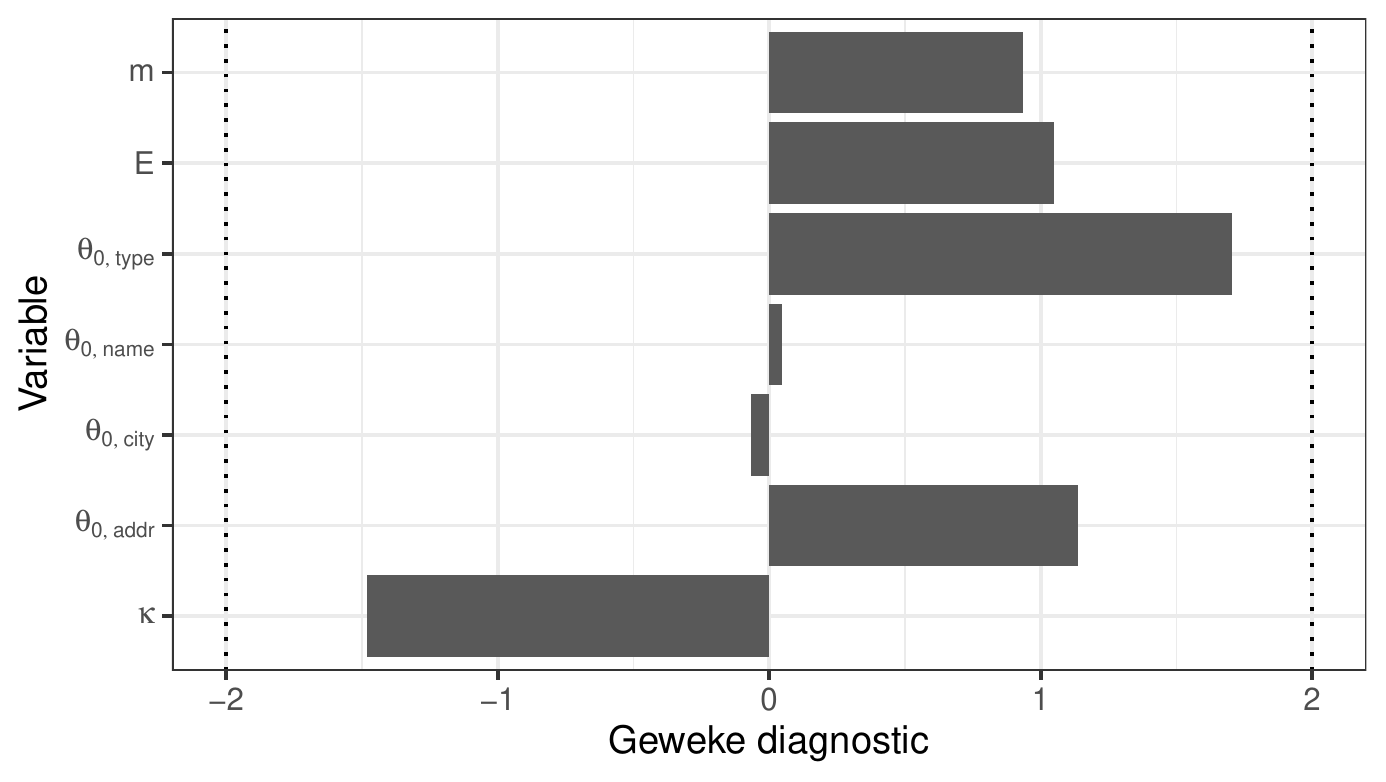} \hfill
\includegraphics[width=0.48\linewidth]{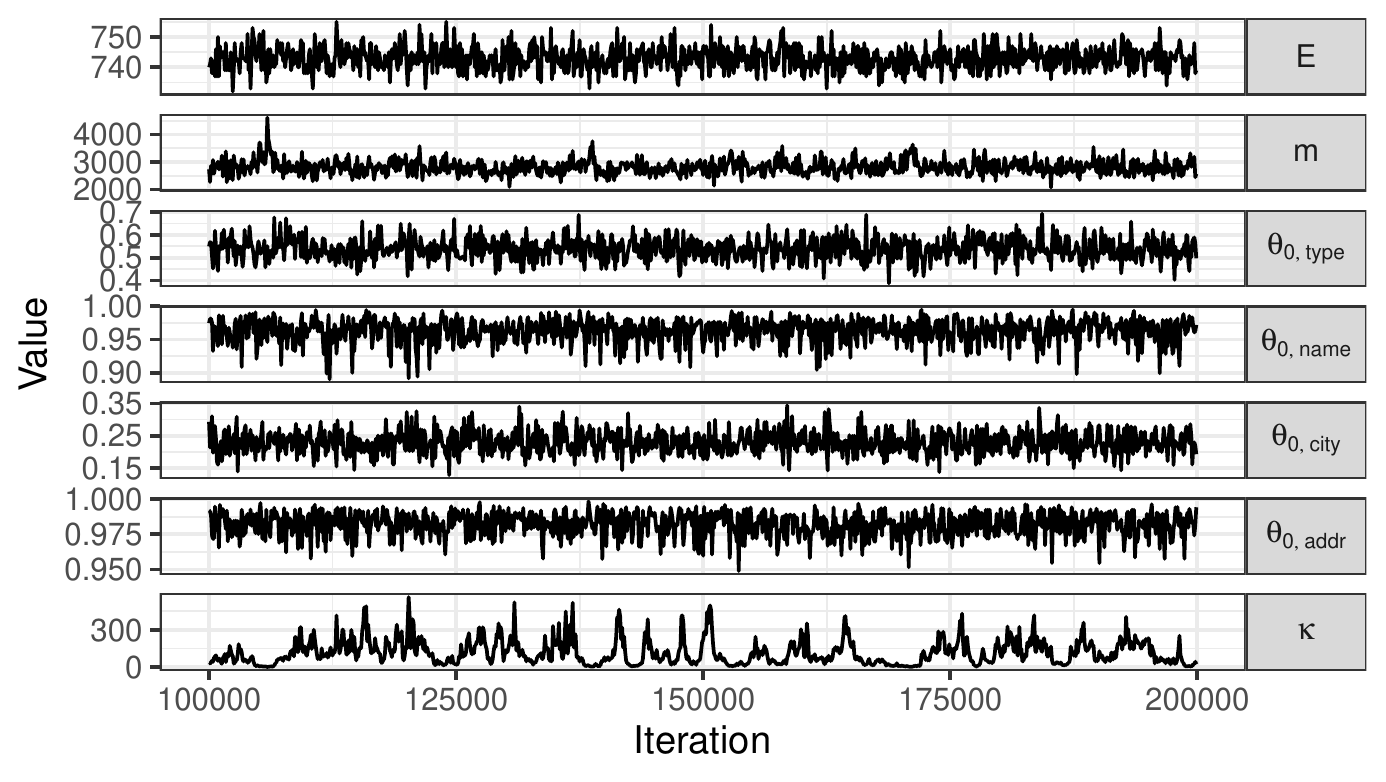} 
\end{fig}

\begin{fig}{\textsf{rest} | \textsf{Coupon} | \textsf{blink}}
\includegraphics[width=0.48\linewidth]{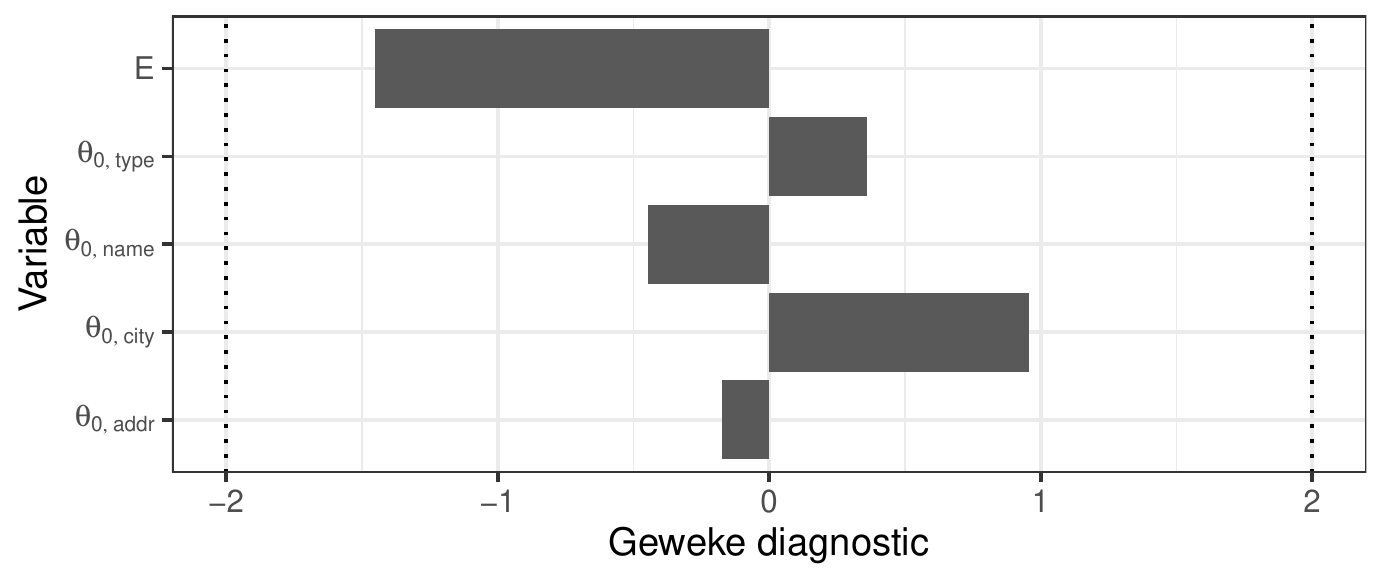} \hfill
\includegraphics[width=0.48\linewidth]{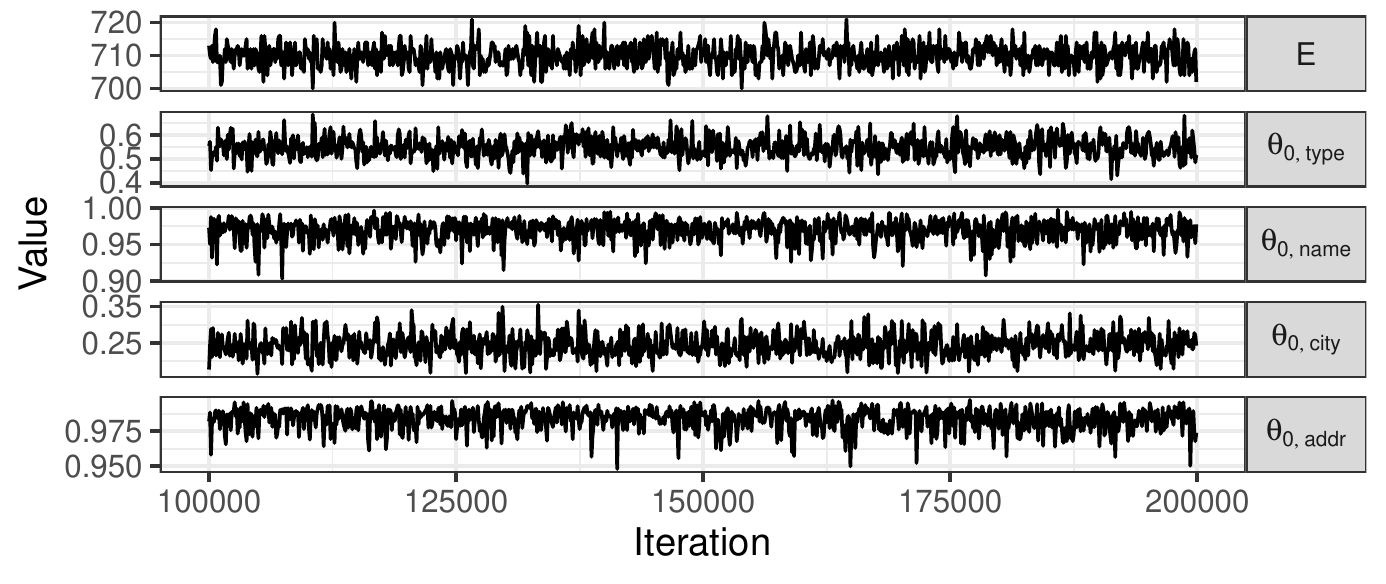}
\end{fig}

\subsection{Comparison with Baseline Models}
Here we present convergence diagnostics for the models fitted in 
Section~\ref{sec:baselines}. 
We present Geweke diagnostic plots and trace plots for a selection of model 
variables for each data set and model. 
Each pair of plots is preceded by a title of the form ``Data set | Model''.
The Geweke diagnostic plot (on the left) depicts a Z-score on the x-axis 
for each variable on the y-axis. 
The Z-score tests for equality of the means of the first 10\% and final 50\% of 
the Markov chain, and is typically expected to be in the range $[-2,2]$ 
\citep{geweke_evaluating_1992}.
The trace plot (on the right) depicts the value of variables (labeled in the 
right panel) for each step in the chain (on the x-axis). 
Note that variable $E$ denotes the number of instantiated entities. 
We replace integer indices for the attributes by named indices. 
For instance, $\theta_{0, \mathrm{city}}$ refers to the distortion probability 
in source $0$ of the attribute called ``city''.

\begin{fig}{\textsf{nltcs} | \textsf{blink}}
\includegraphics[width=0.48\linewidth]{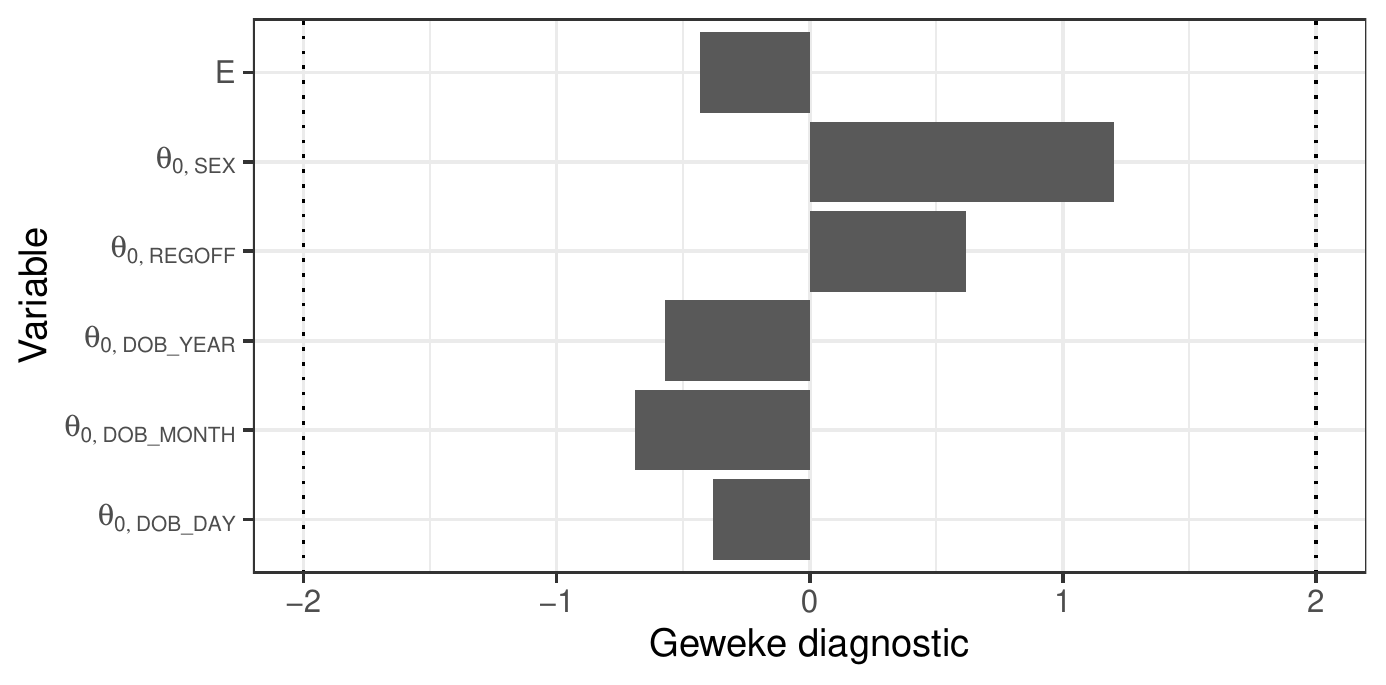} \hfill
\includegraphics[width=0.48\linewidth]{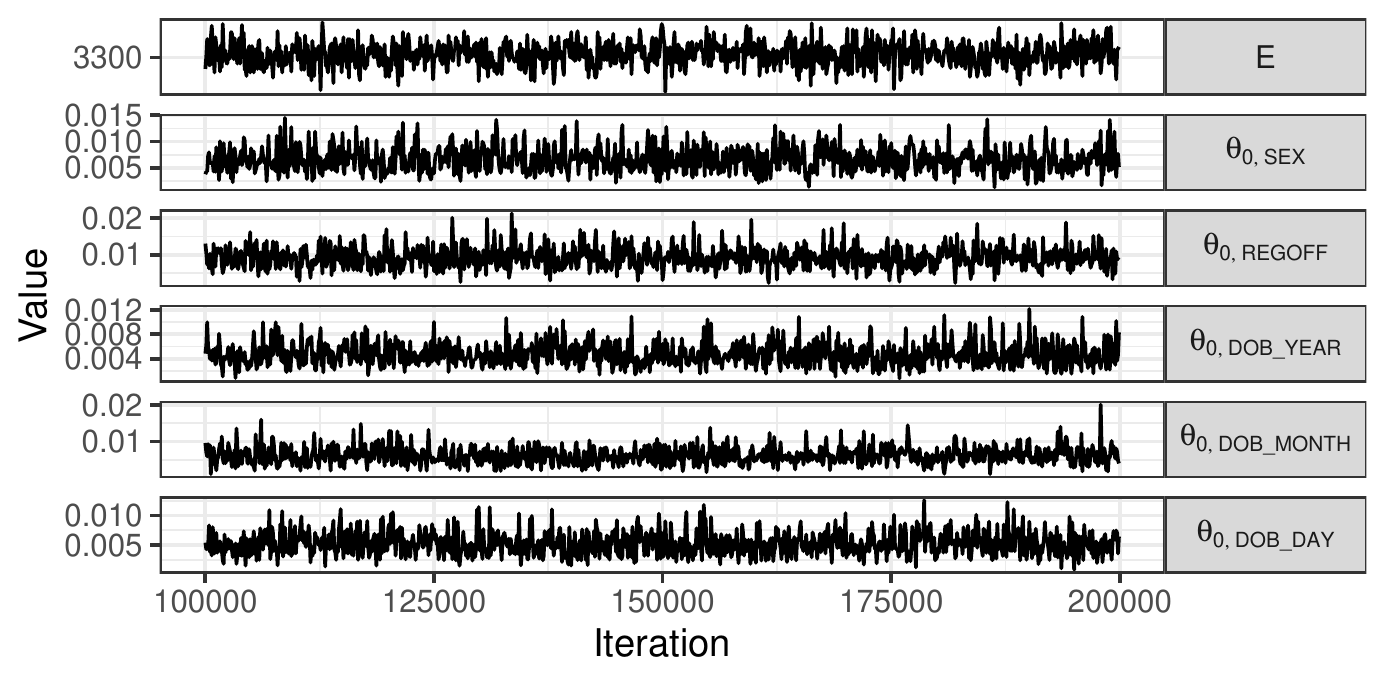} 
\end{fig}

\begin{fig}{\textsf{RLdata} | \textsf{blink}}
\includegraphics[width=0.48\linewidth]{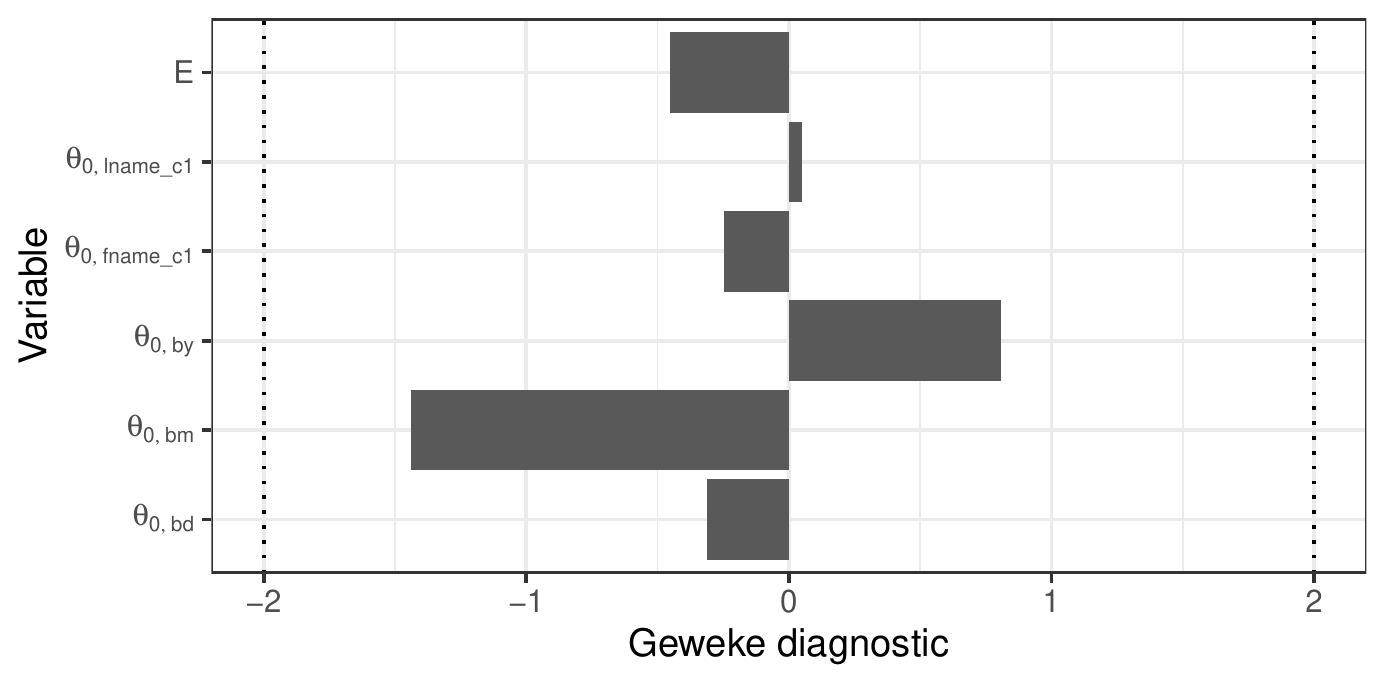} \hfill
\includegraphics[width=0.48\linewidth]{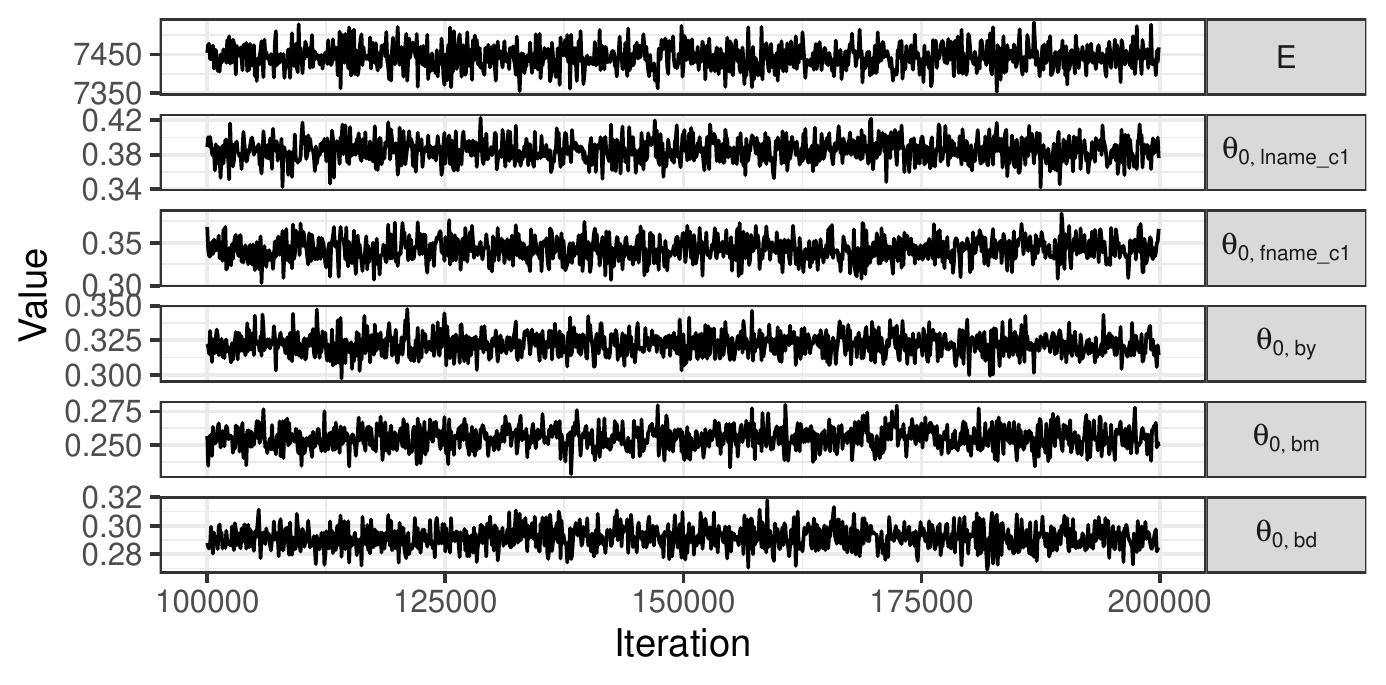}
\end{fig}

\begin{fig}{\textsf{cora} | \textsf{blink}}
\includegraphics[width=0.48\linewidth]{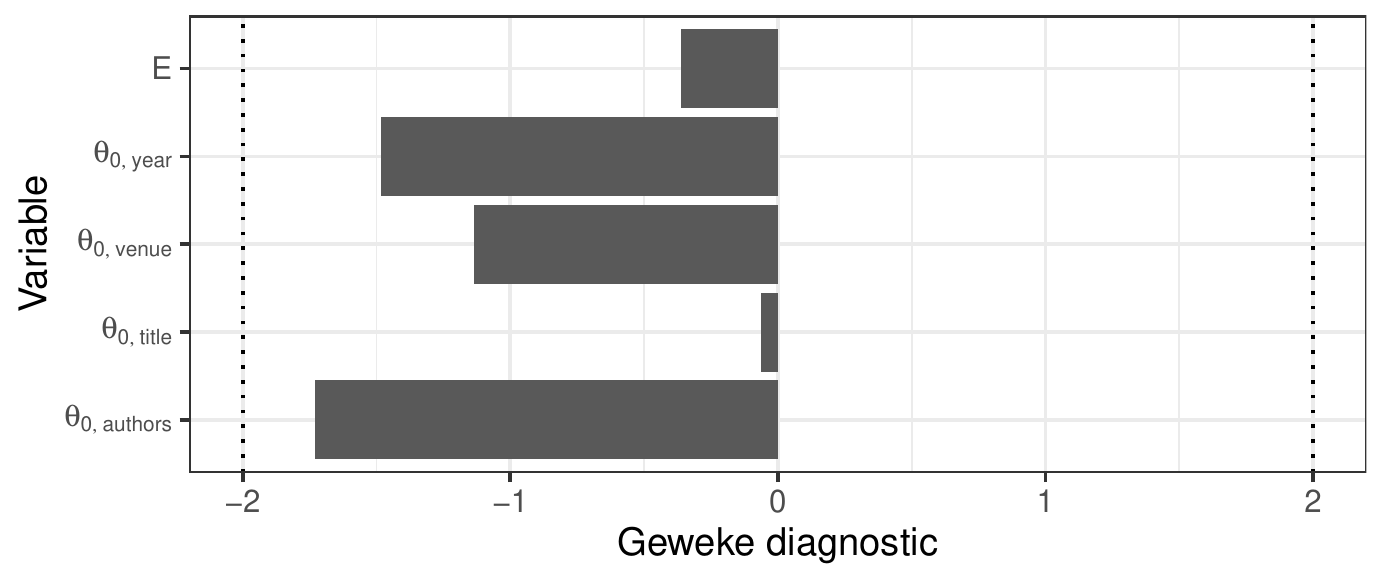} \hfill
\includegraphics[width=0.48\linewidth]{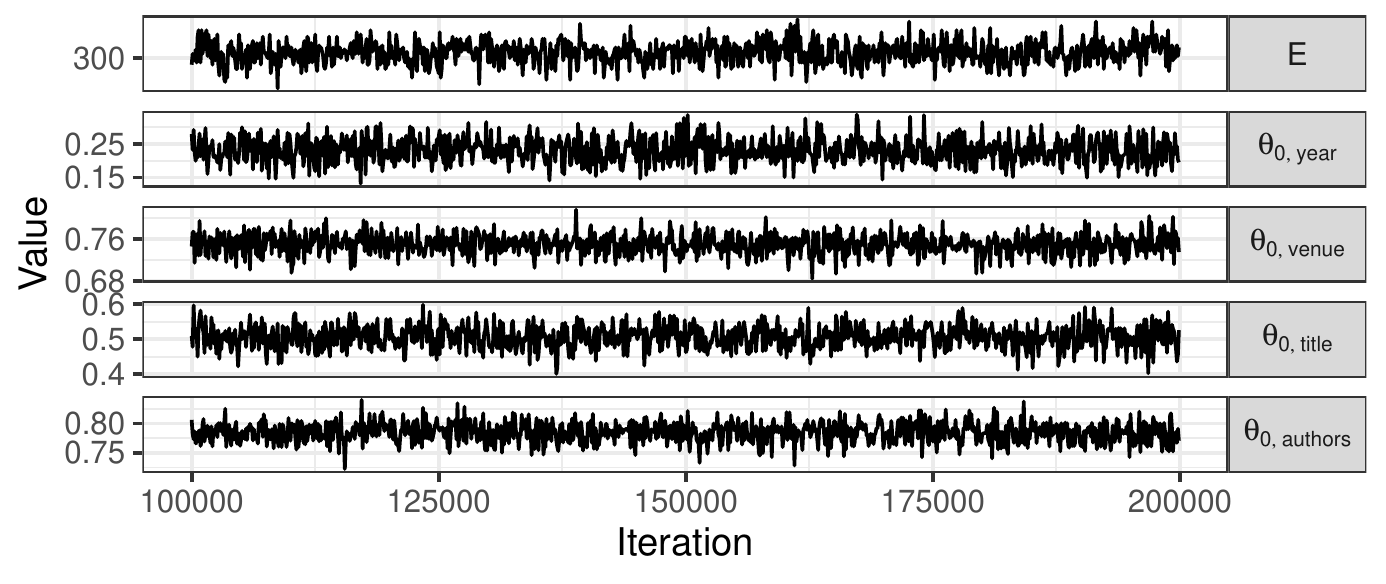} 
\end{fig}

\begin{fig}{\textsf{rest} | \textsf{blink}}
\includegraphics[width=0.48\linewidth]{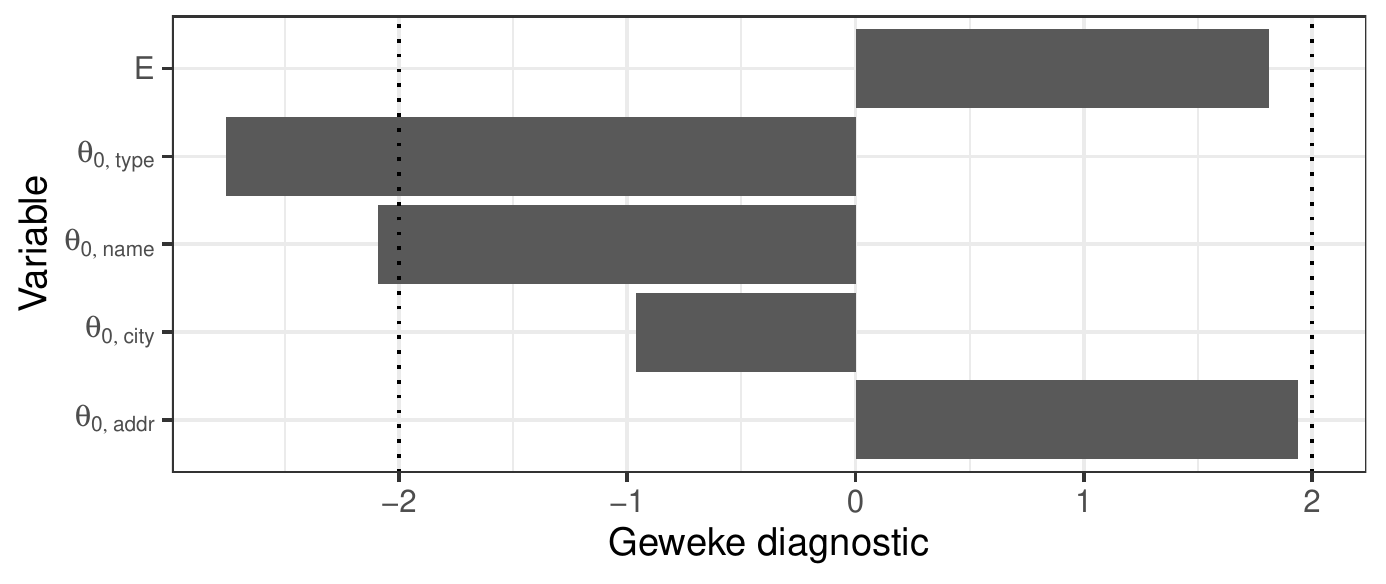} \hfill
\includegraphics[width=0.48\linewidth]{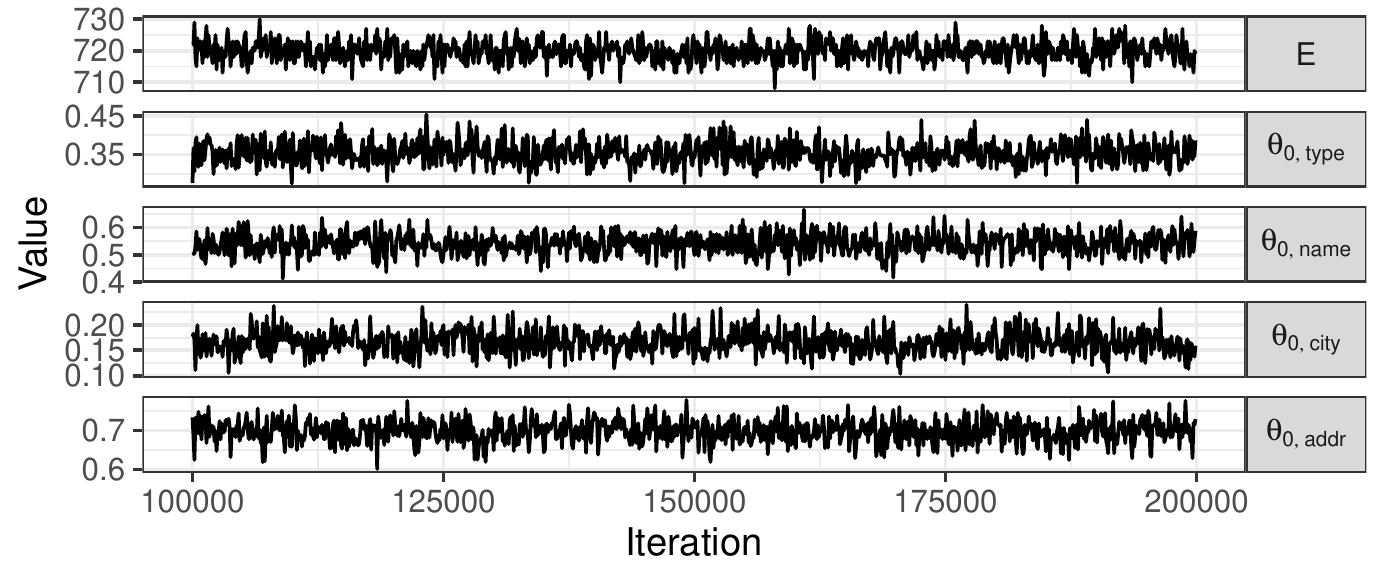}
\end{fig}

\begin{fig}{\textsf{nltcs} | \textsf{Ours}}
\includegraphics[width=0.48\linewidth]{nltcs_ours_coupon_Mon_Jun_20_23_38_18_2022_geweke-plot} \hfill
\includegraphics[width=0.48\linewidth]{nltcs_ours_coupon_Mon_Jun_20_23_38_18_2022_trace-plot} 
\end{fig}

\begin{fig}{\textsf{RLdata} | \textsf{Ours}}
\includegraphics[width=0.48\linewidth]{RLdata10000_ours_coupon_Mon_Jun_20_23_38_24_2022_geweke-plot} \hfill
\includegraphics[width=0.48\linewidth]{RLdata10000_ours_coupon_Mon_Jun_20_23_38_24_2022_trace-plot} 
\end{fig}

\begin{fig}{\textsf{cora} | \textsf{Ours}}
\includegraphics[width=0.48\linewidth]{cora_ours_coupon_Fri_Jun_24_00_10_49_2022_geweke-plot} \hfill
\includegraphics[width=0.48\linewidth]{cora_ours_coupon_Fri_Jun_24_00_10_49_2022_trace-plot} 
\end{fig}

\begin{fig}{\textsf{rest} | \textsf{Ours}}
\includegraphics[width=0.48\linewidth]{restaurant_ours_coupon_Tue_Jun_21_00_05_10_2022_geweke-plot} \hfill
\includegraphics[width=0.48\linewidth]{restaurant_ours_coupon_Tue_Jun_21_00_05_10_2022_trace-plot}
\end{fig}

\begin{fig}{\textsf{nltcs} | \textsf{Sadinle}}
\includegraphics[width=0.48\linewidth]{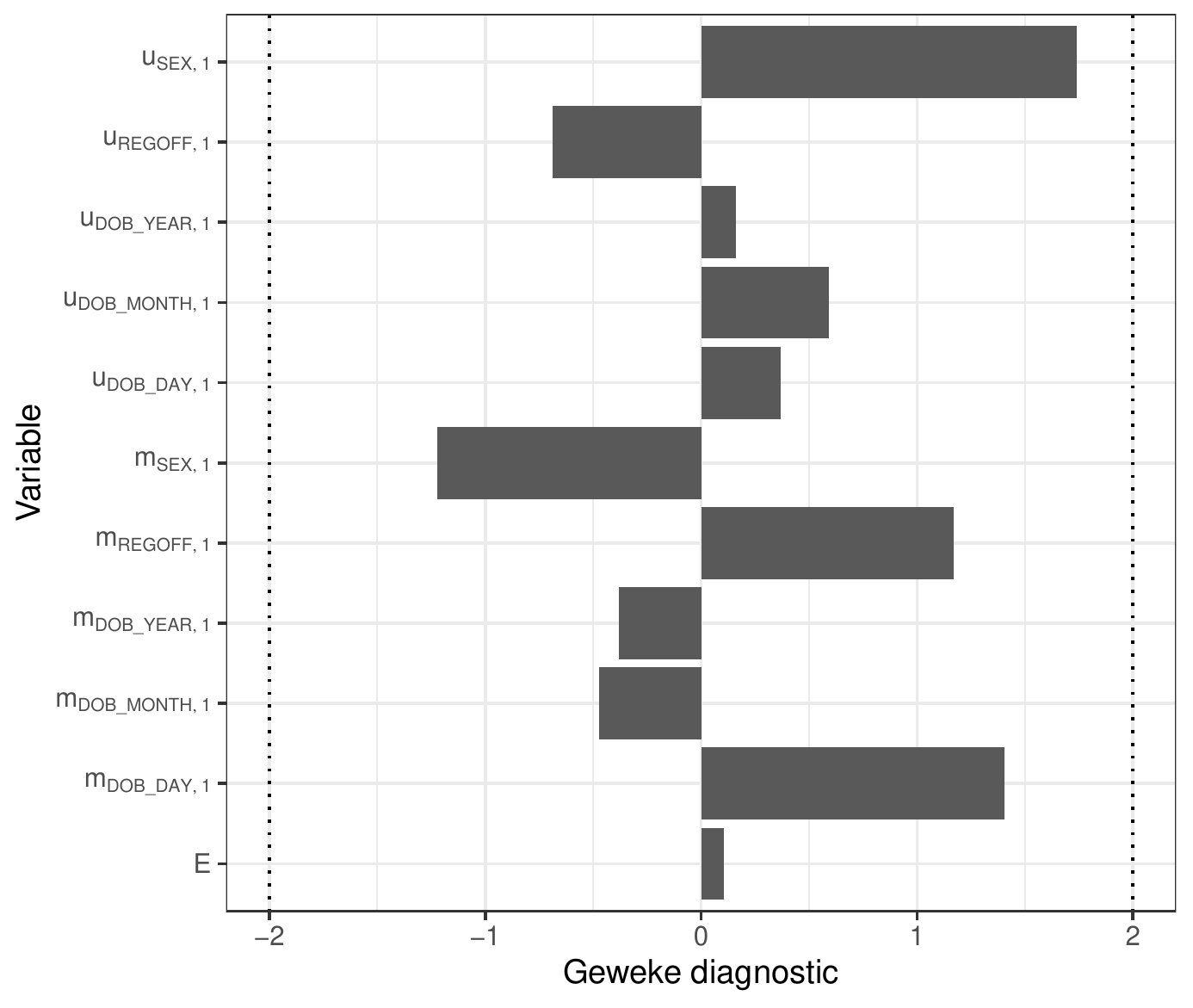} \hfill
\includegraphics[width=0.48\linewidth]{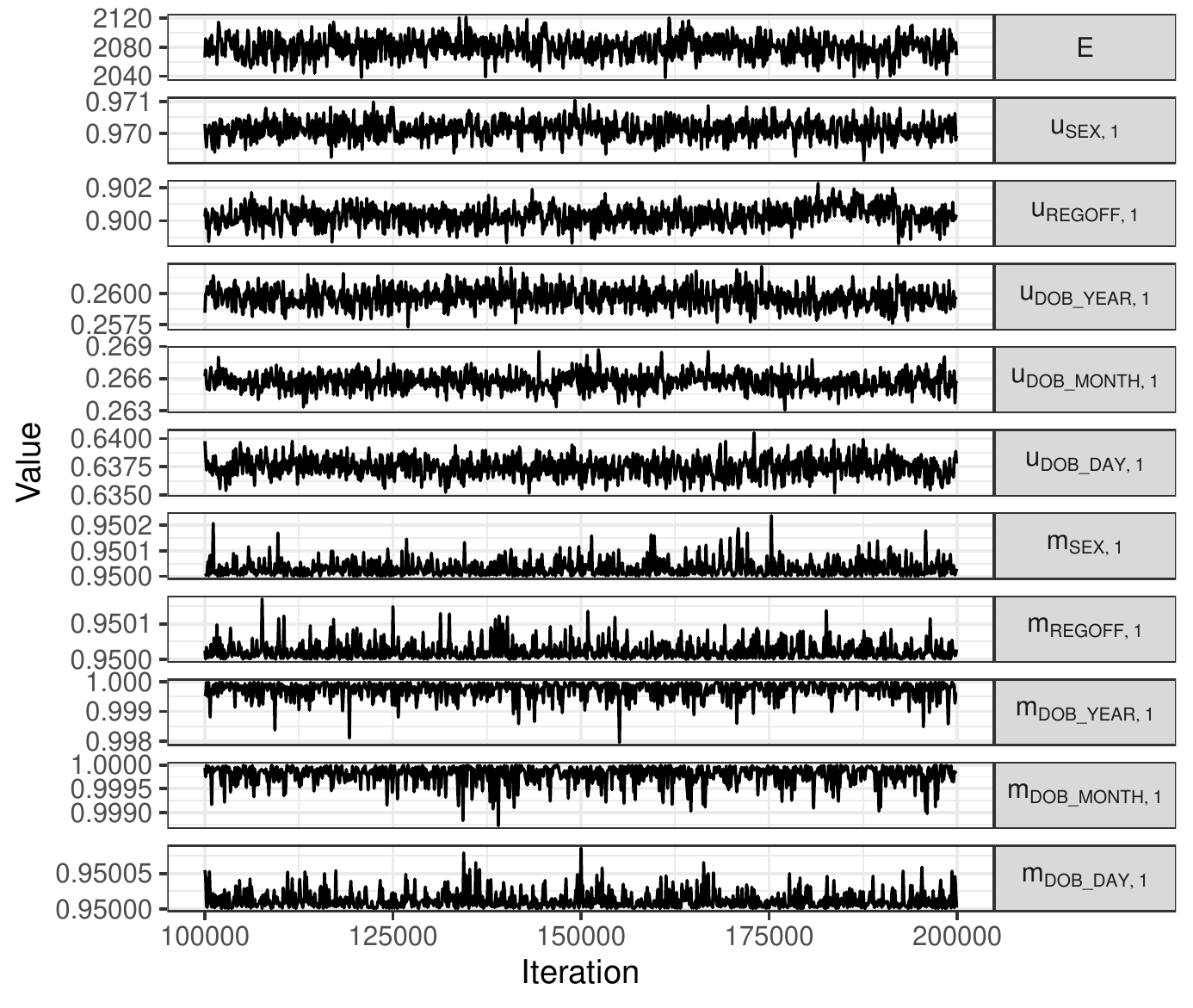}
\end{fig}

\begin{fig}{\textsf{RLdata} | \textsf{Sadinle}}
\includegraphics[width=0.48\linewidth]{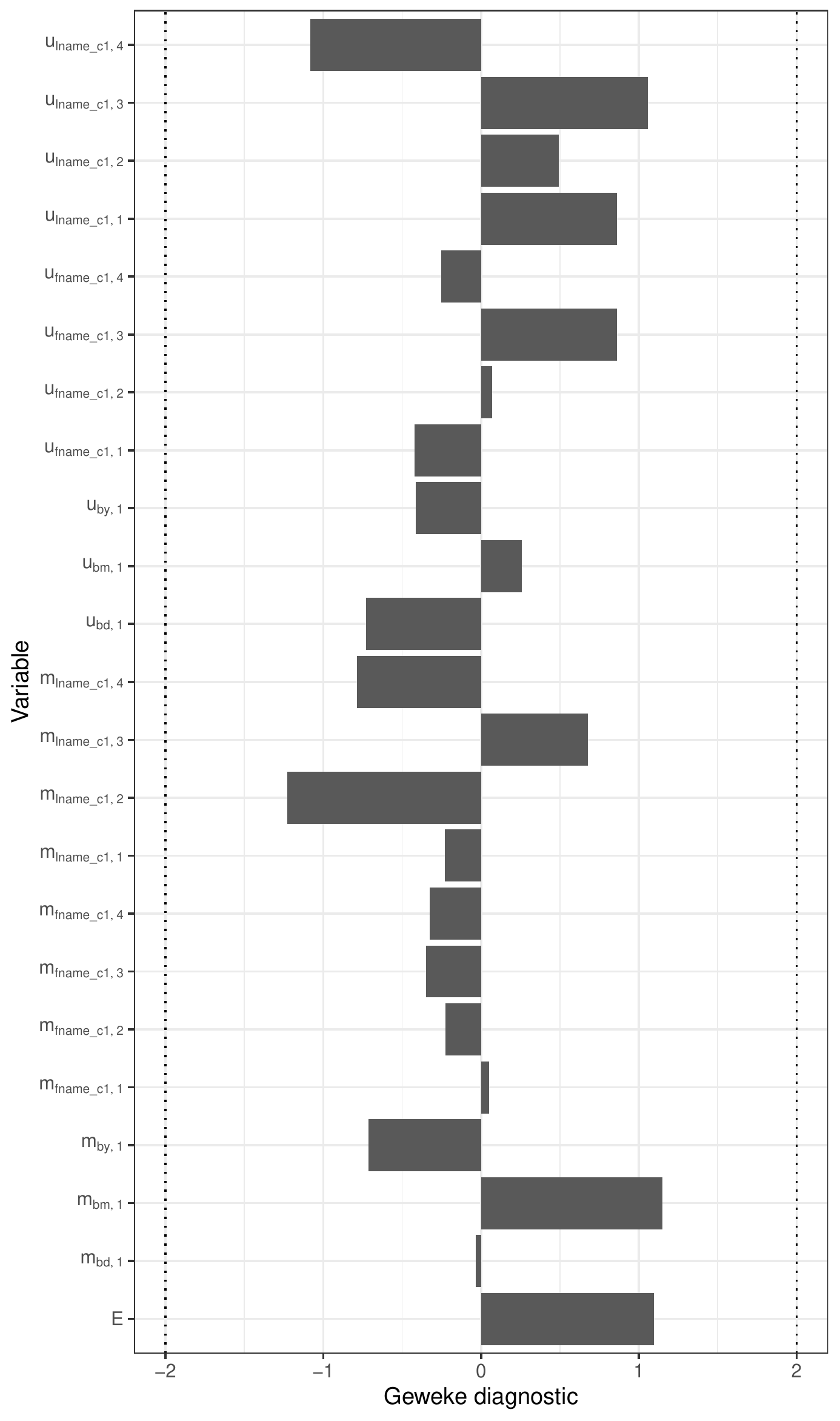} \hfill
\includegraphics[width=0.48\linewidth]{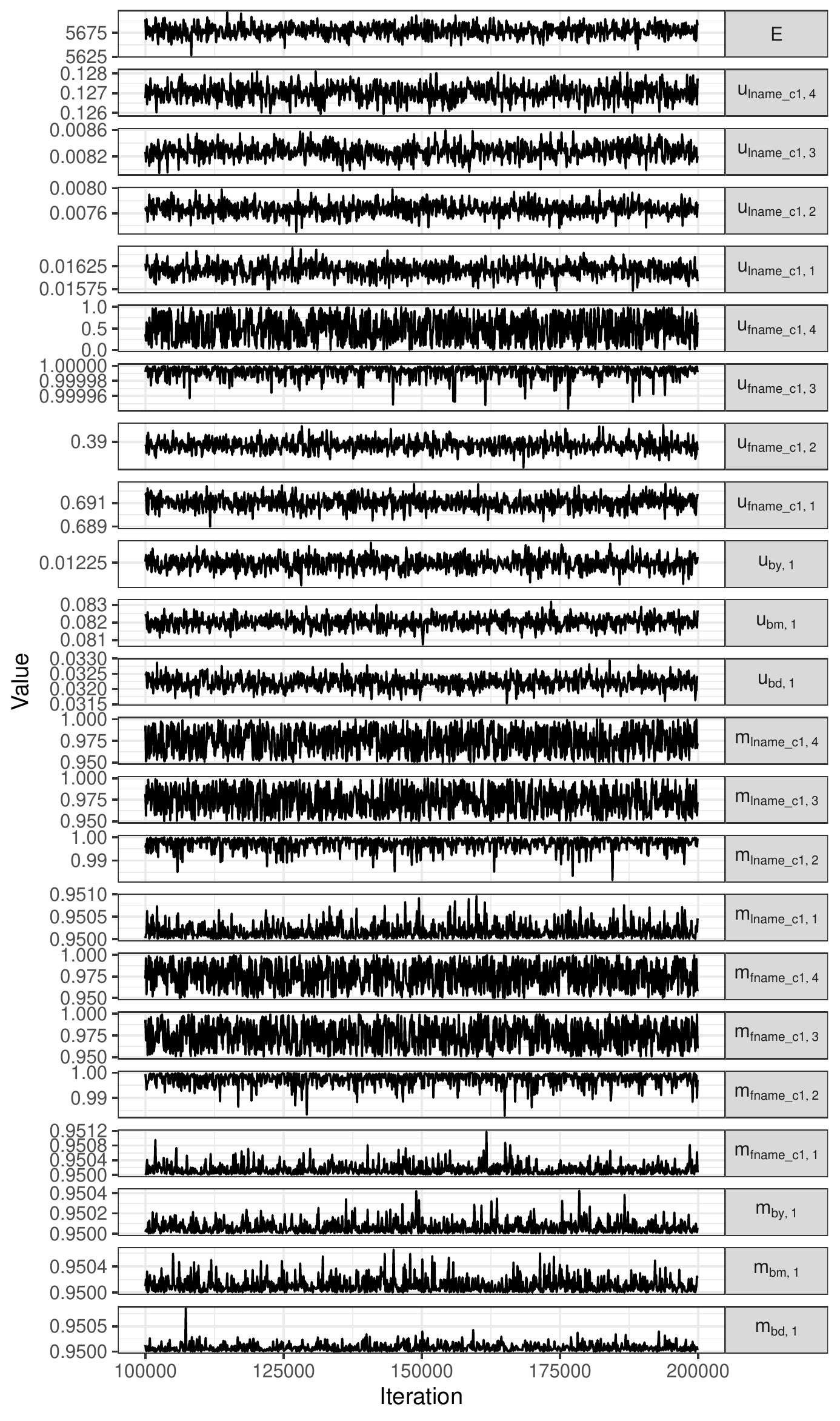} 
\end{fig}

\begin{fig}{\textsf{cora} | \textsf{Sadinle}}
\includegraphics[width=0.48\linewidth]{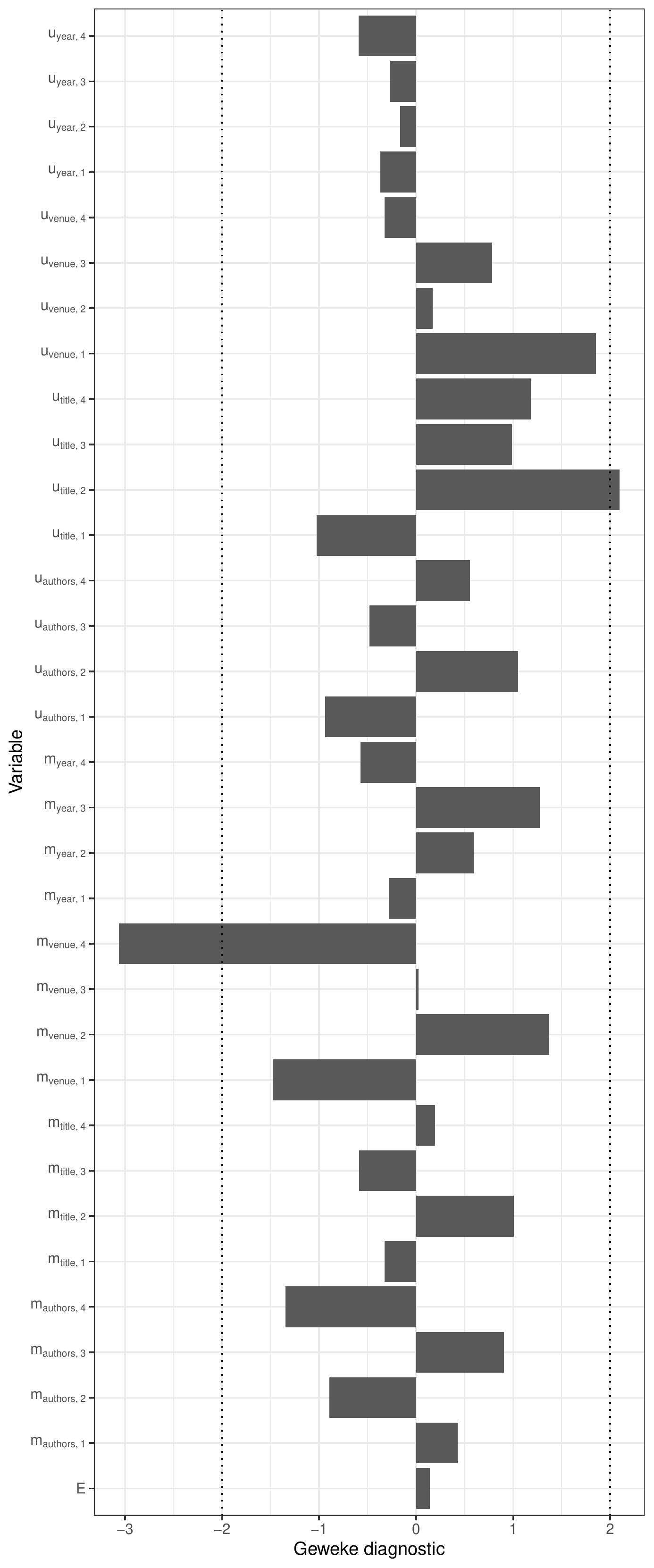} \hfill
\includegraphics[width=0.48\linewidth]{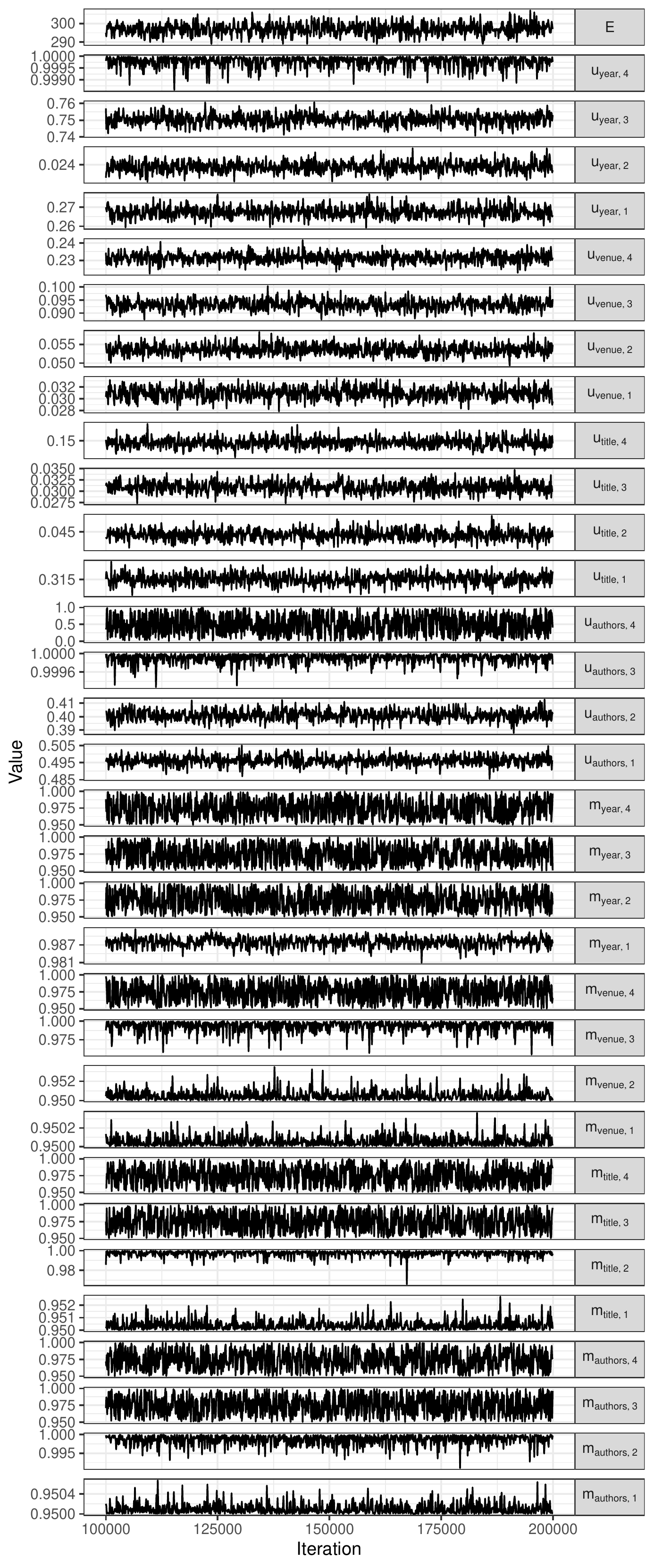} 
\end{fig}

\begin{fig}{\textsf{rest} | \textsf{Sadinle}}
\includegraphics[width=0.48\linewidth]{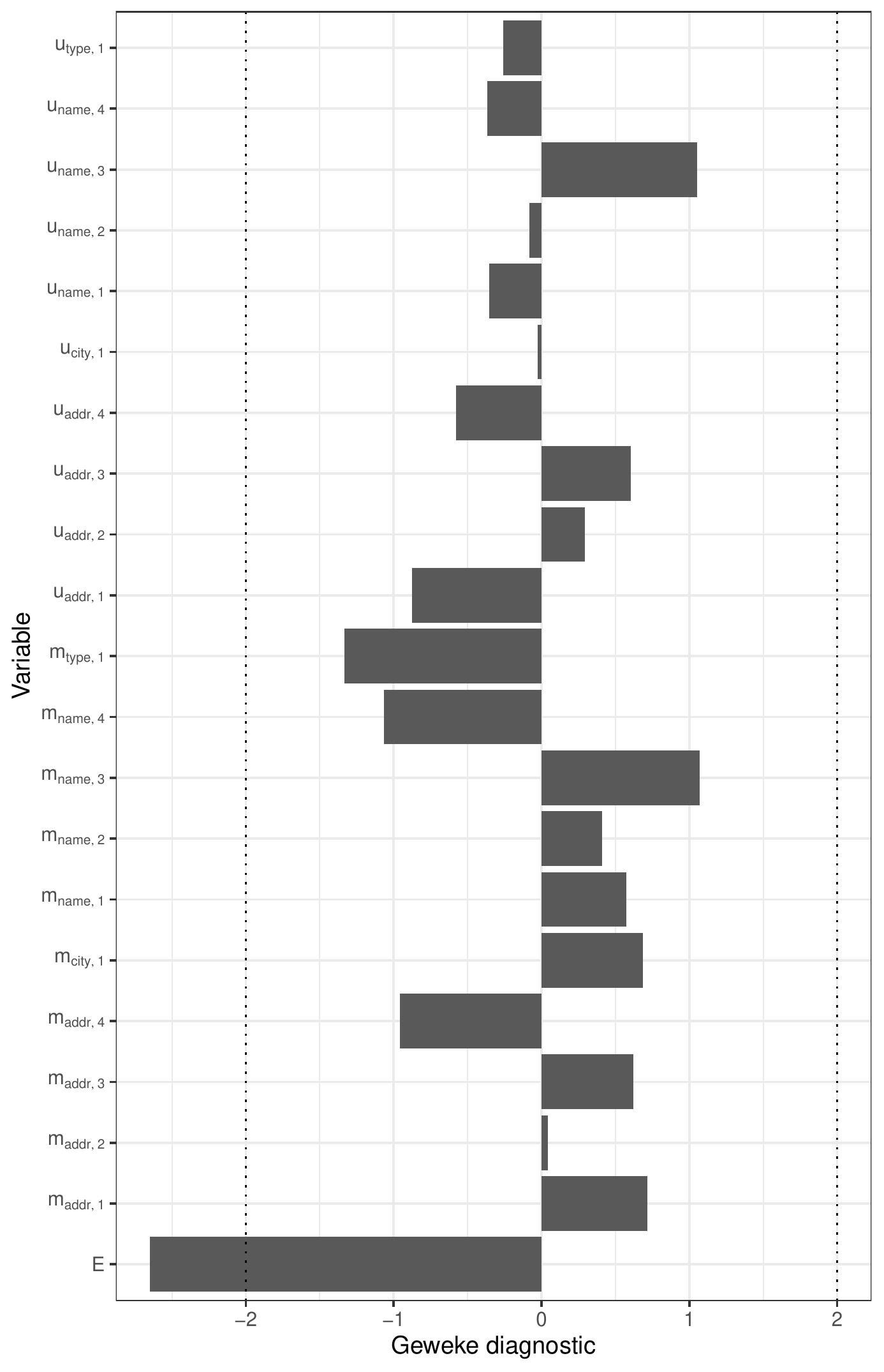} \hfill
\includegraphics[width=0.48\linewidth]{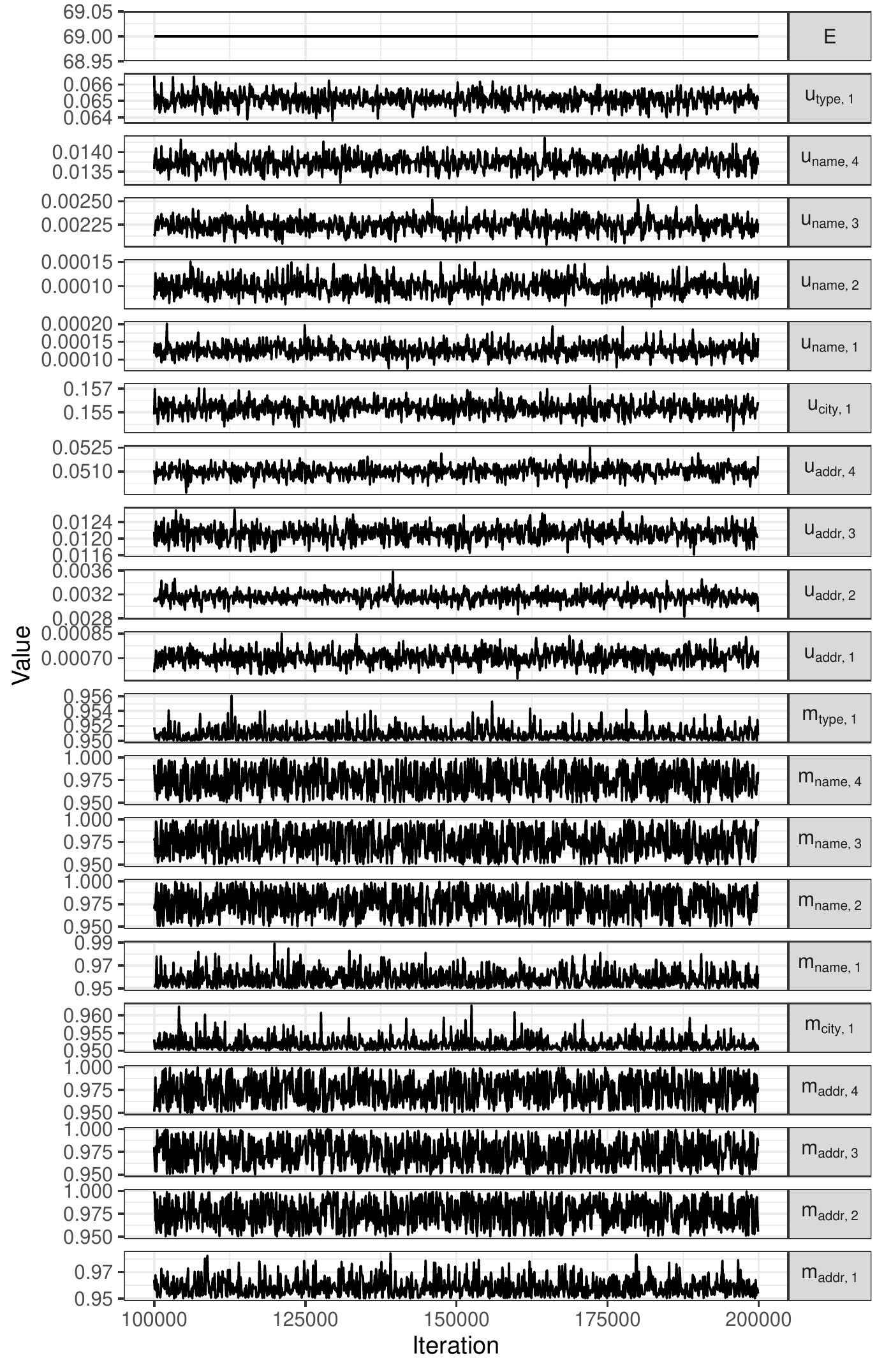}
\end{fig}

\end{document}

%% file: figures/ewens-pitman-crp.pdf_tex
%% Creator: Inkscape 1.2 (dc2aedaf03, 2022-05-15), www.inkscape.org
%% Accompanies image file 'ewens-pitman-crp.pdf' (pdf, eps, ps)
\begingroup%
  \makeatletter%
  \providecommand\color[2][]{%
    \errmessage{(Inkscape) Color is used for the text in Inkscape, but the package 'color.sty' is not loaded}%
    \renewcommand\color[2][]{}%
  }%
  \providecommand\transparent[1]{%
    \errmessage{(Inkscape) Transparency is used (non-zero) for the text in Inkscape, but the package 'transparent.sty' is not loaded}%
    \renewcommand\transparent[1]{}%
  }%
  \providecommand\rotatebox[2]{#2}%
  \newcommand*\fsize{\dimexpr\f@size pt\relax}%
  \newcommand*\lineheight[1]{\fontsize{\fsize}{#1\fsize}\selectfont}%
  \ifx\svgwidth\undefined%
    \setlength{\unitlength}{426bp}%
    \ifx\svgscale\undefined%
      \relax%
    \else%
      \setlength{\unitlength}{\unitlength * \real{\svgscale}}%
    \fi%
  \else%
    \setlength{\unitlength}{\svgwidth}%
  \fi%
  \global\let\svgwidth\undefined%
  \global\let\svgscale\undefined%
  \makeatother%
  \begin{picture}(1,0.25586854)%
    \lineheight{1}%
    \setlength\tabcolsep{0pt}%
    \put(0,0){\includegraphics[width=\unitlength,page=1]{ewens-pitman-crp.pdf}}%
    \put(0.11986755,0.02811604){\color[rgb]{0,0,0}\makebox(0,0)[ct]{\lineheight{1.25}\smash{\begin{tabular}[t]{l}Entity 1\end{tabular}}}}%
    \put(0.35745927,0.02811604){\color[rgb]{0,0,0}\makebox(0,0)[ct]{\lineheight{1.25}\smash{\begin{tabular}[t]{l}Entity 2\end{tabular}}}}%
    \put(0.64475856,0.02811604){\color[rgb]{0,0,0}\makebox(0,0)[ct]{\lineheight{1.25}\smash{\begin{tabular}[t]{l}Entity $k$\end{tabular}}}}%
    \put(0.88072269,0.02811604){\color[rgb]{0,0,0}\makebox(0,0)[ct]{\lineheight{1.25}\smash{\begin{tabular}[t]{l}New entity\end{tabular}}}}%
    \put(0.5,0.02811604){\color[rgb]{0,0,0}\makebox(0,0)[ct]{\lineheight{1.25}\smash{\begin{tabular}[t]{l}$\cdots$\end{tabular}}}}%
    \put(0.50847035,0.14748556){\color[rgb]{0,0,0}\makebox(0,0)[ct]{\lineheight{1.25}\smash{\begin{tabular}[t]{l}\Large $\cdots$\end{tabular}}}}%
    \put(0.11875,0.14523474){\color[rgb]{0,0,0}\makebox(0,0)[ct]{\lineheight{1.25}\smash{\begin{tabular}[t]{l}$\displaystyle\frac{N_1-\sigma}{N+\alpha}$\end{tabular}}}}%
    \put(0.35625004,0.14523474){\color[rgb]{0,0,0}\makebox(0,0)[ct]{\lineheight{1.25}\smash{\begin{tabular}[t]{l}$\displaystyle\frac{N_2-\sigma}{N+\alpha}$\end{tabular}}}}%
    \put(0.64374989,0.14523474){\color[rgb]{0,0,0}\makebox(0,0)[ct]{\lineheight{1.25}\smash{\begin{tabular}[t]{l}$\displaystyle\frac{N_E-\sigma}{N+\alpha}$\end{tabular}}}}%
    \put(0.88124994,0.14523474){\color[rgb]{0,0,0}\makebox(0,0)[ct]{\lineheight{1.25}\smash{\begin{tabular}[t]{l}$\displaystyle\frac{\alpha+E \sigma}{N+\alpha}$\end{tabular}}}}%
    \put(0,0){\includegraphics[width=\unitlength,page=2]{ewens-pitman-crp.pdf}}%
  \end{picture}%
\endgroup%

%% file: figures/plate-diagram.tex
\begin{tikzpicture}
  \tikzset{
    hatch distance/.store in=\hatchdistance,
    hatch distance=8pt,
    hatch thickness/.store in=\hatchthickness,
    hatch thickness=2pt
  }

  \makeatletter
  \pgfdeclarepatternformonly[\hatchdistance,\hatchthickness]{flexible hatch}
  {\pgfqpoint{0pt}{0pt}}
  {\pgfqpoint{\hatchdistance}{\hatchdistance}}
  {\pgfpoint{\hatchdistance-1pt}{\hatchdistance-1pt}}%
  {
    \pgfsetcolor{\tikz@pattern@color}
    \pgfsetlinewidth{\hatchthickness}
    \pgfpathmoveto{\pgfqpoint{0pt}{0pt}}
    \pgfpathlineto{\pgfqpoint{\hatchdistance}{\hatchdistance}}
    \pgfusepath{stroke}
  }
  \makeatother

  \tikzstyle{main}=[circle, minimum size=9mm, thick, draw=black!80, node distance=18mm]
  \tikzstyle{param}=[rectangle, minimum size=8mm, thick, draw=black!80, node distance=18mm]
  \tikzstyle{connect}=[-latex, thick]
  \tikzstyle{shortconnect}=[-latex, thin]
  \tikzstyle{plate}=[rectangle, inner xsep=4.5mm, inner ysep=4.5mm, yshift=1mm, draw=black, rounded corners=6pt]
  \tikzstyle{platelab}=[anchor=north east, xshift=-0.4mm, yshift=-0.4mm]

  \node[main, pattern=flexible hatch, pattern color=black!10] (x) {$x_{ia}$};
  \node[main] (lambda) [left of=x, xshift=-4mm] {$\lambda_{i}$};
  \node[main] (pi) [left of=lambda] {$\vec{\pi}$};
  \node[main] (sigma) [above of=pi] {$\sigma$};
  \node[main] (alpha) [below of=pi] {$\alpha$};
  \node[param, fill=black!10] (zeta1) [left of=sigma, yshift=-5mm] {$\zeta^{(1)}$};	
  \node[param, fill=black!10] (zeta0) [left of=sigma, yshift=5mm] {$\zeta^{(0)}$};
  \node[param, fill=black!10] (chi1) [left of=alpha, yshift=-5mm] {$\chi^{(1)}$};	
  \node[param, fill=black!10] (chi0) [left of=alpha, yshift=5mm] {$\chi^{(0)}$};
  \node[main] (z) [below of=x] {$z_{ia}$};
  \node[param] (omega) [right of=z] {$\omega_{ia}$};
  \node[main] (y) [right of=omega, xshift=2mm] {$y_{ea}$};
  \node[main, fill=black!10] (s) [left of=z, xshift=-4mm] {$\varsigma_{i}$};
  \node[main] (G) [below of=y] {$G_{a}$};
  \node[main] (H) [above of=y] {$H_{ea}$};
  \node[main] (rho) [right of=H, yshift=5mm] {$\rho_{a}$};
  \node[param, fill=black!10] (tau1) [right of=rho, yshift=-5mm] {$\tau_a^{(1)}$};
  \node[param, fill=black!10] (tau0) [right of=rho, yshift=5mm] {$\tau_a^{(0)}$};
  \node[param] (psi) [right of=H, yshift=-5mm] {$\vec{\psi}_{a}$};
  \node[param, fill=black!10] (upsilon) [right of=G, yshift=-5mm] {$\upsilon_{a}$};
  \node[param, fill=black!10] (phi) [right of=G, yshift=5mm] {$\phi_{a}$};
  \node[main] (theta) [below of=z, yshift=-8mm] {$\theta_{sa}$};
  \node[param, fill=black!10] (beta1) [right of=theta, yshift=-5mm] {$\beta_{sa}^{(1)}$};	
  \node[param, fill=black!10] (beta0) [right of=theta, yshift=5mm] {$\beta_{sa}^{(0)}$};
  
  \path (beta0) edge [connect] (theta)
  (beta1) edge [connect] (theta)
  (theta) edge [connect] (z)
  (s) edge [connect] (z)
  (z) edge [connect] (x)
  (lambda) edge [connect] (x)
  (pi) edge [connect] (lambda)
  (y) edge [connect] (H)
  (H) edge [connect] (x)
  (tau1) edge [connect] (rho)
  (tau0) edge [connect] (rho)
  (rho) edge [connect] (H)
  (psi) edge [connect] (H)
  (y) edge [connect] (x)
  (upsilon) edge [connect] (G)
  (phi) edge [connect] (G)
  (y) edge [connect] (psi)
  (sigma) edge [connect] (pi)
  (alpha) edge [connect] (pi)
  (chi0) edge [connect] (alpha)
  (chi1) edge [connect] (alpha)
  (zeta0) edge [connect] (sigma)
  (zeta1) edge [connect] (sigma)
  (y) edge [connect] (omega)
  (omega) edge [connect] (z)
  (G) edge [connect] (y);

  \node[plate, fit=(y) (H)] (plate-e) {};
  \node[platelab] at (plate-e.north east) (e) {\footnotesize$e \in 1 \ldots E$};
  \node[plate, fit=(z) (x) (lambda) (s) (omega)] (plate-i) {};
  \node[platelab] at (plate-i.north east) {\footnotesize$i \in 1 \ldots N$};
  \node[plate, fit=(beta0) (beta1) (theta)] (plate-s) {};
  \node[platelab] at (plate-s.north east) {\footnotesize$s \in 1 \ldots S$};
  \node[plate, fit=(phi) (beta0) (beta1) (x) (omega) (plate-e) (plate-s) (psi) (rho) (tau1) (tau0)] (plate-a) {};
  \node[platelab] at (plate-a.north east) {\footnotesize$a \in  1 \ldots A$};
\end{tikzpicture}